%% file: thesisarxiv.tex
\documentclass[10pt,a4paper,twoside,english]{report}
\usepackage{a4wide}
\usepackage{amsmath}
\usepackage{amsfonts}
\usepackage{amscd}
\usepackage[matrix,frame,arrow]{xypic}
\usepackage[all,curve,arc]{xy}

\usepackage{graphicx}
\usepackage{ifpdf}

\ifpdf
\usepackage[pdfpagemode = UseOutlines,bookmarks=true,bookmarksopen=false]{hyperref}
\hypersetup{pdfauthor={Matthew Palmer},	pdftitle={Relativistic quantum information theory and quantum reference frames}, pdfsubject={PhD thesis}}
\else
\usepackage[hypertex,colorlinks = true, linkcolor=dgreen, urlcolor = dgreen,citecolor=dgreen]{hyperref} %LINKS IN DVI
\fi

\usepackage{caption}
\usepackage{subfig} %% AFTER caption

\numberwithin{equation}{section}
\numberwithin{figure}{section}
\numberwithin{table}{section}
\newcommand{\figref}[1]{Figure~\ref{#1}}
\newcommand{\secref}[1]{Section~\ref{#1}}
\newcommand{\appref}[1]{Appendix~\ref{#1}}
\newcommand{\chref}[1]{Chapter~\ref{#1}}
\newcommand{\tabref}[1]{Table~\ref{#1}}
\newcommand{\Ssref}[1]{\S\ref{#1}}
\newcommand{\Eeqref}[1]{Equation~\eqref{#1}}
\newcommand{\comment}[1]{}
\newcommand{\abs}[1]{\left\vert#1\right\vert}
\newcommand{\set}[1]{\left\{#1\right\}}
\newcommand{\ket}[1]{\left\vert#1\right\rangle}
\newcommand{\bra}[1]{\left\langle#1\right\vert}
\newcommand{\kb}[2]{\vert#1\rangle\!\langle#2\vert}
\newcommand{\bk}[2]{\left\langle#1\vert#2\right\rangle}
\newcommand{\Bigbk}[2]{\Big\langle #1\Big\vert#2\Big\rangle}
\newcommand{\bigbk}[2]{\big\langle #1\big\vert#2\big\rangle}
\newcommand{\ex}[1]{\langle#1\rangle}
\newcommand{\proj}[1]{\kb{#1}{#1}}

\newcommand{\eps}{\varepsilon}
\newcommand{\half}{\frac12}
\newcommand{\ot}{\!\otimes\!}
\newcommand{\di}{\mathrm{d}}
\newcommand{\one}{\mbox{$1 \hspace{-1.0mm} {\bf l}$}}
\newcommand{\ii}{\mathrm{i}}
\newcommand{\Ee}{\mathrm{e}}

\newcommand\RF{\text{\tiny{RF}}}
\newcommand\SG{\text{\tiny{SG}}}

\newcommand{\m}[1]{\mathcal{#1}}

\newcommand{\Hi}{\mathcal{H}}

\newcommand{\M}{\mathcal{M}}

\DeclareMathOperator{\Tr}{Tr}

\DeclareMathOperator{\diag}{diag}

\DeclareMathOperator{\op}{Q}

\DeclareSymbolFont{symbolsC}{U}{txsyc}{m}{n}
\DeclareMathSymbol{\lambdabar}{\mathord}{symbolsC}{111}

\newcommand\be{\begin{equation}}
\newcommand\ee{\end{equation}}
\newcommand\bbm{\begin{bmatrix}}
\newcommand\ebm{\end{bmatrix}}

\usepackage[square, comma, numbers, sort]{natbib}
\begin{document}

\pagenumbering{none}
\include{titlenoim}
\date{October 10, 2013}
\maketitle

\vspace*{\fill}
\begin{center}
\copyright{} Matthew C.\;Palmer, 2013.

Version: December 17, 2013, arXiv
\end{center}
%\begingroup
%\let\pagenumbering\relax
%\mbox{}\pagebreak

%\endgroup

\pagenumbering{roman}
\include{abstract}
\newpage
\input{acknowledgements}
\newpage
\input{arxivnote}
\phantomsection
\cleardoublepage
%\addcontentsline{toc}{chapter}{Table of contents}
\pdfbookmark[0]{Table of contents}{table of contents}

\tableofcontents
\listoffigures

%%%%%%%%%%%%%%%%%%%%%%%%%%%%%%%%%%%%%%%%%%%%%%%%%%
\cleardoublepage
\pagenumbering{arabic}
\include{Intro}

\cleardoublepage
\graphicspath{{LQiCST/}}
\include{LQiCST/Qubits_curvedspacetime}

\cleardoublepage
%\graphicspath{{SGWKB/}}
\include{SGWKB/relspinmeas0.5}

\cleardoublepage
\include{LQiCSTfront}

%%%%%%%%%%%%%%%%%%%%%%%%%%%%%%%%%%%%%%%%%%%%%%%%%%%%%%%%%%
%%%%%%%%%%%%%%%%%%%%%%%%%%%%%%%%%%%%%%%%%%%%%%%%%%%%%%%%%%
\cleardoublepage
\graphicspath{{CoQRF/}}

%%%%%Chopped up QRF%%%%%%%%%%%%%%%%%%%%%%%%%%%%%%%%%%%%%%%%

\input{CoQRF/Pieces/CoQRF10_6Sec1}
%
%%%%S2 10.4 structure%%%%%%%%%%%%%%%%%%%%%%%%%%%%%%
\subsection{External versus internal, and implicated versus non-implicated reference frames}
\input{CoQRF/Pieces/CoQRF10_6Sec2part2}

\input{CoQRF/Pieces/CoQRF10_6Sec2part3}
\subsection{Group-invariant states and the encoding map}\label{sec-quantRF}
\input{CoQRF/Pieces/CoQRF10_6Sec2part4}

\input{CoQRF/Pieces/CoQRF10_6Sec2part5}

\input{CoQRF/Pieces/CoQRF10_6Sec2part6}

\input{CoQRF/Pieces/CoQRF10_6Sec2part6.5}

\input{CoQRF/Pieces/CoQRF10_6Sec2part7}
\subsection{Quantum reference frame states}

\input{CoQRF/Pieces/CoQRF10_6Sec2part8}
\subsection{Dequantisation and effective decoherence of the quantum system\label{sec-introdequant}}
\input{CoQRF/Pieces/CoQRF10_6Sec2part9}

%%%% S3 on - Common %%%%%%%%%%%%%%%%%%%%%%%%%%%%%%

\input{CoQRF/Pieces/CoQRF10_6Sec3on}

%%%%%%%%%%%%%%%%%%%%%%%%%%%%%%%%%%%%%%%%%%%%%%%%%%%%%%%%%%
%%%%%%%%%%%%%%%%%%%%%%%%%%%%%%%%%%%%%%%%%%%%%%%%%%%%%%%%%%

\cleardoublepage
\input{outlook}

\cleardoublepage
\phantomsection
\addcontentsline{toc}{chapter}{Bibliography}\label{Ch-bib}
\input{thesisarxiv.bbl}

%\bibliography{../../all}

\cleardoublepage
\appendix
\setcounter{section}{0}

%\graphicspath{{LQiCST/}}
\input{LQiCST/QST_App}

\cleardoublepage
\graphicspath{{CoQRF/}}
\input{CoQRF/CoQRF105_App}

\end{document}

%% file: titlenoim.tex
\title{Relativistic quantum information theory\\ and quantum reference frames}
\author{Matthew C. Palmer\\
\normalsize A thesis submitted in fulfilment of the requirements\\
\normalsize for the degree of Doctor of Philosophy\\
\normalsize School of Physics\\
\normalsize University of Sydney%\\
%\vspace{4cm}\\
%\ifpdf\includegraphics[height=4cm]{University_of_Sydney.pdf}
%\else\includegraphics[height=4cm]{University_of_Sydney.eps}\fi\\
%\vspace{-9cm}
}
%\\
%\normalsize 2013}

%{\center
%%\ifpdf
%%\includegraphics[height=50mm]{Figure7_6.pdf}
%%\else
%%\fi
%}

%\date{today}

%\begin{center}
%%\large
%
%%\normalsize
%
%%\Large {title}
%%\normalsize
%%\medskip
%
%
%
%
%\end{center}

%\newpage
%\maketitle

% \nocite{*}
%\renewcommand{\baselinestretch}{1.3} 

%% file: abstract.tex
\phantomsection
\addcontentsline{toc}{chapter}{Abstract}
\begin{center}

\vspace*{10mm}

{\Large \bf Abstract}

\end{center}

This thesis is a compilation of research in relativistic quantum information theory, and research in quantum reference frames. The research in the former category concerns the fundamentals of quantum information theory of localised qubits in curved spacetimes. This part of the thesis details how to obtain from field theory a description of a localised qubit in curved spacetime that traverses a classical trajectory. The particles to provide the physical realisations of a localised qubit are photons and massive spin-$\half$ fermions, e.g.\;electrons. We use a high frequency WKB approximation of the Maxwell field and Dirac field, respectively, to obtain integral curves for the particle, and equations governing the evolution of the two-dimensional quantum state and its absolute phases. The quantum information theory is then developed by defining a relativistic measurement formalism, and then constructing algorithms for path superpositions with interferometry, and entanglement and teleportation. This provides a foundation for the approximation of classical particle qubits in curved spacetime, as well as providing a complete covariant quantum information theory for describing localised qubits in curved spacetimes.

Subsequently, the measurement formalism for massive spin-$\half$ fermions is formalised by deriving from field theory the quantum observable for a Stern--Gerlach measurement of a fermionic qubit moving relativistically with respect to the Stern--Gerlach apparatus. Using again the WKB limit, the interaction of the fermion field with the electromagnetic field of the Stern--Gerlach apparatus demonstrates spin-dependent deflection of trajectories with the spin quantisation axis matching the operator derived from the relativistic transformation properties of electromagnetic fields. This provides justification from relativistic field theory of the appropriate interaction and relativistic transformation properties of a fermion and a Stern--Gerlach magnet.

The second vein of research of this thesis regards what behaviour a relativity principle may have in the context of quantum reference frames. A relativity principle in a physical theory dictates how the description of a physical system and its dynamics change under a change in coordinates or reference frame. This research explores the consequences of performing this change of reference frame in the quantum reference frame framework. The scenario involves a quantum `system', with degrees of freedom and quantities defined relationally using an additional quantum system which acts as a reference frame. There is also a second quantum reference frame which is uncorrelated with the system or the first quantum reference frame. In order to change over to using the second reference frame to define the relational quantities for the `system', the two frames must become correlated. The quantum reference frames are quantum systems, so a quantum measurement of the two reference frames is required in order to accomplish this correlation. Due to the imperfect ability of the quantum reference frames to act as frames, this measurement, and subsequent discarding of the first reference frame, results in decoherence on the `system' quantum degrees of freedom. In this derivation the frames are treated as physical systems, but there is an alternative description of this change of reference frames procedure in which the reference frames are treated as external, static background elements. The decoherence then occurs to the `system' without interaction with any other degrees of freedom. This is a type of `intrinsic decoherence', which has been proposed as a semiclassical phenomenon of quantum gravity that arises due to the inherently quantum nature of space.

%% file: acknowledgements.tex
\phantomsection
\addcontentsline{toc}{chapter}{Acknowledgements}

\begin{center}

\vspace*{3cm} {\Large \bf Acknowledgements }
\end{center}
I would like to dearly thank:

Stephen Bartlett, Hans Westman, Florian Girelli, and Maki Takahashi, in so many capacities. Hannah Kennelly. My parents and siblings for their immense support; Karen Palmer for proofreading. The Sydney crew\comment{ Felix Lawrence, Stephen Dekker, Sahand Mahmoodian, George Brawley, Parry Chen,  Ben Fulcher, Hugo Dupree, Scott Brownless, Andrew Darmawan, Joel Wallman, Courtney Brell}, including the quantum, photonics, complex systems, and astro people, picking out Felix Lawrence, Joel Wallman, and Maki again for additional thanks regarding thesis advice. Daniel Terno, Terry Rudolph. Tim Ralph. The UQ crew\comment{ Jacques Pienaar, Nathan Walk}.
The Manly High crew, picking out Flynn Pettersson as PhD advice-giver. James Erickson and Ben Fulcher for providing electronic and global friendship and company.

Thank you also to all my other friends, colleagues, and family who have provided support or company in some way.

I thank my thesis examiners for their constructive comments and insight.

%% file: arxivnote.tex
%\phantomsection
%\addcontentsline{toc}{chapter}{Note for arXiv version}

\begin{center}

\vspace*{3cm} {\Large \bf Note for arXiv version}
\end{center}
This thesis contains work found in the quant-ph arXiv preprints \href{http://arxiv.org/abs/1108.3896}{1108.3896}, \href{http://arxiv.org/abs/1208.6434}{1208.6434}, and \href{http://arxiv.org/abs/1307.6597}{1307.6597}.

\chref{sec-LQiCST} contains joint work with Maki Takahashi and Hans Westman from \href{http://arxiv.org/abs/1108.3896}{1108.3896}, published as \cite{PTW11}. The chapter contains corrections and clarifications to the version of the work in \cite{PTW11}. The chapter also contains some major additions beyond the published paper. These are: a section providing an introduction to general relativity \S\ref{sec-GRintro}; sections regarding how to determine the transformation of a state along a trajectory, \Ssref{sec-statetransformation}, \Ssref{sec-fermionstatetransformation}, and \Ssref{sec-photonstatetransformation}; a paragraph about Wigner rotations \Ssref{fermionwignerrot}; and numerical calculations for the COW neutron gravitational interferometry experiment including \tabref{fig-phaseresults}, in \S\ref{neutronphase}. I thank MT and HW for feedback and suggestions regarding these additions.

\chref{ch-RSO} consists of the paper \href{http://arxiv.org/abs/1208.6434}{1208.6434}, published as \cite{PTW12}, with minor corrections and adjustments. This also was joint work with MT and HW.

\chref{ch-2} is my own. I thank MT for comments on a draft, and MT and HW for discussions.

\chref{ch-CoQRF} is an updated version of \href{http://arxiv.org/abs/1307.6597}{1307.6597}, \cite{PGB13}, and contains research done with Florian Girelli and Stephen Bartlett.

%% file: Intro.tex
\chapter{Introduction}

Quantum mechanics and general relativity are both extremely successful theories. However, the theories each have a limited domain of applicability which cannot adequately describe extreme phenomena where both quantum and gravitational effects are important. There is research into developing a fundamentally new theory that will combine the phenomena and experimental predictions from both of these existing theories: a theory of quantum gravity. This would be a theory of microscopic matter in gravitational settings that is (a) consistent with quantum mechanics and general relativity in their domains, as well as (b) providing novel predictions for potentially observable phenomena not explained in the existing physical theory. In this work it seems extremely difficult, given the immense theoretical overhead, to derive predictions for new realistic phenomena, and so far no complete theory exists.

However, this is not the only way to explore physics at the boundary of quantum mechanics and general relativity. There is also a top-down approach in which one seeks semiclassical phenomena from modified simpler theories. This is in order to gain an intuition for phenomena in semiclassical gravity scenarios, which in turn may motivate the structure and predictions in quantum gravity theories. This was the approach taken for the research contained in this thesis. The models in this thesis combine quantum mechanics with select elements of special and general relativity in order to derive phenomena expected to be the semiclassical and most accessible novel effects of a quantum theory of gravity. Quantum information theory is at the core of the research in this thesis, the latter involving development of an operational theory for finite dimensional systems in relativistic scenarios, and using finite dimensional systems to develop decoherence effects that may occur at a semiclassical level in quantum gravity.

The contribution to physics this thesis offers is in two strains of research. The first concerns the fundamentals of quantum information theory in curved spacetimes. The fundamental basis of quantum information theory is the simplification of a quantum field theory to non-relativistic situations in which the details of the field can be ignored to the extent that one can extract from it a finite-dimensional Hilbert space. For the purposes of many experiments this is a good approximation. For example, in many scenarios one can use the simplifying model of a point-like electron with a two-dimensional Hilbert space for its spin. This quantum information theory is developed in flat space, but is applied to describe experiments in the gravitational field of the Earth. The epistemological intuition then is that this picture of particles with finite Hilbert spaces is still a valid approximation in weak gravitational fields: a quantum information theory should emerge in this limit from quantum field theory on curved spacetime. How do we formally obtain this simplification, and what are its limitations and new phenomena?

This research is ostensibly in the area of `relativistic quantum information theory', but it is quite distinct to other research in this community. Rather than the study of particle production and entanglement degradation for quantum fields, this research constructed quantum information theory for localised qubits in curved space time consisting of photons or massive spin-$\half$ fermions.

In the second chapter of this thesis, this philosophy of deriving simplified results from a limit of field theory was applied to measurement of spin in a relativistic setting. This was done in order to obtain justification from relativistic quantum theory on the correct interaction and relativistic transformation properties of a charged fermion with a Stern--Gerlach magnet. This is because there has been debate in the literature regarding what spin operator to use in relativistic measurements of spin. The purpose of the spin operator in this context should be to determine how to relate a measurement direction in the rest frame of a measurement apparatus to the spin eigenstates of the measurement for a spin particle in relativistic motion with respect to the apparatus. The method by which spin is measured involves interaction with a magnetic field, and this has a specific transformation between Lorentz frames which differs to how other proposals of spin operator transform.

The second strain of research is in quantum reference frames. Quantum reference frames provide a practical way to send quantum information between parties who lack a shared frame of reference. They also provide an analogue for how quantum space may behave.

Spacetime geometry in general relativity is dynamical, with a relationship between the distribution of matter and the curvature of space. This concept combined with the quantum nature of matter indicates the possibility that spacetime also has a quantum nature. There is then a question of how quantum matter will interact with this quantum space. Quantum reference frames provide a way to study this type of interaction, as they model the uncertainty of frames or coordinates, as well as allowing the quantum physics to be independent of the choice of classical frames or coordinates, considered to be unphysical. This idea of background independence is also at the core of general relativity, so quantum reference frames provide a simple toy theory for combining quantum mechanics with the dynamic nature of spacetime.

There has been research regarding the behaviour of quantum reference frames after measurement, and what correspondence this might have with elements of quantum gravity. That research has involved analyses of the behaviour of a single quantum reference frame. In my research I considered how quantum reference frames might behave in a relativity principle. A relativity principle in a physical theory dictates how the description and dynamics of a physical system change under a change of reference frame. For the research in this section we compared the description of a quantum system when using one quantum reference frame to the description when using a second quantum reference frame, but in order to use the second reference frame a measurement of the two reference frames is required. This research details the decoherence that results from changing the description from one reference frame to another. A connection with intrinsic decoherence is made when one considers to describe this decoherence to the quantum state when the reference frames are not treated as quantum states, but as external non-dynamic quantities. One would then witness decoherence of an isolated quantum system, so called `intrinsic decoherence,' due to the inherent quantum nature of space. This effect is a proposed semi-classical effect of quantum gravity.

\section{Structure of the thesis}

The thesis from this point consists primarily of published and submitted papers. The next three chapters detail the research regarding construction of a quantum information theory of relativistic localised qubits starting from field theory. The bulk of the research is in the `Localised qubits in curved spacetimes' paper in \chref{sec-LQiCST}. Directly following this is the paper deriving the spin operator corresponding to spin measurement of a relativistic charged massive fermion from field theory, in \chref{ch-RSO}. \chref{ch-2} is a short summary and discussion of the ways in which spin is represented in relativistic settings. This concludes the first part of the thesis. The work on the decoherence obtained due to changing quantum reference frames constitutes \chref{ch-CoQRF}. Following this is the conclusion for the thesis, \chref{ch-summary}. The combined bibliography for all chapters follows on page~\pageref{Ch-bib}. The appendices of the thesis consist of the appendix for `Localised qubits in curved spacetimes', \appref{ch-app3}, and the appendix for `Changing quantum reference frames', \appref{ch-app5}. \appref{ch-app3} contains some additional mathematical details about spinors, and a proof regarding the mathematical requirements of a transformation law for polarisation on non-geodesic null trajectories. \appref{ch-app5} contains an analysis of Balanced Homodyne Detection.

%% file: LQiCST/Qubits_curvedspacetime.tex
%%%%%%%%%%%%%%%%%%%%%%%%%%%%%%%%%%%%%%%%%%%%%%%%%%
%
%\begin{document}
%\bibliographystyle{ieeetr}
%
\chapter{Localised qubits in curved spacetimes}
\label{sec-LQiCST}
%{\center{ Matthew C Palmer, Maki Takahashi, and Hans F Westman.}}
%\author{\\\\
%Matthew C. Palmer$^1$, Maki Takahashi$^1$, and Hans F. Westman$^{1,2,3}$\\
%{\small \it $^{1}$School of Physics, The University of Sydney, Sydney, NSW 2006, Australia}\\%
%{\small \it $^{2}$Centre for Time, The University of Sydney, Sydney, NSW 2006, Australia}\\%
%{\small \it $^{3}$Perimeter Institute for Theoretical Physics, Waterloo, Ontario N2L 2Y5, Canada}\\[2mm]%
%}

%\date{{\small \today }}
%
%\maketitle

%%%%%%%%%%%%%%%%%%%%%%%%%%%%%%%%%%%%%%%%%%%%%%%%%
%\begin{abstract}
\subsection*{Abstract}
We provide a systematic and self-contained exposition of the subject of localised qubits in curved spacetimes. This research was motivated by a simple experimental question: if we move a spatially localised qubit, initially in a state $|\psi_1\rangle$, along some spacetime path $\Gamma$ from a spacetime point $x_1$ to another point $x_2$, what will the final quantum  state $|\psi_2\rangle$ be at point $x_2$? This chapter addresses this question for two physical realisations of the qubit: spin of a massive fermion and polarisation of a photon. Our starting point is the Dirac and Maxwell equations that describe respectively the one-particle states of localised massive fermions and photons. In the WKB limit we show how one can isolate a two-dimensional quantum state which evolves unitarily along $\Gamma$. The quantum states for these two realisations are represented by a left-handed 2-spinor in the case of massive fermions and a four-component complex polarisation vector in the case of photons. In addition we show how to obtain from this WKB approach a fully general relativistic description of gravitationally induced phases. We use this formalism to describe the gravitational shift in the Colella--Overhauser--Werner 1975 experiment. In the non-relativistic weak field limit our result reduces to the standard formula in the original paper. We provide a concrete physical model for a Stern--Gerlach measurement of spin and obtain a unique spin operator which can be determined given the orientation and velocity of the Stern--Gerlach device and velocity of the massive fermion. Finally, we consider multipartite states and generalise the formalism to incorporate basic elements from quantum information theory such as quantum entanglement, quantum teleportation, and identical particles. The resulting formalism provides a basis for exploring precision quantum measurements of the gravitational field using techniques from quantum information theory.
%\end{abstract}

%%%%%%%%%%%%%%%%%%%%%%%%%%%%%%%%%%%%%%%%%%%%%%%%%%%

\newpage

%%%%%%%%%%%%%%%%%%%%%%%%%%%%%%%%%%%%%%%%%%%%%%%%%%%

%\newpage
\setcounter{section}{-1}
\section{Notation and conventions\label{notation}}
%%%%%%%%%%%%%%%%%%%%%%%%%%%%%%%%%%%%%%%%%%%%%%%%%%
We use the following index notation:
\begin{itemize}
\item[-] $\mu,\nu,\rho,\sigma,\ldots$ denote spacetime tensor indices
\item[-] $I,J,K,L,\ldots=0,1,2,3$ denote tetrad indices.
\item[-] $i,j,k,l,\ldots=1,2,3$ for spatial components of the tetrad (the `triad')
\item[-] $A,B,C,D,\ldots=1,2$ for spinor indices
\item[-] $A',B',C',D',\ldots=1,2$ for conjugate spinor indices
\end{itemize}
The Minkowski metric is defined as $\eta_{\mu\nu}=\diag(1,-1,-1,-1)$. We generally use natural units where $c=\hbar=1$, and in addition we set the charge of a proton to $e=1$.

We use the Weyl representation for the Dirac $\gamma$-matrices
\begin{eqnarray*}
\gamma^I=\left(\begin{array}{cc}0&\sigma^I\\ \bar{\sigma}^I&0\end{array}\right)
\end{eqnarray*}
where $\sigma^I=(1,\sigma^i)$ and $\bar{\sigma}^I=(1,-\sigma^i)$, and $\sigma^i$ are the usual Pauli matrices. Writing this object in spinor notation we have $\sigma^I=\sigma^I_{\ AA'}$ and $\bar{\sigma}^I=\bar{\sigma}^{IA'A}$. In order to interpret the spatial parts $\sigma^i_{\ AA'}$ and $-\bar{\sigma}^{iA'A}$ as the Pauli matrices we use the convention in \cite{Bailin,DHM2010}: for $\bar{\sigma}^I=\bar{\sigma}^{IA'A}$ the primed index is the row index and the unprimed index is the column index, and the opposite assignment occurs for $\sigma^I=\sigma^I_{\ AA'}$. In spinorial notation $\sigma^I$ is not an operator on the space of spinors to itself; rather, an operator $\hat{A}$ carries an index structure $A_{A}^{\ B}$ or $A_{\ B}^{A}$. Throughout this chapter we will switch between the implicit index notation $\hat{A}$ and $A_{A}^{\ B}$ or $A_{\ B}^{A}$.

%\newpage

%%%%%%%%%%%%%%%%%%%%%%%%%%%%%%%%%%%%%%%%%%%%%%%%%%%%%%%%%%%%%%%%%
\section{Introduction}
%%%%%%%%%%%%%%%%%%%%%%%%%%%%%%%%%%%%%%%%%%%%%%%%%%%%%%%%%%%%%%%%%
This chapter will provide a systematic and self-contained exposition of the subject of localised qubits in curved spacetimes with the focus on two physical realisations of the qubit: spin of a massive fermion and polarisation of a photon. Although a great amount of research has been devoted to quantum field theory in curved spacetimes, e.g.\;\cite{Birrell,WaldQFT,Mukhanov}, the quantum information theory of black holes \cite{TernoBH05,MOD12,HarlowHayden13,AMPS13,NVW13,LLT13,Susskind13}, and also more recently to relativistic quantum information theory in the presence of particle creation and the Unruh effect \cite{Alsing-DiracUnruh,Louko,Fuentes,FuentesBerry,Martinez}, the literature about localised qubits and quantum information theory in curved spacetimes is relatively sparse \cite{TerashimaUeda03,ASK,Pienaar11}. In particular, we are aware of only three papers, \cite{TerashimaUeda03,ASK,BDT11}, that deal with the following question: if we move a spatially localised qubit, initially in a state $|\psi_1\rangle$, along some spacetime path $\Gamma$ from a point $p_1$ in spacetime to another point $p_2$, what will the final quantum state $|\psi_2\rangle$ be at point $p_2$? This, and other relevant questions, were given as open problems in the field of relativistic quantum information by Peres and Terno in \cite[p.19]{PeresTerno04}. The formalism developed in this chapter will be able to address such questions, and will also be able to deal with the basic elements of quantum information theory such as entanglement and multipartite states, teleportation, and quantum interference.

The basic object in quantum information theory is the qubit. Given a Hilbert space of some physical system, we can physically realise a qubit as any two-dimensional subspace of that Hilbert space. However, such physical realisations will in general not be localised in physical space. Furthermore, it is debated whether the qubits provided by such nonlocal systems are useful realisations \cite{Dowker11,Bradler11,Montero11comment,Calmet12,Friis13}. In this chapter we shall restrict our attention to physical realisations that are well-localised in physical space so that we can approximately represent the qubit as a two-dimensional quantum state attached to a single point in space. From a spacetime perspective a localised qubit is then mathematically represented as a sequence of two-dimensional quantum states along some spacetime trajectory corresponding to the worldline of the qubit.

In order to ensure relativistic invariance it is then necessary to understand how this quantum state transforms under a Lorentz transformation. However, as is well-known, there are no finite-dimensional faithful unitary representations of the Lorentz group \cite{Wigner} and in particular no two-dimensional ones. The only faithful unitary representations of the Lorentz group are infinite dimensional (see e.g.\;\cite{KimNoz}).  Hence, these cannot be taken to mathematically represent a qubit, i.e.\;a two-level system. Naively it would appear that a formalism for describing localised qubits which is both relativistic and unitary is a mathematical impossibility.

In the case of flat spacetime the Wigner representations \cite{Wigner,Weinberg} provide unitary and faithful but infinite-dimensional representations of the Lorentz group. These representations make use of the symmetries of Minkowski spacetime, i.e.\;the full inhomogeneous Poincar\'e group which includes rotations, boosts, and translations. The basis states $|p,\sigma\rangle$ are taken to be eigenstates of the four momentum operators (the generators of spatio-temporal translations) $\hat{P}^\mu$, i.e.\; $\hat{P}^\mu|p,\sigma\rangle=p^\mu|p,\sigma\rangle$ where the symbol $\sigma$ refers to some discrete degree of freedom, perhaps spin or polarisation. One strategy for obtaining a two-dimensional (perhaps mixed) quantum state $\rho_{\sigma\sigma'}$ for the discrete degree of freedom $\sigma$ would be to trace out the momentum degree of freedom. But as shown in \cite{PeresScudoTerno02,PeresTerno03b,PeresTerno04,BartlettTerno05} this density operator does not have covariant transformation properties.  The mathematical reason, from the theory presented in this chapter, is that the quantum states for qubits with different momenta belong to {\it different} Hilbert spaces. Thus, the density operator $\rho_{\sigma\sigma'}$ is then a mixture of states which belong to different Hilbert spaces. The operation of `tracing out the momenta' is neither physically meaningful nor mathematically motivated.\footnote{This will be discussed in more detail in Chapters \ref{ch-RSO} and \ref{ch-2}.}

Another strategy for defining qubits in a relativistic setting would be to restrict to momentum eigenstates $|p,\sigma\rangle$. The continuous degree of freedom $P$ is then fixed and the remaining degrees of freedom are discrete. In the case of a photon or fermion the state space is two dimensional and this can then serve as a relativistic realisation of a qubit. This is the strategy in \cite{TerashimaUeda03,ASK} where the authors develop a theory of transport of qubits along worldlines. However, when we go from a flat spacetime to curved we lose the translational symmetry and thereby also the momentum eigenstates $\ket{p,\sigma}$. The only symmetry remaining is local Lorentz invariance which is manifest in the tetrad formulation of general relativity. Since the translational symmetry is absent in a curved spacetime it seems difficult to work with Wigner representations which rely heavily on the full inhomogeneous Poincar\'e group. The use of Wigner representations therefore needs further justification as they do not exist in curved spacetimes.

In this chapter we shall refrain from using the infinite-dimensional Wigner representation, only translating results to the Wigner representation for comparison and for familiarity for some readers. Since our focus is on qubits physically realised as polarisation of photons and spin of massive fermions our starting point will be the field equations that describe those physical systems, i.e.\;the Maxwell and Dirac equations in curved spacetimes. Using the WKB approximation we then show in detail how one can isolate a two-dimensional Hilbert space and determine an inner product, unitary evolution, and a quantum state in a Lorentz covariant formalism. Our procedure reproduces the results of \cite{TerashimaUeda03,ASK}, and can be regarded as an independent justification and validation.

Notably, possible gravitationally induced global phases \cite{Anandan,Stodolsky,AudretschLammerzahl,Sakurai,Werner,Alsing}, which are absent in \cite{TerashimaUeda03,ASK}, are automatically included in the WKB approach. Such a phase is irrelevant if only single trajectories are considered. However, quantum mechanics allows for more exotic scenarios such as when a single qubit is simultaneously transported along a superposition of paths. In order to analyse such scenarios it is necessary to determine the gravitationally induced phase difference. We show how to derive a simple but fully general relativistic expression for such a phase difference in the case of spacetime Mach--Zehnder interferometry. Such a phase difference can be measured empirically \cite{Colella} with neutrons in a gravitational field. See \cite{Anandan,Mannheim,VarjuRyder} and references therein for further details and generalisations. The formalism developed in this chapter can easily be applied to any spacetime, e.g.\;spacetimes with frame-dragging.

This chapter aims to be self-contained and we have therefore included necessary background material such as the tetrad formulation of general relativity, the connection 1-form, spinor formalism and more (\Ssref{sec-refframes} and \ref{secspinornotation}). For example, the absence of global reference frames in a curved spacetime has a direct bearing on how entangled states and quantum teleportation in a curved spacetime are to be understood conceptually and mathematically. We discuss this in \secref{secQIinCST}.

%\tblue{The necessary background material for the work in this chapter will be provided in \secref{sec-refframes}. This includes the tetrad formulation of general relativity, the connection 1-form, and spinor formalism. For example, the absence of global reference frames, discussed in \Ssref{sec-refframes}, has a direct bearing on the conceptual and mathematical understanding of entangled states and quantum teleportation in a curved spacetime, discussed in \secref{secQIinCST}. }

%%%%%%%%%%%%%%%%%%%%%%%%%%%%%%%%%%%%%%%%%%%%%%%%%%%%%%%%%%%%%%%%
\section{An outline of methods and concepts}
%%%%%%%%%%%%%%%%%%%%%%%%%%%%%%%%%%%%%%%%%%%%%%%%%%%%%%%%%%%%%%%%
In this section we provide a general outline of the main ideas and concepts needed to understand the topic of localised qubits in curved spacetimes.

%%%%%%%%%%%%%%%%%%%%%%%%%%%%%%%%%%%%%%%%%%%%%%%%%%%%%%%%%%%%%%%%
\subsection{Localised qubits in curved spacetimes}\label{sec:localizedqubits}
%%%%%%%%%%%%%%%%%%%%%%%%%%%%%%%%%%%%%%%%%%%%%%%%%%%%%%%%%%%%%%%%
Let us now make precise the concept of a localised qubit. As a minimal characterisation, a {\em localised qubit} is understood in this chapter as any two-level quantum system which is spatially well-localised. Such a qubit is effectively described by a two-dimensional quantum state attached to a single point in space. From a spacetime perspective the history of the localised qubit is then a sequence (i.e.\;a one-parameter family) of two-dimensional quantum states $|\psi(\lambda)\rangle$ each associated with a point $x^\mu(\lambda)$ on the worldline of the qubit parameterised by $\lambda$. In this chapter we will focus on qubits represented by the spin of an electron and the polarisation of a photon and show how one can, by applying the WKB approximation to the corresponding field equation (the Dirac or Maxwell equation), extract a two-level quantum state associated with a spatially localised particle.

The sequence of quantum states $\ket{\psi(\lambda)}$ must be thought of as belonging to {\it distinct} Hilbert spaces $\m H_{x(\lambda)}$ attached to each point $x^\mu(\lambda)$ of our trajectory. The situation is identical to that in differential geometry where one must think of the tangent spaces associated with different spacetime points as mathematically distinct: since the parallel transport of a vector along some path from one point to another is path dependent there is no natural identification between vectors of one tangent space and the other.\footnote{The formal structure of this is the `vector bundle'.} The parallel transport, for any type of object, is simply a sequence of infinitesimal Lorentz transformations acting on the object and it is this sequence that is in general path dependent. Thus, if we are dealing with a physical realisation of a qubit whose state transforms non-trivially under the Lorentz group, as is the case for the two physical realisations that we are considering, we must also conclude that in general it is not possible to compare quantum states associated with distinct points in spacetime. As we shall see in sections \ref{secfermionQS} and \ref{sec-photonIP}, Hilbert spaces for different {\it momenta} $p^{\mu}(\lambda)$ of the particle carrying the qubit must also be considered distinct. The Hilbert spaces will therefore be indexed as $\mathcal{H}_{x,p}$, and so along a trajectory there will be a family of Hilbert spaces $\m H_{(x,p)(\lambda)}$.

\begin{figure}[h]
\begin{center}
\ifpdf
\includegraphics[width=8cm]{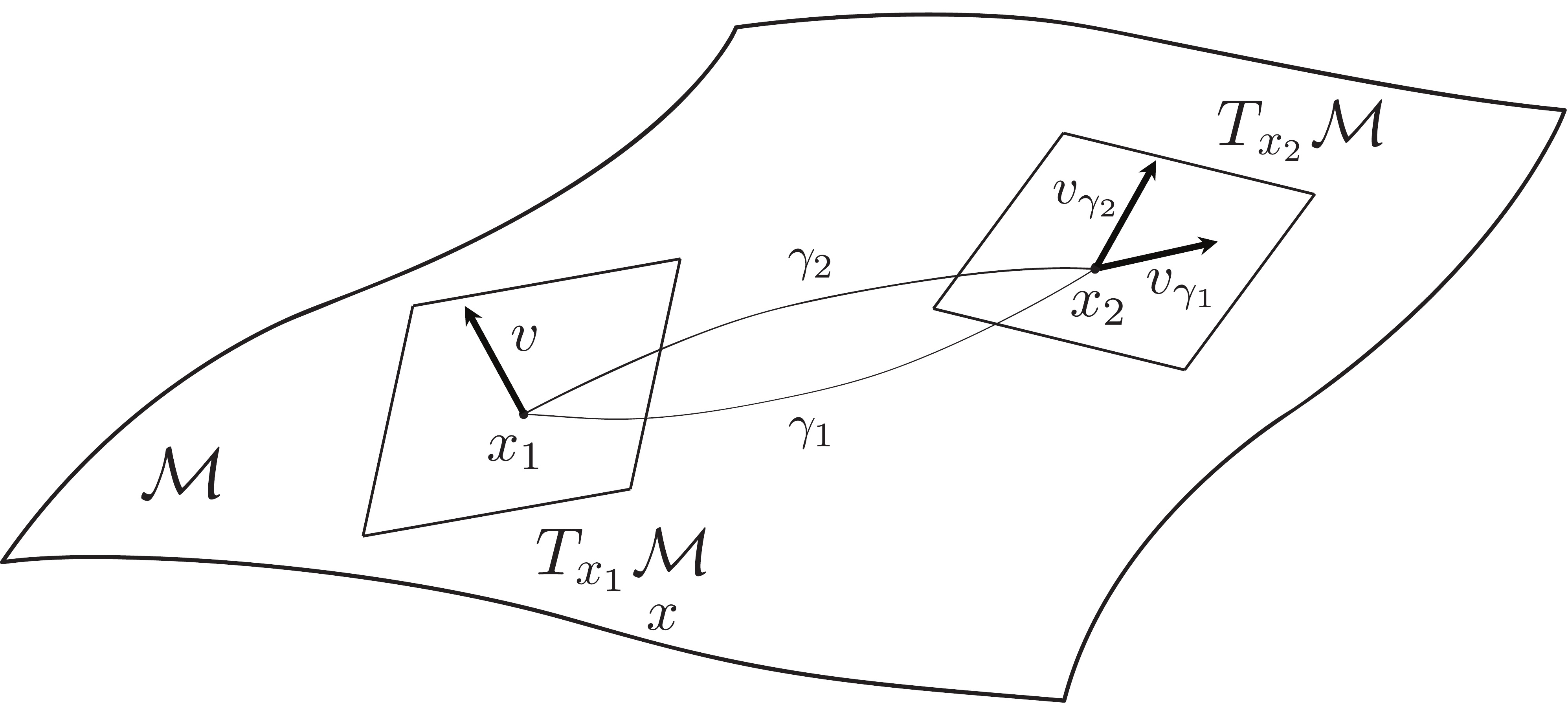}
\includegraphics[width=8cm]{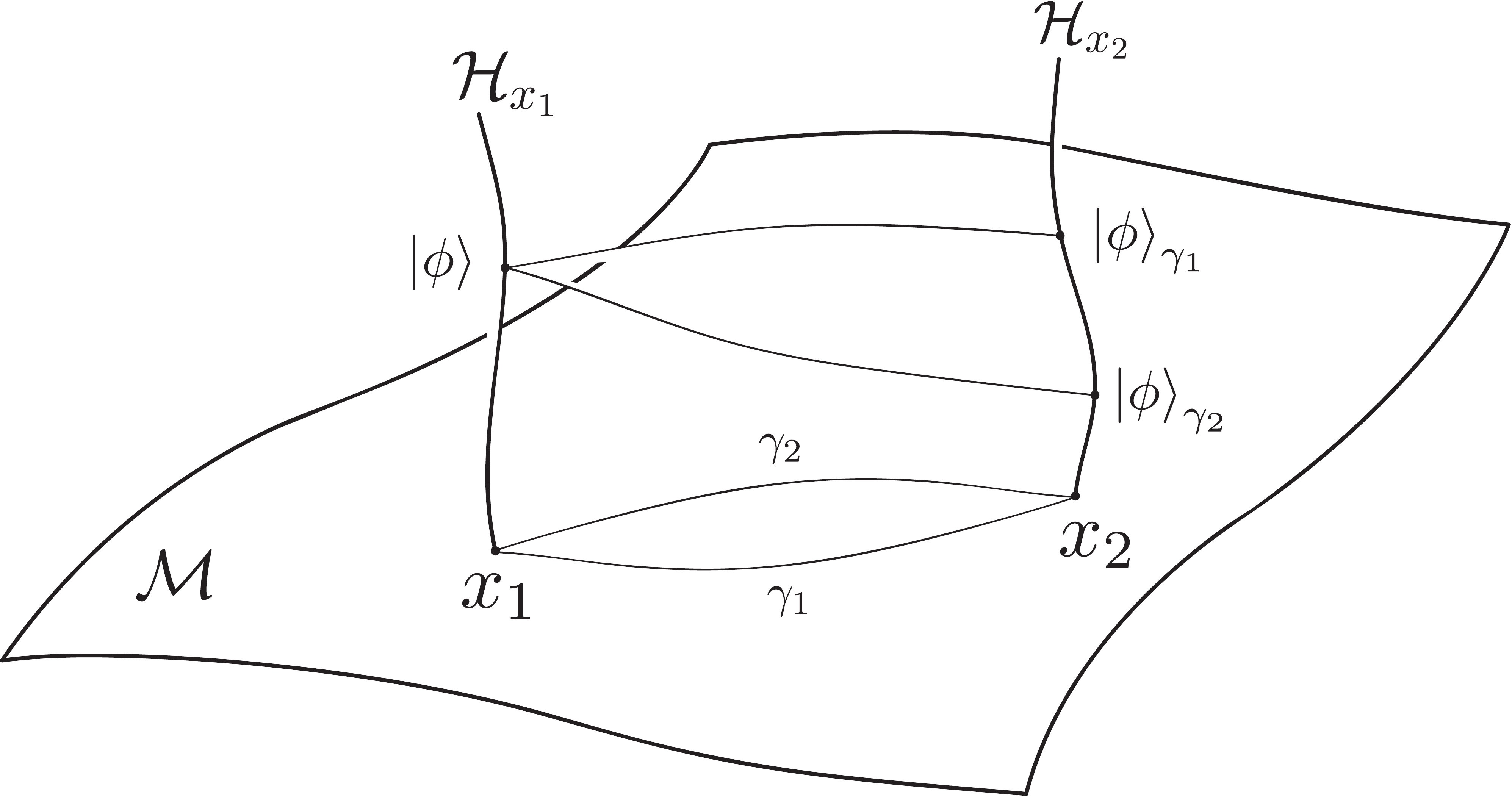}
\else
\includegraphics[width=8cm]{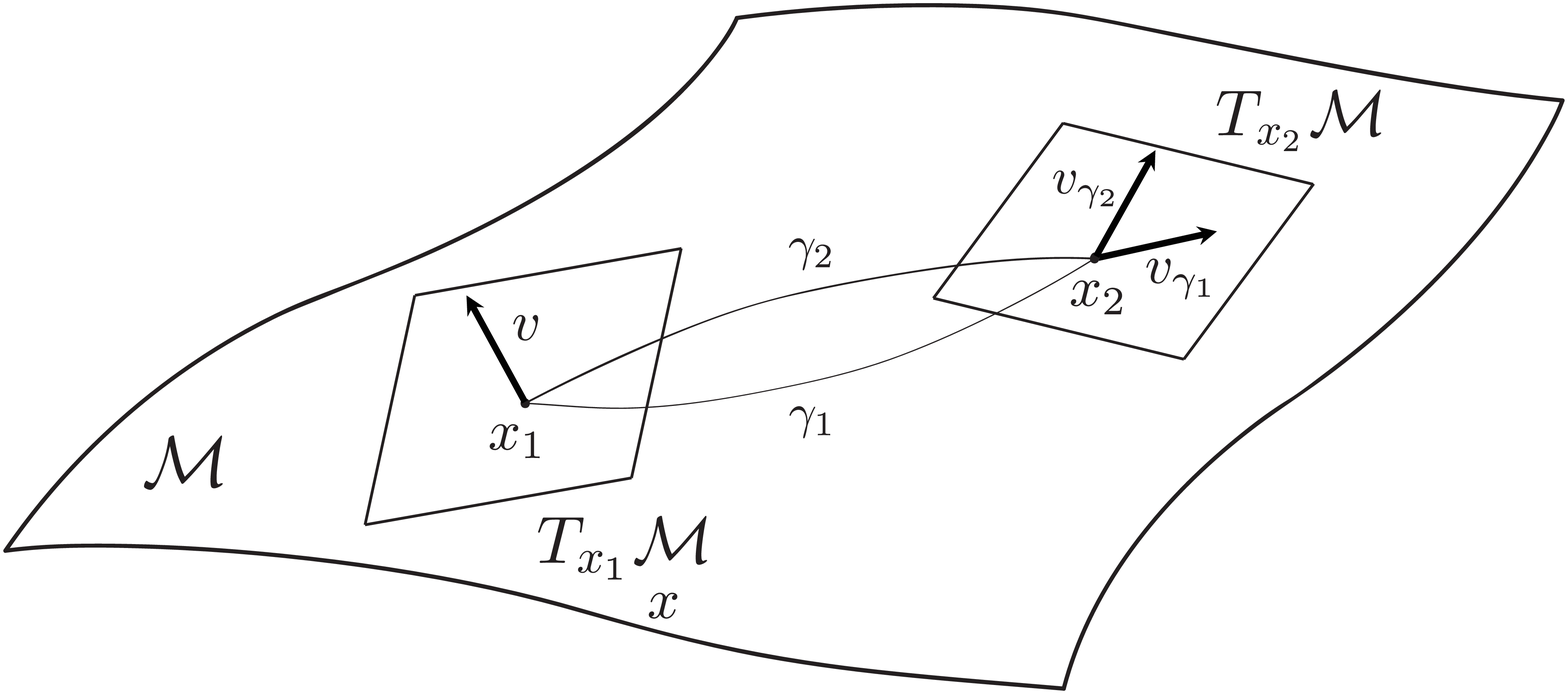}
\includegraphics[width=8cm]{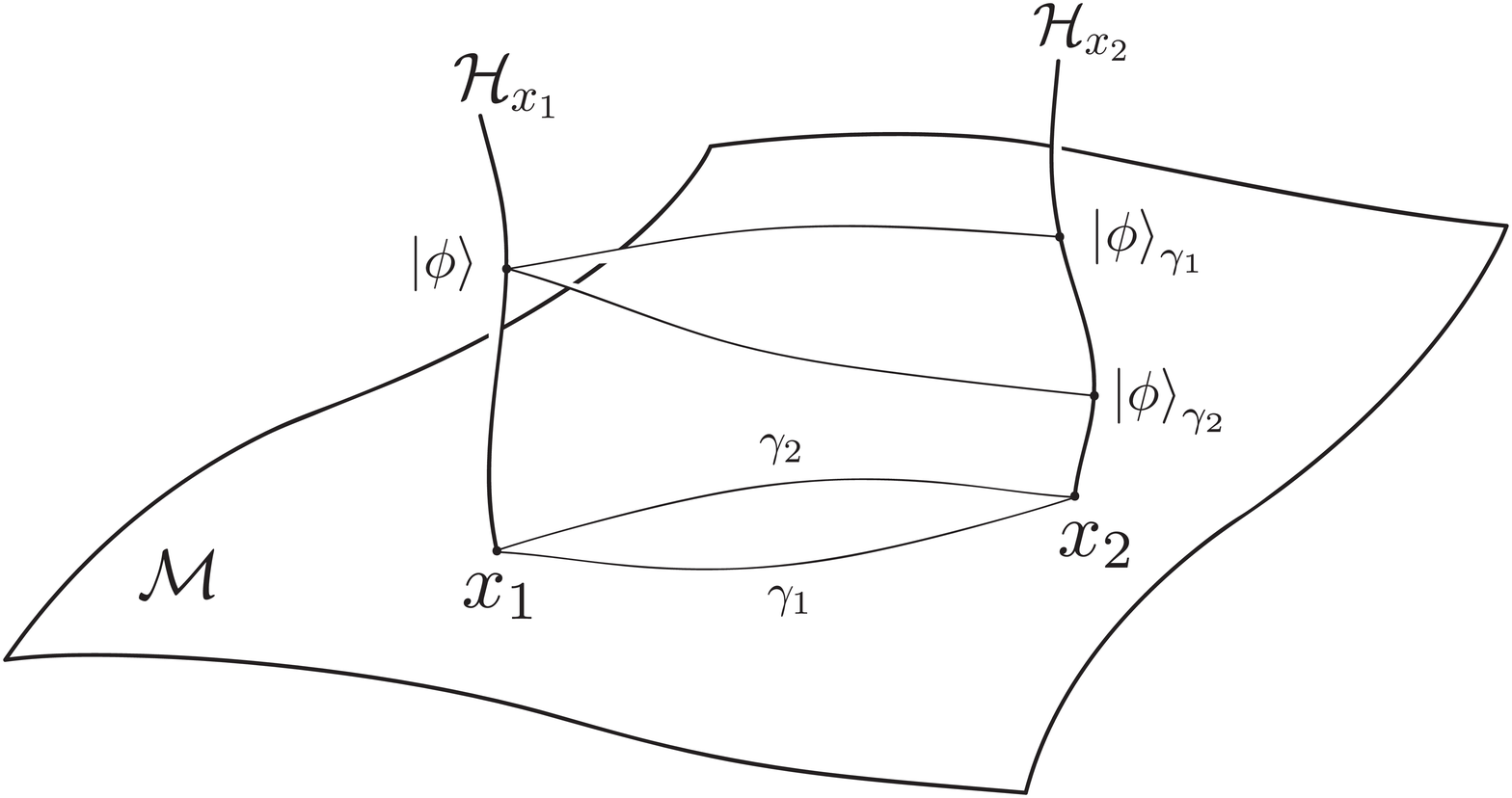}
\fi
\end{center}
\caption[Parallel transport of vectors in tangent spaces versus Hilbert spaces]{If we parallel transport a vector $v$ from point $x_1$ to $x_2$ along two distinct trajectories $\Gamma_1$ and $\Gamma_2$ in curved spacetime we generally obtain two distinct vectors $v_{\Gamma_1}$ and $v_{\Gamma_2}$ at $x_2$. Thus, no natural identification of vectors of one tangent space and another exists in general and we need to associate a distinct tangent space for each point in spacetime. The same applies to quantum states and their Hilbert spaces: the state at a point $x_2$ of a qubit moved from a point $x_1$ would in general depend on the path taken, and hence the Hilbert space for each point is distinct.  The Hilbert spaces $\m H_{x_1}$ and $\m H_{x_2}$ are illustrated as vertical `fibres' attached to the spacetime points $x_1$ and $x_2$.}
\label{comparevector}
\end{figure}

The ambiguity in comparing separated states has particular consequences: It is in general not well-defined to say that two quantum states associated with distinct points in spacetime are the {\it same}. Nor is it mathematically well-defined to ask how much a quantum state has ``really" changed when moved along a path. Nevertheless, if two initially identical states are transported to some point $x$ but along two distinct paths, the difference between the two resulting states is well-defined, since we are comparing states belonging to the same Hilbert space (see \figref{comparevector}).

There are also consequences for how we interpret basic quantum information tasks such as
quantum teleportation: When Alice ``teleports" a quantum state over some distance to Bob we would like to say that it is the {\it same} state that appears at Bob's location. However, this will not have an unambiguous meaning. An interesting alternative is to instead use the maximally entangled state to {\it define} what is ``the same'' quantum state for Bob and Alice, at their distinct locations. We return to these issues in \S\ref{teleportation}.

In a strict sense a localised qubit can be understood as a sequence of quantum states attached to points along a worldline. We will however relax this notion of localised qubits slightly to allow for path superpositions as well. More specifically, we can consider scenarios in which a single localised qubit is split up into a spatial superposition, transported simultaneously along two or more distinct worldlines, and made to recombine at some future spacetime region so as to produce quantum interference phenomena (see \Ssref{secPhase}). We will still regard these spatial superpositions as localised if the components of the superposition are each localised around well-defined spacetime trajectories.

%%%%%%%%%%%%%%%%%%%%%%%%%%%%%%%%%%%%%%%%%%%%%%%%%%%%%%%%%%
\subsection{Physical realisations of localised qubits\label{localqubits}}
%%%%%%%%%%%%%%%%%%%%%%%%%%%%%%%%%%%%%%%%%%%%%%%%%%%%%%%%%%
The concepts of a classical bit and a quantum bit (cbit and qubit for short) are abstract concepts in the sense that no importance is usually attached to the specific way in which we physically realise the cbit or qubit. However, when we want to manipulate the state of the cbit or qubit using external fields, the specific physical realisation of the bit becomes important. For example, the state of a qubit, physically realised as the spin of a massive fermion, can readily be manipulated using an external electromagnetic field, but the same is not true for a qubit physically realised as the polarisation of a photon.

The situation is no different when the external field is the gravitational one. In order to develop a formalism for describing transport of qubits in curved spacetimes it is necessary to pay attention to how the qubit is physically realised. Without knowing whether the qubit is physically realised as the spin of a massive fermion or the polarisation of a photon, for example, it is not possible to determine how the quantum state of the qubit responds to the gravitational field. More precisely: gravity, in part, acts on a localised qubit through a sequence of Lorentz transformations which can be determined from the trajectory along which it is transported and the gravitational field, i.e.\;the connection one-form ${\omega_{\mu}}^I_{\ J}$. Since different qubits can constitute different representations under the Lorentz group, the  influence of gravity will be representation dependent. This is not at odds with the equivalence principle, which only requires that the qubits are acted upon with the same Lorentz transformation.

%%%%%%%%%%%%%%%%%%%%%%%%%%%%%%%%%%%%%%%%%%%%%%%%%%%%%%%%%
\subsection{Our approach}
%%%%%%%%%%%%%%%%%%%%%%%%%%%%%%%%%%%%%%%%%%%%%%%%%%%%%%%%%
Our starting point will be the one-particle excitations of the respective quantum fields. These one-particle excitations are described by complexified {\it classical} fields $\Psi$ or $A_{\mu}$, which are governed by the classical Dirac or Maxwell equation, respectively. Our goal is to formulate a mathematical description for localised qubits in curved spacetime. Therefore we must find a regime in which the spatial degrees of freedom of the fields are suppressed so that the relevant state space reduces to a two-dimensional quantum state associated with points along some well-defined spacetime trajectory. Our approach is to apply the WKB approximation to these field equations (sections \ref{fermsemiclassapprox} and \ref{photongeoapprox}) and study spatially localised solutions. In this way we can isolate a two-dimensional quantum state that travels along a classical trajectory.

In the approach that we use for the two realisations, we start with a general wavefunction for each field, decomposed as
\begin{eqnarray*}
\phi_A(x)=\psi_A(x)\varphi(x)\Ee^{\ii\theta(x)}\ \mbox{ or }\ A_{\mu}(x)=\psi_{\mu}(x)\varphi(x)\Ee^{\ii\theta(x)}
\end{eqnarray*}
where the two-component spinor field $\phi_A(x)$ is the left-handed component of the Dirac field $\Psi$ and is sufficient for describing a fermion \cite{DHM2010}, and $x^{\mu}$ is some coordinate system. The decompositions for the two fields are similar: $\theta(x)$ is the phase, $\varphi(x)$ is the real-valued envelope, and $\psi_A(x)$ or $\psi_{\mu}(x)$ are fields that encode the quantum state of the qubit in the respective cases. These latter objects are respectively the normalised two-component spinor field and normalised complex-valued polarisation vector field. Note that we are deliberately using the same symbol $\psi$ for both the two-component spinor $\psi_{A}$ and the polarisation 4-vector $\psi_{\mu}$ as it is these variables that encode the quantum state in each case.

The WKB limit proceeds under the assumptions that the phase $\theta(x)$ varies in $x$ much more rapidly than any other aspect of the field and that the wavelength of the phase oscillation is much smaller than the spacetime curvature scale.  Expanding the field equations under these conditions we obtain:
\begin{itemize}
\item a field of wavevectors $k_{\mu}(x)$ whose integral curves satisfy the corresponding classical equations of motion;
\item a global phase $\theta$, determined by integrating $k_\mu$ along the integral curves;
\item transport equations that govern the evolution of  $\psi_A$ and $\psi_\mu$ along this family of integral curves;
\item a conserved current which will be interpreted as a quantum probability current.
\end{itemize}
The assumptions of the WKB limit by themselves do not ensure a spatially localised envelope $\varphi(x)$, and therefore do not in general describe localised qubits. In sections \ref{fermionlocalization} and \ref{photonlocalization} we add further assumptions that guarantee that the qubit is localised during its transport along the trajectory. The spatial degrees of freedom are in this way suppressed and we can effectively describe the qubit as a sequence of quantum states, encoded in the objects $\psi_A(\tau)$ or $\psi_\mu(\lambda)$. These objects constitute non-unitary representations of the Lorentz group. As we shall see in \S\ref{sec-unitarity}, unitarity is recovered once we have correctly identified the respective inner products. Notably, the Hilbert spaces $\m H_{(x,p)(\lambda)}$ we obtain are labelled with both the position and the momentum of the localised qubit.

Finally, since the objects $\psi_A(\tau)$ and $\psi_\mu(\lambda)$ have been separated from the phase $\Ee^{\ii\theta(x)}$, the transport equations for these objects do not account for possible gravitationally induced global phases. We show how to obtain such phases in  \S\ref{secPhase} from the WKB approximation. Thus, with the inclusion of phases, we have provided a complete, Lorentz covariant formalism describing the transport of qubits in curved spacetimes. Hereafter it is straightforward to extend the formalism to several qubits in order to treat multipartite states, entanglement and teleportation (\S\ref{secQIinCST}), providing the basic ingredients of quantum information theory in curved spacetimes.

%%%%%%%%%%%%%%%%%%%%%%%%%%%%%%%%%%%%%%%%%%%%%%%%%%%%%%%%%%%%%%%%
\section{Issues from quantum field theory and the domain of applicability\label{sec-domain}}
%%%%%%%%%%%%%%%%%%%%%%%%%%%%%%%%%%%%%%%%%%%%%%%%%%%%%%%%%%%%%%%%
The formalism describing qubits in curved spacetimes presented in this chapter has its specific domain of applicability and cannot be taken to be empirically correct in all situations. One simple reason for this is that the current most fundamental theory of nature is not formulated in terms of localised qubits but instead involves very different objects such as quantum fields. There are four important issues arising from quantum field theory that restrict the domain of applicability:
\begin{itemize}
\item the problem of localisation;
\item particle number ambiguity;
\item particle creation;
\item the Unruh effect.
\end{itemize}
Below we discuss these issues and indicate how they restrict the domain of applicability of the formalism of this chapter. It is worth noting that quantum field theory in curved spacetimes is itself limited in scope, away from quantum gravity scenarios such as extremely high energies or curvature ($10^{-15}\mathrm m$ or $10^{-23}\mathrm s$) \cite{Birrell}. The domain of applicability we obtain is well within this scope.

%%%%%%%%%%%%%%%%%%%%%%%%%%%%%%%%%%%%%%%%%%%%%%%%%%%
\subsection{The localisation problem\label{sec-localisation}}
%%%%%%%%%%%%%%%%%%%%%%%%%%%%%%%%%%%%%%%%%%%%%%%%%%%
The formalism of this chapter concerns spatially localised qubits, with the wavepacket width being much smaller than the curvature scale. However, it is well-known from quantum field theory that it is not possible to localise one-particle states to an arbitrary degree \cite{Knight61}. For example, localisation of massive fermions is limited by the Compton wavelength $\lambda_c=h/mc$ \cite{Newtonwigner}. More precisely, any wavefunction constructed from exclusively positive frequency modes must have a tail that falls off with radius $r$ slower than $\Ee^{-r/\lambda_c}$ \cite{Hegerfeldt85}. However, this is of no concern if we only consider wavepackets with a width much larger than the Compton wavelength. This consequently restricts the domain of applicability of the material in this chapter. In particular, since the width of the wavepacket is assumed to be much smaller than the curvature scale (see \S\ref{sec-particlenumberambiguity}), the localisation theorem means that we cannot deal with extreme curvature scales of the order of the Compton wavelength.

A similar problem exists also for photons. Although the Compton wavelength for photons is ill-defined, it has also been shown that they must have non-vanishing sub-exponential tails \cite{Hegerfeldt,Birula,Keller05,BBBB09}.

Given these localisation theorems it is not strictly speaking possible to define a localised wavepacket with compact support. However, for the purpose of this chapter we will assume that most of the wavepacket is contained within some region, smaller than the curvature scale, and the exponential tails outside can safely be neglected in calculations. We will assume from here on that this is indeed the case.

%%%%%%%%%%%%%%%%%%%%%%%%%%%%%%%%%%%%%%%%%%%%%%%%%%%%%%%%%%
\subsection{Particle number ambiguity\label{sec-particlenumberambiguity}}
%%%%%%%%%%%%%%%%%%%%%%%%%%%%%%%%%%%%%%%%%%%%%%%%%%%%%%%%%%
One important lesson that we have learned from quantum field theory in curved spacetimes is that a natural notion of particle number is in general absent; see e.g.\;\cite{WaldQFT}. It is only under special conditions that a natural notion of particle number emerges. Therefore, for arbitrary time-dependent spacetimes it is not in general possible to talk unambiguously about the spin of {\it one} electron or the polarisation state of {\it one} photon as this would require an unambiguous notion of particle number. This is important in this chapter because a qubit is realised by the spin of one massive fermion or polarisation of one photon.

The particle number ambiguity can be traced back to the fact that the most fundamental mathematical objects in quantum field theory are the quantum field operators and not particles or Fock space representations. More specifically, how many particles a certain quantum state is taken to represent depends in general on how we expand the quantum field operators in terms of annihilation and creation operators $(\hat{a}_i,\hat{a}^\dagger_i)$:
\begin{eqnarray*}
\hat{\phi}(x)=\sum_i \bar{f}_i\hat{a}_i+f_i\hat{a}^\dagger_i
\end{eqnarray*}
which in turn depends on how the complete set of modes (which are solutions to the corresponding classical field equations) is partitioned into positive and negative frequency modes $(f_i,\bar{f}_i)$. In particular, the number operator $\hat{N}\equiv \sum_i\hat a^\dagger_i\hat a_i$ depends on the expansion of the quantum field operator $\hat{\phi}(x)$.

There are then two issues with particle number. The first is that particle number $N$ is defined globally, so there is no localised definition of particle content. The second is that field mode expansion can be done in an infinitude of distinct ways (e.g.\;Minkowski or Rindler) related by Bogoliubov transformations \cite{Haag55,Birrell,CliftonHalvorson01,EarmanFraser06}. So even globally there is ambiguity of particle content depending on the field mode decomposition chosen. Particle number is therefore ill-defined. Since we base our approach on the existence of well-defined {\em one-particle} states for photons and massive fermions, the particle number ambiguity seems to raise conceptual difficulties. %The second is that even globally, the Stone-von Neumann theorem regarding unitary equivalence of field mode expansions fails for quantum fields as they have an infinite number of degrees of freedom (Haag's theorem).

We will now argue from the equivalence principle we can recover an unambiguous definition of particle number for spatially localised states. Consider first vanishing external fields and thus geodesic motion (we will turn to non-geodesics in the next section). In a pseudo-Riemannian geometry, for any sufficiently small spacetime region we can always find coordinates such that the metric tensor is the Minkowski metric $g_{\mu\nu}\stackrel{*}{=}\eta_{\mu\nu}$, where $\stackrel{*}{=}$ denotes a non-covariant equality, i.e.; true in one choice of coordinates, and the affine connection is zero $\Gamma^{\rho}_{\mu\nu}\stackrel{*}{=}0$. However, this is true also for a sufficiently narrow strip around any extended spacetime trajectory, i.e.\;there exists an extended open region containing the trajectory such that $g_{\mu\nu}\stackrel{*}{=}\eta_{\mu\nu}$ and $\Gamma^{\rho}_{\mu\nu}\stackrel{*}{=}0$ \cite{WeinbergGC}. Thus, as long as the qubit wavepacket is confined to that strip it can be described as travelling in a flat spacetime.\footnote{Curvature cannot be set to zero in this process \cite[\S1.6]{MTW}, but we will see in \S\ref{secFermion} and \S\ref{sec-Photons} that for wavelengths much smaller than the curvature, the WKB limits of the field equations allow the effect of curvature to drop out. See \cite{Audretsch81,ASK} for treatments of this effect.} In particular, the usual free Minkowski modes $\Ee^{\pm ip\cdot x}$ form a complete set of solutions to the wave equation for wavepackets localised within the strip. Using these modes we can then define positive and negative frequency, as would be detected by inertial observers. Choosing these inertial observers as the preferred frames in which to determine particle number, the notion of particle number thus becomes well-defined. Thus, if we restrict ourselves to qubit wavepackets that are small with respect to the typical length scale associated with the spacetime curvature, the particle number ambiguity is circumvented and it becomes unproblematic to think of the classical fields $\Psi(x)$ and $A_\mu(x)$ as describing one-particle excitations of the corresponding quantum field.

%%%%%%%%%%%%%%%%%%%%%%%%%%%%%%%%%%%%%%%%%%%%%%%%%%%%%%%%%%%%%%%%%%%%%
\subsection{Particle creation and external fields}
%%%%%%%%%%%%%%%%%%%%%%%%%%%%%%%%%%%%%%%%%%%%%%%%%%%%%%%%%%%%%%%%%%%%%

Within a strip as defined in the previous section, the effects of gravity are absent and therefore there is no particle creation due to gravitational effects for sufficiently localised qubits. If the trajectory $\Gamma$ along which the qubit is transported is non-geodesic, non-zero external fields need to be present along the trajectory. For charged fermions we could use an electromagnetic field. However, if the field strength is strong enough it might cause spontaneous particle creation and we would not be dealing with a single particle and thus not a two-dimensional Hilbert space. As the formalism of this chapter presupposes a two-dimensional Hilbert space, we need to make sure that we are outside the regime where particle creation can occur.

When time-dependent external fields are present, the normal modes $\Ee^{\pm ip\cdot x}$ are no longer solutions of the corresponding classical field equations and there will in general be no preferred way of partitioning the modes $(f_i,\bar f_i)$ into positive and negative frequency modes. Therefore, even when we confine ourselves to within the above mentioned narrow strip, particle number is ambiguous.

This type of particle number ambiguity can be circumvented with the help of asymptotic `in' and `out' regions in which the external field is assumed to be weak. In the scenarios considered in this chapter there will be a spacetime region $\mathcal{R}_\text{\it prep.}$ in which the quantum state of the qubit is prepared, and a spacetime region $\mathcal{R}_\text{\it meas.}$ where a suitable measurement is carried out on the qubit. The regions are connected by one or many timelike paths along which the qubit is transported. The regions $\mathcal{R}_\text{\it prep.}$ and $\mathcal{R}_\text{\it meas.}$ are here taken to be macroscopic but still sufficiently small such that no tidal effects are detectable, and so special relativity is applicable. We allow for non-zero external fields in these regions and along the trajectory, though we assume that external fields (or other interactions) are weak in these end regions so that the qubit is essentially free there. This means that in $\mathcal{R}_\text{\it prep.}$ and $\mathcal{R}_\text{\it meas.}$ we can use the ordinary Minkowski modes $\Ee^{ip\cdot x}$ and $\Ee^{-ip\cdot x}$ to expand our quantum field. This provides us with a natural partitioning of the modes into positive and negative frequency modes and thus particle number is well-defined in the two regions $\mathcal{R}_\text{\it prep.}$ and $\mathcal{R}_\text{\it meas.}$. For our purposes we can therefore regard (approximately) the regions $\mathcal{R}_\text{\it prep.}$ and $\mathcal{R}_\text{\it meas.}$ as the asymptotic `in' and `out' regions of ordinary quantum field theory.

If we want to determine whether there is particle creation we simply `propagate' (using the wave equation with an external field) a positive frequency mode (with respect to the free Minkowski modes in $\mathcal{R}_\text{\it prep.}$) from region $\mathcal{R}_\text{\it prep.}$ to $\mathcal{R}_\text{\it meas.}$. We are not concerned with what form a mode takes outside these regions and strip. In region $\mathcal{R}_\text{\it meas.}$ we then see whether the propagated mode has any negative frequency components (with respect to the free Minkowski modes in $\mathcal{R}_\text{\it meas.}$). If negative frequency components are present we can conclude that particle creation has occurred (see e.g.\;\cite{PeskinSchroeder}). This will push the physics outside our one-particle-excitation formalism and we need to make sure that the strength of the external field is sufficiently small so as to avoid particle creation.

One also has to avoid spin-flip transitions in photon radiation processes such as gyromagnetic emission, which describes radiation due to the acceleration of a charged particle by an external magnetic field, and the related Bremsstrahlung, which corresponds to radiation due to scattering off an external {\it electric} field \cite{Bordovitsyn,Melroseqpd1,Melroseqpd2}. For the former, a charged fermion will emit photons for sufficiently large accelerations and can cause a spin flip and thus a change of the quantum state of the qubit.   Fortunately, the probability of a spin-flip transition is much smaller than that of a spin conserving one, which does not alter the quantum state of the qubit \cite{Melroseqpd2}. In this chapter we assume that the acceleration of the qubit is sufficiently small so that we can ignore such spin-flip processes.

%%%%%%%%%%%%%%%%%%%%%%%%%%%%%%%%%%%%%%%%%%%%%%%%%%%%%%%%%%%%%%%%
\subsection{The Unruh effect}
%%%%%%%%%%%%%%%%%%%%%%%%%%%%%%%%%%%%%%%%%%%%%%%%%%%%%%%%%%%%%%%%
Consider the case of flat spacetime. A violently accelerated particle detector could click (i.e.\;indicate that it has detected a particle) even though the quantum field $\hat{\phi}$ is in its vacuum state. This is the well-known Unruh effect \cite{Unruh,Birrell,Louko}. What happens from a quantum field theory point of view is that the term for the interaction between a detector and a quantum field allows for a process where the detector gets excited and simultaneously excites the quantum field. This effect is similar to that when an accelerated electron excites the electromagnetic field \cite{Akhmedov}. A different way of understanding the Unruh effect is by recognising that there are two different timelike Killing vector fields of the Minkowski spacetime: one generates inertial timelike trajectories and the other generates orbits of constant proper acceleration. Through the separation of variables of the wave equation one then obtains two distinct complete sets of orthonormal modes: Minkowski modes and Rindler modes, corresponding respectively to each Killing field. The positive Minkowski modes have negative frequency components with respect to the Rindler modes and it can be shown that the Minkowski vacuum contains a thermal spectrum with respect to a Rindler observer.

In order to ensure that our measurement and preparation devices operate `accurately', their acceleration must be small enough so as not to cause an Unruh type effect.
%
%%%%%%%%%%%%%%%%%%%%%%%%%%%%%%%%%%%%%%%%%
\subsection{The domain of applicability}
%%%%%%%%%%%%%%%%%%%%%%%%%%%%%%%%%%%%%%%%%
Let us summarise. In order to avoid unwanted effects from quantum field theory we have to restrict ourselves to scenarios in which:
\begin{itemize}
\item the qubit wavepacket size is much smaller than the typical curvature scale (to ensure no particle number ambiguity);
\item in the case of massive fermions, because of the localisation problem the curvature scale must be much larger than the Compton wavelength;
\item there is at most moderate proper acceleration of the qubit (to ensure no particle creation or spin-flip transition due to external fields);
\item there is at most moderate acceleration of preparation and measurement devices (to ensure negligible Unruh effect).
\end{itemize}
For the rest of the chapter we will tacitly assume that these conditions are met.

%%%%%%%%%%%%%%%%%%%%%%%%%%%%%%%%%%%%%%%%%%%%%%%%%%%%%%%%%%%%%%%%%
\section{Reference frames and connection 1-forms\label{sec-refframes}}
%%%%%%%%%%%%%%%%%%%%%%%%%%%%%%%%%%%%%%%%%%%%%%%%%%%%%%%%%%%%%%%%%

The notion of a local reference frame, which is mathematically represented by a tetrad field $e^\mu_I(x)$, is essential for describing localised qubits in curved spacetimes. This section provides an introduction to some of the elements of general relativity and to the mathematics of tetrads with an eye towards their use for quantum information theory in curved spacetime. The hurried reader may want to skip to \S\ref{secFermion}. A presentation of tetrads can also be found in \cite[App. J]{Carroll}.

%%%%%%%%%%%%%%%%%%%%%%%%%%%%%%%%%%%%%%%%%%%%%%%%%%%%%%%%%%%%%%%%%
\subsection{Curved spacetime\label{sec-GRintro}}
%%%%%%%%%%%%%%%%%%%%%%%%%%%%%%%%%%%%%%%%%%%%%%%%%%%%%%%%%%%%%%%%%

To set the scene of curved spacetime, this subsection provides a very quick introduction to some of the elements of general relativity \cite{WeinbergGC,MTW,Wald,Hartlegravity}. General relativity is a theory of gravity in which spacetime is curved and the free-fall trajectories (geodesics) are the straight lines in this geometry.

To begin with, we will have some spacetime coordinates $x^\mu$, with $\mu=t,x,y,z$ with which we can map out the positions of matter. The geometry of spacetime is determined by the distribution of this matter via the Einstein Field Equations
\[
R_{\mu\nu}(x)-\half g_{\mu\nu}(x) R(x)+g_{\mu\nu}(x)\Lambda=8\pi GT_{\mu\nu}(x),
\]
a tensor of nonlinear equations relating the distribution of energy (including matter) $T_{\mu\nu}$ to the geometry of the manifold $R$, $R_{\mu\nu}$, $\Lambda$ and $g_{\mu\nu}$ (the Ricci curvature scalar, Ricci curvature tensor, the cosmological constant, and the metric tensor) at each point $x$. In this chapter we will not be determining solutions of the space time and its geometry from the distribution of matter. Instead we will presume there is a well-defined field of metric tensors $g_{\mu\nu}(x)$ which satisfies the Einstein Field Equations. Given the metric, what we are concerned with in this chapter is the movement of quantum fields on a manifold with geometry described by $g_{\mu\nu}(x)$.

Except at exceptional points such as singularities, for a point $x$ on the manifold we can compute infinitesimal elements $\di x^\mu(x)$, each of which is a vector. These infinitesimal elements, combined with the metric tensor $g^{\mu\nu}(x)$, determine the infinitesimal path length between two neighbouring points in the spacetime: $\di s^2=g_{\mu\nu}\di x^\mu\di x^\nu$ where $g_{\mu\nu}$ is the inverse of $g^{\mu\nu}$ in that $g_{\mu\nu}g^{\nu\rho}=\delta_\mu^\rho$. The set of $\di x^\mu$ at each point $x$ also provides a frame in whose basis we can write vectors. In components, contravariant spacetime vectors are $V^\mu=(V^t,V^a)=(V^t,V^x,V^y,V^z)$, and the covariant vector is written $V_\mu=g_{\mu\nu}V^\nu$.

In \S\ref{sec-connectiononeform} we construct the tetrad $e_\mu^I$ to provide an orthonormal basis, the so-called Lorentz frame at each point. We use the $-2$ signature for general relativity.\footnote{This is less common for working with vectors, because the spatial hypersurfaces have negative components, but it is more sensible for work with spinors, since the abstract and geometric methods for raising and lowering spinor indices produce the same spinor with the same sign (see \cite{Wald} and appendix section \ref{spinhalfLG}).} Therefore, in this basis we have the Minkowski metric $\eta^{IJ}\sim\diag(1,-1,-1,-1)\sim \eta_{IJ}$ at each point. Lorentz contravariant vectors are written $V^I=(V^0,V^i)=(V^0,V^1,V^2,V^3)$, and the corresponding covariant vector is written $V_I=\eta_{IJ}V^J=(V^0,-V^1,-V^2,-V^3)=(V_0,-V_i)$.

After introducing reference frames and tetrad frames in more detail in the following two sections, in \S\ref{sec-connectiononeform} we then describe how vectors map between tangent spaces. In order to do this we require some additional geometric quantities which can be derived from the metric field, in particular the Christoffel symbols:
\be
\Gamma_{\mu\nu}^\rho=\half g^{\rho\sigma}\left(\partial_\mu g_{\sigma\nu}+\partial_\nu g_{\sigma\mu}-\partial_\sigma g_{\mu\nu}\right).\label{eq-affine}
\ee
We then obtain the covariant derivative, which produces the parallel transport of a vector along a geodesic trajectory, and the Fermi--Walker equation, governing the torque-free transport of a vector along an accelerated trajectory.

%\tred{The core principle of `general relativity' is the principle of general equivalence, which states that for a small enough object, the effect of gravity is indistinguishable from an non-gravitational acceleration: That inertial and gravitational mass are equivalent \cite{Hartlegravity}.}

%%%%%%%%%%%%%%%%%%%%%%%%%%%%%%%%%%%%%%%%%%%%%%%%%%%%%%%%%%%%%%%%%
\subsection{The absence of global reference frames}
%%%%%%%%%%%%%%%%%%%%%%%%%%%%%%%%%%%%%%%%%%%%%%%%%%%%%%%%%%%%%%%%%
One main issue that arises when generalising quantum information theory from flat to curved spaces is the absence of a global reference frame. On a flat space manifold one can define a global reference frame by first introducing, at an arbitrary point  $x_1$, some orthonormal reference frame, i.e.\;we associate three orthonormal spatial vectors $(\hat x_{x_1},\hat y_{x_1},\hat z_{x_1})$ with the point $x_1$. In order to establish a reference frame at some other point $x_2$ we can parallel transport each of the three vectors to that point. Since the manifold is flat the three resulting orthonormal directions are independent of the path along which they were transported. Repeating this for all points $x$ in our space we obtain a unique field of reference frames $(\hat{x}_x,\hat{y}_x,\hat{z}_x)$ defined for all points $x$ on the manifold.\footnote{In this chapter we will implicitly always work in a topologically trivial open set. This allows us to ignore topological issues, e.g.\;the fact that not all manifolds will admit the existence of an everywhere non-singular field of reference frames.} Thus, from an arbitrarily chosen reference frame at a single point $x_1$ we can erect a unique {\em global reference frame}.

However, when the manifold is curved no unique global reference frame can be established in this way. The reference frame obtained at point $x_2$ by the parallel transport of the reference frame at $x_1$ is in general dependent on the path along which the frame was transported. Thus, in general there is no path-independent way of constructing global reference frames. Instead we have to accept that the choice of reference frame {\em at each point} on the manifold is completely arbitrary, leading us to the notion of local reference frames.

To illustrate this situation and its consequences in the context of quantum information theory in curved space, consider two parties, Alice and Bob, at separated locations. First we turn to the case where the space is flat and the entangled state is the singlet state. The measurement outcomes will be anticorrelated if Alice and Bob measure along the {\em same} direction. In flat space the notion of `same direction' is well-defined. However, in curved space, whether two directions are `the same' or not is a matter of pure convention, since the direction obtained from parallel transporting a reference frame from Alice to Bob is path dependent. Thus, the phrase `Alice and Bob measure along the {\em same} direction' does not have an unambiguous meaning in curved space.

With no natural way to determine that two reference frames at separated points have the same orientation, we are left with having to keep track of the arbitrary local choice of reference frame at each point. The natural way to proceed is then to develop a formalism that will be reference frame {\em covariant}, with the empirical predictions (e.g.\;predicted probabilities) of the theory required to be manifestly reference frame {\em invariant}. The formalism obtained in this chapter meets these two requirements.

%%%%%%%%%%%%%%%%%%%%%%%%%%%%%%%%%%%%%%%%%%%%%%%%%%%%%%%%%%%%%%%%%%%%%%%%
\subsection{Tetrads and local Lorentz invariance \label{tetradllinvar}}
%%%%%%%%%%%%%%%%%%%%%%%%%%%%%%%%%%%%%%%%%%%%%%%%%%%%%%%%%%%%%%%%%%%%%%%%

The previous discussion was in terms of a curved space and a spatial reference frame consisting of three orthonormal spatial vectors. However, in this chapter we consider curved spacetimes, and so we have to adjust the notion of a reference frame accordingly. We can do this by simply including the 4-velocity of the spatial reference frame as a {\em fourth} component $\hat t_x$ of the reference frame. Thus, in relativity a reference frame $(\hat{t}_x,\hat{x}_x,\hat{y}_x,\hat{z}_x)$ at some point $x$ consists of three orthonormal spacelike vectors and a timelike vector $\hat t_x$.

Instead of using the cumbersome notation $(\hat t_x,\hat{x}_x,\hat{y}_x,\hat{z}_x)$ to represent a local reference frame at a point $x$ we adopt the compact standard notation $e^\mu_I(x)$. Here $I=0,1,2,3$ labels the four orthonormal vectors of this reference frame such that $e^\mu_0\sim\hat{t},e^\mu_1\sim\hat{x},e^\mu_2\sim\hat{y}$, and $e^\mu_3\sim\hat{z}$, and $\mu$ labels the four components of each vector with respect to the coordinates on the curved manifold. The object $e^\mu_I(x)$ is called a {\em tetrad field}. This object represents a field of arbitrarily chosen orthonormal basis vectors for the tangent space for each point in the spacetime manifold $\m M$. This orthonormality is defined in spacetime by
\begin{eqnarray*}
g_{\mu\nu}(x)e^\mu_I(x)e^\nu_J(x)=\eta_{IJ}
\end{eqnarray*}
where $g_{\mu\nu}$ is the spacetime metric tensor and $\eta_{IJ}$ is the local flat Minkowski metric. Furthermore, orthogonality implies that the determinant $e=\det( e^\mu_I)$ of the tetrad as a matrix in $(\mu,I)$ must be non-zero. Thus there exists a unique inverse to the tetrad, denoted by $e^I_\mu$, such that $e^I_\mu e^\mu_J=\eta^I_J = \delta^I_J$ or $e^I_\mu e^\nu_I=g^{\nu}_{\mu}=\delta^{\nu}_{\mu}$. Making use of the inverse $e^I_\mu$ we obtain
\begin{eqnarray*}
g_{\mu\nu}(x)=e^I_\mu(x) e^J_\nu(x)\eta_{IJ}.
\end{eqnarray*}
Therefore, if we are given the inverse reference frame $e^I_\mu(x)$ for all spacetime points $x$ we can reconstruct the metric $g_{\mu \nu}(x)$. The tetrad $e^\mu_I(x)$ can therefore be regarded as a mathematical representation of the geometry.

As stressed above, on a curved manifold the choice of reference frame at any specific point $x$ is completely arbitrary. Consider then {\it local}, i.e.\;spacetime-dependent, transformations of the tetrad $e^\mu_I(x)\rightarrow e'^{\mu}_I(x)=\Lambda_I^{\ J}(x)e^\mu_J(x)$ that preserve orthonormality;
\begin{equation}
\eta_{IJ}=g_{\mu \nu}(x)e'^{\mu}_I(x) e'^{\nu}_J(x)\\=g_{\mu \nu}(x)\Lambda_I^{\ K}(x)e^\mu_K(x)\Lambda_J^{\ L}(x)e^\nu_L(x)=\eta_{KL}\Lambda_I^{\ K}(x)\Lambda_J^{\ L}(x)\label{etaLorentzinvariance}.
\end{equation}
The transformations $\Lambda_I^{\ J}(x)$ are recognised as local Lorentz transformations, leaving $\eta_{IJ}$ invariant. Given that the matrices $\Lambda_I^{\ J}(x)$ are allowed to depend on $x^\mu$, so that different transformations can be performed at different points on the manifold,  the reference frames associated with different points are therefore allowed to be changed in an uncorrelated manner. However for continuity reasons we will restrict $\Lambda_I^{\ J}(x)$ to local {\em proper} Lorentz transformations, i.e.\;members of $SO^{+}(1,3)$.

The inverse tetrad $e^I_\mu$ transforms as $e^I_\mu\rightarrow e'^I_\mu=\Lambda^I_{\ J}e^J_\mu$ where $\Lambda^I_{\ K}\Lambda_J^{\ K}=\delta^I_J$. We now see that the gravitational field $g_{\mu\nu}$ is invariant under these transformations:
\begin{equation}
g'_{\mu\nu}=\eta_{IJ}e'^I_\mu e'^J_\nu=\eta_{IJ}\Lambda^I_{\ K} e^K_\mu \Lambda^J_{\ L}e^L_\nu=\eta_{IJ}\Lambda^I_{\ K}\Lambda^J_{\ L}e^K_\mu e^l_\nu=\eta_{KL} e^K_\mu e^L_\nu=g_{\mu\nu}.\label{gLorentzinvariance}
\end{equation}

Therefore, all tetrads related by a local Lorentz transformation $\Lambda^I_{\ J}(x)$ represent the same geometry $g_{\mu\nu}$. Thus, by switching from a metric representation to a tetrad representation we have made manifest {\em local Lorentz invariance}.

As stated earlier it will be useful to formulate qubits in curved spacetime in a reference frame covariant manner. To do so we need to be able to represent spacetime vectors with respect to the tetrads and not the coordinates. A spacetime vector $V$ expressed in terms of the coordinates will carry the coordinate index $V^{\mu}$. However, the vector could likewise be expressed in terms of the tetrad basis, in this case  $V^{\mu} = V^{I}e^{\mu}_{I}$ where $V^{I}$ are the components of the vector in the tetrad basis given by $V^I=e^I_\mu V^\mu$. We  can therefore work with tensors represented either in the coordinate basis labelled by Greek indices $\mu, \nu, \rho,\text{ etc}$ or in the tetrad basis where tensors are labelled with capital Roman indices $I, J, K,\text{ etc}$. The indices are raised or lowered either with $g^{\mu\nu}$ or with $\eta^{IJ}$ depending on the basis.\footnote{See Notation and conventions,  \secref{notation}.} We will switch between tetrad and coordinate indices freely throughout this chapter.

%%%%%%%%%%%%%%%%%%%%%%%%%%%%%%%%%%%%%%%%%%%%%%%%%%%%%%%%%%%%%%%%
\subsection{The connection 1-form\label{sec-connectiononeform}}
%%%%%%%%%%%%%%%%%%%%%%%%%%%%%%%%%%%%%%%%%%%%%%%%%%%%%%%%%%%%%%%%

In order to define a covariant derivative and parallel transport one needs a connection. When this connection is expressed in the coordinate basis, which is in general neither normalised nor orthogonal, this is referred to as the affine connection $\Gamma^\rho_{\mu\nu}$, given in \eqref{eq-affine}. Alternatively if the connection is expressed in terms of the orthonormal tetrad basis it is called the {\em connection one-form} $\omega_{\mu\ J}^{\ I}$. To see this, consider the parallel transport of a vector $V^\mu$ along some path $x^\mu(\lambda)$ given by the equation
\be
\frac{DV^\mu}{D\lambda}\equiv\frac{\di V^\mu}{\di \lambda}+\frac{\di x^\nu}{\di\lambda} \Gamma^\mu_{\nu\rho}V^\rho\equiv0.\label{eq-coordVPT}
\ee
where $\lambda$ is some arbitrary parameter. The vector $V^\mu$ in the tetrad basis is expressed as $V^\mu=V^I e^\mu_I$. We can now re-express the parallel transport equation in terms of the tetrad components $V^I$:
\begin{align*}
\frac{D(e^\mu_IV^I)}{D\lambda}\equiv & \frac{\di(e^\mu_IV^I)}{\di\lambda}+ \frac{\di x^\nu}{\di\lambda}\Gamma^\mu_{\nu\rho}e^\rho_IV^I\\
=&e^\mu_I\left(\frac{\di V^I}{\di\lambda}+\frac{\di x^\nu}{\di\lambda}\left[e^I_\rho\partial_\nu e^\rho_J+\Gamma^\sigma_{\nu\rho}e^I_\sigma e^\rho_J\right]V^J\right).
\end{align*}
Thus, if we define
\begin{eqnarray*}
\omega_{\nu\ J}^{\ I} \equiv e^I_\rho\partial_\nu e^\rho_J+\Gamma^\sigma_{\nu\rho}e^I_\sigma e^\rho_J,
\end{eqnarray*}
the equation for the parallel transport of the tetrad components $V^I$ can be written as
\be
\frac{DV^I}{D\lambda}\equiv\frac{\di V^I}{\di\lambda}+ \frac{\di x^\nu}{\di\lambda}\omega_{\nu\ J}^{\ I}V^J=0.\label{eq-vecPT}
\ee
The object $\omega_{\nu\ J}^{\ I}$ is called the {\em connection 1-form} or {\em spin-$1$ connection} and is merely the affine connection $\Gamma^\mu_{\nu\rho}$ expressed in a local orthonormal frame $e^\mu_I(x)$. It is also called a {\em Lie-algebra -valued 1-form} since, when viewed as a matrix $(\omega_\nu)^{I}_{\ J}$, it is a 1-form in $\nu$ of elements of the Lie algebra $\mathfrak{so}(1,3)$. The connection 1-form encodes the spacetime curvature  but unlike the affine connection it transforms in a covariant way (as a covariant vector, or in a different language, as a 1-form) under coordinate transformations, due to it having a single coordinate index $\nu$. However, as can readily be checked from the definition, it transforms {\em inhomogeneously} under a change of tetrad $e^I_\mu(x)\rightarrow \Lambda^I_{\ J}(x)e^J_\mu(x)$:
\begin{eqnarray}
\omega_{\mu\ J}^{\ I} \rightarrow\omega_{\mu\ J}^{\prime\ I}=\Lambda^I_{\ K}\Lambda_J^{\ L}\omega_{\mu\ L}^{\ K}+\Lambda^I_{\ K}\partial_\mu \Lambda_J^{\ K}.\label{connectiontransform}
\end{eqnarray}
The inhomogeneous term $\Lambda^I_{\ K}\partial_\mu \Lambda_J^{\ K}$ is present only when the rotations depend on the position coordinate $x^\mu$ and ensures that the parallel transport $\frac{DV^I}{D\lambda}$ transforms properly as a contravariant vector under local Lorentz transformations.

\subsection{Vector transport\label{sec-statetransformation}}

The parallel transport equations \eqref{eq-coordVPT} and \eqref{eq-vecPT} with $\lambda\to\tau$, and $V^\mu$ or $V^I$ spacelike, govern the precession of a gyroscope spin $V$ moving along a geodesic timelike trajectory $x^\mu(\tau)$ with velocity $u^\mu=\di x^\mu/\di \tau$ \cite{Hartlegravity}. Alternatively, with $u_\mu(\lambda)$ light-like, so that $x^\mu(\lambda)$ is a null trajectory, these equations describe the parallel transport of a vector along the trajectory that a ray of light could take.

There is a third transport equation worth introducing: the Fermi--Walker derivative \cite{WeinbergGC,MTW}, which governs the torque-free transport of a vector $V^I$ along an {\it accelerated} timelike trajectory, with proper acceleration $a^J:=\frac{D u^J}{D\tau}$;
\be
\frac{D^{FW}V^I}{D\tau}\equiv \frac{DV^I}{D\tau}+(u^I a_J-a^I u_J)V^J=0\label{eq-VFW}.
\ee
If proper acceleration is zero, the parallel transport equation is recovered.

%\subsection{Calculating the state transformation\label{sec-statetransformation}}

These transport equations describe the infinitesimal transformation $R^I_{\;J}\di\lambda$ or $R^I_{\;J}\di\tau$ to vectors transported along a trajectory, where $R^I_{\;J}$ is an element of the Lie algebra. If we want to determine the final state for $V^I$ along the trajectory $x^\mu(\lambda)$, we would want to solve the transport derivative defined by \eqref{eq-vecPT} or \eqref{eq-VFW}, i.e.\;solve the differential equation $\di V^I/\di\lambda=R^I_{\;J}(\lambda)V^J$ to produce a relation $V^I(\lambda_\text{end})=\mathrm T^I_{\;J}V^J(\lambda_\text{start})$. This transformation operator $\mathrm T^I_{\;J}$ is given by
\[
\mathrm T^I_{\;J}=\m P\exp\left[\int_{\lambda_\text{start}}^{\lambda_\text{end}}R^I_{\;J}(\lambda)\di\lambda \right],
\]
the path-ordered exponential of compositions of the infinitesimal operators, where $\exp$ is the matrix exponential and $\m P$ is the path-ordering operator \cite{Weinberg,PeskinSchroeder}. For example, for the Fermi--Walker transport this is
\[
\mathrm T^I_{\;J}=
\m T\exp\left[\int_{\tau_\text{start}}^{\tau_\text{end}}
\left(- u^\nu\omega_{\nu\ J}^{\ I}-(u^I a_J-a^Iu_J)\right)\di\tau \right]\label{eq-vecstatetransformation}
\]
where the time-ordering operator $\m T$ is used for timelike or lightlike trajectories \cite{Weinberg,PeskinSchroeder}. These ordering operators are needed in the general case because the operators corresponding to the transformation at $\lambda_1$ may not necessarily commute with those at $\lambda_2$ in the series expansion of the exponential. The transformation operator obtained depends on the choice of tetrads at the ends of the trajectory, so it is important to be clear regarding the orientation of preparation and measurement apparatuses at these locations in order to obtain unambiguous results.

%%%%%%%%%%%%%%%%%%%%%%%%%%%%%%%%%%%%%%%%%%%%%%%%%%%%%%%%%%
\section{The qubit as the spin of a massive fermion\label{secFermion}}
%%%%%%%%%%%%%%%%%%%%%%%%%%%%%%%%%%%%%%%%%%%%%%%%%%%%%%%%%%

A specific physical realisation of a qubit is the spin of a massive fermion such as an electron. An electron can be thought of as a spin-$\half$ gyroscope, where a rotation of $2\pi$ around some axis produces the original state but with a minus sign. Such an object is usually taken to be represented by a four-component Dirac field, which constitutes a reducible spin-$\half$ representation of the Lorentz group. However, given that we are after a qubit and therefore a two-dimensional object, we will work with a two-component Weyl spinor field $\phi_A(x)$, with $A=1,2$, which is the left-handed component of the Dirac field (see \appref{secspinornotation}).\footnote{We could work instead with the right-handed component, but this would yield the same results.} We shall see that this is sufficient for describing a massive fermion \cite{DHM2010}. The Weyl spinor itself constitutes a finite-dimensional faithful -- and therefore {\it non-unitary} -- representation of the Lorentz group \cite{KimNoz} and one may therefore think that it could not mathematically represent a quantum state. As we shall see, unitarity is recovered by correctly identifying a suitable inner product.

We will begin by considering the Dirac equation in curved spacetime minimally coupled to an electromagnetic field. We rewrite this Dirac equation in second-order form (called the van der Waerden equation) where the basic field is now a left-handed Weyl spinor $\phi_A$. This equation is then studied in the WKB limit which separates the spin from the spatial degrees of freedom. We then localise this field along a classical trajectory to arrive at a transport equation for the spin of the fermion which forms the physical realisation of the qubit. We find that this transport equation corresponds to the Fermi--Walker transport of the spin along a non-geodesic trajectory plus an additional  precession of the fermion's spin due to the presence of local magnetic fields. We will see that from the WKB approximation a natural inner product for the two-dimensional vector space of Weyl spinors emerges. Furthermore, we will see in \secref{secfermionQS} that in the rest frame of the qubit the standard notion of unitarity is regained. It is also in this frame where the transport equation is identical to the result obtained in \cite{TerashimaUeda03}.

%%%%%%%%%%%%%%%%%%%%%%%%%%%%%%%%%%%%%%%%%%%%%%%%%%%%
\subsection{The WKB approximation}\label{fermsemiclassapprox}
%%%%%%%%%%%%%%%%%%%%%%%%%%%%%%%%%%%%%%%%%%%%%%%%%%%%

Before we begin our analysis of the Dirac equation in the WKB limit we refer the reader to  \appref{secspinornotation} for notation and background material on spinors. This material is necessary for the relativistic treatment of massive fermions. Our approach is based on a version of the WKB approximations in \cite{Howison05}.

%%%%%%%%%%%%%%%%%%%%%%%%%%%%%%%%%%%%%%%%%%%%%%%%%
\subsubsection{The minimally coupled Dirac field in curved spacetime}
%%%%%%%%%%%%%%%%%%%%%%%%%%%%%%%%%%%%%%%%%%%%%%%%%
Fermions in flat spacetime are governed by the Dirac equation $\ii\gamma^\mu\partial_\mu\Psi=m\Psi$. Since we are dealing with curved spacetimes we must generalise the Dirac equation to include these situations. This is done as usual through minimal coupling by replacing the partial derivatives by covariant derivatives. The covariant derivative of a Dirac spinor is defined by \cite{Nakahara}
\begin{eqnarray}
\nabla_\mu\Psi=(\partial_\mu-\frac{\ii}{2}\omega_{\mu IJ}S^{IJ})\Psi\label{spinhalfcovarderiv}
\end{eqnarray}
where $S^{IJ}=\frac \ii4[\gamma^I,\gamma^J]$ are the spin-$\half$ generators of the Lorentz group and $\gamma^{I}$ are the Dirac $\gamma$-matrices which come with a tetrad rather than a tensor index. The gravitational field enters through the spin-1 connection $\omega_{\mu IJ}$. We assume that the fermion is electrically charged and include an electromagnetic field $F_{IJ}$  by minimal coupling so that we can consider accelerated trajectories. The Dirac equation in curved spacetime minimally coupled to an external electromagnetic potential field $A_\mu$ is then given by
\begin{eqnarray}
\ii\gamma^\mu D_\mu\Psi=m\Psi\label{curveddiraceq}
\end{eqnarray}
where we define the $U(1)$ covariant derivative as $D_\mu=\nabla_\mu-\ii eA_\mu$.

%%%%%%%%%%%%%%%%%%%%%%%%%%%%%%%%%%%%%%%%%%%%%%%%%%%%%%%%%
\subsubsection{The van der Waerden equation: an equivalent second order formulation}
%%%%%%%%%%%%%%%%%%%%%%%%%%%%%%%%%%%%%%%%%%%%%%%%%%%%%%%%%
In order to proceed with the WKB approximation it is convenient to put the Dirac equation into a second-order form. This can be done by making use of the Weyl representation of the $\gamma$-matrices (see \ref{secspinornotation} for further details). In this representation the $\gamma$-matrices take on the form
\begin{eqnarray*}
\gamma^I=\begin{pmatrix}0&\sigma^I_{\ AA'}\\ \bar{\sigma}^{IA'A}&0\end{pmatrix}.
\end{eqnarray*}
The Dirac equation then splits into two separate equations
\begin{subequations}
\begin{eqnarray}
\ii\bar{\sigma}^{\mu A'A}D_\mu\phi_A&=m\chi^{A'}\label{left}\\
\ii\sigma^{\mu}_{\ AA'}D_\mu\chi^{A'}&=m\phi_A\label{right}
\end{eqnarray}
\end{subequations}
with $\bar{\sigma}^{\mu A'A}\equiv e^{\mu}_I\bar{\sigma}^{IA'A}$ and $\sigma^{\mu}_{\ A'A}\equiv e^{\mu}_I\bar{\sigma}^{I}_{\ A'A}$, and $\Psi=(\phi_A,\chi^{A'})$, where $\phi_A$ and $\chi^{A'}$ are left- and right- handed 2-spinors respectively. Solving for $\chi^{A'}$ in \Eeqref{left} and inserting the result into \eqref{right} yields a second-order equation called the {\it van der Waerden equation} \cite{SakuraiAQM}
\be
\sigma^\mu_{\ AA'}\bar{\sigma}^{\nu A'B}D_\mu D_\nu\phi_B+m^2\phi_A=0
\ee
which, with 2-spinors, provides a physically equivalent formulation of massive fermions to the Dirac equation and 4-spinor formalism \cite{DHM2010}. We can rewrite this equation in the following way
\begin{eqnarray}
0&=&\sigma^\mu_{\ AA'}\bar{\sigma}^{\nu A'B} D_\mu D_\nu\phi_B+m^2\phi_A\notag\\
&=&\sigma^\mu_{\ AA'}\bar{\sigma}^{\nu A'B}\left( D_{\{\mu} D_{\nu\}}+ D_{[\mu} D_{\nu]}\right)\phi_B+m^2\phi_A\notag\\
&=&g^{\mu\nu} D_\mu D_\nu\phi_A-\ii L^{\mu\nu\ B}_{\ \ A}(\mathfrak{R}_{\mu\nu B}^{\ \ \ \ C}-\ii e\delta_B^{\ C}F_{\mu\nu})\phi_C+m^2\phi_A\label{vdWexpanded}
\end{eqnarray}
where we have used that $2 D_{[\mu} D_{\nu]} = [D_\mu, D_\nu]$ and $2 D_{\{\mu} D_{\nu\}} = \{D_\mu, D_\nu\}$, and $\sigma^{\{ \mu}\bar\sigma^{\nu\} } = g^{\mu\nu}$. We identify $F_{\mu\nu}\equiv2\nabla_{[\mu} A_{\nu]}$ as the electromagnetic tensor and $\mathfrak{R}_{\mu\nu A}^{\ \ \ \ B}\phi_{B}:=2\nabla_{[\mu}\nabla_{\nu]}\phi_{A}$ as a spin-$\half$ curvature 2-form associated with the left-handed spin-$\half$ connection $\frac{\ii}{2}\omega_{\mu IJ}L^{IJ\ B}_{\ \ A}$, where $\hat{L}^{\mu\nu} = e^{\mu}_I e^{\nu}_J\hat{L}^{IJ} = \frac \ii2\sigma^{[\mu},\bar{\sigma}^{\nu]}$ are the left-handed spin-$\half$ generators related to the Dirac four-component representation by  $\hat{S}^{\mu\nu}=\hat{L}^{\mu\nu}\oplus \hat{R}^{\mu\nu}$. We have tacitly assumed here that the connection is torsion-free. Torsion can be included (at least in the case of vanishing electromagnetic field) and will slightly modify the way the spin of the qubit changes when transported along a trajectory. We refer the reader to \cite{Audretsch,Anandan94,Bergmann} for further details on torsion.

%%%%%%%%%%%%%%%%%%%%%%%%%%%%%%%
\subsubsection{The basic ansatz}
%%%%%%%%%%%%%%%%%%%%%%%%%%%%%%%
The starting point of the WKB approximation is to write the left-handed two-spinor field $\phi_A$ as
\begin{eqnarray*}
 \phi_A(x)=\varphi_A(x) \Ee^{\ii\theta(x)/\epsilon}
\end{eqnarray*}
with $\theta(x)$ real, and $\varphi_A(x)$ complex spinor-valued. We then study the van der Waerden equation in the limit $\epsilon\rightarrow0$, where $\epsilon$ is a convenient expansion parameter. Physically this means that we are studying solutions for which the phase is varying much faster than the complex spinor-valued amplitude $\varphi_A$. In the high frequency limit $\epsilon\rightarrow0$ the fermion will not `feel' the presence of a finite electromagnetic field. We are therefore going to assume that as the frequency increases the strength of the electromagnetic field also increases. We thus assume that the electromagnetic potential is given by $\frac1\epsilon A_\mu$. $\epsilon$ is to be thought of as a `dummy' parameter whose only role is to identify the different orders in an expansion. Once the different orders have been identified the value of $\epsilon$ in any equation can be set to 1.

%%%%%%%%%%%%%%%%%%%%%%%%%%%%%%%%%%%%%%%%%%%%%%%%%%%%%%%%%%%%%%%%%%%%%%%%
\subsubsection{The van der Waerden equation in the WKB limit}
%%%%%%%%%%%%%%%%%%%%%%%%%%%%%%%%%%%%%%%%%%%%%%%%%%%%%%%%%%%%%%%%%%%%%%%%
Rewriting the van der Waerden equation in terms of the new variables $\varphi_A$ and $\theta$, and collecting terms of similar order in $\frac{1}{\epsilon}$, yields
\begin{equation}
g^{\mu\nu}\nabla_\mu \nabla_\nu\varphi_A-\ii L^{\mu\nu\ B}_{\ \ A}\mathfrak{R}_{\mu\nu B}^{\ \ \ \ C}\varphi_C+\frac{\ii}{\epsilon}(2k^\mu\nabla_\mu\varphi_A+\varphi_A\nabla_\mu k^\mu+\ii eF_{\mu\nu}L^{\mu\nu\ B}_{\ \ A}\varphi_B)
-\frac{1}{\epsilon^2}k_\mu k^\mu\varphi_A+m^2\varphi_A=0\label{vdWWKB}
\end{equation}
where we define the momentum/wavevector as the gauge invariant quantity $k_\mu=\nabla_\mu\theta-eA_\mu$.

If we assume that both the typical scale $\ell$ over which $\varphi_A$ varies and the curvature scale $\m R$ are large compared to the scale $\lambdabar$ over which the phase varies (which is parameterised by $\epsilon$), the first two terms of \eqref{vdWWKB} can be neglected. In the WKB limit the mass term represents a large number and is therefore treated as a ${1}/{\epsilon^2}$ term. The remaining equations are then
\begin{subequations}
\begin{align}
&2k^\mu\nabla_\mu\varphi_A+\varphi_A\nabla_\mu k^\mu+\ii eF_{\mu\nu}L^{\mu\nu\ B}_{\ \ A}\varphi_B =0\label{transncons}\\
&k^\mu k_\mu-m^2=0\label{disprel}.
\end{align}
\end{subequations}
%
%%%%%%%%%%%%%%%%%%%%%%%%%%%%%%%%%%%%%%%%%%%%%%%%%%%%%%%%%%%
\subsubsection{Derivation of the spin transport equation and conserved current \label{spinortransportderiv}}
%%%%%%%%%%%%%%%%%%%%%%%%%%%%%%%%%%%%%%%%%%%%%%%%%%%%%%%%%%%
The dispersion relation \eqref{disprel} implies that $k$ is timelike. Furthermore, by taking the covariant derivative of the dispersion relation and assuming vanishing torsion
\begin{eqnarray*}
\nabla_\nu(k^\mu k_\mu-m^2)&=&2k^\mu\nabla_\nu k_\mu=2k^\mu\nabla_\nu(\nabla_\mu\theta-eA_\mu)\\
&=&2(k^\mu\nabla_\mu k_\nu+ek^\mu F_{\mu\nu})=0
\end{eqnarray*}
we readily see that the integral curves of $u^\mu(x)\equiv k^\mu(x)/m$, defined by $\frac{\di x^\mu}{\di\tau}=u^\mu$, satisfy the classical Lorentz force law
\begin{eqnarray}
m\frac{D^2x^\mu}{D\tau^2}+e\frac{\di x^\nu}{\di\tau} F_\nu^{\ \mu}=0\label{FWvelocity}.
\end{eqnarray}
where $a^\mu\equiv\frac{D^2x^\mu}{D\tau^2}=\frac{\di x^\nu}{\di\tau}\nabla_\nu u^\mu$ and $u^\mu u_\mu=1$. Thus, the integral curves of $k^\mu$ are classical particle trajectories.

To see the implications of the first equation \eqref{transncons} we contract it with $k_\mu\bar{\sigma}^{\mu A'A}\bar{\varphi}_{A'}$, where $\bar\varphi_{A'}=\overline{\varphi_A}$ is the complex conjugate dual spinor (see \appref{ch-app3}), and add the result to its conjugate. Simplifying this sum with the use of \eqref{FWvelocity} and the identity (\cite[Eqn (2.85) p19]{DHM2010})

\begin{eqnarray*}
\bar{\sigma}^{KA'A}L^{IJ\ B}_{\ \ A}=\frac{\ii}{2}(\eta^{KI}\bar{\sigma}^{JA'B}-\eta^{KJ}\bar{\sigma}^{IA'B}-\ii\epsilon^{KIJ}_{\ \ \ \ L}\bar{\sigma}^{LA'B})
\end{eqnarray*}
yields
\begin{eqnarray}
\nabla_\mu (\varphi^2)k^\mu+\varphi^2\nabla_\mu k^\mu=0\label{contracted}
\end{eqnarray}
where $\varphi^2\equiv u_\mu\bar{\sigma}^{\mu A'A}\bar{\varphi}_{A'}\varphi_A$. \Eeqref{contracted}  can also be rewritten as
\begin{eqnarray}
\nabla_\mu(\varphi^2k^\mu)=0\label{diraccurrent}
\end{eqnarray}
which tells us that we have a conserved energy density $j^\mu\equiv\sqrt{-g}\varphi^2k^\mu$, \footnote{$\varphi^2$ has dimension $L^{-3}$.} with $g=\det g_{\mu\nu}$.

Secondly, \eqref{contracted} yields $\nabla_\mu k^\mu=-(2k^\mu\nabla_\mu\varphi)/\varphi$ and when this is inserted back into \eqref{transncons} we obtain
\begin{eqnarray*}
2k^\mu\nabla_\mu\psi_A+\ii eF_{\mu\nu}L^{\mu\nu\ B}_{\ \ A}\psi_B =0.
\end{eqnarray*}
By making use of the integral curves $x^\mu(\tau)$ we obtain the ordinary differential equation
\begin{eqnarray}
\frac{D\psi_A}{D\tau}+\ii\frac e{2m}F_{IJ}L^{IJ\ B}_{\ \ A}\psi_B =0\label{fermionundecipheredtransport}
\end{eqnarray}
where $\frac{D\psi_A}{D\tau}=\frac{\di\psi_A}{\di\tau}-\frac \ii2u^\mu\omega_{\mu IJ}L^{IJ\ B}_{\ \ A}\psi_B$ is the spin-$\half$ parallel transport. \Eeqref{fermionundecipheredtransport} governs the evolution of the normalised spinor $\psi_A\equiv\varphi_A/\varphi$ along integral curves. Below $\psi_A$ will assume the role of the qubit quantum state.

%%%%%%%%%%%%%%%%%%%%%%%%%%%%%%%%%%%%%%%%%%%%%%%%%%%%%%%%%%%%%%
\subsection{Qubits, localisation and transport}\label{fermionlocalization}
%%%%%%%%%%%%%%%%%%%%%%%%%%%%%%%%%%%%%%%%%%%%%%%%%%%%%%%%%%%%%%
The aim of this chapter is to obtain a formalism for localised qubits. However, the WKB approximation does not guarantee that the fermion is spatially localised, i.e.\;the envelope $\varphi(x)$ need not have compact support in a small region of space. In addition, even if the envelope initially is well-localised there is nothing preventing it from distorting and spreading, and becoming delocalised. We therefore need to make additional assumptions beyond the WKB approximation to guarantee the initial and continued localisation of the qubit. As pointed out in \S\ref{sec-particlenumberambiguity}, by restricting ourselves to localised envelopes we avoid the particle number ambiguity and can interpret the Dirac field as a one-particle quantum wavefunction.

%%%%%%%%%%%%%%%%%%%%%%%%%%%%%%%%%%%%%%
\subsubsection{Localization\label{fermionlocalizationsubsec}}
%%%%%%%%%%%%%%%%%%%%%%%%%%%%%%%%%%%%%%
Before we begin let us be a bit more precise as to what it means for a qubit to be `localised'. In order to avoid the particle number ambiguity we know that the wavepacket size $\m L$ has to be much less than the curvature scale $\m R$. We also know from quantum field theory that it is not possible to localise a massive fermion to within its Compton wavelength $\lambda_\text{com}\equiv h/mc$ using only positive frequency modes (see \S\ref{sec-localisation}). Mathematically we should then have $\lambda_\text{com}<\m L \ll \m R$ where $\m L$ is the packet length in the rest frame of the fermion. If $\lambda_\text{com}\sim \m R$ the formalism of this chapter will not be empirically correct.

How well-localised a wavepacket is, is determined by the support of the envelope. Strictly speaking we know from quantum field theory that a localised state will always have exponential tails which cannot be made to vanish using only positive frequency modes. However, the effects of such tails are small and for the purpose of this chapter we will neglect them and assume that the wavepacket has compact support.

The equation that governs the evolution of the envelope within the WKB approximation is the continuity equation \eqref{diraccurrent}:
\begin{eqnarray*}
\nabla_\mu (u^\mu \varphi^2(x))=0,
\end{eqnarray*}
with $u^\mu=k^\mu/m$. If we assume that the divergence of the velocity field $u^\mu$ is zero, i.e.\;$\nabla_\mu u^\mu=0$, the continuity equation reduces to
\begin{eqnarray*}
\nabla_\mu (u^\mu \varphi^2(x))=u^\mu\nabla_\mu \varphi^2+\varphi^2\nabla_\mu u^\mu=u^\mu\nabla_\mu \varphi^2=0,
\end{eqnarray*}
or, using the integral curves of $u^\mu$,
\begin{eqnarray*}
\frac{\di \varphi^2}{\di\tau}=0.
\end{eqnarray*}
Thus, the shape of the envelope in the qubit's rest frame remains unchanged during the evolution. However, because of the uncertainty principle \cite{PeresQM}, if the wavepacket has finite spatial extent it cannot simultaneously have a sharp momentum, and therefore the divergence in velocity cannot be exactly zero. We can then relax the assumption, since the only thing that we need to guarantee is that the final wavepacket is not {\em significantly} distorted compared to the original one. Since $\nabla_\mu u^\mu$ measures the rate of change of the rest-frame volume $\frac1V\frac{\di V}{\di\tau}$ \cite{MTW}  we should require that
\begin{eqnarray*}
\langle\nabla_\mu u^\mu\rangle \ll\frac{1}{\tau_{_\Gamma}}
\end{eqnarray*}
where $\ex{\nabla_\mu u^\mu}$ is the weighted average value of $|\nabla_\mu u^\mu|$ over the envelope, and $\tau_{_\Gamma}$ the proper time along some path $\Gamma$  assumed to have finite length. If we combine this assumption of negligible divergence with the assumption that the envelope is initially localised so that the wavepacket size is smaller than the curvature scale, we can approximately regard the envelope as being rigidly transported while neither distorting nor spreading during its evolution.

To further suppress the spatial degrees of freedom we need also an assumption about the two-component spinor $\psi_A(x)$. This variable could vary significantly within the localised support of the envelope $\varphi(x)$. However, as we want to attach a single qubit quantum state to each point along a trajectory we need to assume that $\psi_A(x)$ only varies along the trajectory and not spatially. More precisely, we assume that $\psi_A(t,\vec x)=\psi_A(t)$ when we use local Lorentz coordinates $(t,\vec x)$ adapted to the rest frame of the particle. This implies that the wavepacket takes on the form
\begin{eqnarray*}
\phi_A(t,\vec{x})=\psi_A(t) \varphi(t,\vec x)\Ee^{\ii\theta(t,\vec x)}.
\end{eqnarray*}
This form is not preserved for all reference frames since in other local Lorentz coordinates $\psi_A$ will have spatial dependence. Nevertheless, if the packet is sufficiently localised and $\psi_A$ varies slowly the wave-packet will approximately be separable in spin and position for most choices of local Lorentz coordinates. With these additional assumptions we have effectively `frozen out' the spatial degrees of freedom of the wavepacket. The spinor $\psi_A$ can now be thought of not as a function of spacetime $\psi_A(x)$ satisfying a partial differential equation, but rather as a spin state $\psi_A(\tau)$ defined on a classical trajectory $\Gamma$ satisfying an ordinary differential equation \eqref{fermionundecipheredtransport}. We can therefore effectively characterise the fermion for each $\tau$ by a position $x^\mu(\tau)$, a 4-velocity $\di x^\mu/\di\tau=u^\mu(\tau)$, and a spin $\psi_A(\tau)$. Once we have identified the spin as a quantum state this will provide the realisation of a localised qubit.

%%%%%%%%%%%%%%%%%%%%%%%%%%%%%%%%%%%%%%%%%%%%%%%%%%%%%%%%%%%%%%%%%
\subsubsection{The physical interpretation of WKB equations\label{sec-spinorphysinterp}}
%%%%%%%%%%%%%%%%%%%%%%%%%%%%%%%%%%%%%%%%%%%%%%%%%%%%%%%%%%%%%%%%%
As discussed in \secref{sec-particlenumberambiguity}, if we restrict ourselves to localised wavepackets we can interpret $\phi_A(x)$ as one-particle excitations of the quantum field. This allows us to interpret the conserved current $j^\mu/m=\sqrt{g}\varphi^2u^\mu$ as the probability current of a single particle. In this way we can provide a physical interpretation of the classical two-component spinor field  $\phi_A(x)$ as a quantum wavefunction of a single particle.

Next, let us examine the transport equation \eqref{fermionundecipheredtransport}. The electromagnetic tensor $F_{IJ}$ that appears in the term $\ii eF_{IJ}\hat L^{IJ}/m$ can be decomposed into a component parallel to the timelike 4-velocity $u^I$ and a spacelike component perpendicular to $u^I$ using a covariant spatial projector $h^I_J =\delta^I_J-u^{I} u_{J}$.
 We can then rewrite $F_{IJ}L^{IJ\ B}_{\ \ A}$ as $(2u_Iu^KF_{KJ}+h_I^{\ K}h_J^{\ L}F_{KL})L^{IJ\ B}_{\ \ A}$.
The first term corresponds to the electric field as defined in the rest frame, $u_Iu^KF_{KJ}$. This will produce an acceleration $u^\mu\nabla_\mu u_I=a_I=-\frac emu^JF_{JI}$ of the fermion as described by the Lorentz force equation \eqref{FWvelocity}.  The second term is recognised as the magnetic field experienced by the particle, i.e.\; the magnetic field as defined in the rest frame of the particle, $B^\text{rest}_{IJ}$.
We thus obtain the transport equation for $\psi_A$;
\begin{eqnarray}
\frac{D\psi_A}{D\tau}-\ii u_Ia_JL^{IJ\ B}_{\ \ A}\psi_B+   \ii\frac e{2m} B^\text{rest}_{IJ}L^{IJ\ B}_{\ \ A}\psi_B=0\label{fermionTP}.
\end{eqnarray}
This has a simple physical interpretation. The third term represents the magnetic precession which is induced by the torque that the magnetic field exerts on the spin. This takes the usual form $\frac \ii2\frac emF_{ij}L^{ij}=\frac \ii2\frac emB_{ij}\half\eps^{ij}_{\ \ k}\sigma^k=-\frac{\ii}{2}\frac em\mathbf B\cdot\mathbf\sigma$ if we express it in a tetrad co-moving with the particle, i.e.\;$e_0^\mu=u^\mu$.

The two first terms represent the spin-half version of the Fermi--Walker derivative, \eqref{eq-VFW}:
\begin{eqnarray}
\frac{D^{FW}\psi_A}{D\tau}\equiv\frac{D\psi_A}{D\tau}-\ii u_Ia_JL^{IJ\ B}_{\ \ A}\psi_B\label{spinhalfFW}.
\end{eqnarray}
The presence of a Fermi--Walker derivative can be understood directly from physical considerations. Heuristically we understand the electron as a spin-$\half$ object, i.e.\;loosely as a quantum gyroscope. The transport of the orientation of an ordinary classical gyro is not governed by the parallel transport equation but rather, it is governed by a Fermi--Walker transport equation. The Fermi--Walker equation arises when we want to move a gyroscope along some spacetime path without applying any external torque \cite{MTW}.\footnote{At first one might think that this is just what the parallel transport equation achieves. However, this is only true for geodesic motion  ($a_I=0$), where the Fermi--Walker and parallel transport equations agree.} We thus identify \eqref{spinhalfFW} as describing torque-free transport of the electron, resulting in the usual Thomas precession of the spin \cite{WeinbergGC}. Finally, the parallel transport term $D\psi_A/D\tau$ encodes the influence of gravity on the qubit, governed by the spin-1 connection $\omega_{\mu\ J}^{\ I}$.

\subsubsection{A summary of the WKB limit}
Let us summarise the results from the previous section.
\begin{itemize}
\item The full wavepacket is written as $\phi_{A}(x) = \psi_{A}(x)\varphi(x)\Ee^{\ii\theta(x)}$.
\item The phase $\theta$ and the vector potential $A_\mu$ define a field of 4-velocities $u_\mu=\frac1m(\nabla_{\mu}\theta-eA_\mu)$.
\item The current $j^\mu/m=\sqrt{g}\varphi^2u^ \mu$ is a conserved  probability density.
\item The integral curves of $u^\mu$ are timelike and satisfy the classical Lorentz equation $ma_\mu=eu^\nu F_{\mu\nu}$.
\item The two-component spinor $\psi_A(\tau)$ defined along some integral curve of $u^\mu$ satisfies the transport equation
\begin{eqnarray}
\frac{D\psi_A}{D\tau}-\ii u_Ia_JL^{IJ\ B}_{\ \ A}\psi_B+ \ii\frac e{2m}h_I^{\ K}h_J^{\ L}F_{KL}L^{IJ\ B}_{\ \ A}\psi_B=0\label{spinhalfFWandBprecession}
\end{eqnarray}
which dictates how the spin is influenced by the presence of an electromagnetic and gravitational field.
\end{itemize}
%

%%%%%%%%%%%%%%%%%%%%%%%%%%%%%%%%%%%%%%%%%%%%%%%%%%%
\subsection{The quantum Hilbert space}\label{fermionQHS}
%%%%%%%%%%%%%%%%%%%%%%%%%%%%%%%%%%%%%%%%%%%%%%%%%%%

The spinor $\psi_A\in W$ (where $W$ is a two dimensional complex vector space) could potentially encode a two dimensional quantum state. However, given that $\psi_{A}$ constitutes a faithful and therefore {\em non-unitary} representation of the Lorentz group this identification might seem problematic. This issue is resolved by identifying a velocity-dependent inner product on the space $W$. In doing so we are able to promote $W$ to a Hilbert space and so regard $\psi_A$ as a quantum state. Let us now show how the two-component spinor $\psi_A$ can be taken as a representation of the quantum state for a qubit, and that it does indeed evolve unitarily.

%%%%%%%%%%%%%%%%%%%%%%%%%%%%%%%%%%%%%%%%%%%%%%%%%%%%%%%%%%%%%%%%%%%%%
\subsubsection{The quantum state and inner product\label{newinnerproduct}\label{secfermionQS}}
%%%%%%%%%%%%%%%%%%%%%%%%%%%%%%%%%%%%%%%%%%%%%%%%%%%%%%%%%%%%%%%%%%%%%
Although the space of two-component spinors $W$ is a two-dimensional complex vector space, it is not a Hilbert space as there is no positive definite sesquilinear inner product defined {\em a priori}. However, in the above analysis of the Dirac  field in the WKB limit the object $I_{u}^{A'A}\equiv u_I\bar{\sigma}^{IA'A}$ emerged naturally. Note that this object is simply the inner product for the Dirac field in the WKB limit and has the appropriate index structure of an inner product for a spinor space (see \ref{appIP}). Thus we take the inner product between two spinors $\psi_A^{1}$ and  $\psi_A^{2}$ to be given by
\begin{eqnarray}
\bk{\psi^{1}}{\psi^{2}}=I_{u}^{A'A}\bar{\psi}^{1}_{A'}\psi^{2}_A=u_I\bar{\sigma}^{IA'A}\bar{\psi}^{1}_{A'}\psi^{2}_A
\label{IP}
\end{eqnarray}
which in the rest frame $u^I=(1,0,0,0)$ takes on the usual form $u_{0}\sigma^{0 A'A}\bar{\psi}^{1}_{A'}\psi^{2}_A = \delta^{A'A}\bar{\psi}^{1}_{A'}\psi^{2}_A$. The connection between Dirac notation and spinor notation can therefore be identified as
\begin{eqnarray*}
|\phi\rangle\sim\phi_A\qquad\langle\phi|\sim I_{u}^{A'A}\bar{\phi}_{A'}.
\end{eqnarray*}

First note that the inner product \eqref{IP} is manifestly Lorentz invariant. This follows immediately from the fact that all indices have been contracted.\footnote{Lorentz invariance can be verified explicitly by making use of $\Lambda^{I}_{\ J}(x)\bar{\sigma}^{JB'B}(x)\bar\Lambda^{A'}_{\ B'}(x)\Lambda^{A}_{\ B}(x)=\bar{\sigma}^{IA'A}$ \cite{DHM2010}.}  Secondly, $I^{A'A}_u$  satisfies all the criteria for an inner product on a complex vector space $W$: Sesquilinearity\footnote{Sesquilinearity is the property that the inner product is linear in its second argument and antilinear in its first.} is immediate, and the positive definiteness follows if $u^I$ is future causal and timelike, since the eigenvalues $\lambda_{\pm}=u^0(1\pm v)$ of $I_{u}^{A'A}$ are strictly positive, where $u^0\equiv(1-v^2)^{-\half}$ and $v$ denotes the speed of the particle as measured in the tetrad frame. Thus, in the WKB limit, $I_{u}^{A'A}$ can be taken to define an inner product on the spinor space $W$ which therefore becomes a Hilbert space. The spinor $\psi_A$ is then a member of a Hilbert space and thus it plays the role of a quantum state. A qubit is then characterised by its trajectory $\Gamma$ and the quantum states $\psi_A(\tau)$ attached to each point along the trajectory.

In \S \ref{sec:localizedqubits} we saw that we need a separate Hilbert space for each spacetime point $x$. However, the inner product is also velocity dependent, or equivalently momentum dependent. Thus, we must also regard states corresponding to qubits with different momenta as belonging to different Hilbert spaces. In particular, we cannot compare or add quantum states with different 4-momenta $p_1\neq p_2$ even if the quantum states are associated with the same position in spacetime. Consequently the Hilbert space of the qubit is labelled not only with its spacetime position but also with its 4-momentum. We therefore denote the Hilbert space as $\m H_{x,p}$. Note that we no longer strictly have a representation of the Lorentz group since an arbitrary Lorentz transformation $\Lambda$ will correspond to a map between distinct Hilbert spaces $\Lambda: \m H_{p}\to \m H_{\Lambda p}$, rather than within a single space. Thus Wigner's theorem; that any faithful unitary {\it representation} of the Lorentz group must be infinite dimensional \cite{Wigner}, remains satisfied.\footnote{We can consider each 2-dimensional Hilbert space individually, or we can consider the family of all Hilbert spaces $\Hi_\text{total}=\bigoplus_{p\in\mathbb R^3} \Hi_p$, an infinite dimensional representation of the Lorentz group. See \chref{ch-2} for more details.}%As described in \S\ref{RSM-newinnerproduct},

%%%%%%%%%%%%%%%%%%%%%%%%%%%%%%%%%%%%%%%%%%%%%%%%%%%%%%%%%%%%%%%%%%%%%
\subsection{The relation to the Wigner formalism\label{fermionwignerrot}}
%%%%%%%%%%%%%%%%%%%%%%%%%%%%%%%%%%%%%%%%%%%%%%%%%%%%%%%%%%%%%%%%%%%%%

As mentioned in the introduction, the Wigner representation is a representation of the Poincar\'e group, which is the Lorentz group in a semidirect product with translations in flat spacetime. Spin states in this representation transform under a Lorentz transformation by `Wigner rotation'
\be
W(\Lambda,p)\ket{\psi}=L^{-1}(\Lambda p)\Lambda L(p)\ket\psi.\label{eq-Wignerrotation}
\ee
A discussion and comparison of the Wigner and Weyl formalisms are given in \chref{ch-2}. The central insight is that the Wigner spin is the covariant spin as written in the rest-frame, $\tilde\psi_A$. The Wigner rotation is thus interpreted as firstly boosting the presentation of the spin to the frame $p$, applying the Lorentz transformation $\Lambda$, then recovering the spin back to the rest frame using the boost $L^{-1}(\Lambda p)$. With these insights we can identify orthonormal bases for covariant spinors, and also reproduce the transport equation for a massive fermion in curved spacetime in the Wigner representation, as in \cite{TerashimaUeda03}.

\subsubsection{Orthonormal bases}

In order to establish the relation between the Weyl spinor and the Wigner representation we first note that the basis
\begin{eqnarray*}
\xi_A = \begin{pmatrix}1\\ 0\end{pmatrix} \qquad \chi_A = \begin{pmatrix}0\\1\end{pmatrix}
\end{eqnarray*}
in which the quantum state is expanded, $\psi_A=\psi_1\xi_A+\psi_2\chi_A$, is an oblique basis and not orthonormal with respect to the inner product $I^{A'A}_u$, i.e.\;$\bk\xi\chi\neq0$ and $\bk\xi\xi\neq1\neq\bk \chi\chi$. One consequence of this is that the transport equation \eqref{spinhalfFW} appears non-unitary as it contains both terms that look Hermitian (e.g.\;$\hat{L}^{ij}=\half\eps^{ij}_{\ \ k}\hat{\sigma}^k$), and terms that look anti-Hermitian (e.g.\;$\hat{L}^{0j}=-\frac \ii2 \hat{\sigma}^j$). However, as will shall see in \S\ref{sec-unitarity} the transport is unitary with respect to the inner product $I_u^{A'A}$.

The connection to the Wigner formalism is seen by re-expressing the quantum state in an orthonormal basis. This is given by
\be
\tilde\xi_A = L(u)^{\ B}_{A}\xi_B \qquad \tilde\chi_A = L(u)^{\ B}_{A}\chi_B\label{eq-spinonbasis}
\ee
where $L(u)_A^{\ B}$ is the spin-$\half$ Lorentz boost \cite{DHM2010}
\begin{eqnarray}
L(u)_{A}^{\;B} = \frac1{\sqrt{2(u_0+1)}}(\delta_I^0+u_I)\sigma^{I\ B}_{\ A}\label{spinhalfLboost}
%\frac1{\sqrt{2(u_0+1)}}(\sigma^{0\ B}_{\ A}+u_I\sigma^{I\ B}_{\ A})
\end{eqnarray}
with the Pauli operators given by $\sigma^{I\ B}_{\ A}=\sigma^0_{\ AA'}\bar\sigma^{IA'B}$. This boost is the spin-$\half$ representation of the spin-1 Lorentz boost
\be
L(u)_{\ J}^{I}=\delta^I_J+2u^I\delta^0_J-\frac1{u_0+1}(\delta^I_0+u^I)(\delta_J^0+u_J)=
\begin{bmatrix}
u_0&u_j\\
u^i&\delta^i_j+\frac1{u_0+1}u^iu_j
\end{bmatrix}\label{spin1Lboost}
\ee
which takes the contravariant vector $\delta^J_0$ to $u^I$, and where $L_I^{\ J}={L^{-1}}^I_{\ J}$. \footnote{The 4-vector expression for the spin-1 boost can be calculated from \eqref{spinhalfLboost}, the Lorentz invariance of $\bar\sigma^{IA'A}$ and \cite[Eq.\;(2.55)]{DHM2010}.}

Orthonormality of the basis \eqref{eq-spinonbasis} follows from the Lorentz invariance of $\bar{\sigma}^{JB'B}$ and the fact that $\xi_A$ and $\chi_A$ are orthonormal with respect to the inner product $\delta^{A'A}$. I.e.:\;orthogonality of $\tilde\xi$ and $\tilde\chi$ can be seen by making use of the invariance of $\bar\sigma^{IA'A}$:
\begin{eqnarray*}
\bigbk{\tilde\xi}{\tilde\chi}&=&\bar L(u)_{A'}^{\ \ C'}\bar\xi_{C'}L(u)_B^{\ D}\chi_Du_I\bar\sigma^{IA'B}
\nonumber\\
&=&\bar\xi_{C'}\chi_D\bar L(u)_{A'}^{\ \ C'}L(u)_B^{\ D}L(u)^{J}_{\ I}\delta_J^0\bar\sigma^{IA'B}=\bar\xi_{C'}\chi_D\delta^{BC'}=0
\end{eqnarray*}
and we can in a similar way demonstrate that $\bigbk{\tilde\xi}{\tilde\xi}=\bk{\tilde\chi}{\tilde\chi}=1$. The components $(\tilde\psi_1,\tilde\psi_2)$ are defined by $\psi_A=\tilde\psi_1\tilde\xi_A+\tilde\psi_2\tilde\chi_A$ and can now be understood as the components $\tilde\psi_A ={ L(u)^{-1}}^{\ B}_{A}\psi_B$ of the spinor in the particle's rest frame.

\subsubsection{The transport equation in the Wigner representation\label{sec-spinWigcomov}}

Given that in the rest frame the basis $(\tilde\xi_A,\tilde\chi_A)$ is indeed orthonormal, it is instructive to also express the Fermi--Walker transport in such a basis. By doing so we will not only see that the evolution is indeed unitary, but in addition we will make contact with the transport equation identified by \cite{TerashimaUeda03} in which the authors made use of infinite-dimensional representations and the Wigner rotations.
%
%Explicitly the spin-$\half$ Lorentz boost as defined above takes the form \cite{DHM2010}
%\begin{eqnarray}
%L_{A}^{\;B}(\beta) = \sqrt{\frac{\gamma+1}{2}}\sigma^{0\ B}_{\ A}+\sqrt{\frac{\gamma-1}{2\beta^2}} \beta_{i}\sigma^{i\ B}_{\ A}\label{spinhalfLboost}
%\end{eqnarray}
%where $\beta^{i}$ is the boost velocity,  $\gamma=(1-\beta^2)^{-\half}$ is its Lorentz factor, and the Pauli operators are given by $\sigma^{I\ B}_{\ A}=\sigma^0_{\ AA'}\sigma^{IA'B}$. The corresponding spin-1 boost (acting on a contravariant vector) is
%\be
%L(u)_{\ J}^{I}=
%\begin{pmatrix}
%\gamma&\gamma\beta_j\\
%\gamma\beta^i&\delta^i_j+\frac{\gamma^2\beta^i\beta_j}{\gamma+1}
%\end{pmatrix}\label{spin1Lboost}
%\ee
%where $\beta_j=\delta_{ij}\beta^i$. \tblue{We want to use $L(u)$ to boost the spinor from a velocity $u$ to 0, therefore, setting $(\gamma,\gamma\beta^i)=u^I$}, substituting $\psi_{A} =  L(u)_{A}^{\;B}\tilde{\psi}_{B}$ into the Fermi--Walker derivative \eqref{spinhalfFW} yields
%
Substituting $\psi_{A} =  L(u)_{A}^{\;B}\tilde{\psi}_{B}$ into the Fermi--Walker derivative \eqref{spinhalfFW} yields
\begin{align*}
\frac{D^{FW}\psi_{A}}{D\tau} &=\frac{\di\psi_{A}}{\di\tau}-\frac \ii2 u^{\mu} \omega_{\mu\;IJ} L^{IJ\ B}_{\ \ A}\psi_{B} - \ii u_{I} a_{J} L^{IJ\ B}_{\ \ A}\psi_{B} \\
&=L(u)_{A}^{\;B} \frac{\di\tilde{\psi}_B}{\di\tau}+ \frac{\di L(u)_{A}^{\;B}}{\di\tau}\tilde{\psi}_B - \ii\left( \frac 12 u^{\mu} \omega_{\mu\;IJ} + u_{I} a_{J}\right) L^{IJ\ B}_{\ \ A}L(u)_{B}^{\;C}\tilde{\psi}_C=0.
\end{align*}
The latter expression can be rearranged to give an evolution equation for the rest-frame spinor
\[
\frac{\di\tilde{\psi}_A}{\di\tau} = \left[-{L(u)^{-1}}_{A}^{\;B} \frac{\di L(u)_{B}^{\;D}}{\di\tau} + \ii\left( \frac 12 u^{\mu} \omega_{\mu\;IJ} +u_{I} a_{J}\right) {L(u)^{-1}}_A^{\ B}L^{IJ\ C}_{\ \ B} L(u)_{C}^{\;D}\right]\tilde{\psi}_D.
\]
One can then simplify this using the identities\footnote{The former can be shown using the Lorentz invariance of $\bar{\sigma}^{IA'A}$ and the definition of ${L^{IJ}}_A^{\ B}$ in terms of $\sigma^I$: see \S\ref{spinhalfLG}. The latter is given in \cite[p9]{Bailin}.} $ \Lambda_{A}^{\ B}\Lambda_{\ C}^{D} {L^{IJ}}_{B}^{\;\;C}=  \Lambda^{I}_{\ K}\Lambda^{J}_{\ L}{L^{KL}}_{A}^{\ D}$ and ${\Lambda^{-1}}_A^{\ B}=\Lambda_{\ A}^B$ to obtain spin-1 boosts to the terms involving $L^{IJ}$. Using now explicit expressions of the spin-$\half$ and spin-1 boosts \eqref{spinhalfLboost} and \eqref{spin1Lboost}, one can cancel many of the terms to yield the result
%\begin{equation}
% \frac{\di\tilde{\psi}_A}{\di\tau} = \frac{\ii\gamma^{2}}{2(\gamma+1)}\beta_{i}\frac{\di\beta_{j}}{\di\tau}\epsilon^{ij}_{\;\;k}\sigma^{k\;B}_{\;A} \tilde{\psi}_{B} +\ii u^{\mu}\left(\frac12  \omega_{\mu\;ij}+\gamma  \omega_{\mu\;0j}\beta_{i}+\frac{\gamma^{2}}{\gamma+1} \omega_{\mu\;il}\beta^{l}\beta_{j} \right)L^{ij\ B}_{\ \ A}\tilde{\psi}_{B}.
%\end{equation}
%where recall $(\gamma,\gamma\beta^i)=u^I$, so
\begin{equation}
 \frac{\di\tilde{\psi}_A}{\di\tau} =\ii\left[ \frac{1}{u^0+1}u_{i}\frac{\di u_j}{\di\tau}+u^{\mu}\left(\frac12  \omega_{\mu\;ij}+ \omega_{\mu\;0j}u_{i}+\frac{1}{u^0+1} \omega_{\mu\;il}u^{l}u_{j} \right)\right]L^{ij\ B}_{\ \ A}\tilde{\psi}_{B}\label{restframetrans}.
\end{equation}
This is the transport equation for the quantum state $\psi_{A}$ expressed in terms of the rest-frame spinor $\tilde{\psi}_{A}$. First we note that the transport is unitary with respect to the standard inner product $\delta^{A'A}$ as it only contains terms proportional to  $\hat{L}^{ij}=\half\eps^{ij}_{\ \ k}\hat{\sigma}^k $. It is however not manifestly Lorentz invariant. Secondly, it is also equivalent to the transport equation derived by \cite{TerashimaUeda03} who used the infinite-dimensional Wigner representations \cite{Weinberg}. We have thus re-derived their result using the Dirac equation in the WKB limit. In addition, we have done so while avoiding the use of momentum eigenstates $|p,\sigma\rangle$, which are strictly speaking not well-defined in a curved spacetime as no translational invariance is present.

Notice that there is no term proportional to the identity $\delta_A^{\ B}$ which would correspond to an accumulation of global phase.  In fact, global phase is missing in \cite{TerashimaUeda03}. On the other hand, as we shall see in \S\ref{secPhase}, these phases are automatically included in the WKB approach adopted in this chapter.

Although the unitarity of the transport becomes manifest when written in terms of the rest-frame spinor it is not necessary to work with \Eeqref{restframetrans}. Once we generalise the notion of unitarity in \S\ref{sec-unitarity} we will see that we can treat the evolution of the quantum state in terms of the manifestly Lorentz covariant Fermi--Walker transport \eqref{spinhalfFW}.

\subsection{Calculating the state transformation\label{sec-fermionstatetransformation}}
Analogously to in \Ssref{sec-statetransformation}, we want to determine the relation between the final state from an initial state $\psi_A(\tau_\text{end})=\mathrm T_A^{\; B}\psi_B(\tau_\text{start})$, transported along a trajectory $x(\tau)$ with the Fermi--Walker transport
\be
\frac{\di\psi_{A}}{\di\tau}=\ii\left(\half u^{\mu} \omega_{\mu\;IJ} + u_{I} a_{J}\right) L^{IJ\ B}_{\ \ A}\psi_{B},
\label{eq-FW}
\ee
with $u_I=\frac{d x^I}{d\tau}$ and $a^I=\frac{Du^I}{D\tau}$. This is a differential equation of the form $\di\psi/\di\tau=i\hat H\psi$, and so the transformation operator $\hat{\mathrm T}$, with the initial condition that $\hat{\mathrm T}=\hat\sigma^0$ for $\tau_\text{end}=\tau_\text{start}$, is given by
\[
\hat{\mathrm T}_\text{Weyl}=\m T\exp\left[\int_{\tau_\text{start}}^{\tau_\text{end}}i\hat H(\tau)\di\tau \right]=
\m T\exp\left[\ii\int_{\tau_\text{start}}^{\tau_\text{end}}
\left(\half u^{\mu} \omega_{\mu\;IJ} + u_I a_J\right)\hat L^{IJ}\di\tau \right]
\]
where $\exp$ is the matrix exponential and $\m T$ is the time-ordering operator. The time ordering of the exponential is required in the general case as evolutions at different times may not commute in the series expansion \cite{Weinberg,PeskinSchroeder}. If one can find a coordinate system so that the infinitesimal evolution is always proportional to a single matrix, then the operators commute and the integral does not require a time ordering operator \cite{Weinberg}.

The $SL(2,\mathbb C)$ generators are $\hat L^{0j}=\frac\ii2 \hat\sigma^{j}$ and $\hat L^{ij}=\frac12\eps^{ij}_{\ \ k}\hat\sigma^k$, so this integral will consist of complex-valued coefficients of the Pauli matrices.

We have also calculated the rest-frame spinor (i.e.\;Wigner) transport, \eqref{restframetrans}. This is also of the form
 \be
 \frac{\di\tilde{\psi}}{\di\tau} = \ii H_{ij}(\tau)\hat L^{ij}\tilde{\psi}\label{eq-rfFW2},
\ee
where $\hat H$ was generically decomposed into real coefficients $H_{IJ}$ of $\hat L^{IJ}$, but only the coefficients of $\hat L^{ij}$ are nonzero in \eqref{eq-rfFW2}. As in \cite{TerashimaUeda03}, the transformation operator will thus be given by
\[
%\psi(\tau_\text{end})=\hat{\mathrm T}\psi(\tau_\text{start}),\qquad
\hat{\mathrm T}_\text{Wigner}=\m T\exp\left[\ii\int_{\tau_\text{start}}^{\tau_\text{end}} H_{ij}(\tau)\hat L^{ij}(\tau)\di\tau \right]
\]
where scalar coefficients of the Pauli matrices in this integrand are imaginary rather than complex-valued, because only the $L^{ij}$ ($SU(2)$) terms entered in \eqref{eq-rfFW2}.

Since the covariant spinor is related to the stopped spinor by the standard boost $L(p)$, we also have two other ways to calculate transformation matrices: The transformation matrix for the rest frame (Wigner) spinor can be obtained from integrating the covariant transport \eqref{eq-FW} along the trajectory to obtain a single final Lorentz transformation $\Lambda$, then applying the standard boosts as per \eqref{eq-Wignerrotation} to obtain a single Wigner rotation for the entire transformation \cite{TerashimaUeda03}:
\[
\tilde\psi(\tau_\text{end})=\hat{\mathrm T}_\text{Wigner}\tilde\psi(\tau_\text{start})={L(p_\text{end})^{-1}}\hat {\mathrm T}_\text{Weyl}L(p_\text{start})\tilde\psi(\tau_\text{start}).
\]
This method would avoid computing the infinitesimal Wigner rotation corresponding to the Lorentz transformation at each point. Conversely, if desired, the transformation for the covariant spinor can be calculated using $\hat{\mathrm T}_\text{Wigner}$ by substituting $\tilde\psi=L^{-1}(p)\psi$:
\[
\psi(\tau_\text{end})=L(p_\text{end})\hat{\mathrm T}_\text{Wigner}L^{-1}(p_\text{start})\psi(\tau_\text{start})=\hat{\mathrm T}_\text{Weyl}\psi(\tau_\text{start}).
\]
There is therefore some freedom in choice of representation, coordinates, and bases in order to obtain the simplest expression for calculation. Note also that the transformation operator will change depending on the choice of initial and final tetrads, even in the Wigner representation. In \secref{sec-measurement} we develop a preparation and measurement formalism so that measurement directions in a lab can be translated into spin observables.

%%%%%%%%%%%%%%%%%%%%%%%%%%%%%%%%%%%%%%%%%%%%%%%%%%%%%%%%%%%%%%%%%%%%%
\section{The qubit as the polarisation of a photon\label{sec-Photons}}
%%%%%%%%%%%%%%%%%%%%%%%%%%%%%%%%%%%%%%%%%%%%%%%%%%%%%%%%%%%%%%%%%

Another specific physical realisation of a qubit is the polarisation of a single photon. This is an important example since it lends itself easily to physical applications. We obtain this realisation via the WKB limit of Maxwell's equations in curved spacetime \cite{MTW,Woodhouse}. The polarisation of a photon is described by a unit spacelike 4-vector $\psi_\mu$ called the polarisation vector \cite{MTW,Rindler,Woodhouse}. Restricting ourselves to localised wavepackets we obtain the description of a photon with definite 4-momentum/wavevector $k^\mu$ and polarisation vector $\psi_\mu(\lambda)$ which is parallel transported along a null geodesic $x^{\mu}(\lambda)$. We will see that in fact $\psi_{\mu}$ contains only two gauge invariant degrees of freedom and thus can be taken to encode the quantum state of a photonic qubit.

Although we consider only geodesic trajectories in this chapter it is possible to consider non-geodesic trajectories. We refer the reader to \appref{jerk} for a discussion of approaches to this problem.  A physically motivated way to obtain non-geodesic trajectories would be to introduce a medium in Maxwell's equations through which the photon propagates. Nevertheless, even without explicitly including a medium, it is easy to include optical elements such as mirrors, prisms, and other unitary transformations as long as their effect on polarisation can be considered separately to the effect of transport through curved spacetime.

%%%%%%%%%%%%%%%%%%%%%%%%%%%%%%%%%%%%%%%%%%%%%%%%%%%%%%%%%%%%%%%%%%%%%
\subsection{Parallel transport from the WKB approximation}\label{photongeoapprox}
%%%%%%%%%%%%%%%%%%%%%%%%%%%%%%%%%%%%%%%%%%%%%%%%%%%%%%%%%%%%%%%%%%%%%
In this section we shall see that the parallel transport equation for the polarisation vector emerges directly from the WKB approximation \cite{MTW,Woodhouse}. Gauge invariance and gauge fixing in the WKB approach are important for properly isolating the quantum state and we have therefore paid attention to this issue.

%%%%%%%%%%%%%%%%%%%%%%%%%%%%%%%%%%%%%%%%%%%%%%%%%%%%%%%%%%%%%%%%%%%%%%%%%%%%%%%%%%%%
\subsubsection{The basic ansatz}
%%%%%%%%%%%%%%%%%%%%%%%%%%%%%%%%%%%%%%%%%%%%%%%%%%%%%%%%%%%%%%%%%%%%%%%%%%%%%%%%%%%%

The WKB approximation for photons follows a procedure similar to that for the Dirac field. First we write the vector potential $A_{\mu}$ as
\begin{eqnarray}
A_\mu=\varphi_\mu \Ee^{\ii\theta/\epsilon}
\label{EMpolarform}.
\end{eqnarray}
As in the case for the Dirac field, the WKB limit is where the phase $\theta$ is oscillating rapidly compared to the slowly varying complex amplitude $\varphi_\mu$. As before, this is expressed through the expansion parameter $\epsilon$. Maxwell's equations can then be studied in the limit $\epsilon\rightarrow 0$. %Although we omit taking the real part of $\varphi_\mu \Ee^{\ii\theta/\epsilon}$ in this section it is implicitly understood that this is done.

%%%%%%%%%%%%%%%%%%%%%%%%%%%%%%%%%%%%%%%%%%%%%%%%%%%%%%%%%%%%%%%%%%%%%%%%%%%%%%%%%%%%
\subsubsection{Gauge transformations in the WKB limit}
%%%%%%%%%%%%%%%%%%%%%%%%%%%%%%%%%%%%%%%%%%%%%%%%%%%%%%%%%%%%%%%%%%%%%%%%%%%%%%%%%%%%
Let us now study the $U(1)$ gauge transformations in terms of the new variables $\theta$ and $\varphi_\mu$. It is clear that not all gauge transformations $A_\mu\rightarrow A_\mu+\nabla_\mu \lambda$ will preserve the basic form $A_\mu=\varphi_\mu \Ee^{\ii\theta/\eps}$. We therefore consider gauge transformations of the form $\lambda=\zeta \Ee^{\ii\theta/\eps}$ where $\zeta$ is a slowly varying function. This class of gauge transformations can be written in the polar form of \eqref{EMpolarform} as
\[
A_\mu\rightarrow A_\mu+\nabla_\mu\lambda= A_\mu+\nabla_\mu(\zeta \Ee^{\ii\theta/\epsilon})=\left(\varphi_\mu+\nabla_\mu \zeta+\frac{\ii}{\epsilon}k_\mu\zeta\right)\Ee^{\ii\theta/\epsilon}
\]
and so $\varphi_\mu\rightarrow \varphi_\mu+\nabla_\mu \zeta+\frac{\ii}{\epsilon}k_\mu\zeta$.

In the limit $\epsilon\rightarrow 0$ note that $\varphi_\mu$ does not behave properly under the gauge transformations of the type that we are considering since the second term blows up. This has no physical significance and is just an artefact of describing the vector potential as being of the specific form \eqref{EMpolarform}. Such a gauge transformation leaves the physics unchanged but will no longer preserve the form of the solution \eqref{EMpolarform} where we have a slowly varying envelope and rapid phase. Because of this it is necessary to further restrict the space of gauge transformations to ``small" gauge transformations $\zeta=-\ii\epsilon\xi$. In that limit we then have
\begin{eqnarray}
\varphi_\mu\rightarrow \varphi_\mu-\ii\epsilon\nabla_\mu \xi+k_\mu\xi \label{varphigauge}
\end{eqnarray}
and so $\varphi_\mu\rightarrow \varphi_\mu+k_\mu\xi+\mathcal{O}(\epsilon)$. %%NOTE!!!
However, as we shall see below, in order to maintain gauge invariance of the equations in all orders of $\epsilon$ it is important to keep both orders of $\epsilon$ in the gauge transformation \eqref{varphigauge}.

%%%%%%%%%%%%%%%%%%%%%%%%%%%%%%%%%%%%%%%%%%%%%%%%%%%%%%%%%%%%%%%%%%%%%%%%%%%%%%%%%%%%
\subsubsection{The gauge condition}
%%%%%%%%%%%%%%%%%%%%%%%%%%%%%%%%%%%%%%%%%%%%%%%%%%%%%%%%%%%%%%%%%%%%%%%%%%%%%%%%%%%%

In the literature we find two suggestions for imposing a gauge. For example, in \cite{MTW} the Lorenz gauge is used, $\nabla_\mu A^\mu=(\nabla_\mu \varphi^\mu+\frac{\ii}{\epsilon}k^\mu \varphi_\mu)\Ee^{\ii\theta/\epsilon}=0$, and in  \cite{Woodhouse} the gauge $k^\mu \varphi_\mu=0$ is imposed so that the complex amplitude $\varphi_\mu$ is always orthogonal to the wavevector $k^\mu$. However, for our purposes neither of these gauge conditions turns out to be suitable. Rather we will work in a gauge where $k_\mu$ and $\varphi_\mu$ are orthogonal up to first-order terms in $\epsilon$, i.e.\;%
\begin{eqnarray*}
\varphi_\mu k^\mu=\epsilon \alpha(x)
\end{eqnarray*}
where $\alpha$ is taken to be some arbitrary function of $x^\mu$.

%%%%%%%%%%%%%%%%%%%%%%%%%%%%%%%%%%%%%%%%%%%%%%%%
\subsubsection{Maxwell's equations in the WKB limit}
%%%%%%%%%%%%%%%%%%%%%%%%%%%%%%%%%%%%%%%%%%%%%%%%
Let us now turn to Maxwell's equations in vacuum:
\begin{eqnarray}
\nabla_\mu F^\mu_{\ \ \nu}=g^{\rho\mu}\nabla_\rho(\nabla_\mu A_\nu-\nabla_\nu A_\mu)=0\label{MW}.
\end{eqnarray}
The equations $\nabla_{[\rho}F_{\mu\nu]}=0$ are mere identities when we work with a vector potential $A_\mu$ rather than the gauge invariant $F_{\mu\nu}\equiv\nabla_\mu A_\nu-\nabla_\nu A_\mu$. If we substitute the ansatz $A_\mu=\varphi_\mu \Ee^{\ii\theta/\epsilon}$ into \eqref{MW} we obtain
\begin{equation}
\square \varphi_\nu-\nabla^\mu \nabla_\nu \varphi_\mu+\frac{\ii}{\epsilon}(2k^\mu\nabla_\mu \varphi_\nu+\varphi_\nu\nabla_\mu k^\mu-k_\nu\nabla_\mu \varphi^\mu-\nabla_\nu(\varphi_\mu k^\mu))-\frac{1}{\epsilon^2}(k^2\varphi_\nu-k_\nu \varphi_\mu k^\mu)=0.\label{MW1}
\end{equation}
Gauge invariance can be a bit subtle in this context so let us make a few remarks. \Eeqref{MW1} is of course invariant under gauge transformations $\varphi_\mu\rightarrow \varphi_\mu-\ii\epsilon\nabla_\mu\xi+k_\mu\xi$ as this is nothing but Maxwell's equations \eqref{MW} rewritten in different variables. However, note that the terms of zeroth, first, and second order (in $1/\epsilon$) of \Eeqref{MW1}:
\begin{subequations}\label{orders}
\begin{align}
&\square \varphi_\nu-\nabla^\mu \nabla_\nu \varphi_\mu\label{zero}\\
&2k^\mu\nabla_\mu \varphi_\nu+\varphi_\nu\nabla_\mu k^\mu-k_\nu\nabla_\mu \varphi^\mu-\nabla_\nu(\varphi_\mu k^\mu)\label{first}\\
&k^2\varphi_\nu-k_\nu \varphi_\mu k^\mu\label{second}
\end{align}
\end{subequations}
are not {\em separately} gauge invariant. This is so because the gauge transformation $\varphi_\mu\rightarrow \varphi_\mu-\ii\epsilon\nabla_\mu \xi+k_\mu\xi$ contains terms of different orders in $\epsilon$. Thus, after a gauge transformation of the second-order term \eqref{second} we end up with first-order terms in $\epsilon$, which then belong to \eqref{first}. Similarly first-order terms in $\epsilon$ in \eqref{first} end up in \eqref{zero}. It is then easy to verify that the entire equation \eqref{MW1} is gauge invariant although the separate terms in \eqref{orders} are not.
\subsubsection{Equations of motions in the gauge $\varphi_\mu k^\mu=\epsilon\alpha$}
Imposing the gauge condition $k^\mu \varphi_\mu=\epsilon\alpha$ on \eqref{MW1} yields the equation
\begin{equation}
\left[\square \varphi_\nu-\nabla^\mu \nabla_\nu \varphi_\mu-\nabla_\nu\alpha+\frac{\ii}{\epsilon}(2k^\mu\nabla_\mu \varphi_\nu+\varphi_\nu\nabla_\mu k^\mu-k_\nu(\nabla_\mu \varphi^\mu-\alpha))-\frac{1}{\epsilon^2}k^2\varphi_\nu\right]\Ee^{\ii\theta/\epsilon}=0.
\end{equation}
We now demand that the solutions for $\varphi_\mu$ be independent of $\epsilon$ in the limit when $\epsilon$ is small. Physically this means that for high frequencies the form of the solutions should be independent of the frequency (parameterised by $\epsilon$). Consequently, each separate order of $\frac{1}{\epsilon}$ in the expansion must be zero. The equations corresponding to the first and second orders then read
\begin{subequations}\label{orders1}
\begin{align}
&2k^\mu\nabla_\mu \varphi_\nu+\varphi_\nu\nabla_\mu k^\mu-k_\nu(\nabla_\mu \varphi^\mu-\alpha)=0\label{firstorder}\\
&k^\mu k_\mu=0\label{secondorder}
\end{align}
\end{subequations}
for $\varphi_\nu\neq0$. The zeroth-order equation is to be thought of as `small' in comparison to the higher order terms in $1/\epsilon$ and is therefore ignored and not imposed as an equation of motion. The second equation \eqref{secondorder} is trivially gauge invariant since $k_\mu$ does not transform. The first equation is only gauge invariant up to first-order terms in $\epsilon$. This can be seen by letting $\alpha$ transform as $\alpha\rightarrow\alpha+k^\mu\nabla_\mu\xi$ under a gauge transformation, making use of \eqref{second}, and the fact that $k_\mu$ satisfies the geodesic equation as shown in \eqref{photonintegralcurves}.

%%%%%%%%%%%%%%%%%%%%%%%%%%%%%%%%%%%%
\subsubsection{The derivation of parallel transport and conserved currents}
%%%%%%%%%%%%%%%%%%%%%%%%%%%%%%%%%%%%

\Eeqref{secondorder} tells us that the wavevector $k_\mu$ is a null vector, and that its integral curves $x^\mu(\lambda)$ defined by $\di x^\mu/\di\lambda\propto k^\mu$ lie on a light cone. Taking the derivative of \Eeqref{secondorder} yields
\begin{equation}
\nabla_\nu (k^\mu k_\mu)=2k^\mu\nabla_\nu k_\mu\equiv2k^\mu\nabla_\nu\nabla_\mu\theta= 2k^\mu\nabla_\mu\nabla_\nu\theta=2k^\mu\nabla_\mu k_\nu=0 \label{photonintegralcurves}
\end{equation}
which tells us that the integral curves are null geodesics.\footnote{We have assumed in \eqref{zero} that the spacetime torsion is zero. A non-zero torsion field could possibly influence the polarisation (see \cite{Bergmann}).} These are expected since we have considered Maxwell's equations in vacuum. Non-geodesic trajectories can be obtained by introducing a medium through which the photon propagates. See \ref{jerk} for a discussion.

Contracting \Eeqref{firstorder} with the complex conjugate $\bar\varphi_v$ and adding to it the complex conjugate of the contraction yields the continuity equation
\begin{eqnarray}
2\bar\varphi^\nu k^\mu\nabla_\mu \varphi_\nu+2\varphi^\nu k^\mu\nabla_\mu \bar\varphi_\nu+2\bar\varphi^\nu \varphi_\nu\nabla_\mu k^\mu=-2\nabla_\mu(\varphi^2k^\mu)=0\label{photoncontinuity}
\end{eqnarray}
where $\varphi^2\equiv-g^{\mu\nu}\bar\varphi_\mu \varphi_\nu$. Note that $\varphi^2$ is gauge invariant up to first-order terms in $\epsilon$, i.e.\;$\varphi^2\rightarrow \varphi^2+\mathcal{O}(\epsilon)$, and therefore also $j^\mu\equiv\sqrt{g}\varphi^2k^\mu$ is gauge invariant to first order. This means that $j^\mu$ is a conserved current in the WKB limit. Since $j^0$ has the units of a probability density\footnote{We recall that $A_\mu$ has dimensions $L^{-1}$ in natural units with $e=1$.} we can interpret $j^\mu$ as a conserved probability density current.

We can also deduce that $\nabla_\mu k^\mu=-\frac{2}{\varphi}k^\mu\nabla_\mu \varphi$ and if we insert this in \Eeqref{firstorder} and define the polarisation vector $\psi_\nu$ through $\varphi_\nu\equiv \varphi\psi_\nu$, we obtain
\begin{align*}
2k^\mu\nabla_\mu \varphi_\nu+\varphi_\nu\nabla_\mu k^\mu-k_\nu(\nabla_\mu \varphi^\mu-\alpha)&=2k^\mu\nabla_\mu \varphi_\nu-\varphi_\nu\frac{2}{\varphi}k^\mu\nabla_\mu \varphi-k_\nu(\nabla_\mu \varphi^\mu-\alpha)\\&=2\varphi k^\mu\nabla_\mu \psi_\nu-k_\nu(\nabla_\mu \varphi^\mu-\alpha)=0
\end{align*}
which implies that
\begin{eqnarray*}
k^\mu\nabla_\mu \psi_\nu=\left(\frac{\nabla_\mu \varphi^\mu-\alpha}{2\varphi}\right)k_\nu.
\end{eqnarray*}
However, since $\alpha$ is arbitrary the whole right-hand side is arbitrary and we can write
\begin{eqnarray}\label{betatrans}
k^\mu\nabla_\mu \psi_\nu=\beta k_\nu.
\end{eqnarray}
The right-hand side is proportional to the wavevector $k_\nu$ and represents an arbitrary infinitesimal gauge transformation of $\psi_{\mu}$. Let us now introduce the integral curves of $u^\mu=k^\mu/E$ given by $\di x^{\mu}/\di \lambda=u^{\mu}$ where $E$ is an arbitrary constant with dimensions of energy. We can then write \Eeqref{betatrans} as
\begin{eqnarray}\label{photonPT}
\frac{D\psi_{\mu}}{D\lambda}=\beta u_{\mu}.
\end{eqnarray}
%s
Thus the transport of the polarisation vector $\psi_\mu$ is given by the parallel transport \eqref{eq-vecPT} along the null geodesic integral curves of $u_{\mu}$, with an arbitrary infinitesimal gauge transformation at each instant.

%%%%%%%%%%%%%%%%%%%%%%%%%%%%%%%%%%%%%%%%%%%%%%%%
\subsection{Localisation of the qubit}\label{photonlocalization}
%%%%%%%%%%%%%%%%%%%%%%%%%%%%%%%%%%%%%%%%%%%%%%%%

As in the fermion case, the WKB approximation is not enough to guarantee either that the wavepacket is localised or that it stays localised under evolution, and again it is not possible to achieve strict  localisation. Indeed it can be proved that a photon must have non-vanishing sub-exponential tails \cite{Hegerfeldt,Birula}. As in the case of fermions, \S\ref{fermionlocalization}, we are going to ignore these small tails and treat the wavepacket as effectively having compact support within some small region much smaller than the typical curvature scale.

The continuity equation \eqref{photoncontinuity} dictates the evolution of the envelope $\varphi(x)$. Divided by the energy as measured in some arbitrary frame it becomes $\nabla_\mu(\varphi^2u^\mu)=0$. Again we see that the assumption $\nabla_{\mu}u^{\mu}=0$ simplifies this equation. However, the interpretation of $\nabla_{\mu}u^{\mu}$ is a bit different. Instead of quantifying how much a spatial volume element is changing (as in \S\ref{fermionlocalization}), it quantifies how much an area element, transverse to $u^{\mu}$ in some arbitrary reference frame, changes \cite{Poisson}:
\begin{eqnarray*}
\nabla_{\mu}u^\mu=\frac1A\frac{\di A}{\di\lambda}
\end{eqnarray*}
where $\lambda$ is an affine parameter defined by $\di x^{\mu}/\di \lambda = u^{\mu}$. In this case we require that $\ex{\nabla_{\mu}u^\mu} \ll 1/\lambda_{\Gamma}$, where ${\lambda_{\Gamma}}$ is the affine length of the trajectory $\Gamma$. Thus, it gives us a measure of the transverse distortion of a wavepacket. For photons there can be no longitudinal distortion since all components, regardless of frequency, travel with the speed of light. Initial localisation and the assumption that $\nabla_{\mu}u^{\mu}\approx0$ therefore guarantee that the wave-packet is rigidly transported along the trajectory.

Once we assume that the polarisation vector $\psi_{I}$ does not vary spatially within the wavepacket we can effectively describe the system as a polarisation vector $\psi_\mu(\lambda)$ for each $\lambda\in\Gamma$. Having effectively suppressed the spatial degrees of freedom of the wavepacket, the polarisation $\psi_\mu$ can thus be thought of as a function defined on a classical trajectory $\Gamma$, satisfying an ordinary differential equation \eqref{photonPT}. A photonic qubit can then be characterised by a position $x^\mu(\lambda)$, a wavevector $k^\mu(\lambda)$, and a spacelike complex-valued polarisation vector $\psi_\mu(\lambda)$.

%%%%%%%%%%%%%%%%%%%%%%%%%%%%%%%%%%%%%%%%%%%%%%%%%%%%%%%%
\subsection{A summary of WKB limit}
%%%%%%%%%%%%%%%%%%%%%%%%%%%%%%%%%%%%%%%%%%%%%%%%%%%%%%%%

To summarise, the WKB approximation yields the following results and equations:
\begin{itemize}
\item The integral curves $x^\mu(\lambda)$ of 4-velocities $u_\mu$ are null geodesics
\item The vector $j^\mu=\sqrt{g}\varphi^2 k^\mu$ is a conserved probability density current (with  $k^\mu=u^\mu E$).
\item The polarisation vector $\psi_\mu$ satisfies $\psi_{\mu} u^\mu=0$ and transforms as $\psi_{\mu}\rightarrow\psi_{\mu}+\upsilon u_\mu$ under gauge transformation up to first-order terms in $\epsilon$.
\item The transport of $\psi_\mu$ is governed by \eqref{photonPT} which is simply the parallel transport along integral curves of $u^\mu$ modulo gauge transformations.
\end{itemize}
We have now established a formalism for the quantum state of a localised qubit which is invariant under $\psi_\mu\rightarrow \psi_\mu+\upsilon u_\mu$ and $\psi_\mu u^\mu=0$ up to first-order terms in $\epsilon$. We shall from this point on  neglect the small terms of order $\epsilon$.

%%%%%%%%%%%%%%%%%%%%%%%%%%%%%%%%%%%%%%%%%%%%%%%%%%%%%%%%%%%%%%%%%%%%%%%%%%
\subsection{The quantum state\label{secphotonQS}}
%%%%%%%%%%%%%%%%%%%%%%%%%%%%%%%%%%%%%%%%%%%%%%%%%%%%%%%%%%%%%%%%%%%%%%%%%%
We now show that the polarisation 4-vector has only two complex degrees of freedom and in fact it can be taken to encode a two-dimensional quantum state. We do this first with a tetrad adapted to the velocity of the photon for simplicity and then with a general tetrad. It is convenient and more transparent to work with tetrad indices instead of the ordinary tensor indices and we shall do so here.

%%%%%%%%%%%%%%%%%%%%%%%%%%%%%%%%%%%%%%%%%%%%%%%5%
\subsubsection{Identification of the quantum state with an adapted tetrad\label{sec-adaptedphotonQS}}
%%%%%%%%%%%%%%%%%%%%%%%%%%%%%%%%%%%%%%%%%%%%%%%%%

Recall from the previous section that we partially fixed the gauge to $u_{I} \psi^{I}=0$. The remaining gauge transformations are of the form $\psi^{I}\rightarrow \psi^{I}+\upsilon u^{I}$. Indeed, if $u_{I} \psi^{I}=0$ we also have that  $u_{I} (\psi^{I}+\upsilon u^{I})=0$ for all complex-valued functions $\upsilon$, since $u^I$ is null.

To illustrate in more detail what effect this gauge transformation has on the polarisation vector we adapt the tetrad reference frame $e^\mu_I$ (defined in \eqref{tetradllinvar}) to the direction of the photon so that $u^\mu\propto e^\mu_0+e^\mu_3$. Notice that there are several choices of tetrads that put the photon 4-velocity into this standard form. The two-parameter family of transformations relating these different tetrad choices are (1) spatial rotations around the $z$-axis and (2) boosts along the $z$-axis.

With a suitable parameterisations of the photon trajectory such that $e^0_\mu(\di x^\mu/\di\lambda)=1$ we can eliminate the proportionality factor and we have $u^\mu=e^\mu_0+e^\mu_3$. In tetrad components $u^I=(1,0,0,1)$ and we see that the tetrad $z$-component $e^\mu_3$ is aligned with the photon's 3-velocity. Since $0=u_{I}\psi^{I}=\psi^{0}-\psi^{3}$ it follows that $\psi^{0}=\psi^{3}=\nu$ and the polarisation vector can be written as
\begin{eqnarray*}
\psi^{I}=\begin{pmatrix}\nu\\\psi^1\\ \psi^2\\\nu\end{pmatrix}.
\end{eqnarray*}
It is clear that a gauge transformation
\begin{eqnarray*}
\psi^I=\begin{pmatrix}\nu\\\psi^1\\ \psi^2\\\nu\end{pmatrix}\rightarrow \psi^I+\upsilon u^I=\begin{pmatrix}\nu+\upsilon\\\psi^1\\ \psi^2\\\nu+\upsilon\end{pmatrix}
\end{eqnarray*}
leaves the two middle components unchanged and changes only the zeroth and third components. The two complex components $\psi^1$ and $\psi^2$, which form the Jones vector \cite{Hecht}, therefore represent gauge invariant true degrees of freedom of the polarisation vector whereas the zeroth and third components represent pure gauge.

We can now identify the quantum state as the two gauge invariant middle components $\psi^1$ and $\psi^2$, where $\psi^1$ is the horizontal and $\psi^2$ the vertical component of the quantum state in the linear polarisation basis:
\begin{eqnarray*}
|1\rangle\sim\left(\begin{array}{c}0\\1\\ 0\\0\end{array}\right),\qquad |2\rangle\sim \left(\begin{array}{c}0\\0\\ 1\\0\end{array}\right)
\end{eqnarray*}
or simply $|A\rangle\sim \delta^I_{A}$ with $A=1,2$. The quantum state is then
\begin{eqnarray*}
\ket{\psi}\sim\psi^{A} = \delta^{A}_{I}\psi^{I}  = \begin{pmatrix}\psi^1\\\psi^2\end{pmatrix}.
\end{eqnarray*}
Note we have deliberately used a notation similar to that used for representing spinors; however, $\psi^{A}$ should not be confused with an $SL(2,\mathbb C)$ spinor.  In order to distinguish $\psi^A$ from $\psi^I$ we will refer to the former as the Jones vector and the latter as the polarisation vector.

%%%%%%%%%%%%%%%%%%%%%%%%%%%%%%%%%%%%%%%%%%%%%%%5%
\subsubsection{Identification of the quantum state with a non-adapted tetrad\label{sec-photonqsident}}
%%%%%%%%%%%%%%%%%%%%%%%%%%%%%%%%%%%%%%%%%%%%%%%%%

\begin{figure}[h]
\begin{center}
\ifpdf
\includegraphics[height=30mm]{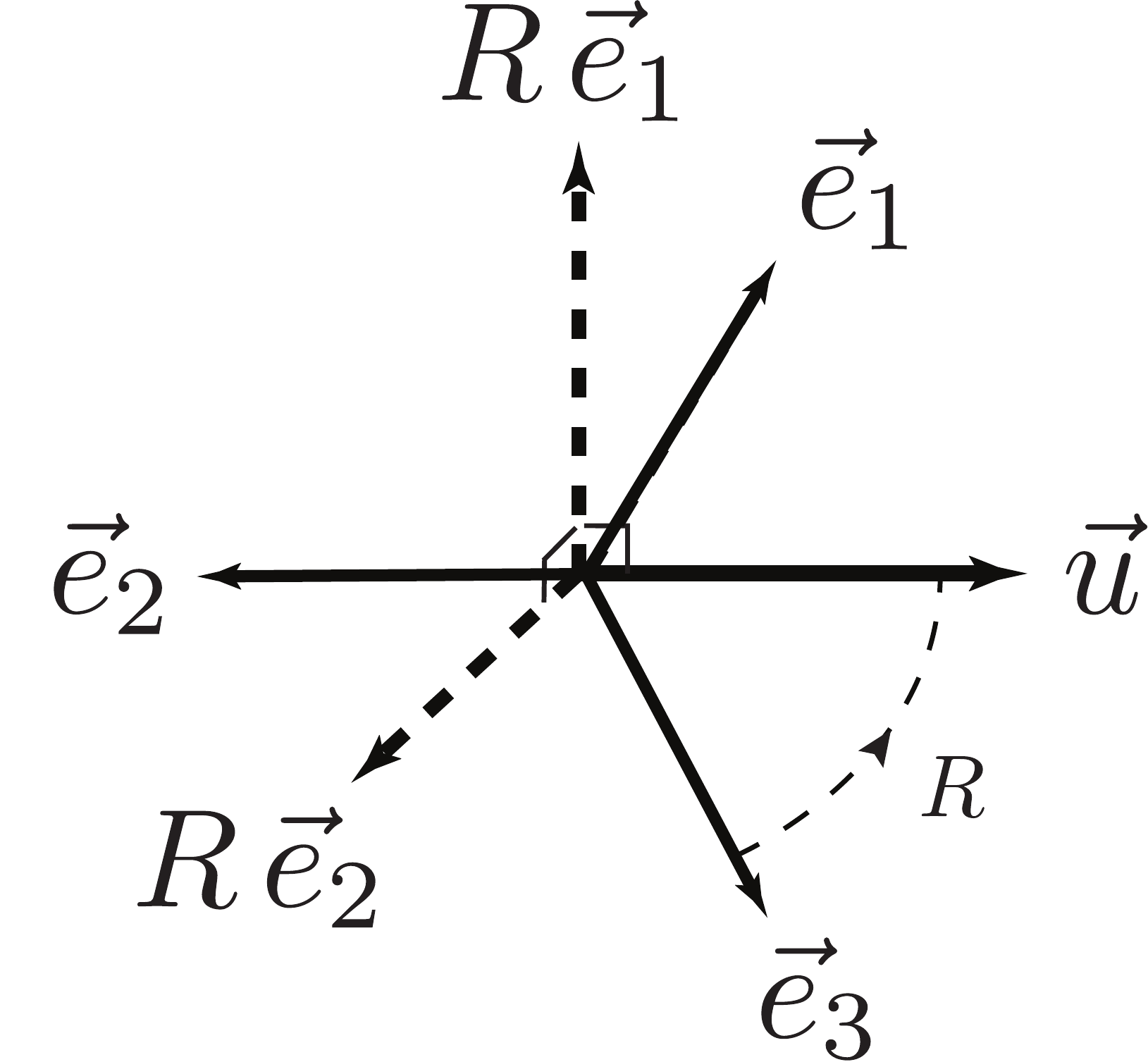}
\else
\includegraphics[height=30mm]{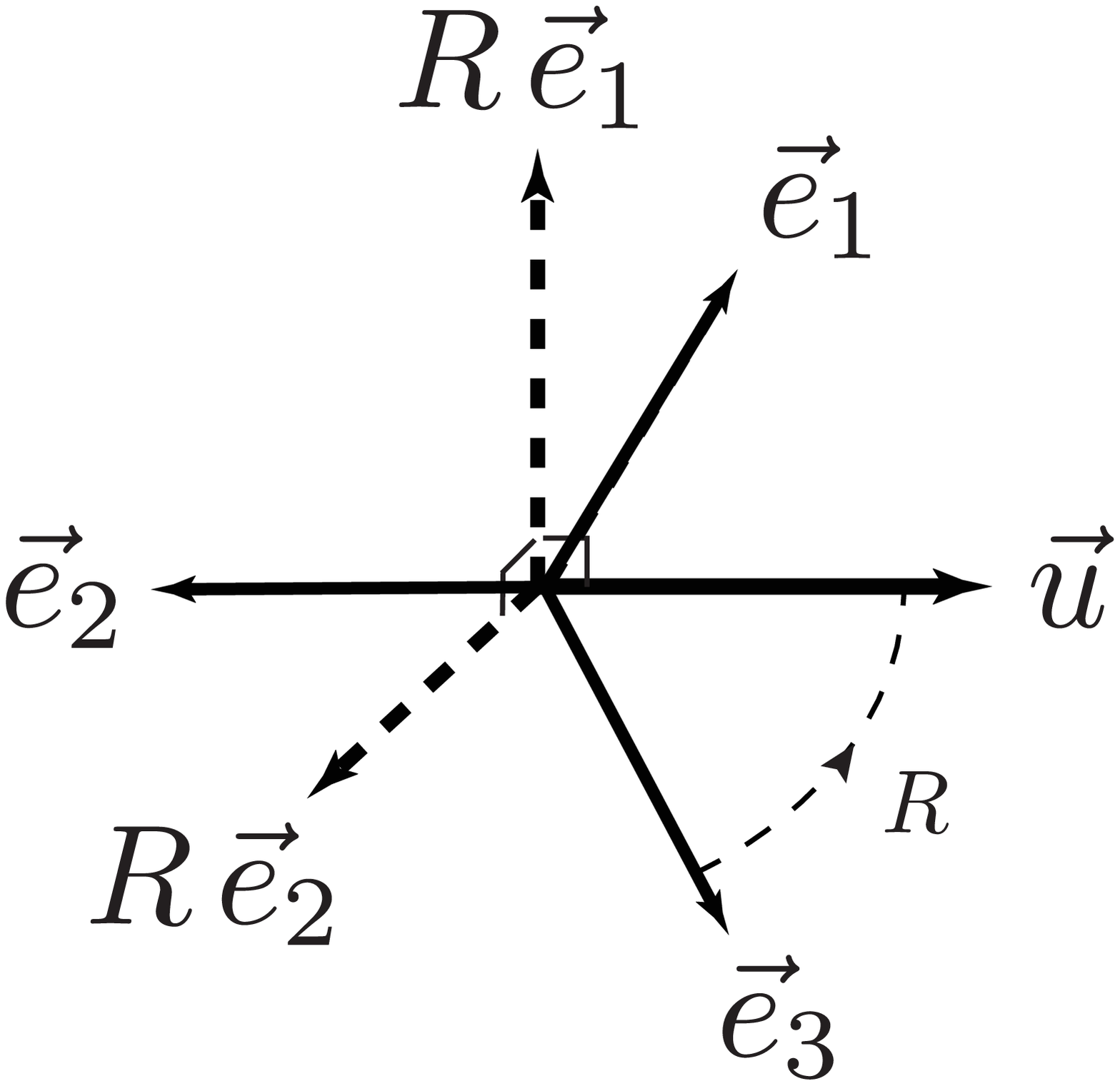}
\fi
\caption[Identifying the polarisation quantum state in a non-adapted tetrad]{The rotation $R$ adapts the spacelike vectors of the tetrad $\vec e_i$ so that the z-axis $\vec e_{3}$ is aligned to the 3-velocity of the photon $\vec u$. A polarisation vector is then in the plane spanned by $R\vec e_1$ and $R\vec e_2$.\label{adaption}}
\end{center}
\end{figure}

In the above discussion we have used an adapted tetrad in order to identify the quantum state. We can write a map for this adaption explicitly, which will provide a generic non-adapted formalism. To adapt one simply introduces a rotation which takes the 4-velocity $u^{I}$ to the standard form \cite{Weinberg}
\begin{eqnarray*}
u^I\to u^{\prime I} = R^{I}_{\ J}u^{J}=\begin{pmatrix}1\\0\\0\\1\end{pmatrix}
\end{eqnarray*}
which results in the tetrad being aligned with the photon's 3-velocity, as illustrated in \figref{adaption}.

Such a rotation is explicitly given by
\be
R^{I}_{\;J}(u)=\delta_0^I\delta_J^0 -\hat r^{I}\hat r_{J}-\frac{u_3}u(P^{I}_{\;J}+\hat r^I\hat r_J) -\sqrt{1-\left(\frac{u_3}u\right)^2}\eps^{I}_{KJ0}\hat r^K\label{eq-photonadaption}
\ee
where $u_3/u=u_\mu e^\mu_3/u=-\cos\theta$ is the angle between the direction of the photon and the z-component of the tetrad, $r^I\equiv \eps^I_{J30}u^J$ is the spatial axis of rotation with $\hat r^{I} \equiv r^{I}/|r|$, and $P^I_{\ J}\equiv \delta^I_J-e^I_0e^0_J$ is the projector onto the spacelike hypersurface orthogonal to the tetrad time axis (see \figref{adaption}).

It is important to stress that there are several other possible choices for this spatial rotation corresponding to different conventions for the linear polarisation basis. Furthermore, the rotation matrix above becomes undefined for $\theta=\pi$ which is unavoidable for topological reasons.

The rotation $R^{I}_{\;J}$ induces a linear polarisation basis $\delta^{A}_{I}R^{I}_{\;J}$. We can now extract the components of the quantum state expressed in this basis as
\begin{eqnarray}
\psi^A= f^A_J\psi^J\text{,\quad with\quad }f^A_J\equiv\delta^{A}_{I}R^{I}_{\;J}.\label{qs-adapt-pol}
\end{eqnarray}
It is clear that the specific linear polarisation basis used here depends on how we have adapted the tetrad to the velocity of the photon. However, regardless of what convention one chooses, the quantum state $\psi^A$ is gauge invariant. Alternatively we could think of the quantum state directly in terms of an equivalence class of polarisation vectors $\psi^I\sim \psi^I+\upsilon u^I$ orthogonal to photon velocity $u^I$.  The advantage of this approach is that once one has developed a gauge invariant formalism one can work solely with the gauge covariant polarisation vector $\psi^I$. This will be addressed below and in \secref{sec-photonunitaritymmt}.

$f^A_I$ from \eqref{qs-adapt-pol} turns out to provide a `diad' frame: The two vectors $f^1_I$ and $f^2_I$ span the two-dimensional space orthogonal to both the photon's 4-velocity $u_I$ and the time component of a tetrad $e^t_I$. If we let $f^I_A$ be the inverse of $f^A_I$ we have that $f^I_Af^A_J = \delta^I_J$ and $f^I_Au_I = f^I_Ae^t_I=0$. In fact, if we define $w_I$ to be a null vector defined by $e^t_I=\half(u_I+w_I)$ \cite{Poisson}, the vectors $u_I$, $w_I$, $f^1_I$ and $f^2_I$ span the full tangent space. This decomposition will be useful when identifying unitary operations in \secref{sec-photonunitaritymmt}.

%%%%%%%%%%%%%%%%%%%%%%%%%%%%%%%%%%%%%%%%%%%%%%%
\subsection{The inner product and Hilbert space\label{newfootnote}\label{sec-photonIP}}
%%%%%%%%%%%%%%%%%%%%%%%%%%%%%%%%%%%%%%%%%%%%%%%

We must identify an inner product on the complex vector space for polarisation so that it can be promoted to a Hilbert space. In the analysis of the WKB limit we found that $j^{I} = -\sqrt{g}k^{I}\varphi^2\eta_{JK}\psi^{J}\bar{\psi}^{K}$ corresponded to a conserved 4-current which was physically interpreted as a conserved probability density current. A natural inner product between two polarisation 4-vectors $\psi^{I}$ and $\phi^{J}$ is then given by
\begin{eqnarray}\label{photonIP}
-\eta_{IJ}\bar\phi^{I}\psi^{J}.
\end{eqnarray}
This form is clearly sesquilinear and positive definite for spacelike polarisation vectors.\footnote{There is no primed index for conjugate terms $\bar\phi^I$ because the vector representation of the Lorentz group is real.} Unlike the case for fermions, the inner product $\eta_{IJ}$ is not explicitly dependent on the photon 4-velocity. However, if we consider the gauge transformation $\psi^{I}\to\psi^{I}+\upsilon_1 u^{I}_{1}$ and $\phi^{I}\to\phi^{I}+\upsilon_2 u^{I}_{2}$ it is clear that unless $u^{I}_{1}=u^{I}_{2}$, i.e.\;$k^{I}_{1}\propto k^{I}_{2}$, the inner product \eqref{photonIP} is not gauge invariant. We conclude that two polarisation vectors corresponding to two photons with non-parallel null velocities do not lie in the same Hilbert space. Furthermore, in order to be able to {\em coherently} add two polarisation states it is  also necessary to have $k_1^I=k_2^I$, i.e.\; the two photons must have the same frequency. Under such conditions the inner product is both Lorentz invariant and gauge invariant. With the inner product \eqref{photonIP} the complex vector space of polarisation vectors is promoted to a Hilbert space which is notably labelled again with both position and 4-momentum $p^I=\hbar k^I$.

In fact, it is a little more subtle. Even though it is not sensible to superpose polarisations for different frequencies on the same trajectory, this does not imply that the polarisation spaces are inequivalent and cannot be factorised into a tensor product (see \cite{PeresTerno03b}). The Hilbert space structure with tensor products of spaces for different frequencies but the same 3-momentum ray would be permitted:
\[
\Hi_\text{total}=\oplus_{\hat{\mathbf p}\in\mathbb R^3}(\mathbb R_{\hat{\mathbf p}}\otimes\Hi_{\mathbf{\hat p}}),
 \]
where $\mathbb R_{\hat{\mathbf p}}$ encodes the frequency $\abs{\mathbf p}$.

The above inner product \eqref{photonIP} reduces to the standard inner product for a two-dimensional Hilbert space. This is best seen through the use of an adapted tetrad. In an adapted frame the inner product of $\psi^I=(\nu,\psi^1,\psi^2,\nu)$ with some other polarisation vector $\phi^I=(\mu,\phi^1,\phi^2,\mu)$ is given by
\begin{eqnarray*}
-\eta_{IJ}\bar{\phi}^I\psi^J=-\bar{\mu}\nu+\bar{\phi}_{1}\psi^1+\bar{\phi}_{2}\psi^2+ \bar{\mu}\nu=\bar{\phi}_1\psi^1+\bar{\phi}_{2}\psi^2=\langle\phi|\psi\rangle.
\end{eqnarray*}
Thus, the standard inner product $\langle\phi|\psi\rangle=\bar{\phi}_{1}\psi^1+\bar{\phi}_{2}\psi^2$ is simply given by $\bk\phi\psi=-\eta_{IJ}\bar{\phi}^I\psi^J$, where we associate
\begin{eqnarray*}
\ket{\psi} \sim \psi^{I}\quad \mbox{and}\quad \bra{\phi} \sim -\bar\phi_{I} = -\eta_{IJ}\bar\phi^{J}.
\end{eqnarray*}
We can now work directly with the polarisation 4-vector $\psi^{I}$ which transforms in a manifestly Lorentz covariant and gauge covariant manner.

%%%%%%%%%%%%%%%%%%%%%%%%%%%%%%%%%%%%%%%%%%%%%%%%%%%%%%%
\subsection{The relation to the Wigner formalism\label{secphotonWigner}}
%%%%%%%%%%%%%%%%%%%%%%%%%%%%%%%%%%%%%%%%%%%%%%%%%%%%%%%

The Wigner rotation $W_A^{\ B}(k,\Lambda)$ on the quantum state represented by the Jones vector which results from the transport of the polarisation vector can be identified in the same way as was done in \S\ref{fermionwignerrot} for fermions. Specifically, this is achieved by determining the evolution of the Jones vector $\psi^{A}$ that is induced by the transport of the quantum state represented by the polarisation 4-vector $\psi^{I}$. Substituting $\psi^{I} = f^{I}_{B} \psi^{B}$ in the transport equation \eqref{photonPT}, $\frac{\di \psi^{I}}{\di \lambda} + u^{\mu}\omega_{\mu \ J}^{\ I}\psi^{J}= \beta u^{I}$, and multiplying by $f^A_I$, we obtain
\begin{eqnarray}
\frac{\di \psi^{A}}{\di \lambda} = - \left(u^{\mu}f_{I}^{A}\omega^{\ I}_{\mu\ J} f^{J}_{B} +f^A_I\frac{\di f^I_B}{\di\lambda}\right)\psi^B + \beta f_{I}^{A} u^{I}.\label{photonqstransport}
\end{eqnarray}
The last term is zero, as $f^A_I$ is the diad frame defined to be orthogonal to $u^{I}$. If we first consider \eqref{photonqstransport} in an adapted tetrad as in \S\ref{sec-adaptedphotonQS} we see that the derivative $\frac{\di f_B^I}{\di\lambda}$ vanishes and the remaining term on the right-hand side can be simplified to
\begin{eqnarray}
\frac{\di \psi^{A}}{\di \lambda} = \ii u^{\mu}\omega_{\mu 12}{\sigma^y}^A_{\ B} \psi^{B}\label{adaptedPQST}
\end{eqnarray}
where we have made use of the antisymmetry of the spin-1 connection in order to introduce the antisymmetric  Pauli Y matrix ${\sigma^y}^A_{\ B}$. \footnote{Note again that this should not be confused with the  $\sigma$-matrices encountered when working with spinors.} \Eeqref{adaptedPQST} is then clearly unitary and helicity preserving as it is proportional to ${\sigma^y}^A_{\ B}$ in the linear polarisation basis. We can now readily identify the infinitesimal Wigner rotation as $W^A_{\ B} = \ii u^{\mu}\omega_{\mu 12} {\sigma^y}^A_{\ B} $, where the rotation angle is $u^{\mu}\omega_{\mu 12}$. In a non-adapted tetrad frame the map $f^I_A=\delta^J_AR_J^{\ I}$ can be seen to put \eqref{photonqstransport} in the form \eqref{adaptedPQST} with a modified spin-1 connection $\omega'^{\ I}_{\mu\ J}$. The Wigner rotation for non-adapted tetrads is then
\begin{eqnarray}
W^A_{\ B}=\ii u^\mu(R^{\;I}_1\partial_\mu R^{2}_{\;I}+R_1^{\;I}\omega_{\mu I}^{\ \ J} R^{2}_{\;J}){\sigma^y}^A_{\ B}.\label{eq-genphotonWigner}
\end{eqnarray}
A general Wigner rotation is understood as the composition of maps $W^A_{\ B}\equiv f_{B}^{I}(\Lambda u) \Lambda_{I}^{\ J}f_{J}^{A}(u)$. It is therefore no surprise that the transport of the polarisation vector induces a Wigner rotation: The action of the gravitational field along a trajectory is simply a sequence of infinitesimal Lorentz transformations which are given by $u^{\mu}\omega_{\mu\ J}^{\ I}$ (\S\ref{localqubits}). The transport of the Jones vector is therefore described by a sequence of infinitesimal Wigner rotations given by \eqref{eq-genphotonWigner}. Notice that the Wigner rotation takes on a form which is not manifestly Lorentz covariant. This is because the Wigner rotation describes a {\it spatial} rotation. This should be contrasted with the manifestly Lorentz covariant representation in terms of parallel transported polarisation vectors.

\subsection{Calculating the state transformation\label{sec-photonstatetransformation}}
As in Sections \ref{sec-statetransformation} and \ref{sec-fermionstatetransformation}, and as in \cite{Alsingphotons} we want to determine the final state transformation for a polarisation quantum state along a null geodesic $\psi(\lambda_\text{end})=\hat{\mathrm T}\psi(\lambda_\text{start})$. For the polarisation 4-vector transport the differential equation is different to previous cases. The polarisation 4-vector is parallel transported with an arbitrary gauge transformation, \eqref{photonPT}. We can write this transport in tensor indices $\mu$ by \eqref{eq-coordVPT}, or in tetrad indices $I$ where the parallel transport is given in terms of the spin-1 connection rather than the affine connection:
\[
\frac{\di \psi_\mu}{\di \lambda}=-u^\nu g^{\rho\sigma}\Gamma^\mu_{\nu\rho}\psi_\sigma+\beta u_\mu,\qquad \frac{\di \psi^I}{\di\lambda}=-u^\nu\omega_{\nu\ J}^{\ I}\psi^J+\beta u^I.
\]
Notice that these have a nonhomogeneous term due to the $\beta u^I$ gauge transformation. The solution to the homogeneous part is
\[
\hat{\mathrm T}_h(\lambda_\text{start};\lambda_\text{end})=\m T\exp\left[\int_{\lambda_\text{start}}^{\lambda_\text{end}}\hat R(\lambda)\di\lambda \right],\qquad \hat{\mathrm T}_h(\lambda_1;\lambda_1):=\hat\sigma^0
\]
where $\exp$ is the matrix exponential and $\m T$ is the time-ordering operator for effects at different times which do not commute \cite{Weinberg,PeskinSchroeder}. The solution to the nonhomogeneous differential equation $\di\psi^I/\di\lambda=R^I_{\ J}(\lambda)\psi^J+\beta(\lambda)u^I(\lambda)$ is \cite{FinanDE}
\[
\psi^I(\lambda_2)={\mathrm T_h}^I_{\ J}(\lambda_\text{start};\lambda_\text{end})\left(\psi^J(\lambda_\text{start})+\int_{\lambda_\text{start}}^{\lambda_\text{end}}\beta(\lambda') {{\mathrm T}^{-1}_h}^J_{\ K}(\lambda_\text{start};\lambda')u^K(\lambda')\;\di\lambda'\right)
\]
and similarly for the transport equation in tensor indices. Recall that the $\beta(\lambda)$ were arbitrary and should give an overall gauge transformation $B\, u^I(\lambda_\text{end})$ on the final polarisation vector. In fact, since $\hat{\mathrm T}_h$ is time ordered, we have that ${\mathrm T_h}^I_{\ J}(\lambda_\text{start};\lambda_\text{end}){{\mathrm T}^{-1}_h}^J_{\ K}(\lambda_\text{start};\lambda')={\mathrm T_h}^I_{\ K}(\lambda';\lambda_\text{end})$. Next, since $\hat{\mathrm T}_h$ is parallel transport, we have that ${\mathrm T_h}^I_{\ K}(\lambda';\lambda_\text{end})u^K(\lambda')=u^I(\lambda_\text{end})$, and so indeed the nonhomogeneous solution is
\begin{align*}
\psi^I(\lambda_2)
%=&{\mathrm T_h}^I_{\ J}(\lambda_\text{start};\lambda_\text{end})\left(\psi^J(\lambda_\text{start})+\int_{\lambda_\text{start}}^{\lambda_\text{end}}\beta(\lambda') {{\mathrm T}^{-1}_h}^J_{\ K}(\lambda_\text{start};\lambda')u^K(\lambda')\;\di\lambda'\right)\\
%=&{\mathrm T_h}^I_{\ J}(\lambda_\text{start};\lambda_\text{end})\psi^J(\lambda_\text{start})+\int_{\lambda_\text{start}}^{\lambda_\text{end}}\beta(\lambda') {\mathrm T_h}^I_{\ J}(\lambda_\text{start};\lambda_\text{end}){{\mathrm T}^{-1}_h}^J_{\ K}(\lambda_\text{start};\lambda')u^K(\lambda')\;\di\lambda'\\
%=&{\mathrm T_h}^I_{\ J}(\lambda_\text{start};\lambda_\text{end})\psi^J(\lambda_\text{start})+\int_{\lambda_\text{start}}^{\lambda_\text{end}}\beta(\lambda') {\mathrm T_h}^I_{\ K}(\lambda';\lambda_\text{end})u^K(\lambda')\;\di\lambda'\\
%=&{\mathrm T_h}^I_{\ J}(\lambda_\text{start};\lambda_\text{end})\psi^J(\lambda_\text{start})+\int_{\lambda_\text{start}}^{\lambda_\text{end}}\beta(\lambda')u^I(\lambda_\text{end})\;\di\lambda'\\
=&{\mathrm T_h}^I_{\ J}(\lambda_\text{start};\lambda_\text{end})\psi^J(\lambda_\text{start})+u^I(\lambda_\text{end})\int_{\lambda_\text{start}}^{\lambda_\text{end}}\beta(\lambda')\;\di\lambda'.
\end{align*}
%
%%
%By gauge fixing, specifically by setting $\beta=0$, one could recover the homogeneous solution.

As in the fermion case in \secref{sec-fermionstatetransformation}, we have also provided the Wigner transport equation \eqref{adaptedPQST} with \eqref{eq-genphotonWigner} for a non-adapted frame:
\[
\frac{\di \psi^{A}}{\di \lambda} = \ii u^\mu(R^{\;I}_1\partial_\mu R^{2}_{\;I}+R_1^{\;I}\omega_{\mu I}^{\ \ J} R^{2}_{\;J}){\sigma^y}^A_{\ B}\psi^{B}.
\]
For calculating polarisation transport it is far easier to use the Wigner formalism: in this formalism the infinitesimal evolution is always proportional to a single matrix, $\hat\sigma^y$. Since all effects thus commute, the integral does not require a time ordering operator \cite{Weinberg}. There are also no gauge degrees of freedom. The transformation is
\[
\psi^A={R_y}(\theta)^A_{\ B}\psi^B,\qquad R_y(\theta)=\bbm\cos\theta&-\sin\theta\\\sin\theta&\cos\theta\ebm,
\]
a rotation around the $y$ axis, with angle
\[
\theta=-\int_{\lambda_\text{start}}^{\lambda_\text{end}} \frac{\di x^\mu}{\di\lambda}\omega_{\mu 12}(\lambda)\;\di\lambda
\]
for the adapted frame. Replace $\omega_{\mu12}\to\left(R^{\;I}_1(\lambda)\frac{\di R^{2}_{\;I}(\lambda)}{\di x^\mu}+R_1^{\;I}(\lambda)\omega_{\mu I}^{\ \ J}(\lambda) R^{2}_{\;J}(\lambda)\right)$ for a non-adapted frame, where $R^I_{\ J}$ are components of the rotation adapting the tetrad \eqref{eq-photonadaption}. In \secref{sec-photonunitaritymmt} we will develop a measurement formalism to complete the mathematics of the covariant formalism, as well as describe the relationship between the spaces of operators in the 4-vector and Jones vector formalisms.

%%%%%%%%%%%%%%%%%%%%%%%%%%%%%%%%%%%%%%%%%%%%%%%%%%%%%%%%%%%%%%%%%%%
\section{Phases and interferometry}\label{secPhase}
%%%%%%%%%%%%%%%%%%%%%%%%%%%%%%%%%%%%%%%%%%%%%%%%%%%%%%%%%%%%%%%%%%%%

So far we have determined the transport of the quantum state of a single qubit along one spacetime trajectory. If we inspect the transport equations \eqref{fermionTP} and \eqref{betatrans} we see that neither one contains a term proportional to the identity ($\delta_A^B$ for fermions and $\delta^I_J$ for photons). Such a term would lead to an overall accumulation of global phase $\Ee^{\ii\theta}\psi_A$ or $\Ee^{\ii\theta}\psi^I$. This leads one to suspect that not all of the possible contributions to the global phase have been taken into account in these transport equations. Indeed this is the case, as can be seen immediately by considering the full wavepacket in the WKB approximation
\begin{eqnarray*}
\Psi_\sigma(x)= \varphi(x)\psi_\sigma(x)\Ee^{\ii\theta(x)}
\end{eqnarray*}
where $\varphi(x)$ is a real-valued envelope and $\sigma=1,2$ for fermions or $\sigma=0,1,2,3$ for photons.\footnote{In the case of a scalar particle (and thus not a qubit) there are still gravitational phases, and here the index $\sigma$ can just be removed.} Clearly there is an additional phase $\theta(x)$ which is not included in $\psi_\sigma(x)$.

Since global phase is unobservable this is of course of no concern if we restrict ourselves to a qubit moving along a single trajectory. However, quantum mechanics allows for more exotic experiments where a single qubit is split up into a spatial superposition, simultaneously transported along multiple distinct paths, and recombined so as to produce quantum interference phenomena. Here it becomes necessary to keep track of the phase difference between the components of the spatial superposition in order to be able to predict the measurement probabilities at the detectors.

In this section we will extend the formalism in this chapter to include gravitationally induced phase difference in experiments involving path superpositions. The formalism will be derived from equations of the WKB approximation together with the assumptions of localisation. With these assumptions, the details of the spatial profile of the qubit become irrelevant, and we can satisfactorily describe the experiment solely in terms of a phase difference $\Delta \theta$ between two quantum states. This phase difference will depend on the spacetime geometry $g_{\mu\nu}$ and the trajectories along which the qubit is simultaneously transported. We show how the various sources for the phase difference can be understood from a wave-geometric picture. We lastly apply the formalism of this chapter to gravitational neutron interferometry \cite{Colella,Sakurai,Werner} and obtain an exact general relativistic expression for the phase difference which in various limits reproduces the results in \cite{Anandan,AudretschLammerzahl,VarjuRyder} in which higher order corrections to the non-relativistic result were proposed.

%%%%%%%%%%%%%%%%%%%%%%%%%%%%%%%%%%%%%%%%%%%%%%%%%%%%%%%%%%%%%%%%%%%%
\subsection{Spacetime Mach--Zehnder interferometry}\label{phasedef}
%%%%%%%%%%%%%%%%%%%%%%%%%%%%%%%%%%%%%%%%%%%%%%%%%%%%%%%%%%%%%%%%%%%%

\begin{figure}[h]
\begin{center}
\ifpdf
\includegraphics[height=60mm]{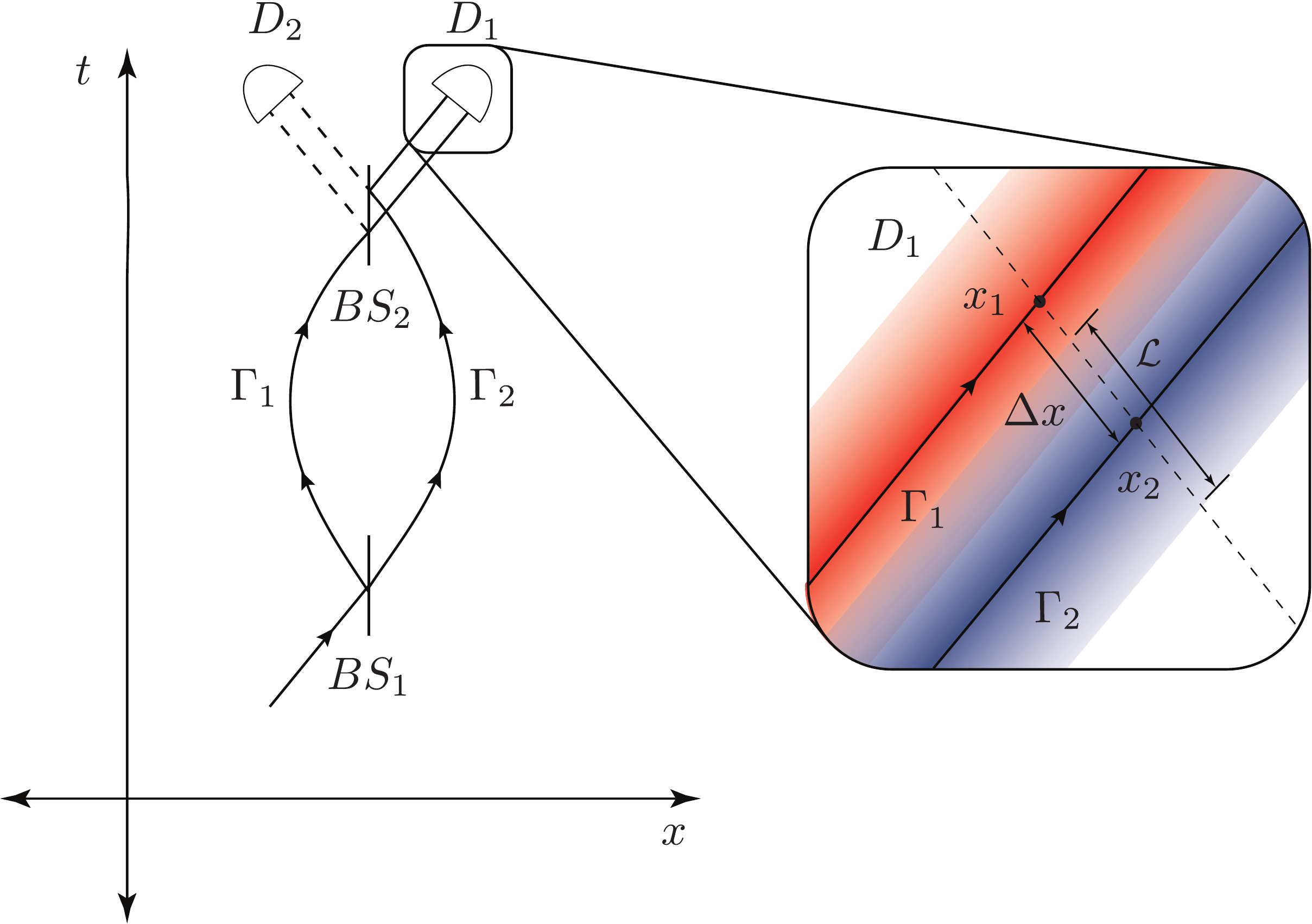}
\else
\includegraphics[height=60mm]{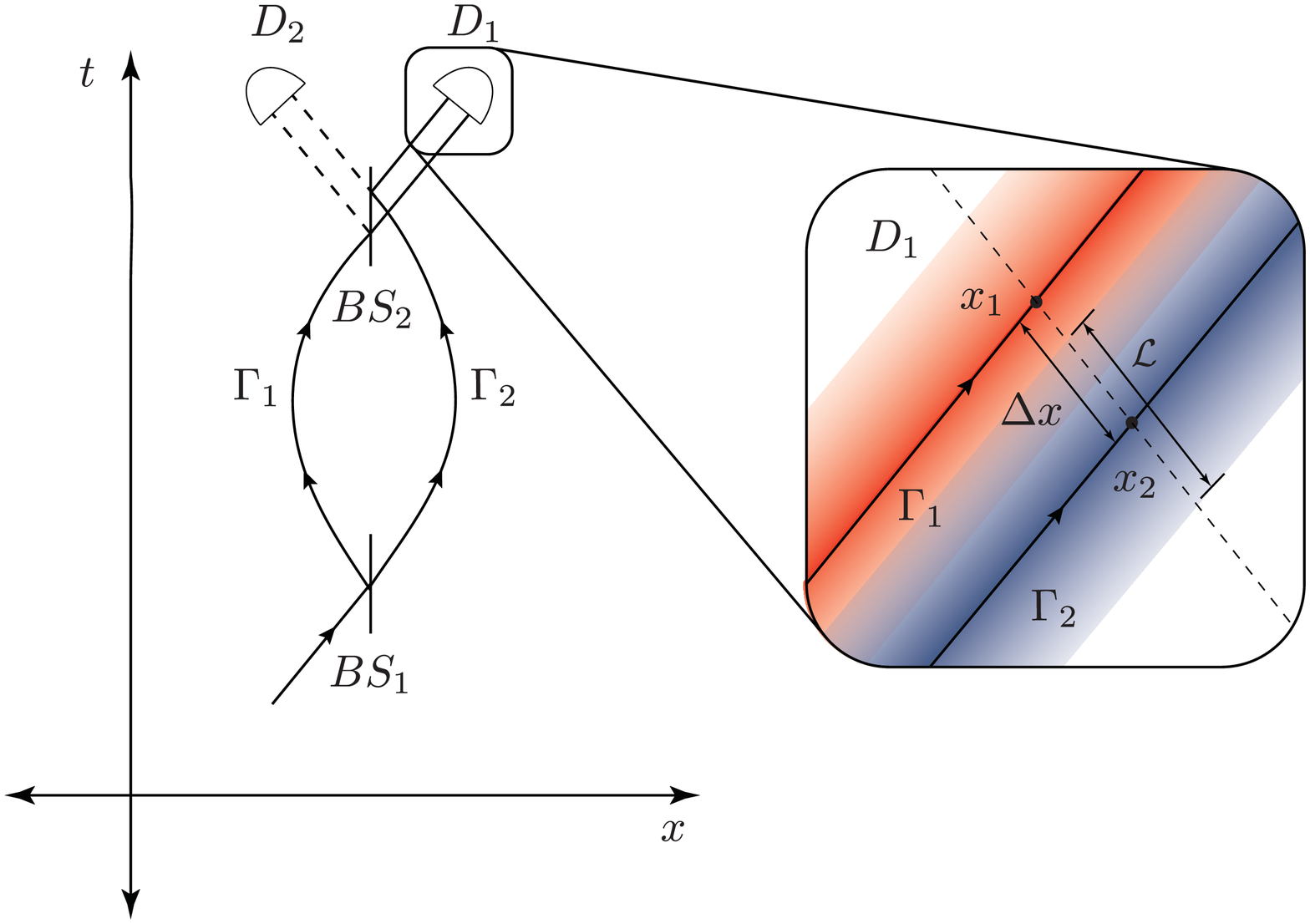}
\fi
\caption[Spacetime Mach-Zehnder interferometer (colour).]{Spacetime figure of a Mach--Zehnder type interferometer, illustrating a single qubit subjected to a beam splitter $BS_1$ resulting in a superposition of the qubit travelling along two distinct spacetime paths $\Gamma_1$ and $\Gamma_2$. In some future spacetime region containing the beam splitter $BS_2$ the components of the spatial superposition are assumed to recombine to produce possible interference phenomena in the detector regions $D_1$ and $D_2$. These regions contain two trajectories, indicating that the times of arrival at the second beam splitter $BS_2$ are not in general the same. The variables $x_{1,2}$ are arbitrary spacetime points in the region $D_1$ along trajectories $\Gamma_1$ and $\Gamma_2$ and are useful for calculating the total phase difference. The red and blue strips represent the spatial extents of the wavepackets along $\Gamma_{1}$ and $\Gamma_{2}$ respectively. These correspond to the length $\m L$ of the wavepacket as measured along the line joining the points $x_{1}$ and $x_{2}$.
\label{fig-spacetimeMZ}}
\end{center}
\end{figure}

We consider, as a concrete example of an interference experiment, standard Mach--Zehnder interferometry. As usual, there is a qubit incident on a beam splitter (e.g.\;a half-silvered mirror) which creates a spatial superposition of the qubit. The two components of the spatially superposed state (each assumed to be spatially well-localised) are then transported along two different paths and later made to interfere using another beam splitter. This produces two output rays each incident on a particle detector, as illustrated in \figref{fig-spacetimeMZ}. The two paths are $\Gamma_1$ and $\Gamma_2$, each integral curves $x^\mu_{1,2}(\lambda)$ of velocities $u_{1,2}^\mu(\lambda)$. We then have two families of Hilbert spaces. A Cauchy surface will intersect the trajectories at $\lambda_1,\lambda_2$, and for any such surface the Hilbert space structure is $\Hi_1(\lambda_1)\oplus\Hi_2(\lambda_2)$ and the norm of the state on the combined Hilbert space is one (see \S\ref{secQIinCST}).

Let us now focus our attention on a small region $D_1$ situated on the right output arm.
There are then two classical paths $\Gamma_1$ and $\Gamma_2$ which arrive at $D_1$, as illustrated in \figref{fig-spacetimeMZ}. Note that in order for us to derive a formalism in terms of quantum states, the wavevectors $k^\mu_1$ and $k^\mu_2$ of $\Gamma_1$ and $\Gamma_2$ must be approximately equal in this region, i.e.\;$k^\mu_1=k^\mu_2=k^\mu$. This is because the Hilbert space of a quantum state is labelled with momentum, as explained in \S\ref{secfermionQS} and \S\ref{sec-photonIP}.

 In general the times of arrival of the two paths $\Gamma_1$ and $\Gamma_2$ at the second beam splitter $BS_{2}$ will differ for the two paths. As we shall see this contributes to the total phase difference between the two packets.

Let $x$ be some suitable local Lorentz coordinate system in region $D_1$. The wavepacket in region $D_1$ is then given by the superposition
\begin{eqnarray}
a\Psi^{(1)}_\sigma(x)+b\Psi^{(2)}_\sigma(x)\label{additionofwavepackets}
\end{eqnarray}
where $\Psi^{(1)}_\sigma(x)$ and $\Psi^{(2)}_\sigma(x)$ are the packets propagated along $\Gamma_1$ and $\Gamma_2$ respectively. $a$ and $b$ are determined from the reflection and transmission coefficients of the various beam splitters in the experiment. In the case of 50-50 beam splitters, $a=b=\frac{\ii}{\sqrt{2}}$ in region $D_1$ (see \figref{fig-spacetimeMZ}). We will ignore any overall global phase factor resulting from reflections.

%%%%%%%%%%%%%%%%%%%%%%%%%%%%%%%%%%%%%%%%%%%%%%%%%%%%%%%%%%%%%%%%%%%%%%%%%%%%%%%%%%%%%%%%%%%%%%%%%%%%%
\subsection{The phase difference from the WKB approximation \label{addingstates}}\label{phasediff}
%%%%%%%%%%%%%%%%%%%%%%%%%%%%%%%%%%%%%%%%%%%%%%%%%%%%%%%%%%%%%%%%%%%%%%%%%%%%%%%%%%%%%%%%%%%%%%%%%%%%%
In order to make empirical predictions in a Mach--Zehnder type interference experiment we must determine explicitly the forms of $\Psi^{(1)}_\sigma(x)$  and $\Psi^{(2)}_\sigma(x)$ in \eqref{additionofwavepackets} in the detector region $D_{1}$. By making use of the field equations in the WKB limit and the localisation assumptions we will see that $\Psi^{(1)}_\sigma(x)$  and $\Psi^{(2)}_\sigma(x)$ will differ by a phase accumulated along the trajectory and a rigid translation/displacement, resulting in an overall phase difference. The derivations differ in the cases of fermions and photons and we will treat them separately.

%%%%%%%%%%%%%%%%%%%%%%%%
\subsubsection{Fermions}
%%%%%%%%%%%%%%%%%%%%%%%%
In the small region $D_1$ the wavepacket in the WKB approximation is given by
\begin{equation}
a\phi^{(1)}_A(x)+b\phi^{(2)}_A(x)=a\varphi_1(x)\psi_A^{(1)}(x)\Ee^{\ii\theta_1(x)}+ b\varphi_2(x)\psi_A^{(2)}(x)\Ee^{\ii\theta_2(x)}\label{superposwavefunctions}
\end{equation}
where $x$ is some local Lorentz coordinate system, and $a$ and $b$ are real-valued coefficients. The functions $\phi_i(x)$, $\psi^{(i)}_{A}(x)$ and $\theta_i(x)$ ($i=1,2$) are defined in \secref{fermsemiclassapprox}. We are now going to successively make use of the equations of the WKB approximation and the localisation assumptions to simplify the expression \eqref{superposwavefunctions} and thereby extract the relative phase difference between the two components in the superposition.

First we use the fact that under the mathematical assumptions detailed in \S\ref{fermionlocalizationsubsec} the envelope will be transported rigidly and will not distort. Therefore, $\varphi_1(x)$ and $\varphi_2(x)$ will differ at most up to a rigid translation and rotation. We assume that the packet is `cigar shaped' and is always oriented in the direction of motion. The final envelopes will then differ at most up to a translation and we can write $\varphi_i(x)=\varphi(x-x_i)$ for some suitable function $\varphi(x)$ and an arbitrary choice of spacetime points $x^\mu_1,x^\mu_2\in D_1$ situated on the trajectories $\Gamma_1,\Gamma_2$ respectively (see \figref{fig-spacetimeMZ}).

If we now assume that
\begin{eqnarray*}
\frac{(x^\mu-x_i^\mu)\nabla_\mu\varphi(x)}{\varphi(x)}\ll1,\qquad i=1,2
\end{eqnarray*}
for all points $x\in\m D_1$, the difference in the envelopes $\varphi_1(x)$ and $\varphi_2(x)$ is negligible. The translational difference in the envelopes can then be neglected and factored out:
\begin{eqnarray}
a\phi^{(1)}_A(x)+b\phi^{(2)}_A(x)\approx\varphi(x)\left(a\psi_A^{(1)}(\Gamma_1)\Ee^{\ii\theta_1(x)}+ b\psi_A^{(2)}(\Gamma_2)\Ee^{\ii\theta_2(x)}\right)\label{locsuperposwavefunctions}
\end{eqnarray}
where $\psi_A^{(i)}(\Gamma_i)$ are determined by integrating the transport equation \eqref{fermionTP} to the points $x_1$, $x_2$, respectively. Thus for the purpose of interferometry the details of the envelope become irrelevant and can be ignored.

We now focus on the phase $\theta_1$ and $\theta_2$. As we pointed out in \secref{newinnerproduct}, in order to coherently add two quantum states it is necessary to assume that the wavevectors of the packets are the same, $k_1^\mu=k_2^\mu=k^\mu$. Therefore in the region $D_1$ we have from the WKB approximation that the phases $\theta_1(x)$ and $\theta_2(x)$ both satisfy the equation
\begin{eqnarray}\label{fermphaseq}
\nabla_\mu\theta=k_\mu+e A_\mu.
\end{eqnarray}
Within the small region  $D_1$ we regard $k_\mu(x)$ and $A_\mu(x)$ as constant and so the partial differential equation \eqref{fermphaseq} has the solution $\theta_i(x)=(k_\mu+e A_\mu)(x^\mu-x^\mu_i)+\theta_i(x_i)$ where $\theta_i(x_i)$ are two integration constants corresponding to the value of $\theta_i(x)$ at the points $x_i$. These integration constants can be determined by integrating \eqref{fermphaseq} along the trajectories $\Gamma_{1,2}$ to the positions $x_{1,2}$ respectively, i.e.\;\begin{eqnarray}\label{fermintphase}
\theta_i(x_i)=\int_{\Gamma_i} (k_\mu+eA_\mu)\di x^\mu+\theta_0
\end{eqnarray}
 where $\theta_0$ is some arbitrary global phase just before the wavepacket was split up by the first beam splitter. Using the above we can rewrite \eqref{locsuperposwavefunctions} as
\begin{eqnarray}\label{waveaddition}
a\phi^{(1)}_A(x)+b\phi^{(2)}_A(x)\approx\varphi(x)\Ee^{\ii\theta_1(x)}\left(a\psi_A^{(1)}(\Gamma_1)+ b\psi_A^{(2)}(\Gamma_2)\Ee^{\ii\Delta\theta}\right)
\end{eqnarray}
where
\begin{eqnarray}
\Delta\theta = (k_\mu+e A_\mu)(x_1^\mu-x^\mu_2)+(\theta_2(x_2)-\theta_1(x_1)).\label{fermionphasedifference}
\end{eqnarray}
It is important to note that this phase difference is independent of $x^\mu\in D_1$ as all dependence on $x$ has been factored out in \eqref{waveaddition}. Furthermore, we note that the choice of $x_1$ and $x_2$ is arbitrary and the phase difference $\Delta\theta$ is also independent of this choice. To see this, consider a different choice of positions, $x_1'=x_1+\delta x_1$ and $x_2'=x_2+\delta x_2$ on $\Gamma_1$ and $\Gamma_2$. This results in a change in the integration constants \eqref{fermintphase} of $\theta_i(x_i)\to\theta_i(x_i)+(k_\mu+eA_\mu)\delta x_i^\mu$ which exactly cancels the change in the term $(k_\mu+eA_\mu)(x_1^\mu-x^\mu_2)$ in \eqref{fermionphasedifference}. Therefore $\Delta\theta$ is independent of the arbitrary positions $x_1$ and $x_2$.

Note that $\Delta\theta$ is not the phase difference determined empirically in a Mach--Zehnder type interference experiment. This is because the transported quantum states $\psi_A^{(1)}(\Gamma_1)$ and $\psi_A^{(2)}(\Gamma_2)$ can contain an additional phase difference induced from their specific evolutions on the Bloch sphere. This transport induced phase difference can be determined from \cite{Bengtsson}
\begin{eqnarray}\label{transphase}
\Ee^{\ii\Delta\theta_\text{Trans}}=\frac{\langle\psi^{(1)}(\Gamma_1)|\psi^{(2)}(\Gamma_2)\rangle}{|\langle\psi^{(1)}(\Gamma_1)|\psi^{(2)}(\Gamma_2)\rangle|},
\end{eqnarray}
which determines complex phase between two quantum states that may not be parallel.\footnote{We note that a precessing spin can induce a reduction of visibility due to differing proper times experienced along each path, similar to the `clock' degree of freedom in \cite{Zych11}.} The region $D_1$ is assumed to be small enough that $\psi^{(1)}_A$ and $\psi^{(2)}_A$ do not vary significantly with changes in $x_1$ and $x_2$. Thus, the {\em total} phase difference $\Delta\theta_\text{Tot}$, which is the quantity that we actually measure in a Mach--Zehnder experiment, is then given by
\begin{eqnarray*}
\Delta\theta_\text{Tot}=\Delta\theta+\Delta\theta_\text{Trans}.
\end{eqnarray*}
This total phase difference $\Delta\theta_\text{Tot}$ can be determined completely from the trajectories $\Gamma_1$ and $\Gamma_2$ and the spacetime geometry $g_{\mu\nu}$ using the transport equation \eqref{fermionTP}. In particular, the phase difference measured by some detector in $D_1$ is independent of the motion of that detector.

Lastly, if we restrict ourselves to measurements that do not probe the spatial profile we can neglect the factor $\varphi(x)\Ee^{\ii\theta_1(x)}$  in \eqref{waveaddition}. All contributions to the phase difference are then contained in $\psi_A^{(\ii)}$ and $\Delta\theta$, and so at $D_1$ we are left with the two-dimensional quantum state
\begin{eqnarray*}
\ket{\psi}_\text{recomb}=a\psi_A^{(1)}(\Gamma_1)+ b\psi_A^{(2)}(\Gamma_2)\Ee^{\ii\Delta\theta}.
\end{eqnarray*}
Therefore the assumptions that led to \eqref{waveaddition} established a formalism for determining the resulting qubit quantum state in region $D_1$ of a Mach--Zehnder type interferometer.

%%%%%%%%%%%%%%%%%%%%%%%%%%%%%
\subsubsection{Photons}
%%%%%%%%%%%%%%%%%%%%%%%%%%%%%

The derivation of the phase difference for photons follows essentially the same path as that for fermions. The starting point is to consider the wavepacket in the small region $D_1$
\begin{equation}
aA_{(1)}^I(x)+bA_{(2)}^I(x)=a\varphi_1(x)\psi^I_{(1)}(x)\Ee^{\ii\theta_1(x)}+ b\varphi_2(x)\psi^I_{(2)}(x)\Ee^{\ii\theta_2(x)}\label{superposwavefunctionsphotons}
\end{equation}
where $x$ is some local Lorentz coordinate system, and $a$ and $b$ are real-valued coefficients. The functions $\varphi_i(x)$, $\psi_{(i)}^{I}(x)$ and $\theta_i(x)$ ($i=1,2$) are defined in \secref{photongeoapprox}. We then make use of the equations of the WKB approximation and the localisation assumptions to simplify the expression \eqref{superposwavefunctionsphotons}. As in the case of fermions, this means that the envelopes $\varphi_1(x)$ and $\varphi_1(x)$ are rigidly transported along their respective trajectories and so they differ at most by a translation i.e.\;$\varphi_i(x)=\varphi(x-x_i)$.

Again, if we assume the change in the envelope is small
\begin{eqnarray*}
\frac{(x^\mu-x_i^\mu)\nabla_\mu\varphi(x)}{\varphi(x)}\ll1,\qquad i=1,2
\end{eqnarray*}
for all points $x\in\m D_1$, the translational difference in the envelopes can be neglected and can be factored out:
\begin{eqnarray*}
aA_{(1)}^I(x)+bA_{(2)}^I(x)\approx\varphi(x)\Ee^{\ii\theta_1(x)}\left(a\psi^I_{(1)}(\Gamma_1)+ b\psi^I_{(2)}(\Gamma_2)\Ee^{\ii\Delta\theta(x)}\right)
\end{eqnarray*}
where $\Delta\theta = \theta_{2}(x)-\theta_{1}(x)$. We then solve the partial differential equation $ \nabla_{\mu}\theta=k_{\mu}$ to determine
\begin{eqnarray}\label{photonphase}
\theta_i(x)=k_\mu(x^\mu-x^\mu_i)+\theta_{i}(x_{i})+\theta_0
\end{eqnarray}
where $\theta_{i}(x_{i}) = \int_{\Gamma_i} k_\mu\di x^\mu$ are again integration constants. Using that $k_{\mu}$ is null and that we are integrating along its integral curves, we have $\int_{\Gamma_i} k_\mu\di x^\mu\equiv0$. Thus, the only contribution to the phase difference is
\begin{eqnarray}
\Delta\theta  =k_\mu(x_1^\mu-x^\mu_2).\label{photonsphasedifference}
\end{eqnarray}
Again note that this phase difference is independent of the position $x^\mu\in D_1$ at which the phase difference is computed. We also have that the phase difference $\Delta\theta$ is independent of the choice of points  $x_1$ and $x_2$. This follows since a change $x^{\mu}_i\to x^{\prime\mu}_i = x^{\mu}_i+\delta x^{\mu}_i=  x^{\mu}_i+\epsilon_{i} k^{\mu}$ leaves $\Delta\theta$ invariant since $k_\mu\delta x_i^\mu=0$.

As in the fermionic case there is also a phase difference $\Delta\theta_\text{Trans}$ defined by \eqref{transphase} related to the transport along the trajectories. What is actually measured in a Mach--Zehnder interference experiment is then
\begin{eqnarray*}
\Delta\theta_\text{Tot}=\Delta\theta+\Delta\theta_\text{Trans}.
\end{eqnarray*}
Just as in the case for fermions we now neglect the spatial part and we end up with the final qubit quantum state at $D_1$
\begin{eqnarray*}
\ket{\psi}_\text{recomb}=a\psi^I_{(1)}(\Gamma_1)+ b\psi^I_{(2)}(\Gamma_2)\Ee^{\ii\Delta\theta(x)}
\end{eqnarray*}
We have now obtained a formalism for describing interference experiments for photons solely in terms of two-dimensional quantum states.
%%%%%%%%%%%%%%%%%%%%%%%%%%%%%
\subsubsection{The recipe for adding qubit states}\label{recipe}
%%%%%%%%%%%%%%%%%%%%%%%%%%%%%%%
Above we have established a formalism for quantum interference phenomena for both fermions and photons in a Mach--Zehnder interference experiment. This description can be summarised by the following recipe for correctly adding the two quantum states:
\begin{enumerate}
\item Transport the quantum states $\psi_\sigma^{(1)}$ and $\psi_\sigma^{(2)}$ to the arbitrary positions $x_1$ and $x_2$ on the respective paths $\Gamma_1$ and $\Gamma_2$ in the recombination region $D_1$ using the appropriate transport equation, either \eqref{fermionTP} for fermions or \eqref{photonPT} for photons.
\item determine the integration constants $\theta_1$ and $\theta_2$. For fermions this is determined by \Eeqref{fermintphase}. For photons this is identically zero.
\item determine the phase difference $\Delta\theta$ using either \eqref{fermionphasedifference} or \eqref{photonsphasedifference}.
\item Finally, the two-dimensional quantum state in region $D_1$ is given by
\begin{eqnarray*}\ket{\psi}_\text{recomb}=a\psi_\sigma^{(1)}+ b\psi_\sigma^{(2)}\Ee^{\ii\Delta\theta}.
\end{eqnarray*}
\end{enumerate}

%%%%%%%%%%%%%%%%%%%%%%%%%%%%%%%%%%%%%%%%%%%%%
\subsection{The physical interpretation of phase in terms of wave geometry}
%%%%%%%%%%%%%%%%%%%%%%%%%%%%%%%%%%%%%%%%%%%%%
We now provide an intuitive wave-geometric picture for the various terms in the phase difference $\Delta\theta$. To do this we will focus on the specific case of fermions. Note however that the essential picture is also applicable to photons and we will comment on photons when necessary. The phase difference for fermions \eqref{fermionphasedifference} is given by
\begin{align}
\Delta\theta=&(k_\mu+eA_\mu)\Delta x^\mu + \int_{\Gamma_2} (k_\mu+eA_\mu)\di x^\mu-\int_{\Gamma_1} (k_\mu+eA_\mu)\di x^\mu \nonumber\\
=& \oint_\Gamma k_\mu \di x^\mu+e\oint_\Gamma A_\mu\di x^\mu
\end{align}
where $\Delta x^\mu\equiv x_1^\mu-x_2^\mu$ and $\Gamma= \Gamma_2+\Gamma_{2\to 1}-\Gamma_1$, where $\Gamma_{2\to 1}$ denotes the straight path going from point $x^\mu_2$ to $x^\mu_1$. The second term in the integral accounts for an Aharonov--Bohm phase. Let us consider the first term. The various contributions to this term are
\begin{eqnarray}
\oint_{\Gamma} k_\mu\di x^\mu = \int_{\Gamma_2} k_\mu\di x^\mu-\int_{\Gamma_1} k_\mu\di x^\mu + k_\mu\Delta x^\mu. \label{massphase}
\end{eqnarray}
The first two terms in the decomposition can each be thought of as representing the accumulation of global phase  along each trajectory, while the third is related to the displacement of the wavepackets. We now show how to interpret these two contributions wave-geometrically.

%%%%%%%%%%%%%%%%%%%%%%%%%%%%%%%%%%%%%%%%%%%%%
\subsubsection{The internal phase shift}
%%%%%%%%%%%%%%%%%%%%%%%%%%%%%%%%%%%%%%%%%%%%%

\begin{figure}[h]
a) $t_1$ \hspace{6cm} b) $t_2$\\
  \centering
\ifpdf
\includegraphics[height=25mm]{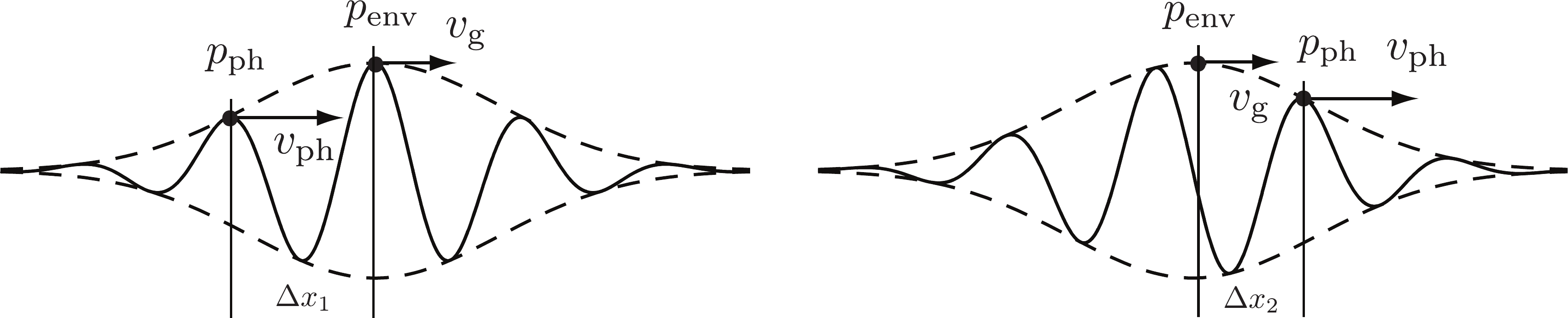}
\else
\includegraphics[height=25mm]{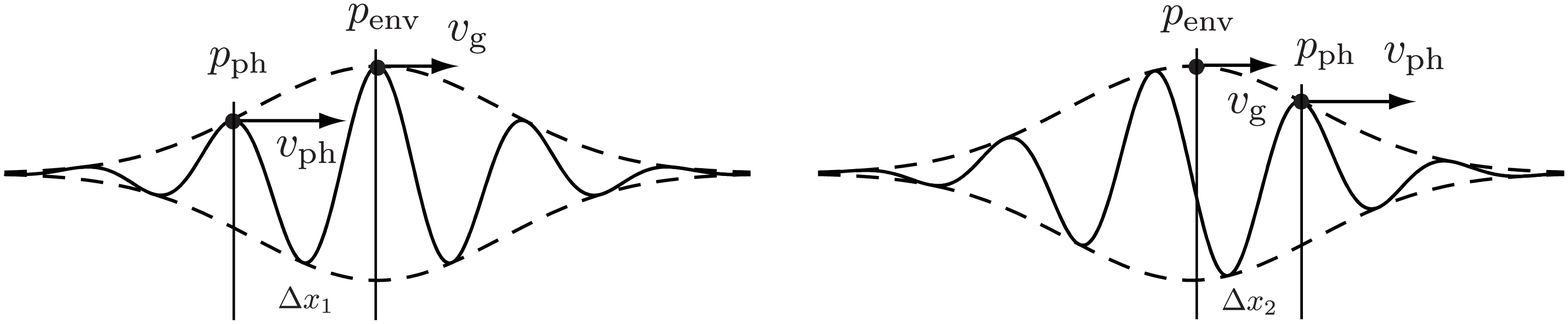}
\fi
\caption[Illustration of internal phase in wave envelopes]{An illustration of the accumulation of internal phase from a time $t_1$ (a) to a later time $t_2$ (b) along a trajectory. The internal phase $\theta_\text{int}=\Delta x_\text{int} /\lambdabar$ is determined by the difference in the offset $\Delta x_\text{int} = \Delta x_{2} - \Delta x_{1}$ of a point $p_\text{ph}$ of constant phase and a point $p_\text{env}$ of constant position on the envelope at the two times $t_1$ and $t_2$. For timelike packet velocities the phase velocity $v_\text{ph}$ is greater than the group velocity $v_\text{g}=c^2/v_\text{ph}<c$ so the internal phase is seen to accumulate along the trajectory.
\label{fig:inthaseoffsets}}
\end{figure}

The first two terms in \eqref{massphase} are integrals of the wavevector $k^\mu$ along the paths $\Gamma_i$, $i=1,2$. If we parameterise the paths with proper time $\frac{\di x^\mu}{\di\tau}=\frac{\hbar}{m}k^\mu$, the integrals become
\begin{eqnarray}\label{pathintphase}
\int_{\Gamma_i} k_\mu\di x^\mu= \int_{\Gamma_i}\di\tau \frac{mc^2}{\hbar}.
\end{eqnarray}
This results in a phase discussed in \cite{Stodolsky,Alsing} and motivated from the relativistic path integral. We can also understand this term in a simple wave-geometric picture. Consider a point $p_{\text{env}}(\tau)$ defined by $\varphi(x)=\text{const}$ which is fixed on the rigidly moving envelope (see \figref{fig:inthaseoffsets}). With respect to some arbitrary reference frame with 4-velocity $n^\mu$, the velocity at which $p_{\text{env}}$ moves in this frame is called the {\em group velocity} $v_\text{g}$ (see e.g.\;\cite{Rindler}), defined by $u^0=e^0_\mu u^\mu=\gamma_{v_\text{g}}=(1-v_\text{g}^2/c^2)^{-1/2}$, and corresponds to the particle's velocity. Secondly, consider a fixed phase point $p_{\text{ph}}$ defined by $\theta(x)=\text{const}$. The speed at which this phase point moves is given by $v_\text{ph} \equiv c^2/v_{\text{g}}$ and is called the {\em phase velocity}. Thus, if $v_{\text{g}}<c$ the points of constant phase move with respect to the wavepacket. It is this difference in velocity that results in the accumulation of the above mentioned path integral phase \eqref{pathintphase}. To see this, first calculate how much distance $\delta x_\text{int}$ is gained by $p_{\text{ph}}$ relative to $p_{\text{env}}$ during some time interval $\di t$ measured in this reference frame. This is given by
\begin{eqnarray*}
\delta x_\text{int} = p_{\text{ph}}-p_{\text{env}} = \left(\frac{c^2}{v_{\text{g}}}-v_{\text{g}}\right)\di t.
\end{eqnarray*}
In order to see how many radians of phase this distance is equivalent to we divide by the reduced wavelength $\lambdabar\equiv \hbar/p =  \hbar/m v_\text{g}$;
\begin{eqnarray*}
\frac{\delta x_\text{int} }{\lambdabar}=\frac{\left(\frac{c^2}{v_{\text{g}}}-v_\text{g}\right)\di t}{\lambdabar}=\frac{\frac{c^2}{v_\text{g}}\gamma^{-2}\di t m\gamma v_\text{g}}{\hbar}=\frac{mc^2}{\hbar}\gamma^{-1}\di t=\frac{mc^2}{\hbar}\di\tau.
\end{eqnarray*}
During a finite period of time we have $\theta_\text{int}\equiv\int\di\tau mc^2/\hbar=\Delta x_\text{int}/\lambdabar$, which is nothing but the path integral phase. We can now interpret the path integral phase as how much the constant phase surfaces have shifted inside the wavepacket. We call this an {\em internal phase shift} $\theta_\text{int}$. When we add two wavepackets it is important to keep track of this phase shift as it may lead to destructive or constructive interference. $\theta_\text{int}$ calculated for each trajectory is simply the integration constants $\theta_{i}(x_{i})$ (e.g.\;\eqref{fermintphase}).

Recall that for photons there was no contribution to the phase from the path integral, i.e.\;the integration constants are $\theta_{i}(x_{i})=0$. From a wave-geometric picture this is due to the fact that the group and phase velocities are equal and therefore $\theta_\text{int} = 0$ .

%%%%%%%%%%%%%%%%%%%%%%%%%%%%%%%%%%%%%%%%%%%%%
\subsubsection{The displacement induced phase difference}
%%%%%%%%%%%%%%%%%%%%%%%%%%%%%%%%%%%%%%%%%%%%%

Let us now provide a wave-geometric interpretation for the third term in \eqref{massphase}, $k_\mu \Delta x^\mu$. First, for simplicity let $\Delta x^\mu = x_1^\mu-x^\mu_2$ be spacelike and orthogonal to some arbitrary unit timelike vector $n^\mu$, i.e.\;$\Delta x^\mu = h^\mu_{\ \nu} \Delta x^\nu$ where $h^\mu_{\ \nu} = \delta^\mu_\nu-n^\mu n_\nu$ projects onto the orthogonal space of $n^\mu$. \footnote{$\Delta x^\mu$ can always be made spacelike orthogonal by changing the arbitrary end points of the trajectories $\Gamma_{1}$ and $\Gamma_{2}$.} $k_\mu \Delta x^\mu$ then simplifies to:
\begin{eqnarray*}
|k_\mu \Delta x^\mu| =|k_\mu h^\mu_{\ \nu} \Delta x^\nu| = k_{\perp} \Delta x_\text{dis} |\cos(\alpha)|=\frac{\Delta x_\text{dis}}{\lambdabar}
\end{eqnarray*}
where $k_\perp = \sqrt{-h^{\mu\nu} k_\mu k_\nu}$ and $\Delta x_\text{dis} = \sqrt{-h_{\mu\nu} \Delta x^\mu\Delta x^\nu}$, and we have used that $|\cos(\alpha)|=1$ since the wavepackets are spatially displaced in the direction of motion, i.e.\;$k_{\perp}^\mu \propto \Delta x^\mu$.

\begin{figure}[h]
  \centering
  \ifpdf
    \includegraphics[height=40mm]{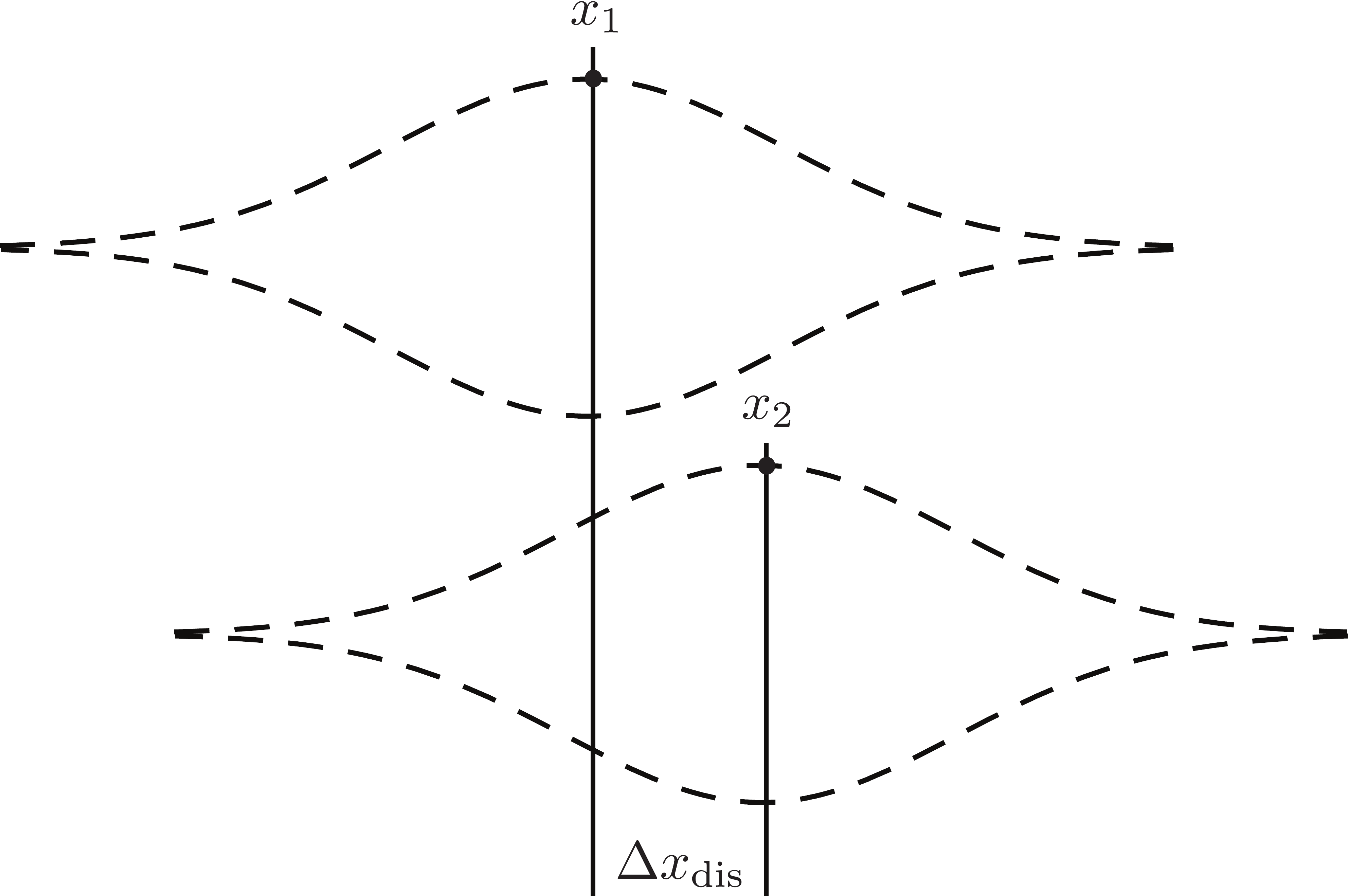}
    \else
  \includegraphics[height=40mm]{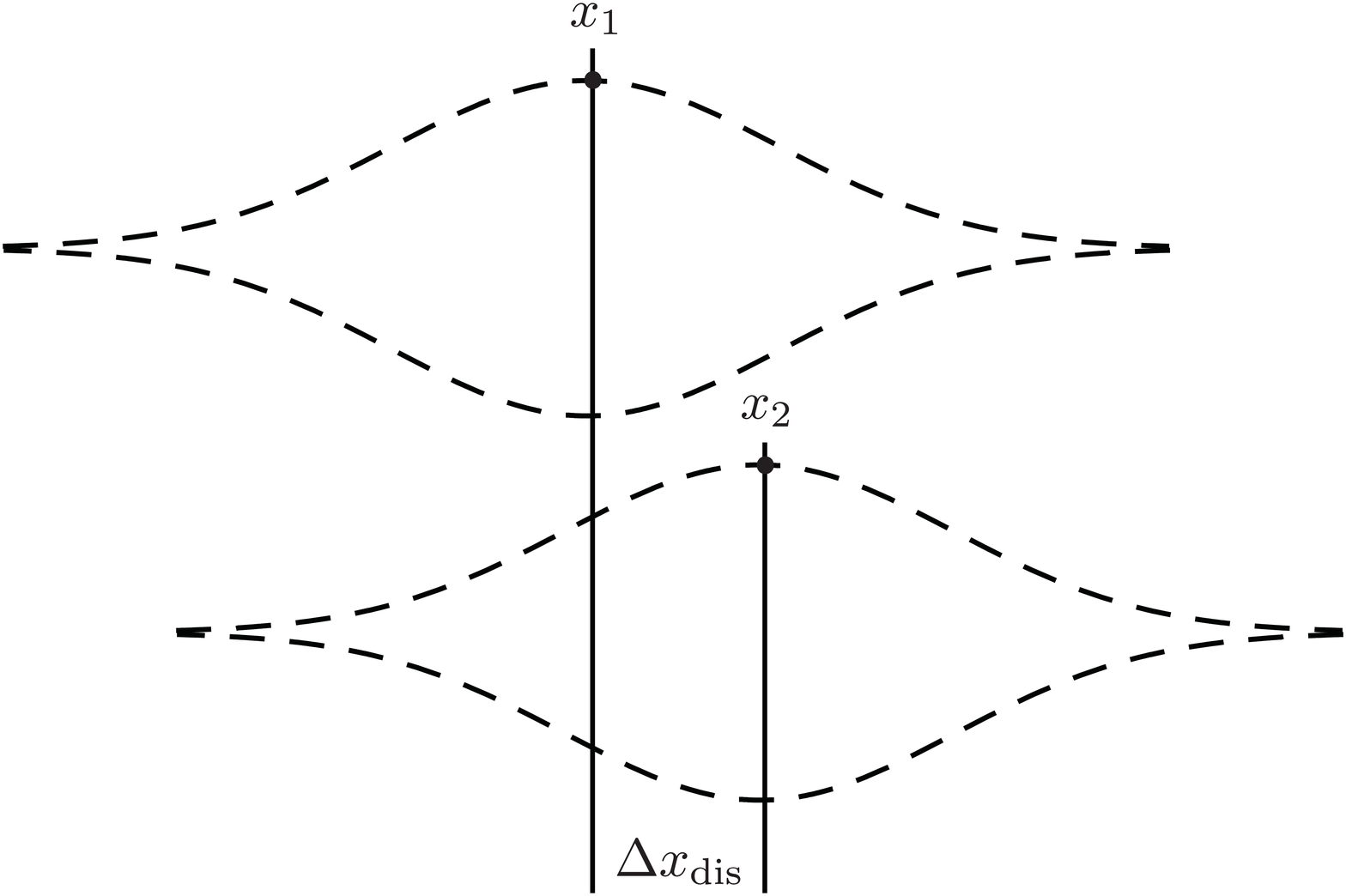}
  \fi
  \caption[Illustration of recombination of envelopes]{Illustration of the recombination of two envelopes in the detector region $D_{1}$. The offset of the classical positions $x_1$, $x_2$ of two wavepackets with the same wavelength $\lambdabar$ produces a displacement phase of $\Delta\theta_\text{dis}=\Delta x_\text{dis}/\lambdabar$.
  \label{fig:Phaseoffsets}}
\end{figure}
This contribution to the phase difference, which is present for both fermions and photons, can therefore be interpreted as the two wavepackets being spatially displaced, as illustrated in \figref{fig:Phaseoffsets} and as argued in \cite{Mannheim}. Note that in order for this {\em displacement induced phase difference} $\Delta\theta_\text{dis}=k_\mu\Delta x^\mu$ to be detectable, the variance in $\Delta x_\text{dis}$ over runs of an interference experiment must be significantly smaller than the wavelength $\lambdabar$. Furthermore, if $\Delta x_\text{dis} \sim \m L$  (see \figref{fig-spacetimeMZ}) then the interference effects will be drastically reduced and when $\Delta x_\text{dis} \geq \m L$ no interference phenomena will be present.\footnote{See also \cite{Zych12}.}

%%%%%%%%%%%%%%%%%%%%%%%%%%%%%%%%%%%%%%%%%%%%%%%%%%%%%%%%%
\subsubsection{Addition of quantum states in the wave-geometric picture}
%%%%%%%%%%%%%%%%%%%%%%%%%%%%%%%%%%%%%%%%%%%%%%%%%%%%%%%%%

\begin{figure}[h]
\centering
\small
\ifpdf
\includegraphics[height=48mm]{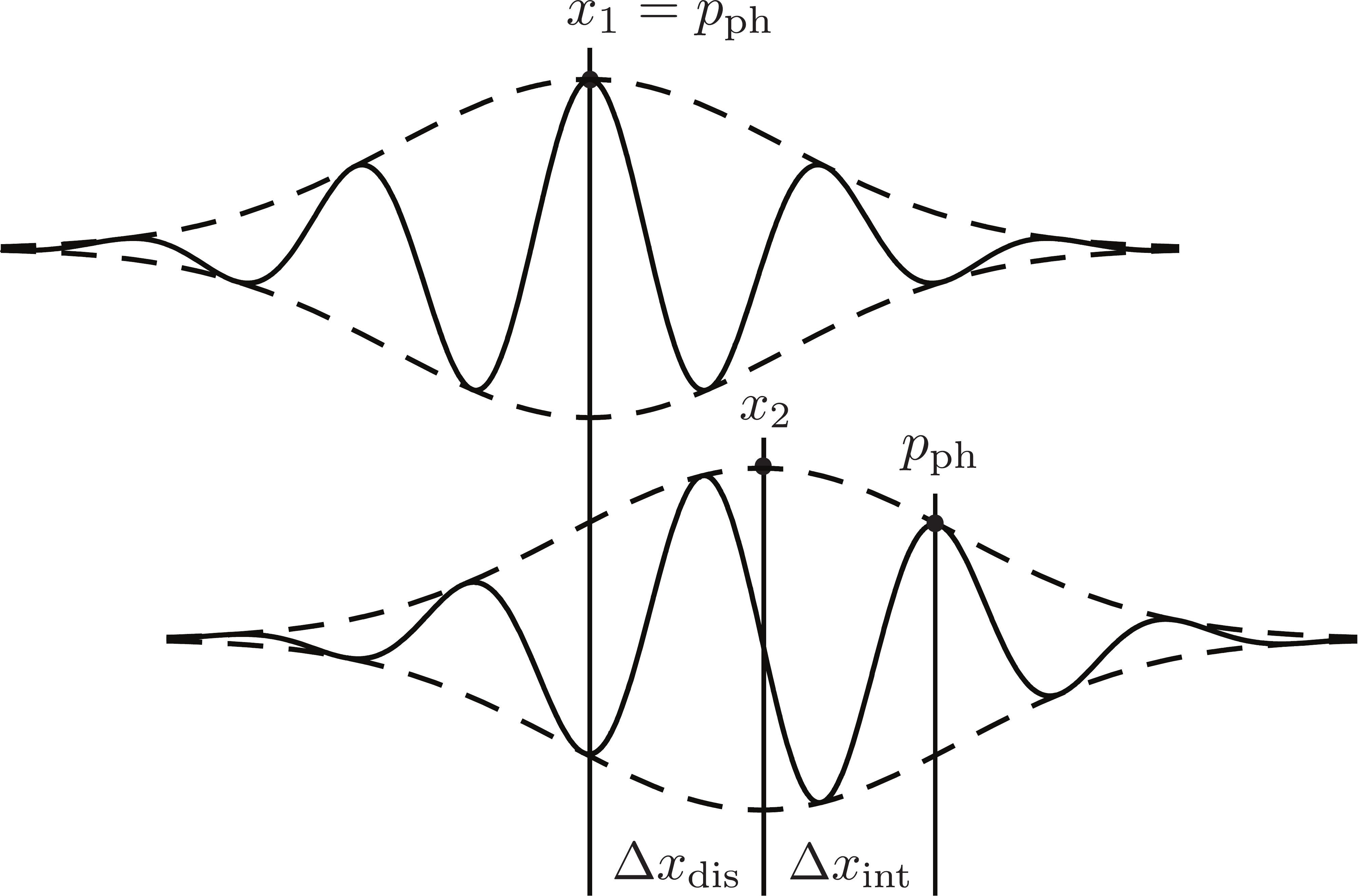}
\else
\includegraphics[height=48mm]{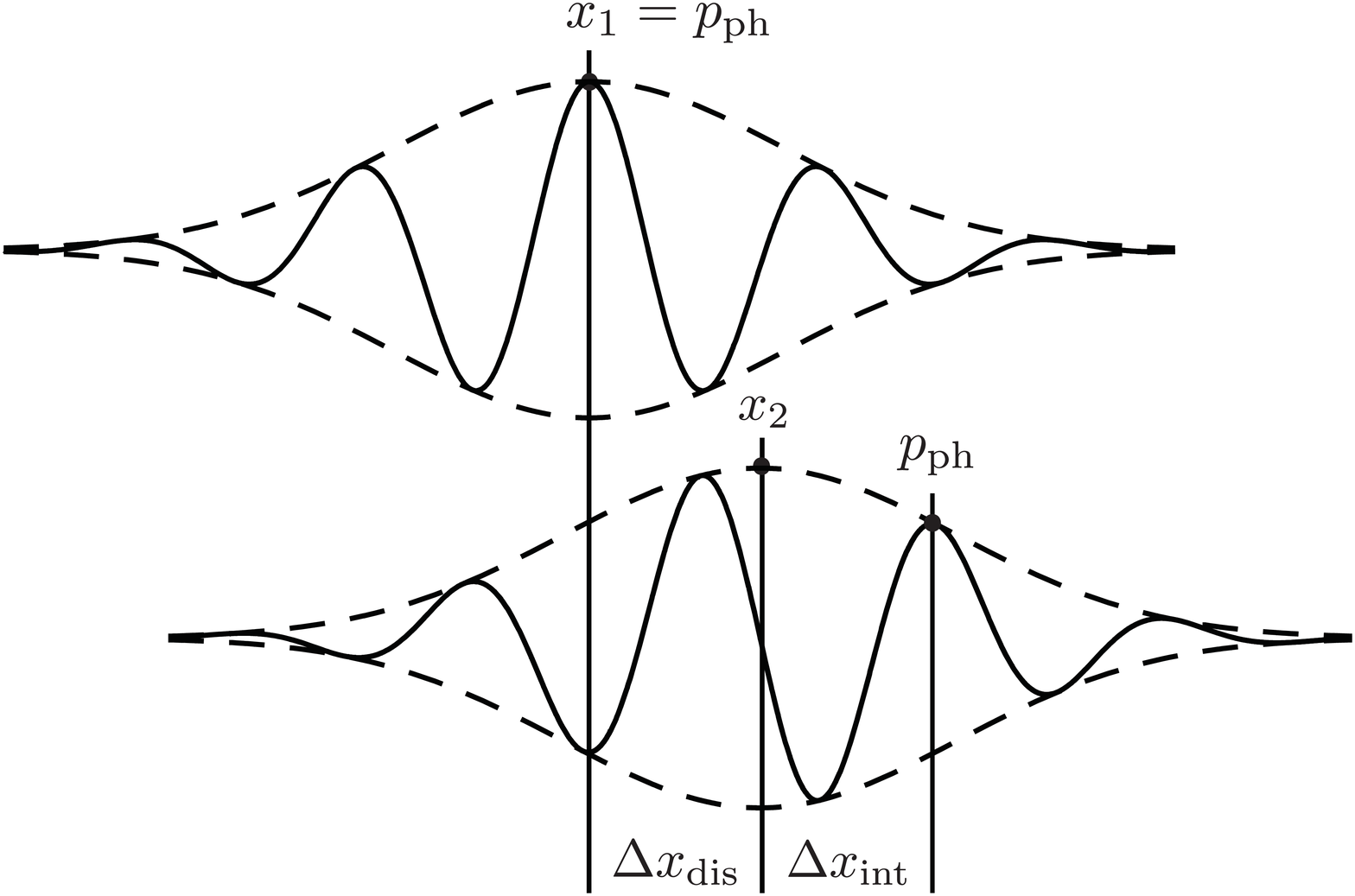}
\fi
\caption[Total phase difference between wavepackets]{From a wave-geometric point of view we can understand the phase difference $\Delta\theta$ as the sum of two contributions: the difference in the {\em internal phase} given by $\Delta\theta_\text{int}=\Delta x_\text{int}/\lambdabar$ and a phase difference $\Delta\theta_\text{dis}=\Delta x_\text{dis}/\lambdabar$ originating from the wavepackets being spatially displaced.\label{fig-spacewaveaddition}}
\end{figure}

The recipe for adding two quantum states \S\ref{recipe} can now readily be understood in terms of wave geometry. In the detector region $D_{1}$ we have two wavepackets whose envelopes, centred at $x_1$ and $x_2$, overlap but are slightly offset  (as in \figref{fig:Phaseoffsets}). Furthermore, each wavepacket has a rapidly oscillating phase that has evolved in a path-dependent way along each of the two distinct trajectories $\Gamma_{1}$ and $\Gamma_{2}$ (as in \figref{fig:inthaseoffsets}). These two effects produce respectively the displacement induced phase difference and the internal phase difference. We then add these wavepackets to obtain the total phase difference $\Delta\theta$. This is illustrated in \figref{fig-spacewaveaddition}. The total phase difference is again  $\Delta\theta_\text{Tot} = \Delta\theta_\text{Trans.}+ \Delta\theta$. The novelty of gravity for interference is that gravitational fields can change the energy and thus wavelength and rate of internal phase accumulation of a particle. In a non-conservative gravitational field, i.e.\;one which does not admit a timelike killing vector field, the components of the superposition can even have unequal frequency at recombination.

%%%%%%%%%%%%%%%%%%%%%%%%%%%%%%%%%%%%%%%%%%%%%%%%%%%%%%%%%%%%%
\subsection{An example: relativistic neutron interferometry}\label{neutronphase}
\begin{figure}[h]
\center
\ifpdf
\includegraphics[height=50mm]{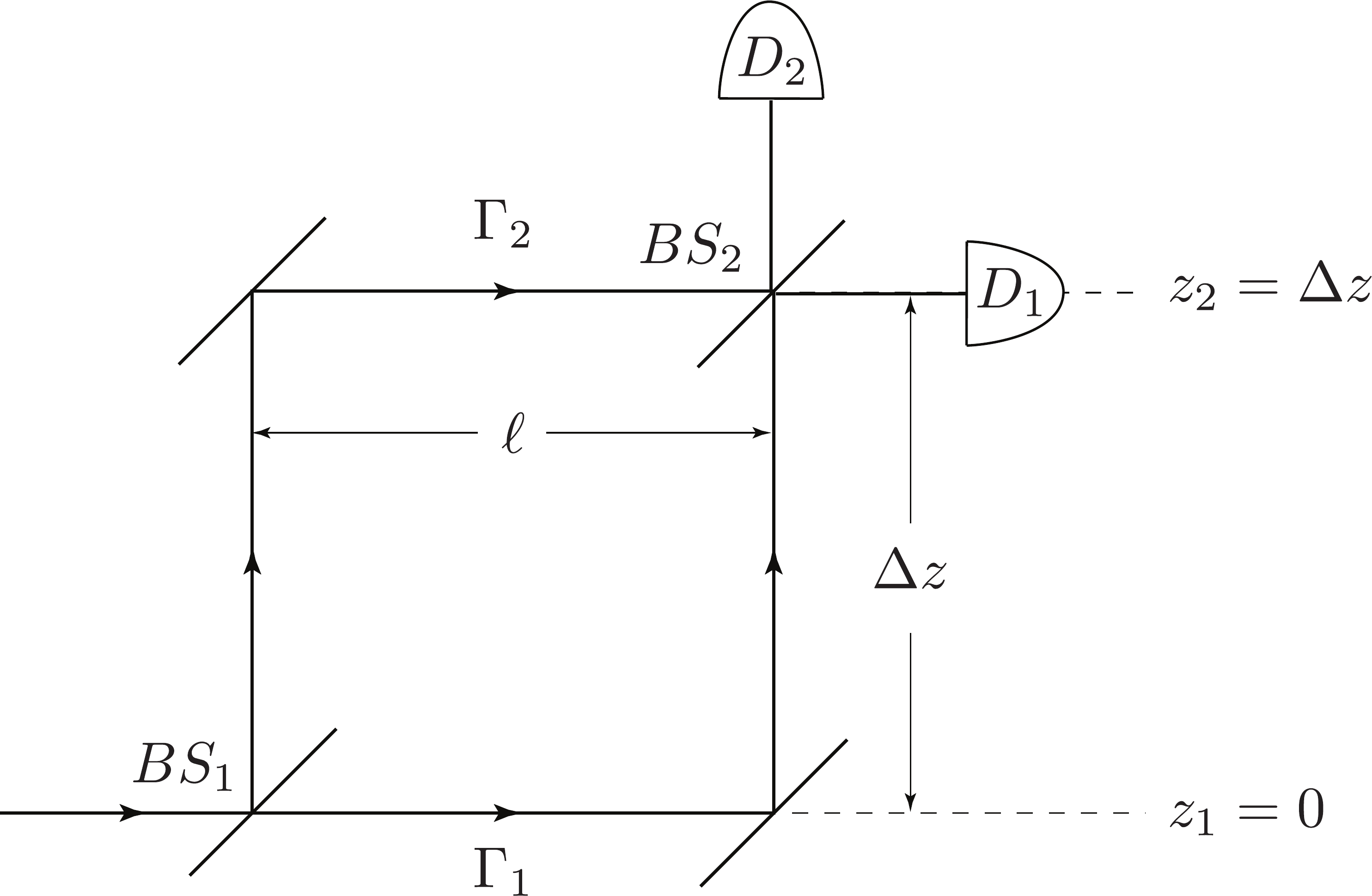}
\else
\includegraphics[height=50mm]{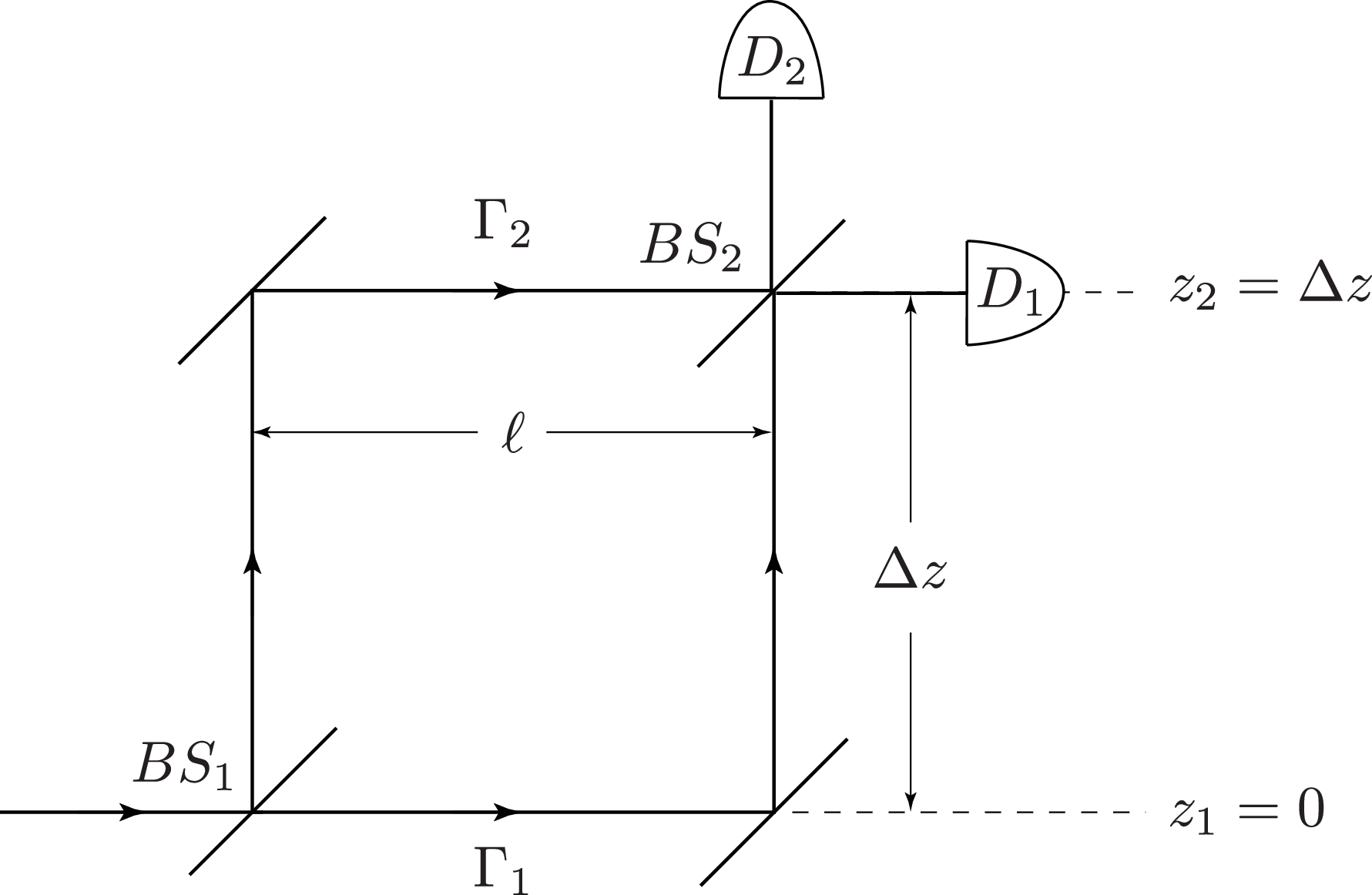}
\fi
\caption[Neutron interferometry schematic]{A Schematic diagram of a neutron interferometer used in the COW experiment. A neutron incident on the first beam splitter $BS_{1}$ is split into a spatial superposition travelling along two distinct paths $\Gamma_{1}$ and $\Gamma_{2}$. We find that $\Gamma_{2}$ accumulates a phase shift with respect to $\Gamma_{1}$ as it is higher in the gravitational field by $\Delta z$. \label{fig-neutroninterferometer}}
\end{figure}

%%%%%%%%%%%%%%%%%%%%%%%%%%%%%%%%%%%%%%%%%%%%%%%%%%%%%%%%%%%%%
As a concrete example for implementing the above recipe for calculating the phase difference we consider the gravitational neutron interferometry experiment illustrated in \figref{fig-neutroninterferometer}, known as the Colella--Overhauser--Werner (COW) experiment \cite{Colella}. The setup is geometrically identical to a Mach--Zehnder interferometer: The wavepacket is as usual split up into a spatial superposition and the respective wavepackets then travel along two distinct paths. The interferometer is oriented such that one path is higher up in the gravitational field relative to the other path. Essentially the two components of the spatial superposition have different speeds and experience two different gravitational potentials, which leads, in the recombination region, to a phase shift. Interference fringes have been observed (see e.g.\;\cite{Colella,Sakurai,Werner}) when the interferometer is rotated in the gravitational field, altering the difference in height of the paths.

In this section we are going to derive the phase difference for this experiment using the relativistic formalism  developed above. The spin of the neutrons is ignored and we treat them as scalar particles with no internal discrete degree of freedom. Therefore there is no need to use the transport equation \eqref{fermionTP}. The effects due to the spin could be included by computing \eqref{transphase}; however the corrections to the overall phase difference are minute, as noted in \cite{VarjuRyder}. From this analysis we will arrive at an exact relativistic result which contains, in certain limits, both approximate relativistic corrections to the COW experiment \cite{Anandan,AudretschLammerzahl,VarjuRyder} as well as the non-relativistic result \cite{Colella,Sakurai,Werner}.

One might represent the gravitational field for this experiment by the Schwarzschild metric. However, since the size of the experimental apparatus is less than a metre and hence small compared to the curvature scale, we can mimic gravity by simply going to an accelerated reference frame. This can be achieved by making use of the Rindler coordinates \cite{MTW,Rindler} in which the flat spacetime metric takes the form\footnote{Note that we could also consider rotating reference frames which would lead to the Sagnac effect \cite{Anandan,Werner,VarjuRyder}, but for simplicity we will stick to the Rindler metric.}
\begin{eqnarray*}
g_{\mu\nu}= \begin{pmatrix}(1+\frac{zg}{c^2})^{2}&0&0&0\\0&-1&0&0\\0&0&-1&0\\0&0&0&-1\end{pmatrix}.
\end{eqnarray*}

Since the Rindler metric is static (i.e.\;independent of $t$) we have a Killing vector $\eta^\mu=(1,0,0,0)$ and hence conserved energy $E\equiv p_\mu\eta^\mu=mc^2\gamma g_{00}$, where $\gamma\equiv(g_{00}-\frac{v^2}{c^2})^{-1/2}$ with $v$ the speed of the neutron as measured in the frame defined by the Killing vector $\eta^\mu$. We can now use this conserved energy to determine the speed $v_2$ of the neutron in the upper path given the speed in the lower path $v_1$, i.e.\;$E_1=mc^2\gamma_1 g_{00}(z_1)=mc^2\gamma_2 g_{00}(z_2)=E_2$. We take the lower path to be at height $z_1=0$, and so if the difference in height is $\Delta z$ we have the height of the top path being $z_2=\Delta z$. Therefore we have the relation $\gamma_2g_{00}(z_2)=\gamma_1$ since $g_{00}(z_1)=1$.

The easiest way to calculate the phase difference is by using the formula
\begin{equation}
\Delta\theta=\oint_{\Gamma} k_\mu\di x^\mu = \int_{\Gamma_2} k_\mu\di x^\mu-\int_{\Gamma_1} k_\mu\di x^\mu + k_\mu\Delta x^\mu\label{COWphase}
\end{equation}
where we have used \eqref{fermionphasedifference} and assumed that $A_{\mu}(x)$ is constant.
This formula contains two arbitrary spacetime points $x_1$ and $x_2$. Here we take these points to be where the trajectories $\Gamma_{1}$ and $\Gamma_{2}$, respectively, hit the second beam splitter $BS_{2}$. Since the spatial positions of these two events are the same in our Rindler coordinate system we have $\Delta x^\mu=(\Delta t,0,0,0)$, where $\Delta t$ is the difference in arrival time. This is given by
\begin{eqnarray*}
\Delta t=\ell\left(\frac{1}{v_1}-\frac{1}{v_2}\right)
\end{eqnarray*}
where $\ell$ is the length of the horizontal legs of the paths. The contribution of the phase difference from the third term is thus
\begin{eqnarray*}
k_\mu\Delta x^\mu=\frac{mc^2\ell}{\hbar}\left(\frac{g_{00}(\Delta z)\gamma_2}{v_1}-\frac{g_{00}(\Delta z)\gamma_2}{v_2}\right)=\frac{mc^2\ell}{\hbar}\left(\frac{\gamma_1}{v_1}-\frac{\gamma_1}{v_2}\right).
\end{eqnarray*}
Let us now turn to the first and second terms in \eqref{COWphase}, representing the internal phase shifts. Since the internal phases accumulated along the vertical components of each path are equal the quantity cancels in the calculation of the phase difference and so it is unnecessary to calculate them. The internal phase shifts of the upper and lower horizontal paths are given by
\begin{eqnarray*}
\theta^{(1)}_\text{int}=\frac{mc^2}{\hbar}\tau_1=\frac{mc^2\ell}{\hbar}\frac{1}{\gamma_1v_1}, \qquad\theta^{(2)}_\text{int}=\frac{mc^2}{\hbar}\tau_2=\frac{mc^2\ell}{\hbar} \frac{1}{\gamma_2v_2}.
\end{eqnarray*}
The phase difference is then given by
\begin{eqnarray*}
\Delta\theta=\theta^{(2)}_\text{int}-\theta^{(1)}_\text{int}+k_\mu\Delta x^\mu=\frac{mc^2\ell}{\hbar}\left(\frac{1}{\gamma_2v_2}-\frac{1}{\gamma_1v_1}+\frac{\gamma_1}{v_1}-\frac{\gamma_1}{v_2}\right)
\end{eqnarray*}
which simplifies to
\begin{eqnarray}\label{exactphasediff}
\Delta\theta=\frac{m\ell\gamma_1}{\hbar}\left(v_1-\frac{v_2}{(1+\frac{\Delta zg}{c^2})^2}\right)
\end{eqnarray}
where we can make the replacement $v_2=c\sqrt{g_{00}\left(1-g_{00}\gamma_1^{-2}\right)}$.

\Eeqref{exactphasediff} is the exact result for the gravitationally induced phase shift in the Rindler metric. This compares to various results in the literature \cite{Anandan,AudretschLammerzahl,Sakurai,Werner,VarjuRyder} for the gravitational effect in the COW experiment, which turn out to be approximations of \eqref{exactphasediff}. It is instructive to take various limits to demonstrate these connections.

Firstly, if we take the weak field limit, $\Delta z g/c^2\ll 1$ we obtain
\begin{eqnarray}
\Delta\theta\approx\frac{m\ell v_1\gamma_1}{\hbar}\left(1-\sqrt{1-2\frac{\Delta zg}{v_1^2}}\right).\label{eq-phaseweakfield}
\end{eqnarray}
If we furthermore take $\Delta zg/v_1^2\ll1$, which corresponds to assuming the relative reduction in velocity is small, $|v_{1}-v_{2}|/v_1\ll1$, and consider only first order terms, we obtain the phase result in \cite{Anandan}. If we instead take the non-relativistic limit ($\gamma_1\approx1$) we obtain the result of \cite{AudretschLammerzahl}. Expanding this result to two orders of $\Delta z g/v_1^2$ gives
\begin{eqnarray}
\Delta\theta\approx\frac{m\ell}{\hbar}\left(\frac{\Delta zg}{v_1}+\frac{\Delta z^{2}g^{2}}{4v_1^3}\right),\label{correction}
\end{eqnarray}
which indicates the `$g^2$' correction term derived in \cite{AudretschLammerzahl}. To leading order in  $\Delta zg/v_1^2$ the non-relativistic limit gives the standard theoretical prediction of the phase difference $\Delta\theta_\text{COW}$ observed in the COW experiment;
\begin{eqnarray}\label{standardphasediff}
\Delta\theta_\text{COW}=\frac{m\Delta z\ell g}{\hbar v_1}.
\end{eqnarray}
There is a reported small discrepancy between measurement and theory \cite{Colella}. However, the error introduced by neglecting corrections in $\Delta z g/v_1^2$ is too small to account for this discrepancy \cite{AudretschLammerzahl}.

The standard result \eqref{standardphasediff} is obtained using a path integral approach (see e.g.\;\cite{Sakurai,Werner,Zych11}). The path integral method allows only for summation over paths which start and end at the same two spacetime points. However, the classical trajectories in this problem in fact do not arrive at the second beam splitter at the same time. Therefore, the standard expression, although a very good approximation in the specific case of the actual experiment under consideration, is not exact even non-relativistically.

Let us examine a COW experiment numerically, presented in \tabref{fig-phaseresults}. From \cite{Sakurai}, the interferometer dimensions are $\Delta z=\ell=3.16\,{\rm cm}$, the neutrons have mass $m=1.67\times10^{-27}\,{\rm kg}$ and speed $v_1=2794\,{\rm m.s}^{-1}$, and $g=9.81\,{\rm m.s}^{-2}$. All results for phase difference $\Delta\theta$ with these data give $55.6\,{\rm rad}$. From \tabref{fig-phaseresults}, the difference between the COW result \eqref{standardphasediff} and the exact result \eqref{exactphasediff} is $1.05\times10^{-6}\,{\rm rad}$,  a difference of 19 parts per billion. The weak field limit is a very good approximation; $10^{-25}$ from the exact result. A nonrelativistic limit is the next best approximation ($10^{-9}$ from the exact result), whereas the small velocity change first order expansion by \cite{Anandan} is less justified, providing only a relatively minor improvement on the COW result.

\begin{table}
\[
\begin{array}{lcccc}
\text{Result} & \text{Approximation} & \text{Ref} & \Delta\text{ from COW} &    \Delta\text{ from Exact}\\
\hline
\text{Exact} & \text{Rindler} &\eqref{exactphasediff} & 1\times10^{-6} & 0 \\
\text{Weak field} & \text{1O }\Delta zg/c^2=3\times10^{-18}  & \eqref{eq-phaseweakfield} & 1\times10^{-6} & -1\times10^{-25} \\
\text{Small $\Delta v$}& \text{wf+1O }\Delta zg/v_1^2=4\times10^{-8} & \text{\cite{Anandan}} & 2\times10^{-9} & -1\times10^{-6}\\
\text{Non relativistic} & \text{wf+0O }\gamma=1+2\times10^{-11} &\text{\cite{AudretschLammerzahl}} & 1\times10^{-6}& -2\times10^{-9}\\
g^2\text{ correction} & \text{nr+2O }\Delta zg/v_1^2 & \eqref{correction} & 6\times10^{-7}& -5\times10^{-7}\\
\text{\cite{Colella}} & \text{nr+1O }\Delta zg/v_1^2 & \eqref{standardphasediff} & 0 & -1\times10^{-6}
\end{array}
\]
%\captionof{figure}{Neutron interferometry numerical results}
\caption[Neutron interferometry numerical results]{\label{fig-phaseresults}Phase results for different approximations starting from the exact result \eqref{exactphasediff} and finishing with the COW approximation \eqref{standardphasediff}. The Approximation column indicates to what order a term is expanded (0O,1O,2O), the numerical value of the term, as well as indicating whether the phase result is based on previous approximations. For example, the `Non relativistic' result is zeroth order (`0O') in $\gamma$, but also possesses all of the Weak field result's approximations (`wf'). The $\Delta$ columns indicate the value of the result subtracting the COW result, and the Exact result, respectively.}
\end{table}

%%%%%%%%%%%%%%%%%%%%%%%%%%%%%%%%%%%%%%%%%%%%%%%%%%%%%%%%%%%%%%%%%%%%%
\section{Elementary operations and measurement formalism\label{sec-unitarity}\label{sec-measurement}}
%%%%%%%%%%%%%%%%%%%%%%%%%%%%%%%%%%%%%%%%%%%%%%%%%%%%%%%%%%%%%%%%%%%%%

In order to develop quantum information theory in curved spacetimes we need to understand how elementary operations such as unitary transformations and state updating are represented within the reference frame covariant formalism of this chapter. This section is dedicated to these issues. In addition we show how Hermitian observables are represented, how to calculate their expectation values, and how to construct explicitly a quantum observable given the measurement direction of a Stern--Gerlach device, or a polariser.

%%%%%%%%%%%%%%%%%%%%%%%%%%%%%%%%%%%%%%%%%%%
\subsection{Fermions}\label{sec-fermionmeasurement}
%%%%%%%%%%%%%%%%%%%%%%%%%%%%%%%%%%%%%%%%%%%

In this section we develop the notion of unitarity, observables and projectors for fermions. The notion of unitarity and observables is not straightforward for two reasons: (1) the inner product is velocity dependent and (2) Hilbert spaces associated with distinct points in spacetime must be thought of as separate. The formalism that we develop addresses these issues in a reference-frame-covariant way.

%%%%%%%%%%%%%%%%%%%%%%%%%%%%%%%%%%%%%%%%%%%
\subsubsection{Unitarity and Hermitian operators}
%%%%%%%%%%%%%%%%%%%%%%%%%%%%%%%%%%%%%%%%%%%

Unitarity is traditionally defined for automorphisms $U:\m H\mapsto\m H$, i.e.\;unitary maps that take elements from one Hilbert space back to the same Hilbert space. The map $U$ is unitary if it satisfies
\begin{eqnarray}
\bk{U\phi}{U\psi}=\bk\phi\psi
\label{ordinaryunitary}
\end{eqnarray}
for all $\phi,\psi\in\Hi$, where the inner product is given by $\bk{\phi}{\psi}=\delta^{A'A}\bar\phi_{A'}\psi_A$. However, for our purposes this definition is too restrictive: we are interested in localised qubits transported along some spacetime trajectory $\Gamma$. The Hilbert spaces associated with the points along $\Gamma$ must be thought of as distinct and therefore a map induced by the transport equation \eqref{fermionTP} cannot be thought of as a map from a Hilbert space to itself. Furthermore, the inner products for the Hilbert spaces depend on the respective 4-velocity. It is then clear that the transformation induced by the transport equation \eqref{fermionTP} is not going to be unitary according to \eqref{ordinaryunitary} as we are not dealing with automorphisms.

Consider therefore a map $U:\m H_1\rightarrow\m H_2$ where $\m H_1$ and $\m H_2$ are two Hilbert spaces on the trajectory $\Gamma$. That a quantum state belongs to $\m H_1$ is indicated by a subscript $\ket{\cdot}_1$ and similarly for $\m H_2$. The `generalised' definition of unitarity then becomes
\begin{eqnarray}
_2\bk{U\phi}{U\psi}_2\ =\ _1\!\bk\phi\psi_1\label{genuni}
\end{eqnarray}
using the inner product $_i\bk{\phi}{\psi}_i=\bar\phi^{(i)}_{A'}I_{u_i}^{A'A}\psi^{(i)}_A$, with $I^{A'A}_{u_i}\equiv u_I^{(i)}\bar\sigma^{IA'A}$, and $\phi^{(i)}_{A},\psi^{(i)}_{A}\in\m H_i$, $i=1,2$. If we adapt the tetrad such that $u^I=(1,0,0,0)$ along the trajectory we see that the inner product $I_u^{A'A}=u_I\bar\sigma^{IA'A}$ becomes the ordinary inner product $\delta^{A'A}$ which is independent of both position and momentum. We would therefore expect the inner product between two quantum states along some trajectory $\Gamma$ to be conserved. To see this let $\phi(\tau)$ and $\psi(\tau)$ represent two quantum states that are Fermi--Walker transported, according to \Eeqref{spinhalfFW} along $\Gamma$. Then we have
\begin{eqnarray}
\frac{\di}{\di\tau}\bk\phi\psi=\frac{D^{FW}}{D\tau}\bk\phi\psi=u_I\bar{\sigma}^{IA'A}\frac{D^{FW}\bar\phi_{A'}}{D\tau}\psi_A+u_I\bar{\sigma}^{IA'A}\bar\phi_{A'}\frac{D^{FW}\psi_A}{D\tau}=0 \quad\label{fermIPcons}
\end{eqnarray}
since the Fermi--Walker derivative of $u^I$ is zero by construction, and the inner product $\bk\cdot\cdot$ is defined using $I_{u(\tau)}^{A'A}$. Strictly speaking, the inner product should be labelled with $\tau$  (i.e.\; $_\tau\bk{\cdot}{\cdot}_\tau$) in order to indicate that we are dealing with different Hilbert spaces. However, for convenience we omit this cumbersome notation.

Let us now consider a more general evolution dictated by a Schr\"odinger equation
\begin{eqnarray}
\frac{D^{FW}\psi_{A}}{D\tau} = \frac{\di\psi_A}{\di\tau}-\ii\left(\frac12\frac{\di x^\mu}{\di\tau}\omega_{\mu IJ}+u_Ia_J\right)L^{IJ\ B}_{\ \ A}\psi_{B} = \ii \op_A^{\ B}\psi_B\label{fermionevolution}
\end{eqnarray}
where $\op_A^{\ B}$ represents some linear operator on $\psi_{B}$. Requiring the inner product to be preserved under the evolution implies that $\op_A^{\ B}$ must for all $\phi,\psi\in\m H$ satisfy
\begin{eqnarray}
u_I\bar{\sigma}^{IA'A}\bar\phi_{A'}\op_A^{\ B}\psi_B-u_I\bar{\sigma}^{IA'A}\bar{\op}_{A'}^{\ B'}\bar\phi_{B'}\psi_A =0 \label{herm}
\end{eqnarray}
or equivalently $\bk{\phi}{\op\psi}=\bk{\op\phi}{\psi}$, which is nothing but the standard definition of a Hermitian operator. Since this must hold for all $\phi_A$ and $\psi_A$ we must have $I_u^{B'A}\op_A^{\ B} = I_u^{A'B}\bar{\op}_{A'}^{\ B'}$.

In spinor notation we can define $\m \op^{A'A}\equiv \op_B^{\ A}I_{u}^{A'B}$ which yields an equivalent definition of Hermiticity for the spinorial object $\m \op^{A'A}$:
\begin{eqnarray*}
\bar{\m \op}^{A'A}=\m \op^{A'A}.
\end{eqnarray*}
An object $\m \op^{A'A}$ satisfying this condition can be written as
\begin{eqnarray}
\m \op^{A'A}=N_I\bar{\sigma}^{IA'A}\label{fermionhermitian}
\end{eqnarray}
for some real-valued coefficients $N_I$. We also have $\op_A^{\ B}=I_{uAA'}\m \op^{A'B}$ where $I_{uAA'}$ is the inverse of $I^{A'A}_u$ defined by $I_{uAA'}I_{u}^{A'B}=\delta_A^B$. The corresponding operator $\op_{A}^{\ B}$ is then given by
\begin{eqnarray}
\op_A^{\ B}&\equiv&I_{uAA'}\m \op^{A'B}=u_I\sigma^I_{\;AB'}N_J\bar{\sigma}^{JB'B}=u_IN_J(\sigma^{[I}\bar{\sigma}^{J]}+ \sigma^{\{I}\bar{\sigma}^{J\}})_A^{\ B}\notag\\
&=&-2\ii u_IN_JL^{IJ\ B}_{\ \;A}+u_{I} N^{I}\delta_A^{\ B}\label{fermionoperator}
\end{eqnarray}
where ${L^{IJ}}_{A}^{\ B}$ are the left-handed $\mathfrak{sl}(2,\mathbb C)$ generators and the term in $\delta_A^{\ B}$ generates changes in global phase.\footnote{Note that one could also have chosen the right-handed representation ${R^{IJ}}_{\ B'}^{A'}$ of the Lorentz group as this would yield the same result.}

It should be noted that the linear operator $\op_{A}^{\ B}$ does not `look' Hermitian when written out in matrix form. For example, $\op_{1}^{\ 2}\neq\bar{\op}_{2}^{\ 1}$.  Rather, it is only the object $\m \op^{A'A}\equiv I_u^{A'B}\op_{B}^{\;A}$ which looks Hermitian in matrix form, i.e.\;$\m \op^{A'A}=\bar{\m \op}^{AA'}$. \footnote{Spinor notation gives the relationship $\overline{\m \op^{A'A}}\equiv\bar{\m \op}^{AA'}\equiv\bar{\m \op}^{A'A}$ between the conjugate and row-column transpose. } The reason for this difference can be clearly seen by expressing $\op_{A}^{\ B}$ in the rest frame of the qubit. In the particle rest frame, Hermitian operators are expressed as $\tilde{\op}_A^{\ B}=N_I\sigma^0_{\ AA'}\bar\sigma^{IA'B}$, and in this case  $\tilde{\op}_{1}^{\ 2}=\bar{\tilde \op}_{2}^{\ 1}$. Thus, from an operator $\hat{\tilde \op}$ which is Hermitian with respect to $\delta^{A'A}$ we can construct another operator $\hat \op$ which is Hermitian with respect to $I^{A'A}_u$ simply by applying a boost, i.e.\;$\op_A^{\ B}=\Lambda_A^{\ C}\tilde{\op}_C^{\ D}\Lambda^{-1B}_D$, where $\Lambda$ is the spin-$\half$ representation of the Lorentz boost that takes $\delta^I_0$ to $u^I$.

A general inner product preserving evolution can therefore be understood as being composed of two pieces. One piece, the Fermi--Walker derivative, dictates how acceleration and the gravitational field affects the quantum state and has therefore a purely geometric character. The Fermi--Walker derivative maps elements between neighbouring Hilbert spaces $\m H_{(x,p)(\tau)}$ and $\m H_{(x,p)(\tau+\delta\tau)}$. The remaining term $\op_A^{\ B}$ encodes possible non-geometric influences on the quantum state and is an automorphism $\op:\m H\rightarrow\m H$. This second term is required to be Hermitian with respect to the inner product $I^{A'A}_u$. \footnote{Hermiticity can alternatively be defined in terms of the partial $\di/\di\tau$ or covariant $D/D\tau$ derivatives but in doing so we would have to modify the definition of a Hermitian operator.}

We have already seen an example of an evolution of the form \eqref{fermionevolution}. In the WKB limit of the minimally coupled Dirac equation we arrived at the transport equation \eqref{fermionTP}, where the linear Hermitian operator took on the form
\begin{eqnarray*}
B_A^{\ B}\equiv -\frac e{2m} B_{IJ}L^{IJ\ B}_{\ \ A}= -\frac e{2m} h_I^{\ K}h_J^{\ L}F_{KL}L^{IJ\ B}_{\ \ A}
\end{eqnarray*}
where $B_{IJ}\equiv h_I^{\ K}h_J^{\ L}F_{KL}$ is the magnetic field experienced by the particle. To see that the magnetic precession term $B_A^{\ \ B}$ is Hermitian with respect to the inner product, we expand the left side of the Hermiticity definition \eqref{herm}:
\begin{eqnarray*}
\bigbk{\phi}{\hat B\psi}-\bigbk{\hat B\phi}{\psi}=-\frac e{2m} \bar\phi\left(u_K\bar{\sigma}^KB_{IJ}\hat L^{IJ}-\bar B_{IJ}\hat{\bar{L}}^{IJ}u_K\bar{\sigma}^K\right)\psi.
\end{eqnarray*}
With $B_{IJ}$ real and making use of the identity $[\bar{\sigma}^K,\hat L^{IJ}]=\ii[\eta^{IJ}\bar{\sigma}^K-\eta^{JK}\bar{\sigma}^I]$ \cite{Bailin}, we have
\begin{align}
\bigbk\phi{\hat B\psi}-\bigbk{\hat B\phi}\psi&=-\frac e{2m} u_K B_{IJ}\bar\phi [\bar{\sigma}^K, \hat L^{IJ}]\psi\\&=-\ii\frac e{2m} \bar\phi\bar{\sigma}^K\psi[u_KB_I^{\ I}-B_K^{\ I}u_I]=0
\label{covhermitcommutator}
\end{align}
since $B_{IJ }u^{J}=0$ and $B_I^{\ I}=0$. The magnetic precession is thus a Hermitian automorphism with respect to the inner product $I_u^{A'A}$.

%%%%%%%%%%%%%%%%%%%%%%%%%%%%%%%%%%%%%%%%%%
\subsubsection{Observables and projective measurements\label{fermeas}}
%%%%%%%%%%%%%%%%%%%%%%%%%%%%%%%%%%%%%%%%%%

Observables are represented by Hermitian operators $\op_A^{\ B}$, which will take the form indicated in \eqref{fermionoperator}. The covariant expression of the expectation value of the observable $\op$ for a spinor $\psi_A$ is given by
\begin{eqnarray}
\langle\psi|\op|\psi\rangle=\bar{\psi}_{A'}N_I\bar\sigma^{IA'A}\psi_A. \label{spacetimemeas}
\end{eqnarray}
Note that in \eqref{spacetimemeas} all indices have been contracted, indicating the expectation value is manifestly a Lorentz invariant scalar and could in principle represent an empirically accessible quantity.

In order to complete the measurement formalism we need to discuss how to determine the post-measurement quantum state. We do this for the simple case of projection-valued measures, however we can easily extend the formalism to generalised measurements. A Hermitian operator has a real eigenvalue spectrum and its normalised eigenstates $|\psi^{(k)}\rangle$ are orthogonal, i.e.
\begin{eqnarray*}
\langle\psi^{(k)}|\psi^{(l)}\rangle=I_{u}^{A'A}\bar{\psi}^{(k)}_{A'}\psi^{(l)}_A=\delta^{kl}.
\end{eqnarray*}
where $k,l=\pm$. The spectral decomposition of an observable $\hat{\op}$  is $\op_{A}^{\ B} = \sum_\pm\lambda_\pm P^{\pm B}_{\ A}$, where the $\lambda_\pm$ are the eigenvalues of $\op_{A}^{\ B}$ and the $P^{\pm B}_{\ A}$ represents the corresponding projector onto the eigenstate $|\psi^{\pm}\rangle$. In spinor notation, the projectors are given by $P^{\pm B}_{\ A}=I_{u}^{A'B}\bar{\psi}^{\pm}_{A'}\psi^{\pm}_A$. A pair of projection operators $P^{\pm B}_{\ A}$ which, together  with the identity operator, span the space of Hermitian observables on $\m H$ can also be written as
\begin{eqnarray}\label{spinprojectors}
P^{\pm B}_{\ A}=\frac{1}{2}(\delta_A^{\ B}\mp 2\ii u_In_JL^{IJ\ B}_{\ \ A})
\end{eqnarray}
with  $n_In^I=1$ and $n_Iu^I=0$. These are the ordinary Bloch sphere projectors but written in a reference frame covariant way. One can then suspect that a measurement of spin along some unit direction can be represented by such projectors. This is indeed the case as we shall see now in the specific case of Stern--Gerlach measurements.

%%%%%%%%%%%%%%%%%%%%%%%%%%%%%%%%%%%%%%%%%%%%%%%%%%%%%%%%%%%%%%%%%%%%%
\subsubsection{The spin operator for a relativistic Stern--Gerlach measurement\label{sec-SternGerlach}}
%%%%%%%%%%%%%%%%%%%%%%%%%%%%%%%%%%%%%%%%%%%%%%%%%%%%%%%%%%%%%%%%%%%%%
To be able to extract empirical predictions from the above formalism we need to determine how $N^{I}$ in \eqref{fermionoperator} corresponds to the relevant parameters defining the experimental setup, e.g.\;the spatial orientation of a Stern--Gerlach magnet. In the literature there exist several proposals for relativistic spin operators and these have been studied for various reasons (see e.g.\;\cite{FoldyWouthuysen,HehlNi,Mashhoon95,Czachor,Ryder98,Ryder99,Ternotworol,Friis-relent}). In this section we are going to be concerned exclusively with constructing a spin observable associated with a relativistic Stern--Gerlach measurement. Notably, the spin operator that we obtain differs from other proposals. This spin operator is investigated in detail in Chapter \ref{ch-RSO}.

In order to obtain the correct relativistic spin observable it is necessary to understand in more detail the physical aspects of the measurement process. In a Stern--Gerlach spin measurement, a particle is passed though an inhomogeneous magnetic field. This causes the wavepacket to separate into two packets of orthogonal spin. A subsequent position measurement then records the outcome. To gain further insight, let us consider this measurement process in the fermion's rest frame where $e^\mu_t=u^\mu$.  In such a frame the stationary qubit is exposed to a magnetic field $B^i$ for a short period of time and it is clear that it is the direction $\frac1BB^i$ (with $B^2\equiv B^iB^j\delta_{ij}$) of the magnetic field that determines what component of the spin we are measuring \cite{PeresQM}.

If the qubit is moving non-relativistically with respect to the Stern--Gerlach device, the spatial direction $\frac1BB^i$ of the magnetic field approximately agrees with the orientation of the Stern--Gerlach device $m^{i}$. However, if the qubit is moving relativistically with respect to the apparatus these directions do not necessarily coincide, nor is their relationship straightforward. We will establish a relation between these two directions, and in doing so we will identify the correct spin observable for a relativistic Stern--Gerlach measurement. In particular the  relativistic spin operator/observable that we obtain depends on the spatial orientation $m^I$ of the apparatus and the 4-velocities $v^I$  and $u^I$ of the apparatus and qubit.

To proceed we first work out an expression for the electromagnetic field $F_{IJ}$ generated by the Stern--Gerlach apparatus. To do that we first introduce the magnetic field 4-vector $M^I=Mm^I$ where $M$ is the magnitude of the Stern--Gerlach magnetic field. We can now define the electromagnetic tensor as
\begin{eqnarray}
F_{IJ}\equiv-\epsilon_{_{LIJK}}v^LM^K.\label{EMfield}
\end{eqnarray}
In the rest frame of the apparatus it takes the form
\begin{eqnarray*}
F_{IJ}\stackrel{*}{=}\begin{pmatrix}0&0\\0&B_{ij}\end{pmatrix}
\end{eqnarray*}
so there is only a magnetic field, and no electric field, generated by the Stern--Gerlach apparatus in its own rest frame (denoted in the equality by $*$). In order to simplify the calculation the gradient $e^{\mu}_{I}\nabla_\mu M$ of the magnetic field strength is assumed to point in the same direction as the magnetic field itself, i.e.\; $e^{\mu}_{I}\nabla_\mu M\propto m_I$. \footnote{Although this is not strictly possible as the magnetic field must satisfy $\nabla_{i}m^{i} = 0$ one can always choose a field which approximately has $e^{\mu}_{I}\nabla_\mu M\propto m_I$ locally \cite{Ballentine}.}

We can now calculate the magnetic field 4-vector $B^I$ corresponding to the magnetic field as measured in the rest frame of the qubit. It is given by $B^I\equiv\half\epsilon^{LIJK}u_LF_{JK}$. Using \eqref{EMfield}, we obtain
\begin{eqnarray*}
B^I=-\half\epsilon^{LIJK}u_L \epsilon_{_{MJKN}}v^MM^N=M^I(v\cdot u)-v^I(M\cdot u)
\end{eqnarray*}
with $\epsilon^{LIJK}\epsilon_{_{MJKN}}=-2(\delta^L_M\delta^I_N-\delta^I_N\delta^L_M)$  and $\epsilon_{_{0123}}=1$ \cite[p87]{MTW}. For spin measurements, the 4-vector $n^I$ is now the normalised qubit rest-frame magnetic field, i.e.
\begin{eqnarray*}
n^I(m,u,v)\equiv\frac{B^I}{B}
\end{eqnarray*}
where $B\equiv\sqrt{-B^IB^J\eta_{_{IJ}}}$, and it is easy to check that $n\cdot u=0$. This expression $n^I$ becomes singular only for unphysical or trivial situations characterised by $u^I$ being null, or $M=0$.

Thus, given the spatial orientation of the Stern--Gerlach apparatus and the 4-velocities of the apparatus and qubit we obtain, using \eqref{fermionoperator}, a relativistic spin operator  given by
\begin{eqnarray*}
\m S_{A}^{\ B} = -2\ii u_I n_J(m,u,v)L^{IJ\ B}_{\ \ A}.
\end{eqnarray*}
The expectation values are calculated using \eqref{spacetimemeas} and the corresponding projectors are given by \eqref{spinprojectors}. We now have a fully relativistic and reference frame invariant measurement formalism for a Stern--Gerlach measurement.

%%%%%%%%%%%%%%%%%%%%%%%%%%%%%%%%%%%%%%%%%%
\subsection{Photons\label{sec-photonunitaritymmt}}
%%%%%%%%%%%%%%%%%%%%%%%%%%%%%%%%%%%%%%%%%%

In this section we develop the notion of unitarity, observables and projectors for photons. The definition of unitarity is in some sense simpler than for fermions as the inner product is not velocity dependent; the difficulty is only in handling the gauge degrees of freedom.

As we have already stated, the polarisation state of a photon can be represented by a spatial complex 4-vector $\psi^I$ orthogonal to the null wavevector $k_\mu$. Furthermore, the corresponding quantum state of a photonic qubit with null velocity $u^I$ was identified as being a member of an equivalence class of polarisation vectors $\psi^I\sim \psi^I+\upsilon u^I$ all orthogonal to $u^I$ (\S\ref{secphotonQS}). With the orthogonality condition and the gauge degree of freedom this space therefore reduced to a two-dimensional Hilbert space on which unitary and Hermitian operators act. We now develop the notion of unitarity, observables and projectors within this four-dimensional formalism.

%%%%%%%%%%%%%%%%%%%%%%%%%%%%%%%%
\subsubsection{Unitarity and Hermitian operators}
%%%%%%%%%%%%%%%%%%%%%%%%%%%%%%%%

Unitarity and Hermiticity are more straightforward with polarisation vectors than spinors because the definition of unitarity $\bk{U\phi}{U\psi} = \bk{\phi}{\psi}$ is in terms of a standard inner product $\eta_{IJ}\bar\phi^I\psi^J$ where $\eta_{IJ}$ is constant. The requirement for unitarity again translates into requiring that the inner product between two polarisation vectors is conserved along trajectories, i.e.
\begin{eqnarray}
\frac{\di \bk{\phi}{\psi}}{\di \lambda} = \frac{D \bk{\phi}{\psi}}{D \lambda} = \frac{D \bar \phi_{I}}{D \lambda}\psi^{I} + \bar\phi_{I} \frac{D \bar \psi^{I}}{D \lambda} = 0\label{photonderivIP}.
\end{eqnarray}

Consider now a Schr\"odinger evolution of the form
\begin{eqnarray}
\frac{D\psi^{I}}{D\lambda}=\beta u^{I} +\ii \op^{I}_{\ J}\psi^{J}
\end{eqnarray}
which is more general than the transport equation \eqref{photonPT}. Substituting into \eqref{photonderivIP}, we get
\begin{eqnarray*}
\bar{\op}^{\ J}_{I}\bar\phi_{J}\psi^{I} -\bar\phi_{I}\op^{I}_{\ J}\psi^{J} = 0
\end{eqnarray*}
where we have used that $\psi^{I}u_{I} = u^{I}\phi_{I} = 0$. Requiring that this hold for all $\phi^{I}$ and $\psi^{I}$ we obtain the standard definition of Hermiticity; $\bar {\op}^{\ \;I}_{J} = \op^{I}_{\ J}$.

In addition to the above we also require that the gauge condition $u_I\psi^I=0$ be preserved. This implies that
\begin{eqnarray*}
\frac{\di(u_{I}\psi^{I})}{\di \lambda} = \frac{D (u_{I}\psi^{I})}{D \lambda} = \frac{D \psi^{I}}{D \lambda}u_{I} = \op^{I}_{\ J}\psi^{J} u_{I} = 0.
\end{eqnarray*}
This condition ensures that Hermitian operators $\op^{I}_{\ J}$ map polarisation vectors into polarisation vectors. Again this should hold for all $\psi^{J}$, so we have the condition $\op^{I}_{\ J} u_{I} \propto u_{J}$. The following two conditions suffice for characterising a general Hermitian operator;
\begin{subequations}
\label{HK0HH}
\begin{align}
\bar{\op}^{\ I}_{J} &= \op^{I}_{\ J},\label{HH} \\
\op^{I}_{\ J} u^{J} & \propto u^{I}. \label{HK0}
\end{align}
\end{subequations}

In order to determine the form of valid operators it is convenient to express the matrix $\op^I_{\ J}$ in terms of a set of basis vectors $\{u^I,w^I,f_1^I,f_2^I\}$ which spans the full tangent space. Recall from \secref{sec-photonqsident} that a diad frame $f_A^I\equiv(f_1^I,f_2^I)$ defines a spacelike two-dimensional subspace orthogonal to two null vectors $u^I,w^I$ with $u^Iw_I=1$. $f_A^If^A_J=h^I_J=\delta^I_J-u^Iw_J-w^Iu_J$ is the metric on the spacelike subspace \cite{Poisson}. One can then define a sixteen-dimensional complex vector space spanned by the outer products of $\{u^I,w^I,f_1^I,f_2^I\}$ with the dual vectors $\{u_J,w_J,f_{1J},f_{2J}\}$. Components of an arbitrary matrix $\op^{I}_{\ J}\in\langle B\otimes\bar B\rangle$ of this space can then be expanded in terms of these sixteen elements.

A valid map $\op^I_{\ J}$ on polarisation vectors must satisfy equations \eqref{HK0HH}. In terms of the sixteen basis elements, no terms in $w^I$ or $w_J$ can exist, since $w^Ik_I\neq0$. Any remaining terms that involve $u^{I}$ or $u_J$ are pure gauge and do not change the polarisation vector. A hermitian operator is therefore, up to gauge, represented as
\begin{eqnarray*}
\op^{I}_{\ J} = a f_1^I f^1_J + \beta f_1^I f^2_J + \bar\beta f_2^I f^1_J + b f_2^I f^2_J
\end{eqnarray*}
where the real numbers $a$ and $b$ and the complex number $\beta$ constitute the four remaining real degrees of freedom. The two conditions in \eqref{HK0HH} thus reduce a $4\times 4$ hermitian matrix to effectively a $2\times 2$ hermitian operator that acts on the transverse spacelike (polarisation) degrees of freedom. Such an operator can be written in terms of the Pauli matrix basis as $\op^{A}_{\ B} = C^a\sigma_{a\ B}^{\ A}$, where the Pauli matrices act on the two dimensional Jones vector. $C^a$ consisting of four coefficients $a=0,1,2,3$ does %FIXED
not transform as a 4-vector and therefore does not have any spatial significance. For the operator $\op_A^{\;B}$ where  $a = 1,2$ or $3$ the eigenbasis corresponds to respectively the diagonal linear polarisation basis, the circular polarisation basis, and the horizontal--vertical polarisation basis of the Jones vector. The relation between the four-dimensional and two-dimensional hermitian operators is $\op^I_{\ J}=f^{I}_{A} \op^{A}_{\ B} f_{J}^{B}$.

%%%%%%%%%%%%%%%%%%%%%%%%%%%%%%%%
\subsubsection{Observables and projective measurements\label{polarizationmeasurement}}
%%%%%%%%%%%%%%%%%%%%%%%%%%%%%%%%
The construction of observables and projectors is identical to that of fermions. Observables are represented by Hermitian operators $\op^{I}_{\ J}$. Let $P_{(k)}^{I}$ represent the eigenvectors of $\op^{I}_{\ J}$. The $P_{(k)}^{I}$ form an orthonormal basis with $\bar{P}_{(k)}^I{P_{(l)}}_I=\delta_{kl}$. The probability $p_{k}$ of getting outcome $\lambda_{k}$ is given in tetrad notation by
\begin{eqnarray}
p_k=|\bar P_{(k)}^I\psi_I|^2\label{eq-polarizeroverlap}.
\end{eqnarray}
The corresponding projector for an eigenvector $P^I_{(k)}$ is $P^I_{\; J} = P^I_{(k)}\bar{P}_{(k)J}$ and the post-measurement state, up to gauge, is given by  $\psi^I\to {\psi'}^{I} = P^{I}_{\ J} \psi^J$.

In this case we have a clean interpretation of projectors $P^I_{k}$ as polariser vectors $P^I$: complex, spacelike normalised vectors orthogonal to photon velocity, $P^Iu_I=0$. Polariser vectors correspond to the physical direction and parameters of an optical polariser: A linear polariser direction is of the form $\Ee^{\ii\theta}P^I$ with $P^I$ real, and a circular polariser is a complex vector $P^I=\frac{1}{\sqrt{2}}(P^I_1+\ii P^I_2)$ with $P_1^IP_2^J\eta_{IJ}=0$ and $\bar P^I_1P^J_1\eta_{IJ}=\bar P^I_2P^J_2\eta_{IJ}=1$. The probability of transmission of a polarisation vector through a polariser is simply the modulus square of the overlap of the polarisation state with the polariser vector \eqref{eq-polarizeroverlap}. Such an overlap clearly does not depend on the tetrad frame used, and indeed all tetrad indices are contracted in \eqref{eq-polarizeroverlap}. The probability $p$ is then manifestly a Lorentz scalar. It is easy to verify that the formalism is invariant under gauge transformations $\psi^I\rightarrow \psi^{I}+\upsilon u^I$ and $P^I\rightarrow P^I+\kappa u^I$. Thus, the probability $p$ is both gauge invariant and Lorentz invariant as should be the case. With this completed measurement formalism it is now possible to work entirely within the covariant polarisation 4-vector formalism.

%%%%%%%%%%%%%%%%%%%%%%%%%%%%%%%%%%%%%%%%%%%%%%%%%%%%%%%%%%%%%%%%%%%%%
\section{Quantum entanglement}\label{secQIinCST}
%%%%%%%%%%%%%%%%%%%%%%%%%%%%%%%%%%%%%%%%%%%%%%%%%%%%%%%%%%%%%%%%%%%%%
Until now we have been concerned with the question of how the quantum state of some specific physical realisation of a single qubit is altered by moving along some well-defined path in spacetime. We shall now show how this formalism can easily be extended to describe entanglement of multiple qubits.

%%%%%%%%%%%%%%%%%%%%%%%%%%%%%%%%%%%%%%%%%%%%%%%%%%%%%%%%%%%%%%%%%%%%%
\subsection{Bipartite states}
%%%%%%%%%%%%%%%%%%%%%%%%%%%%%%%%%%%%%%%%%%%%%%%%%%%%%%%%%%%%%%%%%%%%%
We have seen that it is necessary to associate a separate Hilbert space with each pair of position and momentum $(x^\mu,p_\mu)$. A single qubit moving along a specific path $x^\mu(\lambda)$ in spacetime will therefore have its state encoded in a sequence of distinct Hilbert spaces associated with the spacetime points along the path.\footnote{Recall that while we can uniquely determine the 4-momentum $p^\mu=m\frac{\di x^\mu}{\di\tau}$ from the trajectory $x(\tau)$ in the case of massive fermions, the same is not true for photons. Due to the arbitrariness of the parametrisation of the null trajectory $x(\lambda)$ we can only determine the null momentum up to a proportionality factor, i.e.\;$p^\mu\propto\frac{\di x^\mu}{\di\lambda}$.} The formalism that we have so far developed determines how to assign a quantum state to each distinct Hilbert space along the path along which we move the qubit. The one-parameter family of quantum states $|\psi(\lambda)\rangle$ is parameterised by some parameter $\lambda$ of the path $x(\lambda)$, and the sequence of Hilbert spaces associated with the path is $\mathcal{H}_{(x,p)(\lambda)}$. The quantum state $|\psi(\lambda)\rangle$ belongs to the specific Hilbert space $\mathcal{H}_{(x,p)(\lambda)}$.

Let us now consider how to generalise the formalism of this chapter to the quantum state of two, possibly entangled, qubits in curved spacetime. Instead of one worldline we will now have two worldlines, $x_1(\lambda_1)$ and $x_2(\lambda_2)$, and consequently instead of one parameter $\lambda$ we now have two independent parameters $\lambda_1$ and $\lambda_2$. Corresponding to each value of $\lambda_1$ and $\lambda_2$ we have two spacetime points and two Hilbert spaces, $\mathcal{H}_{(x_1,p_1)(\lambda_1)}$ and $\mathcal{H}_{(x_2,p_2)(\lambda_2)}$. It is therefore clear that the quantum state describing the two qubits is mathematically described by a quantum state $|\psi(\lambda_1,\lambda_2)\rangle\in\mathcal{H}_{(x_1,p_1)(\lambda_1)}$ which belongs to the tensor product Hilbert space $\otimes\mathcal{H}_{(x_2,p_2)(\lambda_2)}$. This can be evaluated at any pair of parameters $(\lambda_1,\lambda_2)$.

In order to calculate statistics (e.g.\;correlation functions) we need to provide an inner product for the tensor product Hilbert space $\mathcal{H}_{(x_1,p_1)(\lambda_1)}\otimes\mathcal{H}_{(x_2,p_2)(\lambda_2)}$. The natural choice is the inner product induced by the inner products for the individual Hilbert spaces.

In the case of fermions the natural choice for the parameterisations $\lambda$ is the proper time $\tau$. The Hilbert spaces $\mathcal{H}_{(x_1,p_1)(\tau_1)}$ and $\mathcal{H}_{(x_2,p_2)(\tau_2)}$ have the inner products given by $I^{A'A}_1=u^1_{I}\bar{\sigma}^{IA'A}$  and $I^{A'A}_2=u^2_{I}\bar{\sigma}^{IA'A}$ where $u^1_I$ and $u^2_I$ are the respective 4-velocities. In our index notation a bipartite quantum state can be represented by an object with two spinor indices $\psi_{A_1 A_2}(x_1(\tau_1),x_2(\tau_2))$. Note however that the indices $A_1$ and $A_2$ relate to two distinct spinor spaces associated with two distinct points $x_1(\tau_1)$ and $x_2(\tau_2)$ and therefore cannot be contracted. The inner product between two bipartite quantum states $|\psi\rangle$ and $|\phi\rangle$ becomes
\begin{eqnarray*}
\bk{\psi}{\phi}_{p_1,p_2}=u^1_I\bar{\sigma}^{I A_1'A_1}u^2_J\bar{\sigma}^{JB_2'B_2}\bar{\psi}_{A_1'B_2'}\phi_{A_1B_2}
\end{eqnarray*}
where $p_1$, $p_2$ are the momenta of the two qubits. In the case of photons the quantum state $\ket{\psi}$ can be represented by a polarisation 4-vector $\psi^I(\lambda)$. A bipartite state $\ket{\psi}$ is then given by a two-index object $\psi^{I_1 I_2}(x_1(\lambda_1),x_2(\lambda_2))$, where $I_1$ and $I_2$ belong to two different tangent spaces and thus cannot be contracted. The requirement that the polarisation vector be orthogonal to the null wavevector generalises to  $u_{I_1}\psi^{I_1I_2} = 0= u_{I_2}\psi^{I_1I_2}$. The inner product between two bipartite quantum states  $|\psi\rangle$ and $|\phi\rangle$ becomes
\begin{eqnarray*}
\bk{\psi}{\phi}_{p_1,p_2} = \eta_{I_1 J_1} \eta_{I_2 J_2} \bar{\psi}^{I_1I_2} \phi^{J_1J_2} .
\end{eqnarray*}
We could also consider bipartite states $\ket{\phi}$ where one component is an electron and the other is a photon. Mathematically this would be represented as $\phi^{I_1}_{A_2}(x_1(\lambda), x_2(\tau))$ and the inner product can be constructed similarly.

Let us now turn to the evolution of bipartite quantum states. The physically available interactions of the qubits are given by local operations. Mathematically this means that the most general evolution of the state vector is given by two separate Schr\"odinger equations:
\begin{eqnarray}
\ii\frac{D^{T}}{D\lambda_1}|\psi(\lambda_1,\lambda_2)\rangle&=&\hat{\op}_1(\lambda_1)\otimes \mathbb{I}|\psi(\lambda_1,\lambda_2)\rangle\label{twoSE1}\\
\ii\frac{D^{T}}{D\lambda_2}|\psi(\lambda_1,\lambda_2)\rangle&=&\mathbb{I}\otimes\hat{ \op}_2(\lambda_2)|\psi(\lambda_1,\lambda_2)\rangle\label{twoSE2}
\end{eqnarray}
where $\hat{\op}_1(\lambda_1)$ and $\hat{\op}_2(\lambda_2)$ are possible local Hermitian operators (as defined in \S\ref{sec-unitarity}) acting on the Hilbert spaces $\mathcal{H}_{(x_1,p_2)(\lambda_1)}$ and $\mathcal{H}_{(x_2,p_2)(\lambda_2)}$. $D^{T}/D\lambda$ denotes the transport law, i.e.\;the Fermi--Walker transport for fermions or the parallel transport for photons.

This mathematical description of the evolution of the quantum state is not standard since we have two Schr\"odinger equations rather than one. However, if we introduce an arbitrary foliation $t(x)$ these two equations can be combined into one Schr\"odinger equation. First we express the parameters as functions of the foliation $\lambda_1=\lambda_1(t)$ and $\lambda_2=\lambda_2(t)$. This allows us to write the quantum state as only depending on one time parameter: $|\psi(t)\rangle=|\psi(\lambda_1(t),\lambda_2(t))\rangle$. The evolution of the quantum state now takes a more familiar form
\begin{eqnarray*}
\ii\frac{D^{T}}{Dt}|\psi(t)\rangle&\equiv&\ii\left(\frac{\di\lambda_1}{\di t} \frac{D^{T}}{D\lambda_1}+\frac{\di\lambda_2}{\di t} \frac{D^{T}}{D\lambda_2}\right)|\psi(\lambda_1(t),\lambda_2(t))\rangle\\
&=&\left(\frac{\di\lambda_1}{\di t}\hat{\op}_1\otimes\mathbb{I}+ \frac{\di\lambda_2}{\di t}\mathbb{I}\otimes\hat{\op}_2\right)| \psi(\lambda_1(t),\lambda_2(t))\rangle\\&=&\hat{\op}|\psi\rangle
\end{eqnarray*}
where $\hat{\op}\equiv\frac{\di\lambda_1}{\di t}\hat{ \op}_1\otimes\mathbb{I}+\frac{\di\lambda_2}{\di t}\mathbb{I}\otimes\hat{\op}_2$ is the total Hamiltonian acting on the full state. For this single Schr\"odinger equation to hold for all paths $x^\mu_1(\lambda_1)$ and $x^\mu_2(\lambda_2)$, and all choices of foliation $t(x)$, and so for all values of $\frac{\di\lambda_1}{\di t}$ and $\frac{\di\lambda_2}{\di t}$, it is necessary that both equations \eqref{twoSE1} and \eqref{twoSE2} hold. Thus, the two mathematical descriptions of the evolution of the quantum state are equivalent when only local operations enter in the evolution.

If the Hamiltonian is not a local one, i.e.\;not of the form $\hat{\op}=a\hat{\op}_1\otimes\mathbb{I}+b\mathbb{I}\otimes\hat{\op}_2$, then it is not possible to cast it into the previous form %FIXED
with two independent evolution equations and it is also necessary to introduce a preferred foliation. However, if all interactions are local the introduction of an arbitrary foliation is not necessary.

The generalisation to multipartite states is straightforward. Furthermore, if we are dealing with identical particles the wavefunction should be symmetrised or antisymmetrised with respect to the particle label, depending on whether we are dealing with bosons or fermions. This will correctly reproduce the Pauli exclusion phenomenon and the Hong--Ou--Mandel bunching phenomenon \cite{HOM}. %fixed
%
%%%%%%%%%%%%%%%%%%%%%%%%%%%%%%%%%%%%%%%%%%%%%%%%%%%%%%%%%%%%%%%%%%%%%
\subsection{State updating and the absence of simultaneity}
%%%%%%%%%%%%%%%%%%%%%%%%%%%%%%%%%%%%%%%%%%%%%%%%%%%%%%%%%%%%%%%%%%%%%
Let us now discuss the issue of state updating for entangled states. Let $\Gamma_1$ and $\Gamma_2$ be two spacetime trajectories along which two qubits are being transported. Furthermore let $x_1(\lambda_1)$ and $x_2(\lambda_2)$ each represent a distinct point on the corresponding trajectory.  Consider now that the two qubits are in the entangled state
\begin{eqnarray*}
\ket{\psi(\lambda_1,\lambda_2)} = \frac{1}{\sqrt 2}\left(\ket{+,\lambda_1}\ket{-,\lambda_2}-\ket{-,\lambda_1}\ket{+,\lambda_2}\right).
\end{eqnarray*}
If a measurement on qubit 1 is carried out at the spacetime point $x_1(\lambda_1)$ with outcome `$+$', the bipartite state has to be updated as follows:
\begin{eqnarray*}
\ket{\psi(\lambda_1,\lambda_2)} \to \ket{+,\lambda_1}\ket{-,\lambda_2}.
\end{eqnarray*}
Note that the value of $\lambda_2$ is left completely arbitrary after the state update. It should therefore be clear that even though the state update is associated with a distinct spacetime point $x_1(\lambda_1)$ on the trajectory $\Gamma_1$ no such point can be identified for $\Gamma_2$. In other words, as long as the local unitary evolution for particle 2 is well-defined, the state updating can be thought of as occurring at any point $x_2(\lambda_2)$ along $\Gamma_2$. The point $x_2(\lambda_2)$ could be in the past, elsewhere, or in the future of $x_1(\lambda_1)$. \footnote{I.e.\;$x_2(\lambda_2)$ has any timelike, lightlike or spacelike relationship with $x_1(\lambda_1)$.} The reason for this freedom in state updating is that a projection operator on particle 1 commutes with any local unitary operator acting on particle 2.

%%%%%%%%%%%%%%%%%%%%%%%%%%%%%%%%%%%%%%%%%%%%%%%%%%%%
\subsection{An example: quantum teleportation}\label{teleportation}
%%%%%%%%%%%%%%%%%%%%%%%%%%%%%%%%%%%%%%%%%%%%%%%%%%%%
As an example of entanglement and state updating let us look at quantum teleportation in a curved spacetime, with an agent Alice teleporting a quantum state to an agent Bob, where each agent is spatially localised, massive, and able to perform local quantum measurements. Although the mathematics is virtually the same as for the non-relativistic treatment, the interpretation is more delicate. In particular, in a curved spacetime the claim that the input state is in some sense the ``same" as the output state seems to lack a well-defined mathematical meaning. However, as we are going to see, in order to carry out the standard teleportation protocol the parties involved must first establish a shared basis in which the entangled state takes on a definite and known form. For example, Alice and Bob could choose the singlet state. This will be called the `canonical' form of the entangled state. Once this shared basis has been established, Alice and Bob have a well-defined convention for comparing quantum states associated with these different points in spacetime. The problems associated with quantum teleportation in curved spacetime are therefore similar to the problems associated with teleportation in flat spacetime when the maximally entangled state is unknown \cite{RudolphSanders01,BRSDialogue,BRS,BRST}.

Consider then three qubits moving along three distinct trajectories $\Gamma_1$, $\Gamma_2$, and $\Gamma_3$. For concreteness assume that the qubits are physically realised as the spins of massive fermions. The tripartite state is then given by
\begin{eqnarray}
\ket{\Upsilon;\lambda_1,\lambda_2,\lambda_3}\in \mathcal{H}_{(x_1,p_1)(\lambda_1)}\otimes\mathcal{H}_{(x_2,p_2) (\lambda_2)}\otimes\mathcal{H}_{(x_3,p_3)(\lambda_3)} \nonumber
\end{eqnarray}
or, written in our index notation, $\Upsilon_{A_1A_2A_3}(\lambda_1,\lambda_2,\lambda_3)$. For the teleportation protocol, Alice will have access to particles 1 and 2 at some stage along the particles' histories, and Bob will have access to particle 3 at some point along the particle's history, independent of Alice's choice.

In order to proceed we define a basis for each one of the three Hilbert spaces $\mathcal{H}_{(x_1,p_1)(\lambda_1)}$, $\mathcal{H}_{(x_2,p_2)(\lambda_2)}$, and $\mathcal{H}_{(x_3,p_3)(\lambda_3)}$ and {\it for all points} along the trajectories $\Gamma_1$, $\Gamma_2$, and $\Gamma_3$. The three pairs of basis vectors are assumed to be orthonormal, and to evolve according to the local unitary evolution (e.g.\;by pure gravitational evolution given by the Fermi--Walker transport \eqref{spinhalfFW}). Therefore, once we have fixed the basis for particle $i$ at one point $x_i(\lambda_i)$ on the trajectory $\Gamma_i$, the basis is uniquely fixed everywhere else along the trajectory, at least where the local unitary evolution is well-defined. It follows that a state can be expressed as a linear combination of these basis states with the components independent of $\lambda_i$, $i=1,2,3$.

We denote these three pairs of (one-parameter families of) orthonormal states as $\phi^{(i)}_{A_i}(\lambda_i)$ and $\psi^{(i)}_{A_i}(\lambda_i)$, with $i=1,2,3$. The orthonormality conditions are explicitly given by
\begin{eqnarray*}
\bigbk{\phi^{(i)};\lambda_i}{\psi^{(i)};\lambda_i}&=&u^{(i)}_{I_i}\bar{\sigma}^{I_iA'_iA_i} \bar{\phi}^{(i)}_{A'_i}(\lambda_i)\psi^{(i)}_{A_i}(\lambda_i)=0\\
\bigbk{\phi^{(i)};\lambda_i}{\phi^{(i)};\lambda_i}&=&u^{(i)}_{I_i}\bar{\sigma}^{I_iA'_iA_i} \bar{\phi}^{(i)}_{A'_i}(\lambda_i)\phi^{(i)}_{A_i}(\lambda_i)=1\\
\bigbk{\psi^{(i)};\lambda_i}{\psi^{(i)};\lambda_i}&=&u^{(i)}_{I_i}\bar{\sigma}^{I_iA'_iA_i} \bar{\psi}^{(i)}_{A'_i}(\lambda_i)\psi^{(i)}_{A_i}(\lambda_i)=1
\end{eqnarray*}
with $i=1,2,3$.

Consider now the specific tripartite three-parameter family of states
\begin{equation}
\Upsilon_{A_1A_2A_3}(\lambda_1,\lambda_2,\lambda_3)= \frac{1}{\sqrt{2}}\left(\alpha\phi^{(1)}_{A_1}(\lambda_1)+ \beta\psi^{(1)}_{A_1}(\lambda_1)\right)\left(\phi^{(2)}_{A_2}(\lambda_2) \phi^{(3)}_{A_3}(\lambda_3)+\psi^{(2)}_{A_2}(\lambda_2)\psi^{(3)}_{A_3} (\lambda_3)\right)\label{tripartite}
\end{equation}
where $\alpha$ and $\beta$ are independent of $\lambda_1$ (since evolution is entirely in the basis vectors) and $|\alpha|^2+|\beta|^2=1$. The maximally entangled state (involving particles $2$ and $3$) only has coefficients equal to $0$ or $1$ which are trivially independent of the parameters $\lambda_2$ and $\lambda_3$. We will call the maximally entangled state in \Eeqref{tripartite} the {\em canonical form}. Other choices of this canonical form are possible but the teleportation protocol used below (see \eqref{Bobunitaries}) will then change accordingly.

As in the flat spacetime description we now proceed to rewrite the state in the Bell basis for the Hilbert space $\mathcal{H}_{(x_1,p_1)(\lambda_1)}\otimes\mathcal{H}_{(x_2,p_2)(\lambda_2)}$:
\begin{eqnarray*}
\Phi^\pm_{A_1A_2}(\lambda_1,\lambda_2)&\equiv&\frac{1}{\sqrt{2}}\left(\phi^{(1)}_{A_1} (\lambda_1)\phi^{(2)}_{A_2}(\lambda_2)\pm\psi^{(1)}_{A_1}(\lambda_1)\psi^{(2)}_{A_2} (\lambda_2)\right)\\
\Psi^\pm_{A_1A_2}(\lambda_1,\lambda_2)&\equiv&\frac{1}{\sqrt{2}}\left(\phi^{(1)}_{A_1} (\lambda_1)\psi^{(2)}_{A_2}(\lambda_2)\pm\psi^{(1)}_{A_1}(\lambda_1)\phi^{(2)}_{A_2} (\lambda_2)\right).
\end{eqnarray*}
In these new bases the state $\ket{\Upsilon}$ reads
\begin{multline*}
\Upsilon_{A_1A_2A_3}(\lambda_1,\lambda_2,\lambda_3)=\frac{1}{2}\big(\Phi^+_{A_1A_2} \left(\alpha\phi_{A_3}+\beta\psi_{A_3}\right)+\Phi^-_{A_1A_2}\left(\alpha\phi_{A_3}- \beta\psi_{A_3}\right)\\ +\Psi^+_{A_1A_2}\left(\beta\phi_{A_3}+\alpha\psi_{A_3}\right)+\Psi^-_{A_1A_2} \left(-\beta\phi_{A_3}+\alpha\psi_{A_3}\right)\big).
\end{multline*}
Alice now performs a Bell basis measurement on the particles $1$ and $2$. It is important to note that the specific physical measurement operation that Alice needs to carry out depends on the bases $(\phi^{(1)}_{A_1},\psi^{(1)}_{A_1})$ and $(\phi^{(2)}_{A_2},\psi^{(2)}_{A_2})$. The basis $(\phi^{(2)}_{A_2},\psi^{(2)}_{A_2})$ is determined by the maximally entangled state. If Alice does not know the maximally entangled state she will not be able to do the correct Bell basis measurement.

The outcome of the Bell basis measurement ($\Phi^+$, $\Phi^-$, $\Psi^+$, or $\Psi^-$) is then communicated to  Bob's side: Bob performs the local unitary operation $U$ on particle 3 (the operation is assumed to act on the state but not on the basis) given by
\begin{eqnarray}\label{Bobunitaries}
U=\left\{\begin{array}{c}\hat\one\ \ \text{if}\ \ \Phi^+\\\hat\sigma_z\ \ \text{if}\ \ \Phi^-\\\hat\sigma_x\ \ \text{if}\ \ \Psi^+ \\\ii\hat\sigma_y\ \ \text{if}\ \ \Psi^-\end{array}\right.
\end{eqnarray}
where $\hat\sigma_x,\hat\sigma_y,\hat\sigma_z$ take the usual form when expressed in the local orthonormal basis $(\phi^{(3)}_{A_3},\psi^{(3)}_{A_3})$. Thus, as was the case for Alice, in order for Bob to know which specific physical operation to carry out, the basis has to be specified. Since the basis is determined by the maximally entangled state, Bob must know the entangled state in order to carry out his operations. Note that if the basis in which Bob applies the operation is incorrect, the state that Bob obtains at the end differs depending on the outcome of the Bell measurement. In the case where Bob implements the operation in the correct basis, Bob's state is given by $\alpha\phi_{A_3}+\beta\psi_{A_3}$.

We should now ask whether we can sensibly view this protocol as a `teleportation' of a quantum state from Alice to Bob in the sense that Bob received the {\em same} state as Alice sent. This hinges on there being a meaningful way of comparing quantum states associated with distinct spacetime points. However, as we have already stressed, if spacetime is curved, sameness of quantum states cannot be established uniquely by parallel transporting one qubit to the other as this would depend on the specific path along which we transport the qubit.

On the other hand, the maximally entangled state, by determining the bases for it to take the canonical form, defines a shared spinor basis for Alice and Bob. If we change the basis, the maximally entangled state would of course change accordingly and would no longer take on the canonical form. Given a shared basis we have a well-defined way of comparing quantum states and in particular a well-defined way to claim that they are the same or not. Thus, when we have a maximally entangled state, there {\it is} a natural way of comparing quantum states associated with distinct spacetime points. It is in using this convention for comparing quantum states that we can claim that Bob did indeed receive the same quantum state, and therefore we can say that the state was in this sense teleported.\footnote{We note that all examples of experimentally produced entangled qubit pairs are produced in localised spatial regions and distributed to the parties. The components of the entangled state will then undergo local unitary evolution along each trajectory.}

In the case of fermions it is also easy to see that the maximally entangled state will also establish a shared reference frame, i.e.\;a shared tetrad. This comes about because from the left-handed spinor $\psi_A$ by means of which the quantum state is expressed we can construct the null {\it Bloch 4-vector} $b^I=\bar\sigma^{IA'A}\bar\psi_{A'}\psi_A$. A maximally entangled state can therefore be loosely understood geometrically as a kind of `non-local connection'.

%%%%%%%%%%%%%%%%%%%%%%%%%%%%%%%%%%%%%%%%%%%%%%%%%%%%%%%%%%%%%%%%%%%%%%%%%%%%
\section{Conclusion, discussion, and outlook}\label{sec8}
%%%%%%%%%%%%%%%%%%%%%%%%%%%%%%%%%%%%%%%%%%%%%%%%%%%%%%%%%%%%%%%%%%%%%%%%%%%%%

Recently there has been increased interest in exploring relativistic quantum information theory in the context of phenomena from quantum field theory such as the Unruh effect and particle number ambiguity \cite{Alsing-DiracUnruh,Fuentes,Martinez}. In contrast, this chapter explored relativistic quantum information in the regime where such effects are negligible and restricted attention to localised qubits for which the particle number ambiguity is circumvented. A localised qubit is understood in this chapter to be any object that can effectively be described by a position and momentum $(x,p)$ and some two-component quantum state $\ket{\psi}$. We obtained a description of localised qubits in curved spacetimes, with the qubits physically realised as the spin of a massive fermion, and the polarisation of a photon.

The original motivation for this  research was to develop a formalism for answering a simple experimental question: if we move a spatially localised qubit, initially in a state $|\psi_1\rangle$ at spacetime point $x_1$, along some classical spacetime path $\Gamma$ to another point $x_2$, what will the final quantum state $|\psi_2\rangle$ be? Rather than working directly with Wigner representations our starting point in answering this question was the one-particle excitations of the quantum fields that describe these physical systems. The one-particle excitations in curved spacetime satisfy respectively the Dirac equation minimally coupled to the electromagnetic field, and Maxwell's equations in vacuum. From these fields we were able to isolate a two-component quantum state and a corresponding Hilbert space.

In the case of fermions, the equation governing the transport of the spin of a fermion consisted of  a spin-$\half$ version of the Fermi--Walker derivative and a magnetic precession term, expressed in a non-orthonormal Hilbert space basis.  This result was expected since an electron can be regarded as a spin-$\half$ gyroscope, and the precession of a classical gyroscope along accelerated trajectories obeys the Fermi--Walker equation. By introducing an orthonormal Hilbert space basis, which physically corresponds to representing the spinor in  the particle's rest frame,  we reproduced the transport equation obtained in \cite{TerashimaUeda03,ASK} which made use of Wigner representations.

We showed by applying the WKB approximation to vacuum Maxwell equations that the polarisation vector of a photon is parallel transported along geodesics and that this corresponds to a Wigner rotation. Furthermore, this rotation is proportional to the spin-1 connection term $\omega_{\mu12}$ when we consider a reference frame where the photon 3-velocity is along the $z$-axis. In this way the effect of spacetime geometry on the quantum state was easily identified.

We worked with faithful finite-dimensional but {\em non-unitary} representations of the Lorentz group, specifically a two-component left-handed spinor $\psi_A$, and polarisation 4-vector $\psi^I$. Nevertheless, by identifying a suitable inner product we obtained a unitary quantum formalism. The advantage of working with non-unitary representations is that the objects which encode the quantum state transform covariantly under actions of the Lorentz group. As a result the transport equations are manifestly Lorentz covariant, in addition to taking on a simple form. The connection to the Wigner formalism was obtained by choosing a reference frame adapted to the particle's 4-velocity, which is a reason why the Wigner formalism is not manifestly Lorentz covariant.

In order to make empirical predictions we need a way to extract probabilities for outcomes from the formalism. Such a measurement formalism was developed for both fermions and photons. The predicted probabilities of outcomes of experiments were shown to be manifestly Lorentz invariant and thus reference frame invariant, resulting in a relativistically invariant measurement formalism. We also derived the specific Hermitian operator corresponding to a Stern--Gerlach measurement, by providing a physical model of this measurement process. In this way a unique spin operator can be identified given the spatial orientation and velocity of the Stern--Gerlach apparatus and the velocity of the particle. Notably this operator does not agree with previous competing proposals \cite{Czachor,Ternotworol,Friis-relent}.

A second advantage of working in terms of the Dirac and Maxwell fields instead of the Wigner representations is that global phases and quantum interference come out automatically from the WKB approximation. By considering spacetime Mach--Zehnder interference experiment we arrived at a general relativistic formula for calculating the gravitationally induced phase difference.  In the specific case of gravitational neutron interferometry we reproduced the existing formulae for the gravitationally induced phase difference as various limits of our formula. Our overall approach, however, provides a general, unified, and straightforward way of calculating phases and interference for any situation.

Finally we generalised this formalism to the treatment of multipartite states, entanglement, and teleportation, thereby extending the formalism to include all the basic elements of quantum information theory.

The Lorentz group played a primary role in the construction of qubits in curved spacetime. This role can be understood in terms of how gravity acts on physical objects. When an object is moved along some path $x^\mu(\lambda)$ in spacetime it passes through a sequence of tangent spaces. These are connected by infinitesimal Lorentz transformations that are determined from the trajectory and the gravitational field (i.e.\;the connection 1-form $\omega_{\mu\ J}^{\ I}$). That is, apart from a possible global phase, this sequence of infinitesimal Lorentz transformations determines how an object is affected by the gravitational field. In particular, if the object has internal degrees of freedom that transform under the Lorentz group, we can determine the effect of gravity on the state of these internal degrees of freedom. For example, this is the explanation for the presence of spin connection terms in the fermion Fermi--Walker transport. It is also an explanation for why the photon Wigner rotation is simply a rotation of the linear polarisation and so respects the helicity of the photon: no Lorentz boost can change frames sufficiently to change the helicity of a photon.

In this chapter we focused on just two physical realisations which constituted non-trivial representations of the Lorentz group. We can nevertheless contemplate other realisations such as composite two-level systems. In order to understand how gravity acts on the qubit state the same general approach applies: One needs to provide a mathematical model of the physical system. Once a model is established one can in principle determine how (if at all) the quantum state transforms under a Lorentz transformation. In addition to this there are other possible gravitational influences on the quantum state such as gravitationally induced phases.

As a concrete example of a physical realisation that would behave very differently to the elementary realisations treated in this chapter, consider a two-level system where the two levels are energy eigenstates $\ket{E_1}$ and $\ket{E_2}$. From ordinary non-relativistic quantum mechanics we know that the total state $\ket{\psi}=a\ket{E_1}+b\ket{E_2}$ will undergo the evolution
\begin{eqnarray*}
\ket{\psi(t)}=a\Ee^{\ii E_1t/\hbar}\ket{E_1}+b\Ee^{\ii E_2t/\hbar}\ket{E_2}.
\end{eqnarray*}
If the composite object is much smaller than the curvature scale and the acceleration is sufficiently gentle to not destroy it we can obtain a fully general relativistic generalisation by simply replacing the Newtonian time $t$ with the proper time $\tau$:
\begin{eqnarray*}
\ket{\psi(\tau)}=a\Ee^{\ii E_1 \tau/\hbar}\ket{E_1}+b\Ee^{\ii E_2 \tau/\hbar}\ket{E_2}.
\end{eqnarray*}
Thus, because of the energy difference we develop a relative phase between the two energy levels which is proportional to the proper time of the trajectory. In principle we can make use of such a two-level system to measure the proper time of a spacetime trajectory.  Since proper time is path dependent we see that the transport of the quantum state is also path dependent, and we can also contemplate possible interference experiments.

The formalism presented in this chapter provides a basis for quantum information theory of localised qubits in curved spacetime. One theoretical application of this is to extend the applicability of clock synchronisation \cite{Josza,Giovannetti01,deBurgh} and reference frame sharing \cite{RudolphSanders01,BRSDialogue} to curved spacetime, and provide an interesting physical scenario for the study of symmetries in quantum mechanics \cite{JonesWiseman,Spekkens}.

This formalism also has potential measurement applications. For example, in the last ten or fifteen years there has been interest in precision measurement of the effects of general relativity. Most recently there has been the experimental confirmation of the predicted frame-dragging effect by Gravity Probe B \cite{gravityprobeB}.\footnote{See also \cite{Brodutch11,BDT11} for a recent theoretical analysis of polarisation rotation due to gravity.} On the other hand, in the same period there has been an increased interest in quantum precision measurements using techniques from quantum information theory \cite{Preskill,Childs,Dariano,Giovannetti04,Chiribella05,Xiang2011}, in particular utilising entangled states. The formalism of this chapter provides a bridge between these developments, providing a solid foundation for considering the effects of gravity on quantum states, and for considering the design of precision measurements of these effects. For localised qubits in curved spacetimes as defined in this chapter, the effect of gravity enters as classical parameters in the unitary evolution of the quantum state. Therefore, one should be able to use these same quantum information theory techniques to increase the precision in measurements of the gravitational field. It is plausible that such an amalgamation of the transformation of the discrete degrees of freedom of a quantum state and the phase accumulation, by increasing the degrees of freedom to be measured, will increase the sensitivity with which possible future sophisticated precision quantum measurements can measure gravitational effects.

For example, spacetime torsion is generally believed, even if non-zero, to be too small to measure with present day empirical methods \cite{Bergmann}. One problem is that torsion, as it is conventionally introduced in Einstein--Cartan theory, does not have any propagating degrees of freedom. Thus, the torsion in a spacetime region is non-zero if and only if the spin density is non-zero there. Experiments to measure torsion thus require objects to pass  {\em though} a material with non-zero spin density to accumulate an effect, while accounting for standard interactions. Needless to say, measuring torsion is then very difficult. However, electrons decouple from matter in the high energy WKB limit and effectively only feel the gravitational field including torsion. Thus, the spin of high energy fermions might carry information about the spacetime torsion.

Finally, on the more speculative side it would perhaps also be interesting to model closed timelike curves \cite{Deutsch,Ralph,Bartlett} within this formalism. It is possible that the relativistic approach to state evolution and bipartite states provides rules or constraints for the manipulation of quantum information in closed timelike curves.

To summarise: in this chapter we have provided a complete account of the transport and measurement of localised qubits, realised as elementary fermions or photons, in curved spacetime. The manifest Lorentz covariance of the formalism allows for a relativistic treatment of qubits, with a perhaps more straightforward interpretation than approaches based on Wigner representations. The treatment of multipartite states, entanglement and interferometry provides a basis for quantum information theory of localised qubits in curved spacetime.

%\section*{Acknowledgements}
%We are indebted to Stephen Bartlett, Daniel Terno, Bruce Yabsley, Don Melrose, Jorma Louko, Florian Girelli, Gerard Milburn and other participants at the conference RQI4 for several stimulating and helpful discussions. We thank Emma Nimmo for help with diagrams. This research was supported by the Perimeter Institute--Australia Foundations (PIAF) program, and the Australian Research Council grant DP0880860.

%%%%%%%%%%%%%%%%%%%%%%%%%%%%%%%%%%%%%%%%%%%%%%%%%%%%%%%%%%%%%%%%
%%%%%%%%%%%%%%%%%%%%%%%%%%%%%%%%%%%%%%%%%%%%%%%%%%%%%%%%%%%%%%%%%

%%%%%%%%%%%%%%%%%%%%%%%%%%%%%%%%%%%%%%%%%%%%%%%%%%%%%%%%%%%%%%%%
%%%%%%%%%%%%%%%%%%%%%%%%%%%%%%%%%%%%%%%%%%%%%%%%%%%%%%%%%%%%%%%%%
%\bibliography{RQI,RQI_RSO}
%%\end{indented}
%\end{document} 

%% file: SGWKB/relspinmeas0.5.tex
%\documentclass[preprint,1p]{elsarticle}
%\journal{Annals of Physics}
%%\documentclass[10pt,a4paper]{article}
%\usepackage{a4wide}
%\usepackage{amsmath}
%\usepackage{amsfonts}
%\usepackage{amssymb}
%\usepackage[hypertex,bookmarks = false, pdfstartview = FitV, colorlinks = false, linkcolor=green, urlcolor = blue]{hyperref}
%
%% MATH ENVIRON----------------------------------------------------
%\newcommand\bpm{\begin{pmatrix}}
%\newcommand\epm{\end{pmatrix}}
%\newcommand\be{\begin{equation}}
%\newcommand\ee{\end{equation}}
%%%%MATHS COMMANDS%%%
%\newcommand{\half}{\frac12}
%\newcommand{\eps}{\varepsilon}
%\newcommand{\ket}[1]{\left\vert{#1}\right\rangle}
%\newcommand{\bra}[1]{\left\langle{#1}\right\vert}
%\newcommand{\kb}[2]{\vert#1\rangle\!\langle#2\vert}
%\newcommand{\bk}[2]{\left\langle#1\vert#2\right\rangle}
%\newcommand{\di}{\mathrm{d}}
%\newcommand{\abs}[1]{\left|#1\right|}
%\newcommand{\m}[1]{\mathcal{#1}}
%\newcommand{\Hi}{\m H}
%\newcommand{\ii}{\mathrm{i}}
%\newcommand{\Ee}{\mathrm{e}}
%\newcommand\MA{\text{\tiny{MA}}}
%\newcommand\RF{\text{\tiny{RF}}}
%\newcommand\SG{\text{\tiny{SG}}}
%\DeclareMathOperator{\Tr}{Tr}
%
%%%%%%%%%%%%%%%%%%%%%%%%%%%%%%%%%%%%%%%%%%%%%%%%%%
%\begin{document}
%\bibliographystyle{ieeetr}
%\biboptions{sort&compress}

\chapter{WKB analysis of relativistic Stern--Gerlach measurements\label{ch-RSO}}

%\author[rvt]{Matthew C Palmer} \ead{m.palmer@physics.usyd.edu.au}
%\author[rvt]{Maki Takahashi}\ead{m.takahashi@physics.usyd.edu.au}
%\author[focal]{Hans F Westman\corref{RSM-cor1}}\ead{hwestman74@gmail.com}
%\cortext[cor1]{Principal corresponding author}
%\address[rvt]{School of Physics, The University of Sydney, Sydney, NSW 2006, Australia}
%\address[focal]{Instituto de F\'isica Fundamental, CSIC, Serrano 113-B, 28006 Madrid, Spain}
%%%%%%%%%%%%%%%%%%%%%%%%%%%%%%%%%%%%%%%%%%%%%%%%%%
%\begin{abstract}
\subsection*{Abstract}
Spin is an important quantum degree of freedom in relativistic quantum information theory. This chapter provides a first-principles derivation of the observable corresponding to a Stern--Gerlach measurement with relativistic particle velocity. The specific mathematical form of the Stern--Gerlach operator is established using the transformation properties of the electromagnetic field. To confirm that this is indeed the correct operator we provide a detailed analysis of the Stern--Gerlach measurement process. We do this by applying a WKB approximation to the minimally coupled Dirac equation describing an interaction between a massive fermion and an electromagnetic field. Making use of the superposition principle we show that the $+1$ and $-1$ spin eigenstates of the proposed spin operator are split into separate packets due to the inhomogeneity of the Stern--Gerlach magnetic field. The operator we obtain is dependent on the momentum between particle and Stern--Gerlach apparatus, and is mathematically distinct from two other commonly used operators. The consequences for quantum tomography are considered.

\newpage

%%%%%%%%%%%%%%%%%%%%%%%%%%%%%%%%%%%%%%%%
\section{Introduction}
%%%%%%%%%%%%%%%%%%%%%%%%%%%%%%%%%%%%%%%%

Over the past decade or so there has been a growing interest in the field of relativistic quantum information \cite{Czachor,Alsing,PeresScudoTerno02,Ternotworol,PeresTerno04,TerashimaUeda03,BartlettTerno05, Caban1,Caban2,Louko,Landulfo,Friis-relent,Fuentes,Brodutch11,BDT11,Pablo1111,PTW11}, the goal being to develop a mathematical framework which combines aspects of both relativity and quantum information theory. The aim of this program is primarily to shed light on the relationship between these two cornerstones of physics but also to investigate possible near future applications in areas such as long range quantum communication in which relativistic effects cannot be neglected (as in \chref{sec-LQiCST}).

One of the necessary features for this program is a relativistic measurement formalism, i.e.\;a recipe for extracting empirical predictions given a measurement setup and a quantum state. First and foremost this formalism is required to be Lorentz invariant in the sense that the predicted statistics should be independent of the reference frame in which we choose to describe the experiment. In order to do this it will be convenient to introduce a notation which is manifestly Lorentz covariant. As result of this we will be required to not only recast Hermitian observables into this relativistic notation but also replace the standard non-relativistic inner product. The advantage of doing this will not only be to develop a relativistic measurement formalism which is manifestly Lorentz invariant but, as we shall see, will also greatly simplify the derivation of a relativistic Stern--Gerlach measurement operator.

This chapter will be concerned with relativistic spin measurements. In the literature there have been several proposals dating back to the 1960's for relativistic spin operators and these have been studied in the context of quantum field theory for various reasons (see e.g.\;\cite{FoldyWouthuysen,Mashhoon95,HehlNi,Ryder98,Ryder99}). More recently, these operators have been used in relativistic quantum information theory to predict measurement statistics for relativistic spin measurements \cite{Czachor,Ternotworol,CzachorWilczewski,Caban1, Caban2,Friis-relent}. The approach of this chapter will not follow these proposals. Rather we will follow a strictly operational approach, where we will expand on results developed in \cite{PTW11, Pablo1111}. Specifically, we will derive the relevant spin operator for a Stern--Gerlach measurement of a relativistic massive fermion. Importantly, our operational approach yields a Hermitian spin operator which is mathematically distinct from these previous proposals.

The outline of this chapter is as follows: We will begin by reviewing the manifestly Lorentz covariant formalism developed in \secref{sec-LQiCST}. We will then derive the relativistic Stern--Gerlach spin observable by modelling a  Stern--Gerlach measurement. Firstly, the specific mathematical form of the Stern--Gerlach operator is established using the transformation properties of the electromagnetic field. Next, to confirm that this is indeed the correct operator we provide a detailed analysis of the Stern--Gerlach measurement process. We do this by applying a WKB approximation to the minimally coupled Dirac equation describing an interaction between a massive fermion and an electromagnetic field. Making use of the superposition principle we show that the $+1$ and $-1$ spin eigenstates of the proposed spin operator are split into separate packets due to the inhomogeneity of the Stern--Gerlach magnetic field. We conclude by discussing the consequences for quantum tomography.

%%%%%%%%%%%%%%%%%%%%%%%%%%%%%%%%%%%%%%%%
\section{Mathematical description of spin qubits \label{RSM-sec-mathdesc}}
%%%%%%%%%%%%%%%%%%%%%%%%%%%%%%%%%%%%%%%%

Before we can describe a quantum mechanical relativistic spin measurement, we must first specify how one can represent spin in such relativistic scenarios, and furthermore specify what the transformation properties under the Lorentz group are for such a representation. Having identified a particular representation we will then need to develop a measurement formalism and identify the form of Hermitian observables.

There are two main ways in which one can represent spin. A common way is to make use of the Wigner representations \cite{Weinberg}, which are in fact infinite dimensional unitary representations of the Poincar\'e group.  This group has the added symmetry of translational invariance, and consequently Wigner basis states $\ket{p,\sigma}$ are labelled with both momentum $p$ and spin $\sigma=1,2$. In this representation the momentum $p$ transforms under the Lorentz group according to $p^\alpha\to \Lambda^\alpha_{\ \beta} p^\beta$, where $\Lambda$ is a general Lorentz transformation and $\alpha, \beta,\ldots = 0,1,2,3$ are spacetime indices, with spacetime metric $\eta^{\alpha\beta} = \mathrm{diag}(1,-1,-1,-1)$. However, the spin component strictly transforms under the {\it Wigner rotations} and constitutes a representation of what is called {\it Wigner's little group}, which is isomorphic to $SU(2)$. Specifically, Wigner's little group consists of the set of Lorentz transformations under which the `standard' momentum $p^{\alpha} = m\delta^\alpha_0$ of a particle with mass $m$ is left invariant \cite{Weinberg}, where $\delta^\alpha_{\beta}$ is the Kronecker delta symbol.

The representation used in this chapter, and in \chref{sec-LQiCST} to which we refer the reader for further details and theoretical background, is distinct from the Wigner representation and de-emphasises the use of the Wigner rotations. Here the mathematical object representing spin is an $SL(2,\mathbb C)$ spinor. This is a two-component complex-valued object $\psi_A$ with index $A=1,2$ that transforms covariantly under the spin-$\half$ representation of the Lorentz group, i.e.\;by $\psi_A\to\Lambda_A^{\ B}\psi_B$. This transformation law differs from the Wigner rotations, which are spatial rotations defined using a preferred frame. The reason why we adopt this alternative but equivalent representation of spin is primarily because we will insist on manifest Lorentz covariance which will greatly simplify our derivation of our relativistic spin operator. However, the results of this chapter are presented in a form in which they can be interpreted in the Wigner representation if desired. A comparison of the Wigner and Weyl representations is presented in \chref{ch-2}.

As this formalism may not be familiar to some, we will first briefly summarise the notation and the key features. Specifically, we will review spinor notation, the definition of an inner product, the modified notion of unitarity, and finally Hermitian operators and observables.

%%%%%%%%%%%%%%%%%%%%%%%%%%%%%%%%%%%%%%%%
\subsection{Representation of spin and spinor notation}
%%%%%%%%%%%%%%%%%%%%%%%%%%%%%%%%%%%%%%%%

A spinor is fundamentally a two-component complex vector $\psi_{A}$ living in a two-dimensional complex vector space $W$. Here spinors are taken to be irreducible representations of $SL(2,\mathbb{C})$, which forms the double cover of the identity component of the Lorentz group, $SO^{+}(1,3)$. As such, $W$ carries a spin-$\half$ representation of the Lorentz group. In the spirit of Relativity we will use a geometric notation similar to that used for tensors. Therefore, a spinor $\psi_{A}$ carries an $SL(2,\mathbb C)$ spinor index $A=1,2$.

Complex conjugation takes a spinor $\psi_{A}\in W$ to a spinor $\overline{\psi_{A}} = \overline{\psi}_{A'}\in\overline{W}$, the conjugate space of $W$. We distinguish the elements of $\overline W$ by placing a prime on the spinor index: $A'$ (as is the notation commonly used in treatments of spinors \cite{Wald,Penrose,Bailin}). The summation of indices then follows the Einstein summation convention: We can only contract when one index appears as superscript and the other as subscript, and only when the indices are either both primed or both unprimed, e.g.\;$\phi_A\psi^A$ and $\xi_{A'} \chi^{A'}$, but not $\bar\phi_{A'} \psi^{A}$. See \appref{secspinornotation} for more details on the mathematics of spinors.

%%%%%%%%%%%%%%%%%%%%%%%%%%%%%%%%%%%%%%%%
\subsection{Lorentz group and $SL(2,\mathbb C)$}
%%%%%%%%%%%%%%%%%%%%%%%%%%%%%%%%%%%%%%%%

 We are concerned with the spin of a massive fermion such as an electron. Such an object is usually taken to be represented by a four-component Dirac field $\Psi(x)$, which constitutes a reducible spin-$\half$ representation of the Lorentz group. At the same time we would like a qubit representation of our spin-$\half$ system, so it is natural to use a two-dimensional object. Such a representation can be found by working with the Dirac field in the Weyl representation. In this representation the Dirac field splits into two 2-component  $SL(2,\mathbb C)$ spinors $\Psi(x) = (\phi_A(x),\chi^{A'}(x))$ which constitute the left and right handed irreducible spin-$\half$ representations of the Lorentz group. We will represent qubits with the two-component left-handed Weyl spinor field $\phi_{A}(x)$. Working instead with the right-handed component $\chi^{A'}$ would yield the same results.

%%%%%%%%%%%%%%%%%%%%%%%%%%%%%%%%%%%%%%%%
\subsubsection{$SL(2,\mathbb C)$}
%%%%%%%%%%%%%%%%%%%%%%%%%%%%%%%%%%%%%%%%

We now turn to the Lorentz group. In the Weyl representation, the Dirac gamma matrices take on the form
\[
\gamma^{\alpha} = \begin{pmatrix}0&\sigma^{\alpha}\\\bar\sigma^\alpha&0\end{pmatrix}
\]
where $\sigma^\alpha\equiv(I,\sigma^i)$ is the Pauli 4-vector and $\bar\sigma^\alpha\equiv(I,-\sigma^i)$. If we use the convention in \cite{Bailin,DHM2010} whereby the primed index for $\bar{\sigma}^\alpha$ is a row index and unprimed is a column index, and the opposite for $\sigma^\alpha$, we can explicitly identify both $\bar\sigma^{0 A'A} = \delta^{A'A}$ and $\sigma^0_{\ AA'}=\delta_{AA'}$ as the $2\times2$ identity matrix, and the spatial parts $-\bar{\sigma}^{iA'A}$ and $\sigma^i_{\ AA'}$ as the usual Pauli matrices. The Weyl representation allows us to extract from the Dirac algebra $\{\gamma^\alpha,\gamma^\beta\}=2\eta^{\alpha\beta}$ the left-handed two-component algebra
\begin{align}
\sigma^\alpha_{\ AA'}\bar\sigma^{\beta A'B}+\sigma^\beta_{\ AA'}\bar\sigma^{\alpha A'B}=2\eta^{\alpha\beta}{\delta}_A^{\ B}\label{RSM-dalg}
\end{align}
where $\delta_{A}^{\ B}$ is the Kronecker delta.

The generators of the group are constructed as $S^{\alpha\beta} = \frac\ii4 [\gamma^{\alpha},\gamma^{\beta}]$ for the 4-component formalism or in spinor notation
\begin{equation}
{L^{\alpha\beta}}_A^{\ B}=\frac\ii4 \left(\sigma^\alpha_{\ AA'}\bar\sigma^{\beta A'B}-\sigma^\beta_{\ AA'}\bar\sigma^{\alpha A'B}\right)\label{RSM-gen}
\end{equation}
for the left-handed 2-spinor.

Note that the Pauli 4-vector is not referred to as an operator in this formalism. An operator for spinors $\psi_B$ instead carries an index structure $\op_{A}^{\ B}$. Wherever possible we will keep indices implicit and use the standard notation $\hat \op$ for operators. Using this notation the components of the generators can be written as
\begin{align*}
\hat L^{0j}=&\frac\ii2 \hat\sigma^{j}, \qquad
\hat L^{ij}=\frac12\eps^{ij}_{\ \ k}\hat\sigma^k
\end{align*}
where $\hat\sigma^{i}$ are the standard Pauli operators.\footnote{While having the same components as the Pauli matrices, note the distinction between the {\it operator} $\hat\sigma^{i}$ which maps a spinor $\psi_{A}\to\phi_{A}$ whereas the object $\bar{\sigma}^{iA'A}$ would map $\psi_{A}\to\phi^{A'}$.} The $\hat L^{0j}$ components generate boosts and the $\hat L^{ij}$ components generate rotations.

The Pauli 4-vector and $\bar\sigma^{\alpha}$ play a special role because they are invariant under Lorentz transformations on all indices, that is \cite{DHM2010},
\begin{equation}
\Lambda^\alpha_{\ \beta}\Lambda^{A}_{\ B}\bar\Lambda^{A'}_{\ B'} \bar{\sigma}^{\beta B'B}=\bar{\sigma}^{\alpha A'A}\label{RSM-LIsigma}
\end{equation}
where $\Lambda^\alpha_{\ \beta}$ is an arbitrary spin-$1$ Lorentz transformation and $\Lambda^{A}_{\ B}$ is the corresponding spin-$\half$ Lorentz transformation, and similarly for $\sigma^\alpha$.

%%%%%%%%%%%%%%%%%%%%%%%%%%%%%%%%%%%%%%%%%%%%%%%%%%%%%%%%%
\subsection{Lorentz invariant measurement formalism\label{RSM-sec-unitarity}\label{RSM-sec-measurement}}
%%%%%%%%%%%%%%%%%%%%%%%%%%%%%%%%%%%%%%%%%%%%%%%%%%%%%%%%%%%%%%%%%%%%%

Up to this point we have discussed spinor notation and the spin-$\half$ representation of the Lorentz group. This structure on its own does not provide a quantum formalism. To achieve a quantum mechanical description we must also introduce an inner product to promote the spinor space $W$ to a Hilbert space $\m H$, and then construct a formalism to extract predictions according to the rules of quantum mechanics. This formalism differs to that used in the Wigner representation since the two component irreducible representation of $SL(2,\mathbb C)$ constitutes a {\it non-unitary} spin-$\half$ representation of the Lorentz group: A spin-$\half$ Lorentz boost (with $\gamma:=(1-\beta^2)^{-\half}$) is given by
\[
\hat L(\beta) = \sqrt{\frac{\gamma+1}{2}}\hat\sigma^{0}+\sqrt{\frac{\gamma-1}{2\beta^2}} \beta_{i}\hat\sigma^{i}.
\]
The eigenvalues of this matrix are $\sqrt{(\gamma+1)/2}\pm\sqrt{(\gamma-1)/2}$. Nevertheless, a notion of unitarity can be recovered by introducing a suitable inner product and not insisting on using representations. The latter is an immaterial effect, whereas the former requires a reformulation of Hermitian operators. We will here simply review the results of the formalism derived in \chref{sec-LQiCST}.

%%%%%%%%%%%%%%%%%%%%%%%%%%%%%%%%%%%%%%%%
\subsubsection{The quantum state and inner product\label{RSM-newinnerproduct}\label{RSM-secfermionQS}}\label{RSM-appIP}
%%%%%%%%%%%%%%%%%%%%%%%%%%%%%%%%%%%%%%%%

In order to promote the spinor space $W$ to a Hilbert space $\m H$ an inner product is required. In spinor notation a sesquilinear inner product requires a spinorial object with index structure $I^{A'A}$. The appropriate object is given by $I_{u}^{A'A}\equiv u_\alpha\bar{\sigma}^{\alpha A'A}$ where $u_{\alpha}$ is the 4-velocity of the particle carrying the spin. For a Wigner state this would be represented by the momentum $p$ in $\ket{p,\psi}$. The inner product between states represented by the two spinors $\psi_A^{1}$ and  $\psi_A^{2}$ is
\begin{eqnarray}
\bigbk{\psi^{1}}{\psi^{2}}=I_{u}^{A'A}\bar{\psi}^{1}_{A'}\psi^{2}_A=u_\alpha\bar{\sigma}^{\alpha A'A}\bar{\psi}^{1}_{A'}\psi^{2}_A
\label{RSM-IP}
\end{eqnarray}
where the connection between Dirac bra--ket notation and spinor notation is identified as
\begin{equation}
|\psi\rangle\sim\psi_A\qquad\langle\psi|\sim I_{u}^{A'A}\bar{\psi}_{A'}.\label{RSM-braket}
\end{equation}
The inner product \eqref{RSM-IP} is Lorentz invariant. This follows immediately from the fact that all indices have been contracted, and that $\bar\sigma^{\alpha A'A}$ is invariant under Lorentz transformation, \eqref{RSM-LIsigma}. This inner product emerges by taking the WKB limit of the Dirac field. In this limit not only do we obtain a well defined inner product on the spinor space and a classical trajectory with velocity $u^{\alpha}$ for a localised wave-packet but in addition a conserved probability current $j^{\alpha}$ that ultimately allows us to apply a quantum mechanical interpretation to the state $\psi_{A}$. This allows us to promote $W$ to a Hilbert space. Given that $I_{u}^{A'A}$ depends on the particle's 4-velocity or equivalently 4-momentum, the corresponding Hilbert space is labelled with momentum $p$:  $\m H_{p}$. With this inner product we have a finite dimensional unitary formalism describing the transformation of spin  $\psi_{A}$ under arbitrary Lorentz transformations. Note that we no longer strictly have a representation of the Lorentz group since an arbitrary Lorentz transformation $\Lambda$ will correspond to a map between distinct Hilbert spaces $\Lambda: \m H_{p}\to \m H_{\Lambda p}$, rather than within a single space.\footnote{We recall that a representation of a group consists of the set of linear operators acting on a single vector space.} By not insisting on using representations we have managed to sidestep Wigner's theorem that any faithful unitary representation of the Lorentz group must be infinite dimensional \cite{Wigner}.

%%%%%%%%%%%%%%%%%%%%%%%%%%%%%%%%%%%%%%%%
\subsubsection{Hermitian operators \label{RSM-sec-hermitian}}
%%%%%%%%%%%%%%%%%%%%%%%%%%%%%%%%%%%%%%%%

Now that we have a well-defined inner product, we can consider the general form of Hermitian operators, and derive the mathematical form of observables. To do this we start with the standard Hermitian property for an operator $\hat \op$:
\begin{align*}
0 &= \bk{\chi}{\op\psi} - \bk{\op\chi}{\psi}\\
   &= I_{u}^{A'A}\bar\chi_{A'} \op_A^{\ B}\psi_B - I_{u}^{A'A} \bar \op_{A'}^{\ B'}\bar\chi_{B'} \psi_A.
\end{align*}
For this to hold for all $\chi_{A}$ and $\psi_{A}$ we must have that $I_{u}^{A'B}\op_B^{\ A} = I_{u}^{B'A} \bar \op_{B'}^{\ A'}$. Making use of \eqref{RSM-dalg} and \eqref{RSM-gen}, and using the self-dual property $\hat L^{\alpha\beta}=\half \ii\eps^{\alpha\beta\gamma\delta}\hat L_{\gamma\delta}$ \cite[Eqn 2.74]{DHM2010} to introduce the {\it Pauli--Lubanski vector} $\hat W^\alpha(p):=\half\eps^{\alpha\beta\gamma\delta}p_\beta \hat L_{\gamma\delta} = \ii p_\beta  \hat L^{\alpha\beta}$, one can then show that a Hermitian operator $\hat \op$ must be of the form
\begin{equation}
\hat \op= 2\ii n_\alpha u_\beta  \hat L^{\alpha\beta}+n_\alpha u_\beta \eta^{\alpha\beta} \hat {\m I}=n_\alpha \left(-\frac{2\hat W^\alpha(p)}{m}+u^{\alpha}\hat{\m I}\right)\label{RSM-fermionoperator}
\end{equation}
with $\hat{\m I}$ the identity, and where each operator is identified by a Lorentz 4-vector $n_\alpha$ of real coefficients. Every Hermitian operator has a real eigenvalue spectrum, and its eigenstates $\ket{\psi^\pm}$ are orthogonal with respect to the inner product $I_u^{A'A}$.

We are now interested in spin observables formed from Hermitian operators \eqref{RSM-fermionoperator}. A spin observable evaluated in the particle's rest frame should reduce to the non-relativistic expression $n_i\sigma^i$ where $n^i$ is the normalised spin measurement direction. This implies that $n^\alpha$ in \eqref{RSM-fermionoperator} is orthogonal to $u^\alpha$, i.e.\;we have the condition
\be
u_\alpha n^\alpha=0.\label{RSM-eq-orthogcond}
\ee
The magnitude of $n^\alpha$ only rescales the eigenvalues of the observable and without loss of generality we can normalise it so that it is spacelike with $n^2=-1$. The Lorentz invariant expectation value of an observable $\hat \op$ for a spinor $\psi_A$ is then given by
\be
\langle\psi|\hat \op|\psi\rangle=-I_u^{A'A}\bar{\psi}_{A'} n_\alpha \frac{2W^{\alpha B}_{\ A} }m \psi_B=-n_\alpha\bar\sigma^{\alpha A'A}\bar\psi_{A'}\psi_A. \label{RSM-spacetimemeas}
\ee
This expression is covariant and has been written in the Weyl representation where we have made use of the relationship $\frac 2m I_u^{A'B}W^{\alpha A}_{\ B}=\bar\sigma^{\alpha A'A}-u^\alpha u_\beta\bar\sigma^{\beta A'A}$. \footnote{For those familiar with the Wigner representation, the expectation value $\bra{k,\psi}\mathbf n\!\cdot\mathbf{\hat\sigma}\ket{k,\psi}$ in the rest frame is written as $-2\bra{p,\psi} n_\alpha \hat W^\alpha(p)/m\ket{p,\psi}$ in a boosted frame, with $\hat W^\alpha(p)/m=\half  L(p)^\alpha_{\ i}\hat\sigma^i$  \cite{Bogoliubov,Czachor,Ternotworol}, where $L(p)$ is the boost relating the frames.} Lorentz invariance follows immediately because all spinor and spacetime indices have been contracted, and all objects transform covariantly.

%%%%%%%%%%%%%%%%%%%%%%%%%%%%%%%%%%%%%%%%
\section{Intuitive derivation of the Stern--Gerlach observable}\label{RSM-SGOP}
%%%%%%%%%%%%%%%%%%%%%%%%%%%%%%%%%%%%%%%%

The task now is to determine the correct spin observable for a Stern--Gerlach measurement in which the Stern--Gerlach apparatus and particle carrying the spin have a relativistic relative velocity. Given the measurement formalism and formation of Lorentz invariant expectation values outlined in the previous section, the problem of determining  a `relativistic Stern--Gerlach spin operator' is reduced to simply determining how $n_{\alpha}$ is related to the direction of the Stern--Gerlach apparatus. This analysis will follow the structure of \secref{secFermion}.

%%%%%%%%%%%%%%%%%%%%%%%%%%%%%%%%%%%%%%%%
\subsection{The non-relativistic Stern--Gerlach experiment\label{RSM-sec-nonrelSG}}
%%%%%%%%%%%%%%%%%%%%%%%%%%%%%%%%%%%%%%%%

Before we consider the relativistic case it is helpful to first review a standard non-relativistic Stern--Gerlach spin measurement represented by the arbitrary non-relativistic Hermitian observable $n_{i}\hat\sigma^{i}$. In this case a particle is passed though an inhomogeneous magnetic field. This causes the wave-packet to separate into two packets of orthogonal spin, after which a position measurement records the outcome. In the rest frame of the particle, the fermion is exposed to a magnetic field $B^\SG_i = |B^\SG| b^\SG_{i}$ for a short period of time, which is the magnetic field as measured specifically in the Stern--Gerlach rest frame, denoted by $\text{SG}$. The direction $b^\SG_{i}$ of the magnetic field defines the quantisation direction of the spin, and the {\it gradient} of the magnetic field $\nabla_{i}|B^\SG|$ determines the rate and direction along which the wave-packet splits into eigenstates of $b^\SG_{i}\hat\sigma^{i}$ \cite{PeresQM}. Therefore in the non-relativistic case the measurement direction $n_{i}$ is simply the direction of the magnetic field in the Stern--Gerlach rest frame, $b^\SG_i$.

%%%%%%%%%%%%%%%%%%%%%%%%%%%%%%%%%%%%%%%%
\subsection{The relativistic Stern--Gerlach experiment}
%%%%%%%%%%%%%%%%%%%%%%%%%%%%%%%%%%%%%%%%

We now turn to the relativistic scenario. In this case the qubit is now moving through the Stern--Gerlach apparatus with relativistic velocity. Viewed in the rest frame of the particle, denoted $\text{RF}$, the measurement process is indistinguishable from the non-relativistic one described in \S\ref{RSM-sec-nonrelSG}. However, in this frame the fermion will experience a {\it transformed} magnetic field $B_{i}^{\RF} = |B^{\RF}|b_{i}^{\RF}$. The measurement direction is now given by $n^i=b^i_\RF$, giving a spin observable of $b_i^\RF\sigma^i$. \footnote{Similarly it is the gradient of the rest frame magnetic field $\nabla_{i}|B^{\RF}|$ which now determines the rate and direction along which the wave-packet splits.} This spin observable is written in the specific frame of the particle rest frame, and due to the transformation properties of the magnetic field, the relationship between $b^\RF_i$ and the orientation of the Stern--Gerlach apparatus is nontrivial. The goal is now to arrive at a covariant expression of the spin observable in terms of the Stern--Gerlach direction $b_{\alpha}^\SG$ and the 4-velocities of the Stern--Gerlach apparatus and the particle.

In order to do this we assume that the electromagnetic field generated consists of purely a magnetic field in the rest frame of the Stern--Gerlach device,
\begin{equation}
F_{\alpha\beta} = -\epsilon_{\alpha\beta\gamma\delta}v^{\gamma}B^{\delta}_\SG\overset{**}{=}\begin{pmatrix}0&0\\0&B_{ij}\end{pmatrix}\label{RSM-SGemfield}
\end{equation}
where $v^{\gamma}$ is the 4-velocity of the Stern--Gerlach device and $B^{\delta}_\SG$ is the magnetic field 4-vector of the Stern--Gerlach device. The double star `$**$' of the right hand side indicates that it has been evaluated explicitly in the frame where $v^{\alpha} \overset{**}{=}(1,0,0,0)$, i.e.\;the Stern--Gerlach rest frame in which the Stern--Gerlach  magnetic 4-vector is given by $B^\delta_\SG\overset{**}{=}(0,B^i_\SG)$.

We can now determine the 4-vector $B_{\RF}^\alpha$ defined by $B^\alpha_\RF\overset*=(0,B^i_\RF)$, where `$*$' indicates evaluation in the particle rest frame, $u^\alpha\overset*=(1,0,0,0)$. The covariant expression for $B^\alpha_\RF$ is given by $B_{\RF}^\alpha\equiv-\frac12\epsilon^{\alpha\beta\gamma\delta}u_\beta F_{\gamma\delta}$, which when inserted into \eqref{RSM-SGemfield} yields
\begin{equation}
B_{\RF}^\alpha=\frac12\epsilon^{\alpha\beta\gamma\delta}u_\beta \epsilon_{\gamma\delta \kappa\lambda}v^\kappa B_\SG^\lambda=B_\SG^\alpha(v\cdot u)-v^\alpha(B_\SG\cdot u)\label{RSM-BRF}
\end{equation}
with $\epsilon^{\alpha\beta\gamma\delta} \epsilon_{\gamma\delta \kappa\lambda}=-2(\delta^\alpha_\kappa\delta^\beta_\lambda-\delta^\alpha_\lambda\delta^\beta_\kappa)$ \cite[p.87]{MTW}, and the notation $a\cdot b\equiv a_\alpha b^\alpha$ indicating a 4-vector scalar product. Considering a spin measurement using this magnetic field, the four-vector $n^\alpha$ in \eqref{RSM-spacetimemeas} is now the normalised direction of the rest frame magnetic field:
\begin{equation}
n^\alpha(m,u,v)\equiv b_{\RF}^\alpha = \frac{B_{\RF}^\alpha}{|B_{\RF}|}
\label{RSM-SGn}
\end{equation}
where $|B_{\RF}|:=\sqrt{-B_{\RF}^\alpha B_{\RF}^\beta\eta_{\alpha\beta}}$, and from \eqref{RSM-BRF} we have $b_{\RF}\cdot u=0$, in agreement with \eqref{RSM-eq-orthogcond}. By \Eeqref{RSM-fermionoperator} the relativistic spin operator is therefore given by
\begin{equation}
\m S_{A}^{\ B}:=-b^\RF_\alpha \frac{{2W^\alpha}_A^{\ B}}m\label{RSM-SG_spin_op},
\end{equation}
and the expectation value of the corresponding measurement is calculated using \eqref{RSM-spacetimemeas}. Expectation values \eqref{RSM-spacetimemeas} are invariant under simultaneous Lorentz transformation of both particle and apparatus, and thus with only the relative velocity between the apparatus and qubit and the spatial orientation of the Stern--Gerlach apparatus, we can calculate the expectation values corresponding to a relativistic Stern--Gerlach spin measurement.

%%%%%%%%%%%%%%%%%%%%%%%%%%%%%%%%%%%%%%%%
\section{WKB analysis of a Stern--Gerlach measurement}\label{RSM-WKB}
%%%%%%%%%%%%%%%%%%%%%%%%%%%%%%%%%%%%%%%%

In \secref{RSM-SGOP} we argued that in the particle rest frame the measurement process is indistinguishable to a non-relativistic one \cite{PeresQM,Ballentine}. From this analysis we saw that it is the direction of the magnetic field, as seen in the particle rest frame, that determines which spin measurement is being carried out. That derivation used standard arguments based on classical relativity and non-relativistic quantum mechanics  (see also \cite{PTW11,Pablo1111}).

In this section we provide a more fundamental analysis of the relativistic Stern--Gerlach measurement process, using the minimally coupled Dirac equation in the WKB limit. Our starting point is an equivalent two component formulation of the Dirac equation called the {\it van der Waerden} equation. The two component field is then expanded, without loss of generality, as a superposition in the eigenbasis of the spin operator \eqref{RSM-SG_spin_op}. Using the linearity of the van der Waerden equation, we can analyse each component of the superposition separately, and identify the classical trajectories of each component. We will use this to show that an inhomogeneous magnetic field results in the splitting of a localised wave-packet with arbitrary spin into two components of orthogonal spin. This analysis singles out \eqref{RSM-SG_spin_op} as the relevant spin operator for a relativistic Stern--Gerlach spin measurement.

%%%%%%%%%%%%%%%%%%%%%%%%%%%%%%%%%%%%%%%%
\subsection{The WKB equations}
%%%%%%%%%%%%%%%%%%%%%%%%%%%%%%%%%%%%%%%%

A qubit physically realised by the spin of a massive fermion is described by the Dirac field, therefore our starting point will be the Dirac equation minimally coupled to the electromagnetic field
\begin{eqnarray}
\ii\gamma^\alpha D_\alpha\Psi=\ii\gamma^\alpha \left(\partial_\alpha-\ii eA_{\alpha}\right)\Psi=m\Psi\label{RSM-curveddiraceq}
\end{eqnarray}
where we define the $U(1)$ covariant derivative as $D_\alpha\equiv\partial_\alpha-\ii eA_\alpha$. $\Psi$ is the Dirac field, $\gamma^{\alpha}$ are the Dirac $\gamma$-matrices and $A_{\alpha}$ is the electromagnetic four potential. Strictly speaking it is the positive frequency solutions of the Dirac equation that describe a one particle state. Furthermore we assume that the strength of the electric field is insufficient to cause particle creation. See \secref{secFermion} for a discussion of spin of a massive fermion as a realisation of a relativistic qubit.

We proceed with the WKB approximation by putting the Dirac equation into a second order form. In the Weyl representation of the Dirac matrices, the field splits into $\Psi=(\phi_A,\chi^{A'})$ \cite{PeskinSchroeder}. The objects $\phi_A$ and $\chi^{A'}$ are each two-component Weyl-spinor fields constituting left- and right-handed spinor representations of $SL(2,\mathbb C)$. In this representation the Dirac equation splits into two separate equations
\begin{subequations}
\begin{align}
\ii\bar{\sigma}^{\alpha A'A}D_\alpha\phi_A&=m\chi^{A'}\label{RSM-left}\\
\ii\sigma^{\alpha}_{\ AA'}D_\alpha\chi^{A'}&=m\phi_A\label{RSM-right}.
\end{align}
\end{subequations}
Solving for $\chi^{A'}$ in \Eeqref{RSM-left}, inserting the result into \eqref{RSM-right}, and rearranging yields a second order equation called the {\it van der Waerden equation} \cite{SakuraiAQM}
\begin{equation}
\eta^{\alpha\beta} D_\alpha D_\beta\phi_A- eF_{\alpha\beta}L^{\alpha\beta\ B}_{\ \ A}\phi_B+m^2\phi_A = 0\label{RSM-vdweqn}
\end{equation}
where $F_{\alpha\beta}=\partial_\alpha A_\beta-\partial_\beta A_\alpha$ is the electromagnetic tensor and we have used that $\hat{L}^{\alpha\beta} = \frac \ii2\sigma^{[\alpha}\bar{\sigma}^{\beta]}$ and $\hat\eta^{\alpha\beta}=\sigma^{\{\alpha}\bar\sigma^{\beta\}}$.

The next step is to consider the van der Waerden equation in the high frequency WKB  limit. In this limit we see that the fermion travels along classical trajectories. We will assume that the field is sufficiently localised for the purposes of the Stern--Gerlach measurement.\footnote{See \chref{sec-LQiCST} for further details on localisation.} The goal then is to show that the wave-packet is split by the Stern--Gerlach magnetic field into two packets of spin corresponding exactly to the eigenstates of the spin operator \eqref{RSM-SG_spin_op}.

Traditional treatments of the WKB approximation begin with an ansatz for the spinor field of the form
\[
\phi_{A}(x) =  \varphi_{A}(x)\Ee^{\ii\theta(x)/\eps}
\]
where $\eps$ is to be thought of as a `dummy' parameter whose only role is to identify the different orders in an expansion. This ansatz is substituted into the van der Waerden equation. One can then expand in the limit $\eps\rightarrow0$; the high frequency limit. Mathematically this corresponds to splitting the field into a rapidly varying phase $\theta$ and a slowly varying envelope $\varphi_{A}$. The phase determines a field of wavevectors $k_{\alpha}\equiv \partial_{\alpha}\theta -e A_{\alpha}$ which in this limit define integral curves along which the envelope is transported.

In the case of a Stern--Gerlach spin measurement we know the initial wave-packet will be split into two wave-packets of orthogonal spin with different wavevectors $k^{\pm}_{\alpha}\equiv \partial_{\alpha}\theta^{\pm} -e A_{\alpha}$. Therefore we must slightly modify the WKB ansatz to
\begin{equation}
\phi_{A} = a \varphi^{+}_{A}\Ee^{\ii\theta^{+}/\eps}+b \varphi^{-}_{A}\Ee^{\ii\theta^{-}/\eps}\label{RSM-wkb_sol}
\end{equation}
where we have, without loss of generality, expanded the spinor field in the eigenbasis of $\m S_{A}^{\ B}$ \eqref{RSM-SG_spin_op} with components $a,b \in\mathbb C$ defined so that $|a|^{2}+|b|^{2} = 1$. The integral curves are determined by the phases  $\theta^{\pm}$ which correspond to the spin eigenstates $\varphi_{A}^{\pm}$. We will see that these components are deflected in two different directions by the inhomogeneous magnetic field.

Using the linearity of the van der Waerden equation we can analyse each component $\phi^{\pm}_{A} = \varphi^{\pm}_{A}\Ee^{\ii\theta^{\pm}/\eps}$ separately. Substituting $\phi^{\pm}_{A}$ into \eqref{RSM-vdweqn} yields
\begin{multline}
\left(\eta^{\alpha\beta}\partial_\alpha \partial_\beta\varphi^{\pm}_A- eF_{\alpha\beta}L^{\alpha\beta\ B}_{\ \ A}\varphi^{\pm}_B+\frac{\ii}{\eps}(2k_{\pm}^\alpha\partial_\alpha\varphi^{\pm}_A+\varphi^{\pm}_A\partial_\alpha k_{\pm}^\alpha) \right.\left.
-\frac{1}{\eps^2}k^{\pm}_\alpha k_{\pm}^\alpha\varphi^{\pm}_A+m^2\varphi^{\pm}_A\right)\Ee^{\ii\theta^{\pm}/\eps} = 0 \label{RSM-vdwwkb}
\end{multline}
where $k^{\pm}_{\alpha}\equiv \partial_{\alpha}\theta^{\pm} -e A_{\alpha}$. It is customary to treat the mass term as $\eps^{-2}$ and we shall do so here. In the $\eps\rightarrow0$ limit, which corresponds to large momentum, we notice that the electromagnetic field term $F_{\alpha\beta}$ has a negligible influence in \eqref{RSM-vdwwkb}. Thus in order for the fermion to `feel' the presence of the electromagnetic field, we will need to treat $F_{\alpha\beta}$ as a $1/\eps$ term.

The WKB approximation proceeds by separating the orders of $\eps$. For our purposes we neglect the lowest order terms and thus obtain the following set of equations
\begin{align}
\frac 1\eps\left(2k_{\pm}^\alpha\partial_\alpha\varphi^{\pm}_A+\varphi^{\pm}_A\partial_\alpha k_{\pm}^\alpha+\ii eF_{\alpha\beta}L^{\alpha\beta\ B}_{\ \ A}\varphi^{\pm}_B\right)&=0, \label{RSM-wkb1}\\
\frac 1{\eps^{2}}\left(k^{\pm}_\alpha k_{\pm}^\alpha-m^2\right)\varphi^{\pm}_A &=0. \label{RSM-wkb2}
\end{align}
The first equation \eqref{RSM-wkb1} will describe the evolution of the spin state of $\varphi_A^\pm$ along a trajectory (as in \chref{sec-LQiCST}). The second equation \eqref{RSM-wkb2} determines the trajectories along which the fermion is transported. As it is, the spin does not couple to the magnetic field, implying that the trajectories cannot be spin-dependent, and thus no spin-dependent deflection of packets can occur. However, if we treat the gradient of the magnetic field as a $\eps^{-2}$ term, we can include such a term in \eqref{RSM-wkb2}:
\begin{align}
\left(k^{\pm}_\alpha k_{\pm}^\alpha-m^2\right)\varphi^{\pm}_A - \eps eF_{\alpha\beta}L^{\alpha\beta\ B}_{\ \ A}\varphi^{\pm}_B&=0. \label{RSM-wkb3}
\end{align}
To zeroth order in $\eps$, \Eeqref{RSM-wkb3} is still the standard dispersion relation, and implies that $k_{\pm}^{\alpha}$ is timelike. However, upon taking the gradient of \eqref{RSM-wkb3}, the second term becomes relevant. It is in this way that the trajectories become spin-dependent, producing the separation of the wave-packet that occurs in a Stern--Gerlach measurement.

%%%%%%%%%%%%%%%%%%%%%%%%%%%%%%%%%%%%%%%%
\subsection{Determining the spin-dependent trajectories\label{RSM-sec-traj}}
%%%%%%%%%%%%%%%%%%%%%%%%%%%%%%%%%%%%%%%%

It is from the integral curves of $\frac{\di x^\alpha_{\pm}}{\di\tau}=u_{\pm}^\alpha(x)\equiv k_{\pm}^\alpha(x)/m$ that we can read off the deflection of the trajectories, where $u^{\alpha}_{+} = u^{\alpha}_{-}$ prior to entering the magnetic field. In order to deduce the implications of \eqref{RSM-wkb3} we first multiply it by $I_{u_\pm}^{A'A}\bar\varphi^{\pm}_{A'}$, obtaining
\begin{align}
\left(k^{\pm}_\alpha k_{\pm}^\alpha-m^2\right)|\varphi^{\pm}|^{2} - \eps eF_{\alpha\beta}I_{u_\pm}^{A'A}L^{\alpha\beta\ B}_{\ \ A}\varphi^{\pm}_B\bar\varphi^{\pm}_{A'}&=0. \label{RSM-wkb4}
\end{align}
Next, we decompose the operator $F_{\alpha\beta}{\hat L^{\alpha\beta}}$ into the electric field $\hat{E}^{\RF}$ and magnetic field $\hat{B}^{\RF}$ operators as measured in the rest frame defined by the initial 4-velocity $u^{\pm}_{\alpha}$. This is given by
\be
\begin{split}
F_{\alpha\beta}\hat L^{\alpha\beta}&=\hat E^\RF+\hat B^\RF\\
& \equiv 2u^{\pm}_{\gamma}u_{\pm}^\alpha F_{\alpha\beta}\hat L^{\gamma\beta}+F_{\alpha\beta}{h_{\pm}}_{\ \gamma}^{\alpha}{h_{\pm}}_{\ \delta}^{\beta} \hat L^{\gamma\delta}\label{RSM-electromagnetic_term}
\end{split}
\ee
where ${h_{\pm}}_{\ \gamma}^{\alpha} \equiv \delta_{\ \gamma}^{\alpha}-u^{\alpha}_{\pm}u^{\pm}_{\gamma}$ is the spacetime projector onto the space orthogonal to the 4-velocity $u_{\pm}^\alpha$. The first term is anti-Hermitian with respect to the inner product $I_{u_\pm}^{A'A}$, whereas the second term is Hermitian.

First consider the magnetic field term $\hat{B}^{\RF}$. Using the self-dual property $\hat L^{\alpha\beta}=\half \ii\eps^{\alpha\beta\gamma\delta}\hat L_{\gamma\delta}$, and substituting the specific form \eqref{RSM-SGemfield} of the electromagnetic field of the Stern--Gerlach apparatus, the expression can be rearranged to give

\begin{equation}
\begin{split}
\hat B^\RF=& F_{\alpha\beta}{h_{\pm}}_{\ \gamma}^{\alpha}{h_{\pm}}_{\ \delta}^{\beta}
\hat L^{\gamma\delta}\\
 =& -2\ii u_{\alpha}B^{\RF}_{\beta}\hat L^{\alpha\beta}\equiv \m |B^\RF|\hat{\m S}\label{RSM-magneticoperator}
\end{split}
 \end{equation}
where $|B_{\RF}|$  is the magnitude of the magnetic field as measured in the rest frame of $u^{\pm}_{\alpha}$ \eqref{RSM-BRF}. We see that the magnetic field operator is in fact the relativistic spin operator \eqref{RSM-SG_spin_op} derived in the previous section multiplied by the field strength. Given that $\varphi^\pm_A$ are defined as eigenstates of this operator, \eqref{RSM-wkb_sol}, we therefore have that
\be
\bra{\psi^{\pm}}\hat{B}^{\RF}\ket{\psi^{\pm}}=\pm |B^{\RF}|\label{RSM-magneticterm}
 \ee
where the quantum state is identified as $\ket{\psi^{\pm}} \sim\varphi^{\pm}_{A}/|\varphi^{\pm}|$ with $|\varphi^\pm|^2\equiv\bar\varphi^\pm_{A'}I_{u_\pm}^{A'A}\varphi_A^\pm$.

Let us now proceed to show that the expectation values of $\hat E_\RF$ in \eqref{RSM-electromagnetic_term} with $\ket{\psi^\pm}$ are zero. Firstly, we define the projector $\hat\Pi^\pm_{B_\RF}\equiv\half(\hat{\m I}\pm\frac1{|B_\RF|^2}\hat B_\RF)$, so we have
\[
\bra{\psi^\pm}\hat E_\RF\ket{\psi^\pm}=\Tr[\hat E_\RF\kb{\psi^\pm}{\psi^\pm}]=\Tr[\hat E_\RF\hat\Pi^\pm_{B_\RF}] =\Tr[\hat E_\RF \half(\hat{\m I}\pm\frac1{|B_\RF|^2}\hat B_\RF)].
\]
Using the Lorentz invariance of the expectation values, we can evaluate the expectation values in the particle rest frame where the operators take on the form $\hat E^\RF\overset*=E_i^\RF\hat\sigma^i$ and $\hat B^\RF\overset*=B_i^\RF\hat\sigma^i$:
\[
\bra{\psi^\pm}\hat E_\RF\ket{\psi^\pm}\overset*=\pm\frac1{|B^\RF|^2}E^\RF_iB_\RF^i.
\]
The electric field vector is given by  $E_\RF^i\overset*=F^{0i}=\eps^{ijk}v_jB_k^\SG$. By \Eeqref{RSM-BRF}, $B_\SG^i$ is a linear combination of $B_\RF^i$ and $v^i$, and so we have $E_\RF^i B^\RF_i=0$. Therefore
\begin{equation}
\bra{\psi^{\pm}}\hat{E}^{\RF}\ket{\psi^{\pm}}= 0.\label{RSM-electricterm}
 \end{equation}

Now substituting \eqref{RSM-magneticterm} and \eqref{RSM-electricterm} into \eqref{RSM-wkb4}, and taking the gradient of the resulting equation, we obtain
\begin{align}
0 & =  \partial_{\alpha}(k^{\pm}_\beta k_{\pm}^\beta)\pm\partial_{\alpha}(|B_{\RF}|)\nonumber\\
& =  2k_{\pm}^\beta\partial_{\beta}k^{\pm}_\alpha + 2ek^{\beta}_{\pm}F_{\beta\alpha}\pm\partial_{\alpha}(|B_{\RF}|)\label{RSM-trajectory}
\end{align}
where we have used that $\partial_{\alpha}(|B_{\RF}|)\sim 1/\eps^{2}$. We see that $\frac{\di x_{\pm}^\alpha}{\di\tau}=u_{\pm}^{\alpha}=k_{\pm}^{\alpha}/m$ must satisfy
\begin{eqnarray}
m\frac{\di^2x_{\pm}^\alpha}{\di\tau^2}+e\frac{\di x_{\pm}^\beta}{\di\tau} F_\beta^{\ \alpha}\pm \frac1m\partial^{\alpha}(|B_{\RF}|)=0\label{RSM-FWvelocity}
\end{eqnarray}
where $\frac{\di^2x_{\pm}^\alpha}{\di\tau^2}=\frac{\di x_{\pm}^\beta}{\di\tau}\partial_\beta u^\alpha_{\pm}$ is the 4-acceleration and $u_{\pm}^\alpha u^{\pm}_\alpha=1$. The first two terms of \eqref{RSM-FWvelocity} are simply the classical Lorentz force law, but in addition to this we have a deflection induced by the non-zero magnetic field gradient $\pm\partial_{\alpha}(|B_{\RF}|)/m$ whose sign depends on whether the spin is parallel or anti-parallel to the magnetic field $B^{\RF}_{\alpha}$.

The implications of \eqref{RSM-FWvelocity} are as follows: prior to measurement we have that  $u_{+}^{\alpha}=u_-^{\alpha}$. The qubit is then exposed to a strongly inhomogeneous electromagnetic field $F_{\alpha\beta}$ for a short period of time. This impulse-like interaction alters the velocity of the respective packets. For an ideal measurement this interaction is short enough that negligible precession of the spin, governed by \eqref{RSM-wkb1}, will occur during this splitting. The end result is the deflection of the $\psi_{A}^+$ component of spin with amplitude $|a|^{2}$ in the direction of the gradient of the magnetic field, and the deflection of the $\psi_A^-$ component with amplitude $|b|^2$ in the opposite direction. A position measurement will then produce the outcome `$+$' with probability $|a|^2$, and `$-$' with $|b|^2$. Thus we conclude that the operator corresponding to relativistic Stern--Gerlach measurement is given by \eqref{RSM-SG_spin_op}.

%%%%%%%%%%%%%%%%%%%%%%%%%%%%%%%%%%%%%%%%
\section{Conclusion and discussion}\label{RSM-conclusion}
%%%%%%%%%%%%%%%%%%%%%%%%%%%%%%%%%%%%%%%%

This chapter provided two distinct ways of identifying the spin observable corresponding to a Stern--Gerlach measurement of a massive fermion where the relative velocity of the particle and Stern--Gerlach apparatus is relativistic. The first approach followed an intuitive argument based on the transformation properties of the electromagnetic field. The second approach was a first-principles approach, starting with the Dirac equation minimally coupled to the electromagnetic field. Using this equation we showed that in the WKB limit the `$+$' spin eigenstate of \eqref{RSM-SG_spin_op} is deflected `up' and the `$-$' spin eigenstate is deflected `down'. We therefore concluded that in the relativistic regime the appropriate spin-operator for a Stern--Gerlach measurement is
\be
 \hat{\m S}(v,p,b^\SG)= - \frac{b^\SG_{\alpha}(v\cdot p)-v_{\alpha}(b^\SG\cdot p)}{\sqrt{(v\cdot p)^2-(b^\SG\cdot p)^2}}\frac{2\hat W^\alpha(p)}m.
\label{RSM-concl-SGspinop}
\ee
Notably the spin operator \eqref{RSM-concl-SGspinop} is momentum-dependent, so that, if the momentum is unknown, it is not possible to determine the expectation value. This has the following implications: Firstly, tracing over momentum of a state written in a tensor product basis of spin and momentum has been used in relativistic quantum information theory to extract the reduced spin density matrix \cite{PeresScudoTerno02,GingrichAdami02,PeresTerno03b,BartlettTerno05,Friis-relent,Choi12}. However, we can see that due to the momentum dependence of the observable \eqref{RSM-concl-SGspinop}, this reduced spin density matrix is not useful for extracting statistics of a relativistic Stern--Gerlach measurement. The usefulness of the reduced spin density matrix is further limited by the fact  that it has no Lorentz covariant transformation properties \cite{PeresScudoTerno02,PeresTerno03b,PeresTerno04,Czachorreply,TernoIRQI,Choi12}.

Secondly, in quantum tomography of spin, one collects measurement data for various measurement directions. One then uses this data to solve for the quantum state. Non-relativistically, the relationship between the quantum state and measurement data is momentum independent, and it is enough to choose three linearly independent directions in order to reconstruct the quantum state. However, in the relativistic Stern--Gerlach case, the experimental data are related to momentum dependent theoretical expectation values of the quantum state, determined by \eqref{RSM-concl-SGspinop}. Thus, if momentum is unknown, three linearly independent directions will not suffice. We leave it as an open question as to what minimal set of measurements is required to reconstruct the state in this relativistic case.

As a final point about the specific form of \eqref{RSM-concl-SGspinop}, we note that in the literature there exist several alternative operators that are used as observables for relativistic spin measurements. Two notable operators that have been proposed in the relativistic quantum information community are $\hat{\m S}'\propto a_{i}\hat W^i$ \cite{Czachor,Friis-relent} and $\hat{\m S}'' \propto a_{i }(\hat W^i-\hat W^0p^i/(p^0+m))$ \cite{Ternotworol,Lee03,LCY04,KimSon,Caban1,Caban2,CRW09,RS09,Landulfo,Moradi10}, where $a_i$ is a parameter determining which measurement is carried out. Although these proposals are Hermitian they are mathematically distinct from \eqref{RSM-concl-SGspinop} and it can be shown that they lead to quantitatively distinct predictions. An intuitive reason for this can be found in \cite{Pablo1111} where the authors show that the directions extracted from the $\hat{\m S}'$ operator do not transform in the same way as a magnetic field. It can also be shown in a similar analysis that the directions extracted from $\hat{\m S}''$ do not transform like a magnetic field.\footnote{Notice that $a_i$ or $b^\SG_\alpha$ perpendicular or parallel to $p^i$ are unchanged measurement directions for all three spin operators.} As a result these proposals cannot be considered to represent a Stern--Gerlach measurement. The question that therefore must be addressed is whether there exists a physical implementation for either of these proposals. However, measurements making use of a coupling of the spin to the electromagnetic field will not yield these spin operators.

%%%%%%%%%%%%%%%%%%%%%%%%%%%%%%%%%%%%%%%%%
%\section*{Acknowledgments}
%%%%%%%%%%%%%%%%%%%%%%%%%%%%%%%%%%%%%%%%%
%
%We would like to thank Stephen Bartlett, Daniel Terno, Aharon Brodutch, and Nikolai Friis for stimulating and helpful discussions. This research was supported by the Perimeter Institute-Australia Foundations (PIAF) program, and the Australian Research Council grant DP0880860. HW was supported by the CSIC JAE-DOC 2011 program.

%
%\section*{References}
%
%\bibliography{RQI_RSO,RQI}
%
%\end{document} 

%% file: LQiCSTfront.tex
\chapter{Additional comments regarding representations of spin\label{ch-2}}

In Chapters \ref{sec-LQiCST} and \ref{ch-RSO}, Wigner rotations and representations of the Lorentz group were briefly mentioned (with material in Sections \ref{notation}, \ref{fermionQHS}, \ref{sec-fermionmeasurement}, \ref{RSM-sec-mathdesc}, and \ref{secspinornotation}, and material regarding state transport in Sections \ref{sec-statetransformation}, \ref{sec-fermionstatetransformation}, and \ref{sec-photonstatetransformation}). The concepts were not discussed in detail and nor was the associated theory emphasised in the chapters. This chapter provides a quick summary and interpretation of the material, compares the representations, and provides some motivation for the predominant use of the Weyl representation in the two previous chapters.

\section{Representations of the Lorentz group}
Spin is an internal degree of freedom that we considered in the previous chapters. The spin quantum state maps to a direction in real space and carries a nontrivial representation of the Lorentz group. Fermion spin carries a spin-$\half$ representation. There are two main forms of representations that one can use. The most common in relativistic quantum information theory is the Wigner representation. This is actually a representation of the Poincar\'e group, so includes a full spin-momentum Hilbert space. The main alternative is the Weyl representation, which is used in Dirac field theory. We use the left-handed 2-component Weyl spinor to represent the spin of a massive fermion \cite{DHM2010}. The field may be formulated in position or momentum coordinates.

\subsection{Weyl representation}
The calculations regarding spin-$\half$ in Chapters \ref{sec-LQiCST} and \ref{ch-RSO} were predominantly using the Weyl representation \cite{Wald,Penrose,Bailin,DHM2010}, a representation of the Lorentz group. In this chapter by `the Weyl representation' we mean the full spin-momentum Hilbert space.

In this representation we have a spinor $\psi_A\in \mathbb C^2$ with index $A=1,2$ which transforms covariantly under the spin-$\half$ representation of $SL(2,\mathbb C)$. That is, a Lorentz transformation $\Lambda$ acts as a linear operator on the spinor;
\be
\Lambda:\psi_A\mapsto\Lambda_A^{\; B}\psi_B
\ee
where the linear operator $\Lambda_A^{\ B}=U(\Lambda)_A^{\ B}$ is an element of a faithful representation $U_A^{\ B}$ of $SL(2,\mathbb C)$; every element of the group has a distinct operator on the space (see \appref{secspinornotation}).

If we take the inner product on the Hilbert space of the spinor to be the quantum mechanical standard $\bk\chi\psi=\bar\chi_{A'}\delta^{A'A}\psi_A=\bar\chi_1\psi^1+\bar\chi^2\psi^2$, the boosts of the representation are nonunitary operators: A spin-$\half$ Lorentz boost \cite{DHM2010}, with boost parameter $\beta$, gamma factor $\gamma=(1-\beta^2)^{-\half}$ and $u^I=(\gamma,\gamma\beta^i)$ is
\[
L_{A}^{\ B}(u) = \frac1{\sqrt{2(u_0+1)}}(\delta_I^0+u_I)\sigma^{I\ B}_{\ A}
\]
where $\sigma^{I\ B}_{\ A}=\sigma^0_{\ AA'}\bar\sigma^{IA'B}$. With $\abs{L\psi}^2=\bar\psi_{B'}\bar L_{\ A'}^{B'}\delta^{A'A}L_A^{\ B}\psi_B$, by direct calculation we have
\[
\bar L_{\ A'}^{B'}\delta^{A'A}L_A^{\ B}=u_I\bar\sigma^{IB'B}%=\gamma(\sigma^{0B'B}+\beta_i\sigma^{iB'B})
\]
and so $\abs{L\psi}^2=\bra\psi u_I\bar\sigma^I\ket\psi=\gamma\abs\psi+\gamma\bra\psi\beta_i\hat\sigma^i\ket\psi$, which in general is not equal to 1. The solution is to define a new inner product for which the spinor has unit norm and the boosts are unitary, namely:
\[
\bk\psi\phi:=\bar\psi_{A'}\,u_I\bar\sigma^{IA'A}\,\phi_A.
\]
This inner product is derived from the Dirac inner product (see \appref{appIP}), and takes the standard quantum mechanics form for the rest frame $p=m(\frac{\di t}{\di\tau},\frac{\di \mathbf x}{\di\tau})=(m,0,0,0)$. The momentum-dependent inner product $I_p$ means that the Hilbert space, defined by the pair $\Hi=(\mathbb C^2,I_p)$ where $\mathbb C^2$ is the vector space and $I_p$ the inner product, is different for each momentum. If we then consider the full representation of the Lorentz group on a relativistic particle, the boosts change momentum, and so boosts move spinors from one spin Hilbert space to another: $\Lambda:\Hi_p\to\Hi_{\Lambda p}$.

Since Hilbert spaces for each spin are distinct, the Hilbert space structure for a spin-momentum wavefunction in this representation is the direct sum of the spin Hilbert spaces $\Hi_p$ over all momentum: $\Hi=\bigoplus_{p\in\mathbb R^3} \Hi_p$.

Additional detail regarding the representation of spin that was used in Chapters \ref{sec-LQiCST} and \ref{ch-RSO} and the Dirac algebra is in \appref{secspinornotation}.

\subsection{Wigner representation\label{sec-compWigner}}
The Wigner representation \cite{Weinberg,TerashimaUeda02,Ternothesis,PeresTerno04,ASK,Friis-relent} is a representation of the Poincar\'e group, which is the Lorentz group in a semidirect product with translations in flat spacetime. One then explicitly constructs spin-momentum states that are eigenstates of the generators of the translations: $\hat P^\mu\psi_{p,\sigma}=p^\mu\psi_{p,\sigma}$ \cite{Weinberg}. Under a Lorentz transformation the momentum transforms as usual: $p\mapsto\Lambda p$. However, in this representation the spin/polarisation components $\sigma$ are made to transform by
\[
U(\Lambda)\psi_{p,\sigma}=\sum_{\sigma'}D_{\sigma'\sigma}(W(\Lambda,P))\psi_{p,\sigma}
\]
determined by a `Wigner rotation'
\be
W(\Lambda,p)=L^{-1}(\Lambda p)\Lambda L(p).\label{eq-Wignerrotation2}
\ee
and where in this case $D_{\sigma'\sigma}(W)$ is the operator for the spin-$\half$ representation of the Wigner rotation $W$. The Wigner rotations are constructed by defining a `standard momentum' $k$ and determining the structure of the subgroup of the Lorentz transformations that leave this momentum invariant, called `Wigner's little group' for $k$. This is an exercise in mathematics and logical deduction, \cite{Weinberg}. For massive particles the standard momentum is usually $k=(m,0,0,0)$ and the little group is all spatial rotations $SO(3)$. The spin-$\half$ representation $D_{\sigma'\sigma}(W)$ is then an element of $SU(2)$. Note that the standard momentum $k$ is an arbitrary choice as long as in the massive case it is timelike, but the `standard Lorentz transformation' $L(p)$ depends on this choice. For massive particles with the usual choice of standard momentum, $L(p)$ is a pure boost which takes $k=(m,0,0,0)$ to $p$.

The outcome is that this representation is not faithful on the spin subspace, since the boosts and rotations of the Lorentz are reduced to a subgroup of transformations. However, these transformations are unitary using standard quantum mechanical inner product, so the inner product for the spin-momentum state is a standard and Lorentz-invariant $(\psi_{p',\sigma'},\psi_{p,\sigma})=\delta_{\sigma'\sigma}\delta^3(p'-p)$ \cite{Weinberg}.

In \cite{TerashimaUeda02,Ternothesis,PeresTerno03b,PeresTerno04,BartlettTerno05,ASK,Friis-relent} the state for a massive fermion is interpreted as $\ket{p}\otimes\ket\sigma$ living in a Hilbert space $\Hi_p\otimes\Hi_\sigma$ which is the tensor product of a momentum Hilbert space with a spin Hilbert space common to spin for all momenta.

\subsection{Comparing the representations\label{sec-comparingreps}}
These representations are both valid ways to describe relativistic localised qubits. The Wigner rotations are a way to obtain a description of relativistic qubits in standard non-relativistic quantum mechanics formalism, that is: with the Lorentz group reduced to spatial rotations on a Hilbert space with the standard inner product. This representation offers the most immediate connection to non-relativistic quantum information theory in that the mathematical structures of the quantum theory are identical.

In comparison, the Weyl representation requires a family of inner products not standard to nonrelativistic quantum mechanics, requiring a redefinition of Hermiticity and unitarity for operators. However, the benefits of the Weyl representation are that spin and polarisation both behave in relativity in the same way as 4-vectors, making it an intuitive representation of spin in which to visualise relativistic qubits, and enabling straightforward calculation of transformations. The covariant representation provides more structure and context for mathematical elements of the theory. This aids in understanding experimental scenarios, greatly simplifies the application of tools of general relativity to quantum information theory, and provides constraints on the mathematical operations that are invalid in relativistic settings. %This latter aspect is specifically the covariant formalism providing rules for preventing misinterpretations of the mathematics, \tred{particularly in regarding states written in one Lorentz frame as being in another Lorentz frames.} \tred{, as is easier to commit with a noncovariant formalism. For example, it is clear that spins with different momenta are inequivalent representations, as they are distinct Hilbert spaces in this representation, not amenable to factorising into a tensor product as in the Wigner representation of spin. ...how is this state viewed in another frame?}

We can relate the two representations of spin, illustrated in \figref{fig-LQFrontLTsreps}. The Weyl spin $\psi_A$ is the spin vector as observed in the frame in which the particle has momentum $p$. In the Wigner representation the spin $\ket\sigma$ is identified as the spin in the frame of the `standard momentum' $k$, i.e.\;the spin as observed in the rest frame of the particle. The Wigner rotation is then the process of applying a boost to the spin $\sigma$ to take it to the momentum $p$, applying the desired Lorentz transformation $\Lambda$, and then boosting the spin back to the standard momentum. The $L(p)$ is therefore the map between representations, mapping between Hilbert spaces $L(p):(\mathbb C^2,I_k)\!=\!\Hi_k\to(\mathbb C^2,I_p)\!=\!\Hi_p$ and rest frame and covariant spins $L(p):\ket{\sigma}\mapsto\psi_A$.

\begin{figure}[h]
\[
\begin{CD}
\psi_A @>\Lambda>> (\psi_\Lambda)_A\\
@V L^{-1}(p) VV  @VV L^{-1}(\Lambda p)V\\
\ket{\sigma} @>>W(\Lambda,p)>\ket{\sigma_\Lambda}
\end{CD}
\]
\caption[Commutative diagram of Weyl and Wigner group actions on spin.]{A commutative diagram comparing the covariant action of the Lorentz transformation $\Lambda$ on the Weyl spinor $\psi_A$ with momentum $p$ (top line) to the Wigner rotation $W(\Lambda,p)$ on the Wigner spin $\ket{\sigma}$ (bottom line). By applying the standard Lorentz boost $L$, a boost that takes $k=(m,0,0,0)$ to momentum $p$, or its inverse, one can move between the Weyl spinor $\psi_A$, and the Wigner spin $\ket\sigma$, which is equivalent to the Weyl spin in the rest frame $k=(m,0,0,0)$.\label{fig-LQFrontLTsreps}}
\end{figure}

%% file: CoQRF/Pieces/CoQRF10_6Sec1.tex
\chapter{Changing quantum reference frames\label{ch-CoQRF}}

\subsection*{Abstract}
We consider the process of changing reference frames in the case where the reference frames are quantum systems.  We find that, as part of this process, decoherence is necessarily induced on any quantum system described relative to these frames.  We explore this process with examples involving reference frames for phase and orientation. Quantifying the effect of changing quantum reference frames provides a theoretical description for this process in quantum experiments, and serves as a first step in developing a relativity principle for theories in which all objects including reference frames are necessarily quantum.

\newpage

\section{Introduction\label{sec-intro}}

Quantum states and dynamics are commonly described with respect to a classical background reference frame. Even defining a basis for the Hilbert space of a quantum systems will in general make reference to a background frame. For example, the state $\ket 0$ for a spin-$1/2$ particle may be defined as the spin parallel to the $z$-axis of a laboratory reference frame, and the $(\ket0+\ket1)/\sqrt2$ state as parallel to the $x$-axis. In place of a classical background frame, one could use a second quantum system prepared in a state that indicates an orientation or alignment of a frame, associated with a group $G$. The basis can then be defined with respect to this `quantum reference frame'.  In this case, quantum information is encoded in degrees of freedom that are independent of the orientation of the laboratory reference frame.

In this chapter we consider the process of changing a quantum reference frame as the first step in understanding the relativity principle for quantum reference frames.  A relativity principle dictates how the description and dynamics of a physical system change when the reference frame used to describe the system is changed. For example, in special relativity, the laws of physics are the same in all inertial frames, with the description related by elements of the Poincar\'e group. Here, we will be investigating how the description of a physical system changes due to a change of inertial quantum reference frame.

A change of reference frame can formally be expressed in two steps.  First, one determines the relationship between the old and new reference frames.  Second, based on this relationship, one defines a transformation on the state of the physical system(s) which changes the state from being with respect to the old frame to now being with respect to the new frame. In a classical theory, all properties of the reference frames can be known to arbitrary precision, and they are unaffected by measurement.  The relationship between two frames is represented by a group element, corresponding to the transformation that takes the first reference frame to one aligned with the second.

Now consider quantum reference frames. A change of quantum reference frame is the process by which a quantum reference frame initially used to define the basis for some quantum system is exchanged for a second quantum reference frame. Quantum reference frames are explicitly physical objects, so we require a physical process to compare and change these reference frames. We must consider what changing a frame means when the frames are quantum. A change of quantum reference frames could have (at least) two interpretations. In the first, there are two reference frames for which the observer knows the orientation. Changing reference frames is simply a matter of discarding the undesired reference frame. The second interpretation is that the orientation of the second reference frame is not initially known. Learning the orientation therefore requires a measurement. Describing this latter change of quantum reference frame is the focus of this chapter. There is a clear experimental importance in such a procedure \cite{BRSDialogue}. For example, in switching from a locked phase or clock reference laser $A$ to an uncorrelated laser $B$, the two lasers must be phase locked.

Quantum reference frames in general will use finite resources, quantified by some parameter such as the Hilbert space dimension of the frame.  If our reference frames describe a continuum of orientations and we restrict the size (e.g., Hilbert space dimension) of the frames, then reference frame states corresponding to different orientations will not be perfectly distinguishable.  This uncertainty of the frame results in decoherence in information encoded using the reference frame. In particular, as we will show, decoherence can result from a change of quantum reference frame. This would be a novel behaviour for a relativity principle.

With a construction and characterisation of a `change of quantum reference frame' procedure, we quantitatively investigate the decoherence resulting from changing physical quantum reference frames.  We interpret this appearance of decoherence in terms of `intrinsic' decoherence, which is a proposed semiclassical phenomenon of quantum gravity arising from fluctuations and fundamental uncertainties in the background space  \cite{Power00,KokYurtsever,Gambini04,Milburn03,GirelliPoulin07,GirelliPoulin08}. This `change of quantum reference frame procedure' can be viewed as a derivation of this `intrinsic' decoherence when the quantum natures of the background space and reference frames are modelled as physical quantum systems.

The structure of the chapter is as follows. \secref{sec:Prelim} reviews the concepts and mathematical formalism of quantum reference frames. \secref{sec-introCoRF} presents the definition of the quantum operation describing a change of quantum reference frame, and an analysis of the properties of this quantum operation for some special cases of quantum reference frame.  We also discuss the significance of the decoherence induced, and what consequences the procedure has. In Sections~\ref{sec-phaseex} and \ref{sec-exampleSU2} we provide examples of the procedure for phase references (characterised by the group $U(1)$), and a Cartesian frame and direction indicator (characterised by $SU(2)$). In \secref{sec-conclusion} we present some concluding remarks.

\section{Preliminaries:  Classical and quantum reference frames\label{sec:Prelim}}
In this section, we review the mathematical tools used in the description of a quantum state relative to a classical or quantum reference frame.  We follow the notation of the review article~\cite{BRS}.

%% file: CoQRF/Pieces/CoQRF10_6Sec2part2.tex
Reference frames in a physical theory can be treated as either external or internal to the physical description. An external frame is a static (i.e., non-dynamical) background resource to which the description of objects is referred.  It is not described as part of the dynamics, and in particular does not interact with physical systems in the theory. We will refer to such reference frames as {\it background} reference frames. Conversely, internal reference frames are dynamical physical objects that are included in the description of the physics.

Let us illustrate the concept of an internal frame with an example of determining the position of a classical particle $S$ on a line. The position $x_S$ of the particle depends on the external coordinates, in particular on the placement of the origin and the choice of scale.  Alternatively, one could use the position $x_{R}$ of one additional reference particle $R$ as a physical token for the origin. The relative position of the initial particle with respect to this reference particle is given by the distance $x_S-x_{R}$. This relative position is a degree of freedom inseparably of both the $S$ particle and $R$ particle. Because this quantity is independent of the background reference frame, or equivalently, is invariant under global translations, we say it is a \emph{relational} quantity.

The physical description when using an internal frame then consists of a system $S$ and reference frame $R$. The important physical quantities are encoded in the relationships between the system and the reference frame, which are invariant under changes of background reference frame. We refer to these as the relational degrees of freedom. 

%% file: CoQRF/Pieces/CoQRF10_6Sec2part3.tex
One should also distinguish whether the reference frame is correlated with the state from one that is uncorrelated. Following \cite{BRSDialogue}, an {\it implicated} reference frame is one which is correlated with the system. A {\it non-implicated} reference frame is uncorrelated with the system, so we have no knowledge of the relations between the system and reference frame. For example, consider a scenario in which a polarised photon is received from a distant source. While the polarisation may be well-defined relative to the reference frame at the source, if the receiving frame is uncorrelated with this, then the information in the state cannot be extracted. We say the receiver's frame is non-implicated.

%% file: CoQRF/Pieces/CoQRF10_6Sec2part4.tex
In this chapter we will use quantum reference frames, which are quantum states that play the role of physical internal reference frames for quantum theories. These quantum states are elements of a separate Hilbert space associated with the quantum reference frame, where different states on this Hilbert space can describe different `orientations' of the quantum reference frame.  To formalise these notions, we will look at how to mathematically describe the manipulation of quantum reference frames and relational quantum degrees of freedom using techniques from the theory of group representations. 

%% file: CoQRF/Pieces/CoQRF10_6Sec2part5.tex
To set up the use of a quantum reference frame for encoding information in relational degrees of freedom, we begin with a background reference frame, and a quantum system $S$ in the state $\rho_S$ with respect to this background frame.  Changes of orientation of this system relative to the classical frame are described by a unitary representation $U_S(g)$ of an element $g$ from a group $G$ which describes all possible changes of orientation.

Next we prepare a quantum state on an additional system $R$ in a quantum reference frame state $\rho_{R}$, also defined with respect to the background reference frame.  This system transforms under a unitary representation $U_R$ of $G$. I.e.,\;the reference frame state breaks a symmetry associated with $G$, which has a representation $U_R\!:\!G\to \m B(\Hi_R)$ on the Hilbert space of the quantum reference frame system $\Hi_R$. We can now consider using $R$ as a reference frame for $S$, i.e., referring to properties of $S$ in relation to those of $R$, independent of the background frame.

To ensure that we are not still making accidental use of the background frame, we can de-implicate it. This de-implication involves decorrelating the compound quantum system $SR$ from the background frame. For a general state $\rho$, this is done by averaging the state over all rotations $g\in G$ using $U(g)$, the unitary representation of $G$ on the total Hilbert space of $\rho$. The resulting map $\m G$ is called the {\it G-twirl} of the state, given by
\be
\m G\left(\m \rho\right)=\int\di\mu(g)\m U(g)[\rho] = \int\di\mu(g) U(g) \rho U(g)^\dag\label{eq-Gtwirl}
\ee
where $\m U(g)[\rho]:=U(g)\rho U(g)^\dag$ is the unitary map of the left action of the group and $\di\mu(g)$ is the group-invariant Haar measure of the group (for example, the $U(1)$ integration measure is $\di\theta/2\pi$).  We restrict our attention to compact Lie groups, where the average is well-defined and bounded.  Note that a $G$-twirled state may be mixed even if the original state $\rho$ was pure.

%% file: CoQRF/Pieces/CoQRF10_6Sec2part6.tex
Applying \eqref{eq-Gtwirl} to $\rho_S\otimes\rho_{R}$, we first note that the unitary representation on the compound Hilbert space $S\otimes R$ is given by $U_S(g)\otimes U_{R}(g)$, constructed from the individual representations of $G$ on $S$ and $R$. We then have
\be
\m G_{S R}\left(\m \rho_S\otimes \rho_{R}\right)=\int\di\mu(g)\m U_S(g)[\rho_S]\otimes\m U_R(g)[\rho_R].\label{eq-encgen}
\ee
This allows us to encode a system $\rho_S$ using a reference frame $\rho_R$. We define a \emph{relational encoding} $\m E_{\rho_R}(\rho_S)$ of $\rho_S$ using a generic reference frame state $\rho_R$ as the $G$-invariant encoded state
\be
\m E_{\rho_R}(\rho_S):=\m G_{S R}\left(\m \rho_S\otimes \rho_{R}\right).\label{eq-encGC}
 \ee 

%% file: CoQRF/Pieces/CoQRF10_6Sec2part6.5.tex
To distinguish $G$-twirled or encoded states from states that may require background reference frames, we will use separate terminology and notation. States $\rho$ that may depend on an implicated background reference frame will be described as `kinematic'. We will call the states that are invariant under G-twirling $\sigma=\m G(\sigma)$ (including G-twirled states $\sigma=\m G(\rho)$) `group-invariant' or `G-invariant'. Reference frames external to a G-invariant state $\sigma$ are always non-implicated and consequently $\sigma$ is well-defined independent of a background reference frame.%}

%% file: CoQRF/Pieces/CoQRF10_6Sec2part7.tex
\subsection{The relational degrees of freedom}

Relational degrees of freedom are those which are independent of any background frame. Given a system state $\rho_S$ and a quantum reference frame $\rho_R$, it is not immediately obvious what the relational degrees of freedom in the $G$-twirled joint state $\m G(\rho_S \otimes \rho_R)$ are. In the following, we will define the Hilbert space subsystems associated with these relational degrees of freedom, following~\cite{BRST,BRS}.  Again, for simplicity of the mathematics, we will consider symmetries corresponding to compact Lie groups such as $U(1)$ and $SU(2)$. However, many of the concepts developed can be directly transferred to general groups and reference frames.

The unitary representation of a compact Lie group on a Hilbert space $\Hi$ consists of a number of inequivalent representations called `charge sectors'. The Hilbert space can be decomposed into a tensor sum of these charge sectors, each labelled by $q$ (for example, $q$ may be total spin in a representation of $SU(2)$ on a collection of spins). Each of these charge sectors may be a {\it reducible} representation, which can be further decomposed into a Hilbert subspace $\m M^{(q)}$ carrying an {\it irreducible} representation (`irrep'), and a `multiplicity space'  $\m N^{(q)}$ which carries the trivial representation and whose dimension indicates how many copies of the irreducible representation exists in the charge sector $q$. The representation on the full Hilbert space then has the structure
\be
\Hi=\bigoplus_{q} \m M^{(q)}\otimes\m N^{(q)},\label{eq-repdecomp}
\ee
where $q$ ranges over all the irreps (charge sectors) of $G$ that are supported on $\Hi$.

The $G$-twirl map \eqref{eq-Gtwirl} is closely related to the representations of the group, in that it averages an input state $\rho$ over the unitary action of every element in the symmetry group. Decomposing this map following \eqref{eq-repdecomp}, we have
\be
\m G(\rho)=\sum_q(\m D_{\m M^{(q)}}\otimes\m I_{\m N^{(q)}})[\Pi^{(q)}\rho{\Pi^{(q)}}^\dag].\label{eq-Gtwirldecomp}
\ee
The terms in this operation are defined as follows.  First, $\Pi^{(q)}$ is the projector onto the subspace $\m M^{(q)}\otimes\m N^{(q)}$, the charge sector $q$. This removes all coherences in the kinematic density operator $\rho$ between the charge sectors. Next, $\m D$ is the complete depolarising channel, which is a trace preserving map that takes every density operator to a scalar multiple of the identity operator on the space $\m M^{(q)}$. This is the effect of an average of rotations on an irrep. Finally $\m I^{(q)}$ is the identity map on the multiplicity subspace $\m N^{(q)}$.

We can now identify the relational degrees of freedom, unaffected by $G$-twirl, as the multiplicity subspaces $\m N^{(q)}$.  The degrees of freedom in the subsystems $\m M^{(q)}$ are defined only with respect to a background frame, and are completely decohered by the $G$-twirl.

%% file: CoQRF/Pieces/CoQRF10_6Sec2part8.tex
\label{sec-groupestates}\label{sec-RFquant}

In this section we describe how to define useful quantum reference frame states.

A reference frame breaks a symmetry by indicating an orientation.  The set of possible orientations is associated with a symmetry group $G$. We begin with a \emph{fiducial state} $\ket{\psi(e)}$, which serves as a quantum reference frame oriented with respect to a background frame and which we choose to associate with the identity $e\in G$. Given this fiducial state we can construct states corresponding to other orientations $g\in G$ by generating the states in the orbit of $\ket{\psi(e)}$ under the group action $U(g)$, yielding $\ket{\psi(g)}:=U(g)\ket{\psi(e)}$ for all $g\in G$.  Such states obey the relation $U(h)\ket{\psi(g)} = \ket{\psi(hg)}$, and we say that they transform \emph{covariantly} under the action of the symmetry group.

Quantum reference frames generally use limited finite resources quantified by some parameter $s_{R}$. A fundamental example of a size parameter is the dimensionality of the Hilbert space $\Hi_{R}$, constraining the number of charge sectors $q_{R}$ under the representation of the group. We define the notation $\ket{s_R;\psi(g)}$ to denote a $G$-covariant state $\psi(g)$ with size parameter $s_R$.  Where it is unnecessary to indicate size, we may suppress the size parameter. The groups considered in the theory of this chapter are compact Lie groups, meaning the reference frames can take one of a continuum of orientations in a closed manifold. With only finite-dimensional representations of such groups, reference frame states for different orientations in a Lie group cannot all be perfectly distinguishable. Consequently, a state will have a mean orientation $g$, but also possess an uncertainty in orientation.

We would like reference frame states to have a well-defined classical limit in which the overlap of states with different orientations becomes zero as the size parameter $s_{R}$ increases to infinity, i.e.,
\begin{equation}
  \lim_{s_R \to \infty} D_{s_R}\abs{\bk{s_R;\psi(g)}{s_R;\psi(h)}}^2 = \delta(gh^{-1}),
\end{equation}
where $\delta(g)$ is the delta function on $G$ defined by $\int\di\mu(g)\delta(g)f(g)=f(e)$ for any continuous function $f$ of $G$ \cite{BRST}, and $D_{s_R}$ is the dimension of the Hilbert space spanned by $\ket{s_R;\psi(h)}$.

In the finite size case, one may wish to maximise the distinguishability of the quantum reference frame used for a Hilbert space size constraint $D_{s_R}$. Distinguishability can be quantified using maximum likelihood or fidelity measures \cite{Chiribella06,BRS,Bisio10}. Since we also want the reference frame states to become ideal in the classical limit, we want this distinguishability to scale with $D_{s_R}$ (see \cite{Chiribella06,Vaccaro08} regarding asymptotic measures). The set of optimal reference frame states $\rho$ for a group $G$ on $D_{s_R}$ dimensions are the \emph{maximum likelihood states} \cite{Chiribella06}, denoted $\ket g$ or $\ket{s_R;g}$. These pure states transform covariantly, and have the property that
\begin{equation}
  \m G(|g\rangle\langle g|) = D_{s_R}^{-1}I\,,\label{eq:uniformGtwirl}
\end{equation}
i.e.\;these have uniform support over their Hilbert space, which will make these states useful in the construction of measurements (POVMs). The form of a maximum likelihood state is specific to the group $G$. In some cases the optimal forms can be directly determined \cite{Bagan01,PeresScudo01,Bagan04,Chiribella04,Chiribella05,Lindner06,Bagan06,BRS}, and we will be using examples of maximum likelihood states in Sections~\ref{sec-phaseex} and \ref{sec-exampleSU2}. 

%% file: CoQRF/Pieces/CoQRF10_6Sec2part9.tex
Equation~\eqref{eq-encGC} describes the quantisation of a reference frame in a Hilbert space $\Hi_{R}$. The reverse of this procedure is the \emph{dequantisation map}~\cite{BRST}
\be
\m R(\sigma_{S R})=D_{s_R}\int\di\mu(g)\bigl[U_S(g^{-1})\otimes\bra{g}_{R}\bigr]\sigma_{S R}\bigl[U_S(g^{-1})^\dag\otimes\ket{g}_{R}\bigr],\label{eq-rec}
\ee
(also called the \emph{recovery map}).  This map describes the measurement of the quantum reference frame on system $R$ against a background reference frame (described by a covariant POVM formed with elements proportional to projectors onto the states $\ket g_R$ on the reference frame), thereby dequantising it, recovering a modified description of the system state. If the reference frame is measured to have orientation `$g$' relative to the background frame, then the orientation of the state is corrected by a rotation $g^{-1}$.

The finite size of a quantum reference token means for symmetries described by compact Lie groups that the token is an imperfect reference frame. Consequently, the use of a quantum reference frame causes an effective decoherence to the information in $\rho_S$. We can describe this decoherence by composing \eqref{eq-encGC} and \eqref{eq-rec} to produce
\be
\m R\circ\m E_{\rho_R}(\rho_S)=D_{s_R}\int\di\mu(g)\bra g\!\rho_R\!\ket g\m U_S(g^{-1})[\rho_S].\label{eq-REmap}
\ee
This map takes the form of a noise map on $\rho_S$, describing a mixing of this state over a distribution of unitaries determined by the distribution $\bra g\rho_R\ket g$.

This recovery map will be useful for identifying the effects of the change of quantum reference frame procedure, and for determining the resulting decoherence in \secref{sec-interpretation}.

%% file: CoQRF/Pieces/CoQRF10_6Sec3on.tex
\section{Change of a quantum reference frame}
\label{sec-introCoRF}

We now consider the central problem of changing quantum reference frames. The operational task for this chapter is defined as follows. An observer possesses a quantum reference frame $A$ and a state $\rho_S$, which are correlated. The observer wishes to use a second reference frame $B$ without knowing anything about its state. The task of the observer is to use the $B$ quantum system as a quantum reference for the system $\rho_S$, and to discard the initial reference frame $A$. This occurs in several experimental guises \cite{BRSDialogue}. For example, in switching phase or clock reference lasers from a locked laser $A$ to an uncorrelated laser $B$, one needs to phase lock the two lasers \cite{PfleegotMandel67PR,PfleegorMandel67PLA,Molmer97PRA,CKR05} (this has been extended to issues in optical teleportation \cite{RudolphSanders01,Fujii03}). In another example, determining the relative phase of two Bose--Einstein condensates \cite{JavanainenYoo96,HostonYou96,WCW96,ATMDKK97,CastinDalibard97,CKR05} can be interpreted as correlating quantum reference frames \cite{BRSDialogue,DB00,Denschlag00}.

We begin this section with a qualitative discussion of the issues regarding measurement when changing reference frames, including an example to illustrate the central ideas. If the reader prefers, this subsection \Ssref{sec-introchangingRFs} can be skipped in favour of the mathematical formulation in \Ssref{sec-genresults}.

\subsection{Changing quantum reference frames:  a qualitative discussion\label{sec-introchangingRFs}}

First, to define a `change of reference frame', suppose we have a system $S$ and implicated reference frame $A$, so that quantities of $S$ are defined using $A$.  Changing quantum reference frames is a procedure that implicates a second reference frame $B$ with $S$, while {\it de}-implicating $A$. The final arrangement should be of the system $S$ with quantities defined with respect to $B$.  We restrict to a static scenario (i.e., a trivial Hamiltonian), and will not consider dynamics in this chapter (see \cite{Poulin07,Angelo12}).

As an example, also investigated in \cite{AharonovKaufherr,ABPSS}, consider a particle $S$ in one dimension, with position defined relative to a reference frame consisting of another particle $A$ which provides an origin.  Introduce a second particle, $B$, which we would like to use as a new reference frame for $S$.  Classically, this seems straightforward:  the position of $S$ described in terms of $B$ will differ by the relative position of the two reference frames, $x_B-x_A$.  (Note that this relational quantity is independent of any background reference frame.) Particle $A$ can be subsequently discarded.

The problem with this naive approach is that we have not learnt of the relationship of $B$ with respect to either $A$ or $S$. In both classical and quantum cases, if the relationship between $A$ to $B$ is initially unknown then a measurement is required in order to determine this relationship. There are two options for doing this. The relationship between $S$ and $B$ can be directly measured, or the relationship between $A$ and $B$ can be measured (giving us the relational quantity $x_B-x_A$ for adjusting the description of $S$). Let us concentrate on the quantum mechanical case now, and first consider a semiclassical configuration in which the $A$ and $B$ reference frames are in position eigenstates and the measurements are ideal projective measurements of relative position. The $S$ state is arbitrary. For the first measurement option, the relative position of $S$ and $B$ is measured. The wavefunction of $S$ may have been encoding more than a position eigenstate, and this state is directly affected by the measurement, as it is projected to a position eigenstate. This behaviour is not consistent with what we expect of a change of reference frame. Rather, the second measurement option is what we will use to change between quantum reference frames. Here the relative position $x_B-x_A$ of the two reference frames $A$ and $B$ is measured, and the system $S$ is not involved. After obtaining a well-defined value of $x_B-x_A$, we can combine this with preexisting knowledge of $x_S-x_A$ to obtain knowledge of $x_S-x_B$, since the associated operators commute. We can now discard the $A$ reference frame and retain a correlation between $S$ and $B$. The new description of the state $S$ will have changed by $x_B-x_A$ due to the difference in position of reference frame $B$ versus $A$, thus accomplishing a change of quantum reference frame.

There are however some subtleties in this procedure. In the above example $A$ and $B$ were position eigenstates and the measurements were projective to these position eigenstates, allowing for arbitrarily good precision in the relational variables. Firstly, we would describe the initial state of $B$ as a completely mixed state because we have minimal prior knowledge of its position. It is this lack of knowledge which necessitates the measurement. Secondly, in general, quantum reference frames will not be able to attain ideal states such as position eigenstates. We will see that the imperfect nature of the $A$ reference frame results in decoherence to the quantum system $\rho_S$ after $A$ is discarded. If the measurement is also only capable of projecting $x_B-x_A$ to a state with finite variance in position, then discarding the frame $A$ yields a system wavefunction correlated with an imperfect reference frame $B$. Dequantising $B$ will then also cause decoherence to the system.

In the next section, we will formalise these concepts and problems, and construct a general framework for describing a change of quantum reference frame.  In particular, because we use quantum states to indicate orientations in a continuous group, in many cases we cannot perfectly distinguish nonorthogonal states for different reference frame orientations. One of the main limiting factors for distinguishability is the dimension of the Hilbert space used for the reference frame. The imperfect distinguishability results in an uncertainty in the orientation given by a quantum reference frame, leading to decoherence when we change the quantum frame used for encoding a quantum system.

\subsection{General results of change of quantum reference frame procedure\label{sec-genresults}}

In this section, we formally develop the change of quantum reference frame procedure and then calculate the final state for a physically relevant class of initial states.

Consider a system $\rho_S$ encoded using a quantum reference frame $\rho_A$, as described in \secref{sec-quantRF}.  The joint system is described by the state $\m E_{\rho_A}(\rho_S)=\m G_{S A}(\rho_S\otimes\rho_A)$. We want to construct a procedure to exchange the reference frame $A$ with a quantum reference frame $B$, defined by a state $\rho_B$, to result in some relational state involving only $\rho_S$ and reference frame $\rho_B$. Both $A$ and $B$ reference frame states are physical objects, so following the discussion in \secref{sec-introchangingRFs}, we will take $B$ in the initial state to be a non-implicated reference frame, i.e., uncorrelated with $S$ and $A$. The initial state in our scenario is then $\m G_{SA}(\rho_S\otimes\rho_A)\otimes\m G_B(\rho_B)$.

Discarding the $A$ frame at this point will simply yield the state $\m G_S(\rho_S)\otimes\m G_B(\rho_B)$. Instead we use an active procedure to change reference frames, which involves a measurement of the relative orientation $h\in G$ between the two reference frames $A$ and $B$. This measurement leads to a correlation in orientation of the two reference frames. Because there was initially correlation between frame $S$ and the system $A$, we obtain correlation between $S$ and $B$. Now we can discard the $A$ reference frame by tracing and use $B$ as the new quantum reference frame. If the reference frames use finite resources, such as finite Hilbert space dimension to indicate orientations, we expect decoherence in the post-measurement state.

\subsubsection{A measurement to determine the relationship between frames\label{sec-demoprop}}

The core element of the procedure to change quantum reference frames is a relational measurement of the two reference frames $A$ and $B$ that determines a relative orientation $h \in G$~\cite{BRSrelational}.  The fact that it is a relational measurement means that it can be made independent of any background reference frame. In the following, we construct the relational POVM and update map for this measurement, and prove key properties of the construction.

The quantum statistics of a relational measurement of the two reference frames $A$ and $B$ are given by a relational POVM $\{E_h|h\in G\}$.  A POVM allows us to calculate the probabilities of the $h$ outcomes for an input state, but here we are equally interested in the post-measurement state.  We therefore construct a family of trace-decreasing completely positive (CP) maps $\m M^h_{AB}$ associated with the POVM elements to determine post measurement states for a given outcome $h$.  (Such maps, which describe the POVM and also the post-measurement update rule, are sometimes called \emph{instruments}.)  We require these operations to be implementable without the use of a background reference frame.

We now define a measurement, as a POVM, satisfying the above conditions. The POVM is designed to determine orientation within the symmetry group, so will be formed from the maximum likelihood states $\ket g$ for the particular symmetry group $G$ of the scenario, using the techniques of \secref{sec-groupestates}.  That is, the maximum likelihood states $\ket g_A$ and $\ket g_B$ for each reference frame system $A$ and $B$ will satisfy the conditions $\m G(|g\rangle_A \langle g|) = D_{s_A}^{-1} I_A$ and $\m G(|g\rangle_B \langle g|) = D_{s_B}^{-1} I_B$, where $D_{s_*}$ are normalisation factors given by the dimensions of the Hilbert space spanned by each projector on $A$ and $B$. (Note, in \cite{BRST}, the fiducial state was normalised such that $\m G(\proj e)=I$.) We then a family of projectors $\Pi_{AB}^{g,h}$ on the two reference frame systems $AB$ given by
\be
\Pi_{AB}^{g,h}= \proj g_A\ot \proj{gh}_B=\m U_{AB}(g)[\proj e_A\ot\proj h_B],\label{eq-proj}
\ee
with $\ket g=U(g)\ket e$. The projector $\Pi^{g,h}_{AB}$ projects onto the state describing an orientation $g\in G$ of the state on $A$ and an orientation $gh\in G$ of the state on $B$.

The projectors are defined with respect to a background frame.  By using a $G$-twirl, we can define relational POVM effects $\{E_h\}$ as
\be
E_h=D_{s_A}D_{s_B}\int \di\mu(g)\Pi_{AB}^{g,h}.
\ee
This measurement satisfies POVM completeness, $\int \di\mu(h)E_h = I_{AB}$. To prove completeness, note that
\begin{align}
\int \di\mu(h)E_h
&=D_{s_A}D_{s_B}\int\di\mu(g)\di\mu(h)\kb gg_A\ot \kb{gh}{gh}_B\nonumber \\
&=\Bigl(D_{s_A}\int\di\mu(g)\kb gg_A\Bigr)\ot \Bigl(D_{s_B} \int\di\mu(h) \kb{h}{h}_B\Bigr) \nonumber \\
&=I_{AB}\,.
\end{align}
with the second line obtained by measure invariance, and the last using the property of maximum likelihood states $\m G(|e\rangle_A \langle e|) = D_{s_A}^{-1} I_A$ and $\m G(|e\rangle_B \langle e|) = D_{s_B}^{-1} I_B$.

With each effect $E_h$, we can define a corresponding CP map $\m M_{AB}^h$ describing both the measurement and subsequent update map in terms of the projectors as
\be
\M^h_{AB}(\rho_{AB})=D_{s_A}D_{s_B}\int\di\mu(g)\Pi^{g,h}_{AB} \;\rho_{AB}\;{\Pi^{g,h}_{AB}}^\dag\label{eq-M}.
\ee
Note that this update map is chosen such that the measurement is repeatable.  As with the POVM, this map can be implemented without the use of a background frame.  We prove this fact by demonstrating that the map is `group-invariant', which means that the measurement map \eqref{eq-M} is invariant under any global rotation $\m U_{AB}(f):=\m U_A(f)\otimes\m U_B(f)$, i.e.\;for any $f\in G$ we have that $\m U_{AB}(f)\circ\m M_{AB}^h\circ\m U_{AB}^\dag(f)=\m M_{AB}^h$ \cite{BartlettWiseman03,GourSpekkens}.

Starting from \eqref{eq-M}, we have
\be
\m U_{AB}(f)\circ\M^h_{AB}\circ\m U_{AB}^\dag(f)[\sigma]=D_{s_A}D_{s_B}\int\di\mu(g)U_{AB}(f)\Pi^{g,h}_{AB} \,U_{AB}^\dag(f)\sigma U_{AB}(f)\,{\Pi^{g,h}_{AB}}^\dag U_{AB}^\dag(f).
\ee
Concentrating on the rotation operators around a projector, from \eqref{eq-proj} we have that
\be
U_{AB}(f)\Pi_{AB}^{g,h}U_{AB}^\dag(f)=\proj{fg}_A\otimes\proj{fgh}_B = \Pi_{AB}^{fg,h}.
\ee
The group invariance of the integration measure in \eqref{eq-M} allows us to redefine $fg\to g$, thereby recovering the original map.

Note that, for a nonabelian symmetry group, this $G$-invariance of the map constrains the construction of the projectors.  If we had instead defined the projectors as $\kb{g}{g}_A\ot\kb{hg}{hg}_B$, the resulting map would not be $G$-invariant except in the special case of $h$ satisfying $hg=gh$ for all $g\in G$ (i.e.\;$h$ in the centre of the group).

\subsubsection{Using the relational measurement to change quantum reference frames\label{sec-specialcases}\label{sec-mainmeasurement}}

We have constructed a quantum operation \eqref{eq-genCQRF} to determine the relative orientation between two quantum reference frames.  In this section we will apply this map to the problem of changing quantum reference frames. Our main result is that when the operation $\m M_{AB}^h$ is applied in order to change the description of $\rho_S$ with a reference frame $A$ to a description with a reference frame $B$, followed by a discarding (tracing out) of the original reference frame $A$, the result is a new relational encoding of $\rho_S$ with $B$, but with the additional effect of decoherence.

Consider the action of the measurement map $\m M_{AB}^h$ given by \Eeqref{eq-M} on generic $G$-invariant states $\sigma_{SAB}$ on systems $S,A,B$ (defined to act as the identity map on $S$).  Because it is $G$-covariant, the map will produce an unnormalised $G$-invariant state on $SAB$. For the purposes of the change of quantum reference frame procedure we want a map from $SAB$ to $SB$, as we want to discard the $A$ reference frame following the measurement. This is done by applying a partial trace over $A$ to the post-measurement state, and the result is a final (unnormalised) $G$-invariant state on systems $S$ and $B$ with correlation between the subsystems. The unnormalised final state for outcome $h$ is
\begin{align}
\Tr_A[\M_{AB}^h(\sigma_{S AB})]=&D_{s_A}D_{s_B}\int\di\mu(g)\bigl[\bra g_A\ot\bra{gh}_B\bigr]\;\sigma_{S AB}\bigl[\ket {g}_A\ot\ket{gh}_B\bigr]\otimes\proj{gh}_B\label{eq-genCQRF}.
\end{align}
The measurement outcome $h$ is a continuous parameter, so we have a probability density function for outcomes $h$ for the measurement of a state $\sigma_{SAB}$ given by
\be
P(h)=\Tr[E_h\sigma_{SAB}]= \Tr[\m M^h_{AB}(\sigma_{SAB})]\,.\label{eq-P}
\ee
The probability density function normalises by $\int P(h)\di\mu(h)=1$ when using the group-invariant Haar measure $\di\mu(g)$.

Consider a relational encoding of a quantum state $\rho_S$ using a quantum reference frame $\ket{\psi(a)}_A$, a pure state with a well-defined orientation $a\in G$.  (For example, this could be the state $\ket{\psi(a)}_A = U(a)\ket e_A$ for $a\in G$, although we do not require that it take the form of a maximum likelihood state.)  No other reference frame is implicated, so we describe the joint $SA$ system by the $G$-twirled state $\m E_{\ket{\psi(a)}_A}(\rho_S)=\m G_{S A}(\rho_S\otimes\proj{\psi(a)}_A)$. We introduce a second reference frame $\rho_B$ which is non-implicated, i.e., uncorrelated with the other two quantum systems, described by the state $\m G_B(\rho_B)$. The full initial state on all components (the system $S$ and both quantum reference frames $A$ and $B$) is then
\be
\sigma_{S AB}=\sigma_{SA}\otimes\sigma_B=\m G_{S A}(\rho_S\otimes\proj{\psi( a)}_A)\otimes\m G_B(\rho_B).\label{eq-uncorinitstate}
\ee

We apply the operation $\sigma_{SAB} \rightarrow \Tr_A[\m M_{AB}^h(\sigma_{SAB})]$ given by \eqref{eq-genCQRF} to this state $\sigma_{SAB}$. This state is group-invariant, satisfying $\m G_{SAB}(\sigma_{SAB})=\sigma_{SAB}$, and $\m M_{AB}^h$ is $G$-covariant, so we can commute the $G$-twirl with the operation, allowing us to write the final state on $SB$ as
\begin{align}
\Tr_A[\M_{AB}^h(\sigma_{SAB})]
=&\m G_{SB}\Bigl[\rho_S\otimes\bigl(D_{s_A}D_{s_B}\int\di\mu(g)\!\abs{\bk g{\psi( a)}_A}^2\!\bra{gh}\!\m G_B(\rho_B)\!\ket{gh}\;\proj{gh}_B\bigr)\Bigr] \nonumber \\
=&\m G_{SB}\Bigl[\rho_S\otimes\bigl(D_{s_A}\int\di\mu(g)\, \abs{\bk g{\psi( a)}_A}^2\;\proj{gh}_B\bigr)\Bigr]\label{eq-finalstate}
\end{align}
where the second line follows from the simplification $\bra{gh}\m G_B(\rho_B)\ket{gh}=\Tr[\m G(\proj{e}) \rho_B]=D_{s_B}^{-1}$, arising from properties of the $G$-twirl and maximum likelihood states. Thus, referring back to \eqref{eq-P}, $P(h)=1$ for states of the form \eqref{eq-uncorinitstate}. We can therefore associate $\Tr[\m M^h_{AB}(\sigma_{SAB})]$  with a trace one normalised state.

The $G$-twirl $\m G_{SB}$ allows us to insert a global rotation $\m U_S(g^{-1})\otimes\m U_B(g^{-1})$ into \eqref{eq-finalstate} without changing the state. We thus obtain the operator $\m U_S(g^{-1})$ on $\rho_S$, and a reference frame state $\proj h_B$.  We can therefore write the final state as a relational encoding
\be
\Tr_A[\m M_{AB}^h(\sigma_{SAB})]=\m G_{SB}\left(\rho_S'\otimes\kb hh_B\right).\label{eq-mainmeas}
\ee
Returning to \eqref{eq-finalstate}, With $\ket{\psi(a)}$ covariant, we have that $\bk g{\psi(a)}_A=\bk{a^{-1}g}{\psi(e)}_A$. Redefining $g\to ag$, we have that the new encoded system state $\rho_S'$ is related to the original system state $\rho_S$ by the composition of maps
\[
\rho_S'=\m F_S^{(A)}\circ\m U_S(a^{-1})[\rho_S]
\]
and the form of the CP map $\m F_S^{(A)}$ is
\be
\m F_S^{(A)}:=D_{s_A}\int\di\mu(g)\abs{\bk{g}{\psi(e)}_A}^2\m U_S(g^{-1}).\label{eq-genCQRFG}
\ee
This CP map is a convex mixture of unitary maps determined by the overlap of a maximum likelihood state with the reference frame state on $A$. Therefore, unless the state $\rho_S$ is $G$-invariant, this map results in decoherence of $\rho_S$ due to the uncertainty in orientation of the reference frame state $\ket{\psi(e)}_A$. Note that the map is independent of the orientation $a$ of reference frame $A$.

Finally, using again the properties of maximum likelihood states we have that $\int\di\mu(g)\abs{\bk g{\psi(e)}_A}^2=D_{s_A}^{-1}$, and therefore the CP map $\m F_S^{(A)}$ is trace-preserving.  We can therefore identify $\Tr_A[\M_{AB}^h(\sigma_{S AB})]$ as being proportional to the normalised (trace one) final state
\be
\sigma_{SB}^h:=\m E_{\ket h_B}\bigl(\m F_S^{(A)}\circ\m U_S(a^{-1})[\rho_S]\bigr).\label{eq-finalnormalised}
\ee
Therefore the procedure maps a relational encoding of $\rho_S$ in terms of $A$ to a normalised relational encoding of $\rho_S'$ in terms of $B$.  Note that this new relational encoding explicitly depends on the measurement outcome $h$, which relates the previous description relative to A to the new description relative to B.

\subsection{Decoherence in the change of quantum reference frame procedure\label{sec-dec2}}

\begin{figure}
\[
\xymatrix{
& \rho_S' \ar[ddr]^{\hbox{Encoding $\m E_{\ket{h}_B}$}}&\\
&{\m E_{\ket h_B}\!\circ\!\m R}&\\
\m E_{\ket{\psi(a)}_A}(\rho_S) \ar[uur]^{\hbox{Recovery $\m R$}} \ar[rr]_{\hbox{Change of frame}}^{\Tr_A[\m M_{AB}^h(*\otimes\sigma_B)]} &&  \m E_{\ket h_B}(\rho_S')
}
\]
\caption[Re-encoding with or without using a background frame]{The procedure \eqref{eq-genCQRF} on states of the form given by \eqref{eq-uncorinitstate} is the same map as $\m E_{\ket h_B}\!\circ\m R$ on an encoding $\m E_{\ket{\psi(a)}_A}(\rho_S)$. The change of quantum reference frame map results in description of the state relative to a  different reference frame without measuring against a background classical frame, whereas the re-encoding map $\m E_{\ket h_B}\!\circ\m R$ changes a reference frame using a background reference frame for the intermediate state $\rho_S'$. \label{fig-CRE}}
\end{figure}
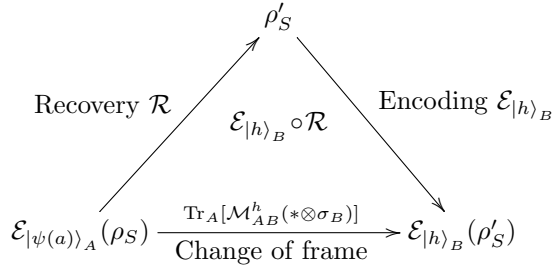

We will now characterise the decoherence of the system $\rho_S$ due to the change of quantum reference frame procedure. As outlined in \secref{sec-introdequant}, for a system $S$ in a relational encoding with a quantum reference frame $R$, the recovery map of \eqref{eq-rec} can be interpreted as measuring the quantum reference frame $R$ against a background classical frame, discarding the quantum reference frame and correcting the orientation of $\rho_S$ if the quantum reference frame is rotated with respect to the background frame. This recovery map applied to a relational encoding of a system $\m E_{\ket{\psi(a)}_R}(\rho_S)$ leads to a noise map on $\rho_S$:
\be
\m R\circ\m E_{\ket{\psi(a)}_R}(\rho_S)=D_{s_R}\int\di\mu(g)\abs{\bk g{\psi(a)}_R}^2\m U_S(g^{-1})[\rho_S].\label{eq-REinterp}
\ee

The form of this map is identical to the form of the decoherence map $\m F_S^{(A)}$ from the change of reference frame procedure in the final encoding, $\m E_{\ket e_B}(\rho_S')$, given by \eqref{eq-genCQRFG}. We then have the equivalence of maps
\be
\m R\circ\m E_{\ket{\psi(a)}_R}\equiv\m F_S^{(R)}\circ\m U_S(a^{-1})\label{eq-REFU}.
\ee
Indeed, as depicted in \figref{fig-CRE}, the re-encoding $\m E_{\ket a_A}(\rho_S)\mapsto\m E_{\ket e_B}(\rho_S')$ achieved by the change of quantum reference frame procedure can also be achieved by concatenating the recovery and encoding maps $\m E_{\ket e_B}\circ\m R(\sigma_{SA})$, apart from the final rotation $\m U_S(h^{-1})$. We can therefore write the final state of the change of quantum reference frame procedure as
\be
\sigma_{SB}^h=\m E_{\ket h_B}\left[(\m R\circ\m  E_{\ket{\psi(a)}_A})[\rho_S]\right],\label{eq-finstateRE}
\ee
an encoded state in which $\rho_S$ is recovered from an initial encoding, and rotated by $h^{-1}$ due to the relative orientation between the old and new reference frames.

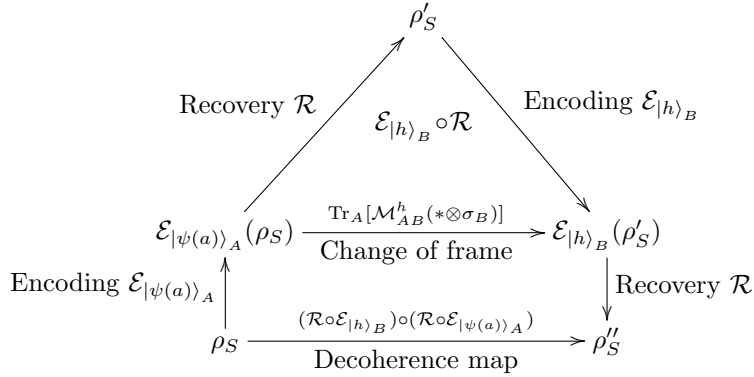
\begin{figure}
\[
\xymatrix{
& \rho_S' \ar[ddr]^{\hbox{Encoding $\m E_{\ket{h}_B}$}}&\\
&{\m E_{\ket h_B}\!\circ\!\m R}&\\
\m E_{\ket{\psi(a)}_A}(\rho_S) \ar[uur]^{\hbox{Recovery $\m R$}} \ar[rr]_{\hbox{Change of frame}}^{\Tr_A[\m M_{AB}^h(*\otimes\sigma_B)]} &&  \m E_{\ket h_B}(\rho_S')\ar[d]^{\hbox{Recovery $\m R$}}\\
\rho_S \ar[rr]_{\hbox{Decoherence map}}^{(\m R\circ\m E_{\ket h_B})\circ(\m R\circ\m E_{\ket{\psi(a)}_A})} \ar[u]^{\hbox{Encoding $\m E_{\ket{\psi(a)}_A}$}}&&\rho_S''}
\]
\caption[Net decoherence map on the system due to the change of reference frame map]{We can add another level to \figref{fig-CRE} whereby the initial state in the change of frame map is encoded from $\rho_S$, and the final state is recovered to $\rho_S''$. We then identify that the change of reference frame map induces a decoherence map on the original $\rho_S$ which is the result of the encoding of the $A$ frame and recovery of the $B$ frame.\label{fig-RERE}}
\end{figure}

In order to determine the net decoherence in \eqref{eq-finstateRE} due to the change of quantum reference frame procedure, we apply the recovery map $\m R$ to $\sigma_{SB}^h$. Using \eqref{eq-rec} and \eqref{eq-finstateRE} we obtain the final system state in terms of a decoherence map on the initial system state,
\begin{equation}
\m R(\sigma_{SB}^h)=\m R\circ\m E_{\ket h_B}(\rho_S')=:\rho_S''.
\end{equation}
We can represent the entire decoherence map on the initial system $\rho_S$ using a commutative diagram in \figref{fig-RERE}. Using our analysis leading to \eqref{eq-REFU}, the net decoherence of $\rho_S$ in the change of quantum reference frame procedure is then exactly the composition of the effective decoherence with rotation correction from the use of the first reference frame, $\m R\circ\m E_{\ket{\psi(a)}_A}[\rho_S]=\m F_S^{(A)}\circ\m U_S(a^{-1})[\rho_S]$, with the effective decoherence from using the second reference frame, $\m R\circ\m E_{\ket h_B}[\rho_S']=\m F_S^{(B)}\circ\m U_S(h^{-1})[\rho_S']$. The final noise map is then
\be
\begin{split}
\rho_S''=\m N[\rho_S]=&(\m R\circ\m E_{\ket h_B})\circ(\m R\circ\m
E_{\ket{\psi(a)}_A})[\rho_S]\\
=&\m F_S^{(B)}\circ\m U_S(h^{-1})\circ\m F_S^{(A)}\circ\m U_S(a^{-1})[\rho_S].
\end{split}\label{eq-fulldecoherence}
\ee
We will use this identification of the decoherence in terms of the encoding and recovery maps in \secref{sec-interpretation} in which we interpret the nature of the decoherence.

\subsubsection{Classical limits\label{sec-classicallimits}}

Recall the notation introduced in \secref{sec-groupestates}, with the states of quantum references frames parameterised by a `size' parameter $s$, and for which $s \to \infty$ describes the classical limit.
Given that decoherence occurs in the change of quantum reference frame procedure due to the uncertainties in orientation of each reference frame state $\ket{s;\psi(g)}$, we want to identify the conditions in which the decoherence in the procedure disappears. For a class of quantum reference frame states $\ket{s;\psi(g)}$ that possess a well-defined classical limit $s\to\infty$ in which uncertainty in orientation disappears, we demonstrate that the change of quantum reference frame procedure has an appropriate classical limit.

To reproduce a classical change of reference frame map, for size parameters $s_A,s_B\to\infty$ we should have the initial relational encoding $\lim_{s_A \to \infty} \m E_{\ket{s_A;a}_A}(\rho_S)$ map to the final relational encoding $\lim_{s_B \to \infty} \m E_{\ket{s_B;ah}_B}(\rho_S)$ with no change to the encoded state $\rho_S$, i.e., no decoherence.  We now show that this is the case.

As $s_A\to\infty$, the overlap of $\ket{s_A;\psi(e)}$ with other orientations in the group becomes zero, i.e., we have
\begin{equation}
   \lim_{s_A \to \infty} D_{s_A}\abs{\bk {s_A;g}{s_A;\psi(e)}_A}^2 = \delta(g)\,,
\end{equation}
where $\delta$ is the Dirac delta function on the group. The decoherence map \eqref{eq-genCQRFG} then becomes the identity map, $\lim_{s_A \to \infty} \m F_S^{(A)}=\m I_S$. The final state is then $\sigma_{S B}^h =\m E_{\ket{s_B;ah}_B}[\rho_S]$ where the size of the reference frame $B$ is determined by the size $s_B$ of the initial $B$ reference frame. This reproduces the required classical limit.  If the $B$ reference frame remains finite, the recovery of this state $\m R(\sigma_{SB}^h)$ will still result in the effective decoherence $\m F_S^{(B)}$. Therefore the classical limit of $A$ of the change of frame procedure corresponds to a relational encoding from a `classical frame' $A$ into a finite-size quantum reference frame $B$. When $s_B$ is also taken to infinity, we must have that the measurement projectors on $B$ have an infinite size, resulting in a `classical' frame on $B$ in the final state. Since this frame has perfect distinguishability, we will avoid any effective decoherence upon de-quantisation from having a relational encoding with a finite reference frame.

Note that when only the $B$ reference frame becomes classical, the decoherence associated with the final decoding $\m F_S^{(B)}$ will disappear, but the decoherence associated with the change of reference frame map $\m F_S^{(A)}$ remains.  This classical limit corresponds to decoding from a quantum frame $A$ into a classical frame $B$ with the decoherence $\m F_S^{(A)}$ associated with the dequantisation of the $A$ frame.

\subsection{Consequences and interpretation of decoherence\label{sec-interpretation}}

In the previous sections, we have developed the mathematical tools to describe the change of a quantum reference frame. Before investigating two examples in Sections~\ref{sec-phaseex} and \ref{sec-exampleSU2}, it is worthwhile to consider at this stage some of the conceptual consequences of the procedure.

As we identified in \secref{sec-dec2}, following the change of quantum reference frame procedure the system in the final encoded state appears to be affected by a form of decoherence.  This decoherence is absent in the classical limit.  In this section we will investigate the properties of the decoherence, the necessity of its existence in a change of reference frame procedure, and consider the consequences for the relativity principle for quantum reference frames, suggesting a connection to a type of \textit{intrinsic decoherence}.

\subsubsection{Properties of the decoherence from changing quantum frames}

First, we pose some questions regarding the properties of the decoherence in the procedure.  Is decoherence necessary when changing a quantum reference frame? Could the decoherence be reduced by changing to a better (more precise) reference frame?

In \secref{sec-dec2} we determined that the decoherence due to changing quantum reference frames with the procedure was $\m F_S^{(A)}$ in \eqref{eq-genCQRFG}; a mixture of unitaries determined by the overlap of the first reference frame state $\ket\psi_A$ with the measurement projector: $\abs{\bk g{\psi(e)}_A}^2$. The states in this overlap are generally not orthogonal unless the reference frame approaches infinite size. Therefore, whenever the reference frames are of finite size (and therefore with a finite asymmetry), decoherence of $\rho_S$ is induced. If the state is recovered from the new reference frame $B$, we also see that there is an additional contribution $\m F_S^{(B)}$ to the net decoherence from reference frame $B$, of the same type as $\m F_S^{(A)}$. As shown in \eqref{eq-fulldecoherence}, we then have that the net decoherence on $\rho_S$ is the composition of the decoherences associated with the original encoding of $\rho_S$ with respect to reference frame $A$, and the final encoding with respect to $B$. As a particular consequence, one cannot change to a larger (more precise) reference frame without strictly increasing decoherence beyond that of the initial encoding. Similarly, in reducing reference frame size, the decoherence will be more substantial than the effective decoherence from encoding using the final reference frame.

\subsubsection{Interpreting the decoherence as an intrinsic decoherence}

Now that we have identified the decoherence as being fundamental to change of quantum reference frames, there is still the question of what consequences this decoherence has for the relativity principle. To this end, we will interpret the decoherence in terms of an intrinsic decoherence.  Intrinsic decoherence is decoherence to a quantum state that occurs without interaction with an environment \cite{Stamp12}.
It has been proposed to occur as a result of fluctuations in the spacetime metric or other aspects of background spacetime due to quantum effects of the spacetime in theories of quantum gravity \cite{Milburn91,Power00,Gambini04,KokYurtsever,GirelliPoulin08}. By internalising parameters into a quantum model, quantum reference frames provide a way to model the effects of a background spacetime. A connection between deformed symmetries of semiclassical gravity and quantum reference frames was demonstrated in \cite{GirelliPoulin07}. Most closely related to quantum reference frame measurement is a model by Milburn for intrinsic decoherence on a quantum state that arises when a translation operator is not exactly known due to the quantisation and uncertainty of the background time parameter~\cite{Milburn03}.

We will interpret the decoherence of the change of quantum reference frame procedure \Ssref{sec-introCoRF} within the spacetime intrinsic decoherence framework introduced above. The change of quantum reference frame procedure is a complete, closed description of the decoherence that occurs to a system $\rho_S$ due to changing between two quantum reference frames. The corresponding description of a change of reference frame when the two reference frames are treated as background frames, so that only the system $\rho_S$ remains quantum mechanical, is that the system experiences a noise map $\m F_S^{(B)}\circ\m U_S(h^{-1})\circ\m F_S^{(A)}\circ\m U_S(a^{-1})$ as an isolated quantum system; i.e.\;in this description the quantum system experiences intrinsic decoherence. Now, running this reasoning in reverse, we have that the change of quantum reference frame map is the self-contained description of this intrinsic decoherence once the quantum nature of the reference frames is included. As such it is an operational derivation of a process that leads to intrinsic decoherence. Note that this particular model consists of an abrupt measurement rather than dynamics or continuous time evolution.

\section{Example: Phase reference \label{sec-phaseex}\label{sec-genAbelgroup}}

In this section, we explore a simple example illustrating the details of the change of quantum reference frame procedure for reference frames associated with an Abelian group.  Specifically, we consider a phase reference, whose orientation corresponds to an element of $U(1)$.  We will pay particular attention to the explicit forms and interpretation of the final state described in \secref{sec-mainmeasurement}, which characterises the effect and describes the decoherence that occurs in the change of quantum reference frame procedure.

The example will be structured as follows. First we will describe the reference frame states we will use and how these allow storage of quantum information in relational degrees of freedom. We will then review the change of quantum reference frame procedure in this Abelian case. We then explicitly calculate the decoherence for the cases where the reference frame is described by a phase eigenstate or coherent state. There will be some comparison of the decoherence for these choices. We will use these results to verify the classical limits of the procedure as described in \secref{sec-classicallimits}. Finally, we comment on the similarity of the relational reference frame measurement to balanced homodyne detection.

\subsection{The representation of $U(1)$ on harmonic oscillators}

We first present the structure of the representation of $U(1)$ on a collection of harmonic oscillators, and how we might encode information in a relational way.  For an Abelian group, group multiplication becomes addition and the identity $e$ can be written as $0$. The unitary group $U(1)$ can be considered as the group of phases $\theta$ with the group multiplication being addition modulo $2\pi$. However, although we can interpret the group elements as phases, we will retain the generic group element notation $g,a,h \in U(1)$ for familiarity. The charge sectors of the representation are subspaces of total photon number. The unitary representation of $U(1)$ on a single mode state is $U(\theta)=\Ee^{\ii \hat n\theta}$ where $\hat n$ is the number operator $\hat n\ket k_\text{Fock}=k\ket k_\text{Fock}$. The $U(1)$ Haar integration measure is $\di\mu(g)=\di g/2\pi$. Therefore, for a single mode harmonic oscillator, the $G$-twirl of a state $\sum_{k=0}^\infty a_k\ket k_\text{Fock}$ is
\begin{equation}
\m G\Bigl(\sum_{k,l=0}^\infty a_ka_l^*\kb kl_\text{Fock}\Bigr)=\int_0^{2\pi}\frac{\di\theta}{2\pi}\Ee^{\ii(k-l)\theta}\sum_{k,l=0}^\infty a_ka_l^*\kb kl_\text{Fock}=\sum_{k=0}^\infty \abs{a_k}^2\kb kk_\text{Fock}
\end{equation}
with the integral giving the constraint $k=l$. The phase information in a single mode state is thus completely decohered. However, if we introduce a second mode, i.e., a second oscillator with distinguishable frequency, we can form the two-mode pure state $\ket{\psi_{SA}}=\sum_{k,l}a_{k,l}\ket{kl}_\text{Fock}$. Written in terms of total photons $2n=k+l$ and difference $2j=k-l$, where $n$ can take any non-negative half-integer value and where $j=-n,-n +1, \dots, n$ \cite{TycSanders}, this becomes $\sum_{n,j}a_{n+j,n-j}\ket{n+j,n-j}_\text{Fock}$. The $G$-twirl on this state is
\begin{align}
\m G_{SA}(\proj\psi_{SA})=&\sum_{n,j}\sum_{m,k}\int \frac{\di\theta}{2\pi}\Ee^{\ii(2n-2m)\theta}\Big(a_{n+j,n-j}\ket{n+j,n-j}_\text{Fock}\Big) \Big(a^*_{m+k,m-k}\bra{m+k,m-k}_\text{Fock}\Big) \nonumber \\
=&\sum_n\Big(\sum_{j,k}a_{n+j,n-j}a_{n+k,n-k}^*\kb{n+j,n-j}{n+k,n-k}_\text{Fock}\Big)
\end{align}
Phase coherence remains within subspaces of total photon number eigenstates, producing a total state that is a mixture over total photon number $2n$ of pure eigenstates $\sum_{j}a_{n+j,n-j}\ket{n+j,n-j}_\text{Fock}$ of total photon number $2n$. With judicious choices of a reference frame state on $A$, a state on $S$ can be relationally encoded into the subspaces of total photon number~\cite{SBRK,BRSDialogue}.

\subsection{Reference frames for $U(1)$}

We define our two reference frames $A$ and $B$ to be single mode harmonic oscillators in group-covariant states $\ket{\psi(gh)}=U(g)\ket{\psi(h)}$. The particular examples we will study are the $U(1)$ maximum likelihood states, and $U(1)$ coherent states, both of which have well-defined size parameters.

\subsubsection{Reference frame $A$ in phase eigenstate\label{sec-BHD}}

The maximum likelihood states (introduced in \secref{sec-groupestates}) for a representation of the $U(1)$ group on a single mode Fock space truncated in maximum photon number $s$ are the bounded-size phase eigenstates with photon number cutoff $s$  \cite{PeggBarnett97}. The phase eigenstate with phase $g$ and size parameter $s$ is given by
\be
\ket{s;g}:=N_s^{-\half}\sum_{k=0}^s\Ee^{\ii k g}\ket k_\text{Fock}\label{eq-phaseeigenstate}
\ee
where $\ket k_\text{Fock}$ is the Fock state with $k$ excitations, and the state normalisation is $N_s=(s+1)=D_s$, the dimension of the Hilbert space.

In addition, as these states satisfy $\m G (\proj {s;g}) = (s+1)^{-1}I_s$, they will also be used to form the projectors \eqref{eq-proj} for measurement.

We now consider the change of quantum reference frame procedure for the $U(1)$ group, using phase eigenstates both for our initial reference frame on $A$ as well as forming the relational measurement.  In this procedure, an initial state $\sigma_{S AB}=\m G_{S A}(\rho_S\otimes\kb{\psi(a)}{\psi(a)}_A)\otimes \m G_B(\rho_B)$ is transformed to the final state for outcome $h$ on $SB$ given by
\be
\sigma_{SB}^h=\Tr_A\bigl[\m M_{AB}^h(\sigma_{SAB})\bigr]=\m E_{\ket{a+h}_B}\bigl(\m F_S^{(A)}[\rho_S]\bigr)\,,
\ee
where we have commuted the rotations $a$, $h$, and the map $\m F_S^{(A)}$ due to $U(1)$ being Abelian. For the state of reference frame $A$ prepared in the bounded-size phase eigenstate $\rho_A=\proj{s_A;a}$ with cutoff $s_A$, the overlap between two phase eigenstates with cutoffs $s$ gives \cite{PeggBarnett89}
\begin{align}
\abs{\bk{s;g}{s;h}}^2=&D_s^{-2}\sum_{k=-s}^s(s+1-\abs k) \Ee^{\ii k(h-g)}\label{eq-U1overlap}\\
=&\frac{1}{(s+1)^2}\frac{1-\cos[(s+1)(h-g)]}{1-\cos[h-g]}\notag.
\end{align}
The measurement of relative orientation $h$ is constructed from a family of projectors \eqref{eq-proj} on the $A$ and $B$ Hilbert spaces. In this example the projectors will be constructed in terms of $U(1)$ maximum likelihood states with size cutoffs $s_A,s_B$. Due to the equally weighted superposition of number states of the phase states in the projectors, a measurement constructed from such a family of projectors resolves the identity on the space of the reference frames $A$ and $B$. The sizes of the projectors is set to be equal to the cutoff of the reference frame states, $s_A$ and $s_B$. The decoherence map \eqref{eq-genCQRF} then takes the form
\begin{align}
\m F_S^{(A)}=&D_{s_A}\abs{\bk{s_A;g}{s_A;0}_A}^2 \m U_S(-g)\label{eq-abelF}\\
=&\frac{1}{(s+1)}\int\frac{\di g}{2\pi}\frac{1-\cos[(s_A+1)g]}{1-\cos g}\m U_S(g^{-1}).\label{eq-mixedU1PEf}
\end{align}

The distribution of unitaries in $g\in G$ is graphed in \figref{fig-mixedU1f} for average photon number $\ex n_A=s_A/2=4$ and $8$. The function is symmetric about $g=0$, at which it is peaked.

We note that the relational measurement has many similarities to \emph{balanced homodyne detection}:  a measurement technique from quantum optics.  We explore this relationship in \appref{app-BHD}.

\subsubsection{Reference frame $A$ in coherent state\label{sec-U1cohstate}}

We also consider reference frame $A$ given by a coherent state
\be
\ket{s_A;g}_{\rm CS}=\Ee^{-s_A^2/2}\sum_{k=0}^\infty \frac{s_A^k\Ee^{\ii kg}}{\sqrt{k!}}\ket k_\text{Fock}.\label{eq-U1CS}
\ee
The coherent state has a well-defined phase $g$ (i.e.\;orientation in $U(1)$) and transforms covariantly under the group: $U(g)\ket{s_A;0}_{\rm CS}=\ket{s_A;g}_{\rm CS}$. It has a size $s_A$ characterised by the square of the mean photon number, $\ex n=s_A^2$. The $G$-twirl of this state gives a Poisson distribution in photon number, with no phase coherence.

Although coherent states are suitable as quantum reference frames, there are challenges to constructing relational measurements using projectors onto these states because $\m G(\proj{s;g})$ is not proportional to the identity.  We therefore restrict to the relational measurement constructed out of phase eigenstate projectors.

Coherent states have non-zero support on all photon numbers $n\to\infty$, so we will use an infinite limit for the size $s_A$ of the projectors on $A$ for this example. The POVM will resolve the identity on the full infinite dimensional Fock space. We will need to keep in mind that the initial $B$ state may also have $s_B\to\infty$, for example, if it is mixture of coherent states, in which case the projectors on $B$ and consequently the post-measurement state on $B$ will have infinite size.

\begin{figure}[h]
\centering
\ifpdf                                                %%%COMMENT FOR SUBMISSION
\includegraphics[height=8cm]{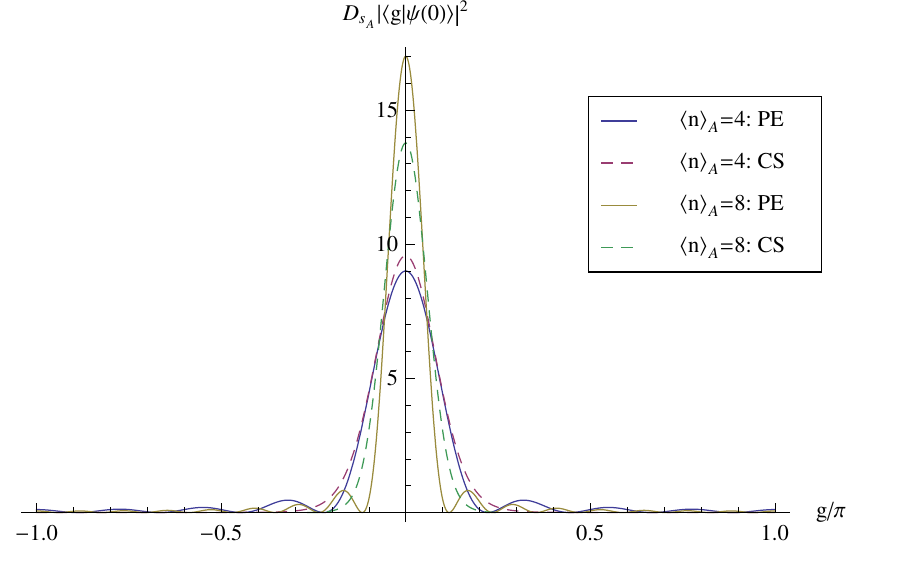}
\else                                                %%%COMMENT FOR SUBMISSION
\includegraphics[height=8cm]{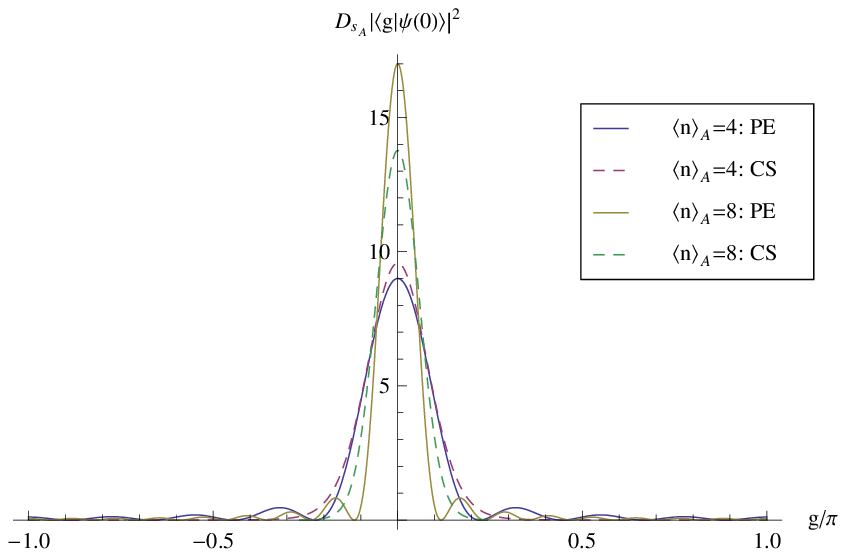}                                                %%%COMMENT FOR SUBMISSION
\fi                                                %%%COMMENT FOR SUBMISSION
\caption[State overlaps for $U(1)$ frames (colour)]{Plotted are the state overlaps $D_{s_A}\abs{\bk g{\psi(0)}}^2$ for reference frame $A$ in a $U(1)$ phase eigenstate (\eqref{eq-U1overlap}, `PE', solid lines) and coherent state (\eqref{eq-U1CSoverlap}, `CS', broken lines), for choices of average photon number $\ex n_A$. This indicates the distribution of unitaries in the decoherence maps $\m F_S^{(A)}$. For small average photon number the decoherence for the coherent states has a narrower peak than the phase eigenstate, but the phase eigenstate becomes more narrowly peaked by $\ex n_A=5$. For calculations the summations for the coherent state overlap were truncated at the 21st terms, accounting for 99.99\% of the support. \label{fig-mixedU1f}}
\end{figure}

The overlap of a coherent state with a phase eigenstate used in the projectors is
\be
\bk{s;g}{t;h}_{\rm CS}=D_s^{-\half}\Ee^{-t^2/2}\sum_{k=0}^s\frac{t^k\Ee^{\ii k(h-g)}}{\sqrt{k!}}
\label{eq-U1CSoverlap}
\ee
where we take the support of the projectors $s\to\infty$.  (Because the POVM has normalisation factors $D_s$, this limit will still result in a well-defined projector.)  The decoherence map \eqref{eq-genCQRFG} is then
\be
\m F_S^{(A)}=\int\di\mu(g) \Ee^{-t^2/2}\sum_{k=0}^s\frac{t^k\Ee^{-\ii kg}}{\sqrt{k!}}\m U_S(-g)\label{eq-mixedU1CSf}
 \ee
The distribution of the unitaries in this decoherence map is plotted in \figref{fig-mixedU1f} for choices of $s_A$, and compared with the corresponding phase eigenstate distribution \eqref{eq-mixedU1PEf} for the same average photon number.

\subsection{Classical limits}

We briefly examine and interpret the results of the change of reference frame procedure for the classical limits of reference frames $A$ and $B$, i.e., when one or both of the size parameters $s_A,s_B$ are taken to infinity.

We will examine the $B$ classical limit first. The decoherence map \eqref{eq-mixedU1PEf} is not dependent on the $B$ frame, so it does not change in the $s_B\to\infty$ limit. The final state then has mixing due to the finite size of the $A$ reference frame. If $B$ is initially in a mixture of a finite size phase eigenstate, then the effect of $s_B\to\infty$ is merely to increase the size of the final reference frame $B$ to its classical limit. For the coherent state example, the final state on $B$ is already an infinite-cutoff phase eigenstate. The interpretation of this limit is a decoding from quantum reference frame to classical frame, with noise accumulated solely due to the encoding with reference frame $A$.

To compute the decoherence map in the limit $s_A\to\infty$ we want to show that the overlap functions \eqref{eq-U1overlap} and \eqref{eq-U1CSoverlap} approach perfect distinguishability. For the phase eigenstate, using \eqref{eq-U1overlap} we can show that in the limit $s_A\to\infty$ the term becomes a delta function
\be
\lim_{s_A \to \infty} D_{s_A}\abs{\bk{s_A;g}{s_A;h}}^2=\lim_{s_A \to \infty}\sum_{k=-s_A}^{s_A}\frac{s_A+1-\abs k}{s_A+1}\Ee^{\ii k(h-g)}= \sum_{k=-\infty}^{\infty}\Ee^{\ii k(h-g)}
=\delta(h-g)\,,\label{eq-U1peoverlaplimit}
\ee
where the denominator is provided by the state normalisation $D_{s_A}=s_A+1$ \eqref{eq-phaseeigenstate} and $\delta$ is normalised in the Haar measure: $\int_0^{2\pi}\delta(g)\frac{\di g}{2\pi}=1$. For coherent states, rather than attempting to directly compute the limit of the overlap, there are existing results we can use: A phase operator can be defined in terms of the states $\ket\theta=\sum_{n=0}^\infty \Ee^{\ii n\theta}\ket n_\text{Fock}$ \cite{WallsMilburn}. These are the same operators that we use in the projectors, so characteristics of phase indicate characteristics of the overlap function \eqref{eq-U1CSoverlap}. Indeed the operator is used to define a phase distribution $P(\theta)=\abs{\bk\theta\psi}^2/2\pi$ for some state $\psi$. Particularly, for $\psi$ a large coherent state $\ket{s,\phi}$, the mean of the phase distribution is $\ex{\theta}=\phi$ and the standard deviation is $\Delta\theta=\frac1{2s}$ \cite{WallsMilburn}. Then, as $s\to\infty$, the phase uncertainty becomes $\Delta\theta\to0$. Therefore we have $\lim_{s\to\infty}\abs{\bk{s,\phi}{\theta}}^2$ is non-zero only for $\theta=\phi$, for which the value is 1.

\section{Example: Cartesian and Direction frames \label{sec-exampleSU2}}

In this section we will consider the change of reference frame procedure for reference frames based on a nonabelian group, $SU(2)$, which describes the orientations of a Cartesian reference frame for three dimensions. We also consider a `direction indicator' state for three dimensions, which, due to rotational invariance around the single indicated direction, is associated with the coset space $SU(2)/U(1)$. We use $SU(2)$ rather than $SO(3)$ so that we can use spin representations.

The representation of $SU(2)$ decomposes a Hilbert space into a tensor sum of charge sectors of total spin $j$, where $j$ is a positive integer or half integer. In general, each of these is a reducible representation which can be further decomposed into a subsystem $\m M_j$ carrying an irreducible representation in a tensor product with a multiplicity subsystem $\m N_j$ which carries the trivial representation. The Hilbert space of a reference frame state would then decompose as $\Hi_A=\bigoplus_{j} \m M_A^{(j)}\otimes\m N_A^{(j)}$ \cite{BRST}.

\subsection{$SU(2)$ fiducial states (Cartesian frame)}

We define our reference frame systems using a Hilbert space $\m H_R=\bigoplus_{j} \m M_R^{(j)}\otimes\m N_R^{(j)}$, with the dimensions of the subsystems $\m M_R^{(j)}$ and $\m N_R^{(j)}$ chosen to be equal.  Such a space carries the regular representation of $SU(2)$, where each irrep $j$ appears with multiplicity equal to its dimension.  Following \cite{BRST}, we define a fiducial Cartesian reference frame state, with truncation parameter $s$, to be
\be
\ket{s;e}:=D_s^{-\half}\sum_{j=0}^s\sqrt{2j+1}\sum_{m=-j}^j \ket{j,m}_\text{N}\otimes\ket{\phi_{j,m}}\label{eq-SU2fidstate}
\ee
which has support on integer spin $j$ charge sectors up to $j=s$. Here, $\ket{j,m}_\text{N}$ is an eigenstate of $J_z$, and these for $m=-j,-j+1,\dots, j$ form a basis for $\m M_A^{(j)}$, denoted by $\ket{\cdot}_\text{N}$. The states $\ket{\phi_{j,m}}$ form a basis $\m N_A^{(j)}$. Together $\sum_{m=-j}^j\ket{j,m}_\text{N}\otimes\ket{\phi_{j,m}}$ forms a state in the spin-$j$ {\it reducible} representation of $SU(2)$ which is maximally entangled between the irreducible representation and multiplicity subspaces. The state normalisation $D_s$ is the dimension of the vector space that $\ket{e_A}$ spans, and is given by $D_s=\sum_{j=0}^s(2j+1)^2=\frac13(2s+1)(2s+3)(s+1)=\binom{2s+3}3$.

For rotations of these states under $SU(2)$ we will use the polar parametrisation:
\[
U(g)=U(\omega,\theta,\phi)=\Ee^{\ii\omega\mathbf n\cdot\mathbf J}
\]
with $\omega$ the rotation angle, $\mathbf n=(\sin\theta\cos\phi,\sin\theta\sin\phi,\cos\theta)$ the axis of rotation, $\frac\phi2,\theta,\omega\in[0,\pi)$, and with the Haar measure given by $\di\mu(g)=\sin^2\frac\omega2\sin\theta\ \di\phi\ \di\theta\ \di\omega/2\pi^2$.

\begin{figure}
\centering
\ifpdf                                                %%%COMMENT FOR SUBMISSION
 \includegraphics[height=6cm]{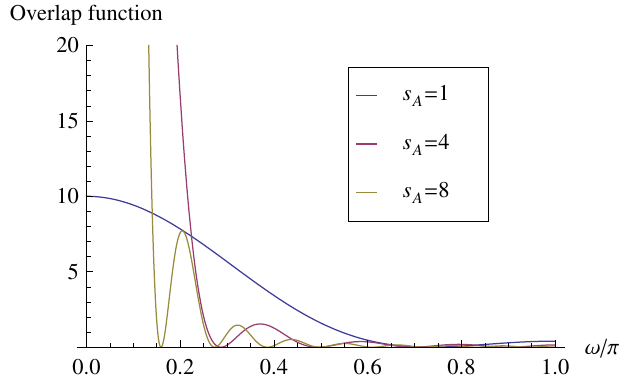}
\includegraphics[bb= 0 0 50 120,clip=true,height=6cm]{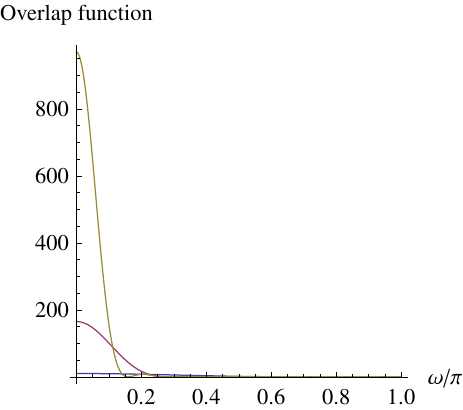}\;\includegraphics[bb= 123 0 133 110,clip=true,height=4.5cm]{SU2FSall.pdf}
\else                                                %%%COMMENT FOR SUBMISSION
 \includegraphics[height=6cm]{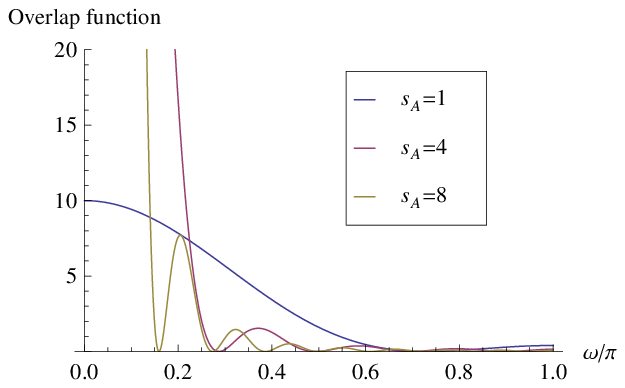}                                                %%%COMMENT FOR SUBMISSION
\includegraphics[bb= 0 0 50 120,clip=true,height=6cm]{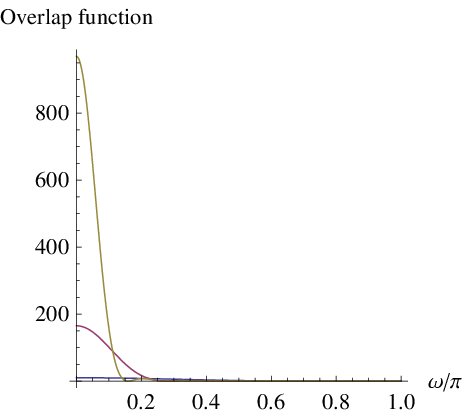}\;\includegraphics[bb= 123 0 133 110,clip=true,height=4.5cm]{SU2FSall.eps}                                                %%%COMMENT FOR SUBMISSION
\fi                                                %%%COMMENT FOR SUBMISSION
\caption[State overlap in decoherence map for $SU(2)$ Cartesian fiducial states (colour)]{Plots of integrand of decoherence map \eqref{eq-mixedSU2GEf} for fiducial state Cartesian reference frames with $s_A=1,4,8$. The fiducial state overlap depends only on the rotation angle $\omega$ and not on the axis of the rotation $\mathbf n=(\sin\theta\cos\phi,\sin\theta\sin\phi,\cos\theta)$. The mixing is symmetric about the identity, identified by $\omega=0$. Plot (a) shows the details of the plot across $\omega$, and (b) shows the full range for small $\omega$, indicating that overlap function becomes extremely peaked near $\omega=0$ even for small $s_A$, since the density of states approaches zero as $\omega\to0$.\label{fig-mixedSU2fPE}}
\end{figure}

For this example, we will use (rotated) fiducial states \eqref{eq-SU2fidstate} to form the measurement projectors \eqref{eq-proj}, with maximum $j$ `cutoffs' $s_A$ and $s_B$ for the projectors on $A$ and $B$, respectively. The overlap function of an unrotated fiducial state with an $SU(2)$-rotated state $\ket{s_A;g}=U(g)\ket{s_A;e}$ of the same size is
\be
\bk{s_A;e}{s_A;g}
=D_{s_A}^{-1}\sum_{j=0}^{s_A}(2j+1) \chi^{(j)}(\omega,\theta,\phi)
\ee
where $\chi^{(j)}(\omega,\theta,\phi)=\cos[(j+\half)\omega]/\cos(\omega/2)$ are the characters of $SU(2)$ \cite{BRST}. Using $\cos[(j+\half)\omega]/\cos(\omega/2)=\sum_{m=-j}^j\Ee^{\ii m\omega}$ and reordering summations (using $\sum_{j=m}^{s_A}=(s_A+1)^2-m^2$) we have
\be
\bk{s_A;e}{s_A;g}
=D_{s_A}^{-1}\sum_{j=0}^{s_A}(2j+1) \sum_{m=-j}^j\Ee^{\ii m\omega}=D_{s_A}^{-1}\sum_{m=-s_A}^{s_A}\Ee^{\ii m\omega}((1+s_A)^2-m^2).\label{eq-su2fiducialoverlap}
\ee
The decoherence map \eqref{eq-genCQRFG} is then
\be
\m F_S^{(A)}=\binom{2s_A+3}3^{-1}\int \frac{\di\omega\di\theta\di\phi}{2\pi^2}\sin^2\Bigl(\frac\omega2\Bigr)\sin\theta\; \Bigl(\sum_{m=-s_A}^{s_A}\Ee^{\ii m\omega}\bigl((1+s_A)^2-m^2\bigr)\Bigr)^2\m U_S(-\omega,\theta,\phi).\label{eq-mixedSU2GEf}
\ee
The state overlap in this map is plotted in \figref{fig-mixedSU2fPE} for choices of $s_A$. The overlap function \eqref{eq-su2fiducialoverlap} is independent of the axis of rotation $\mathbf n$ of $g$, depending only on the rotation angle $\omega$.

\subsubsection{Classical limits of reference frame states}

Again, we can verify several classical limits:  the limit in which the $A$ reference frame becomes infinitely large; when the $B$ reference frame becomes infinitely large; and when both become infinitely large.

We can write the overlap function \eqref{eq-su2fiducialoverlap} as
\be
\bk{s_A;e}{s_A;g}=D_{s_A}^{-1}\Bigl((1+s_A)^2+\frac{\di^2}{\di\omega^2}\Bigr)\sum_{k=-s_A}^{s_A} \Ee^{\ii k\omega}
\ee
Then for the limit $s_A\to\infty$, we have, since $D_{s_A}\sim s_A^3$,
\be
\lim_{s_A\to\infty}D_{s_A}^{-1}\Bigl((1+s_A)^2+\frac{\di^2}{\di\omega^2}\Bigr)\sum_{k=-s_A}^{s_A}\Ee^{\ii k\omega}= D_{s_A}^{-1}\frac2{\omega^2}\delta(\omega)\label{eq-SU2FSoverlaplimit}
\ee
and in addition we enforce the normalisation condition $\bk{s_A;e}{s_A;e}=1$ for all $s_A$.

We can replace one inner product in the decoherence map \eqref{eq-mixedSU2GEf} with \eqref{eq-SU2FSoverlaplimit} to obtain $\frac2{\omega^2}\delta(\omega)\bk g e_A\simeq \frac2{\omega^2}\delta(\omega)$ in the integrand. Now integrating over $\omega$, the unitary is constrained to $\m U_S(0,\theta,\phi)=\m I$, and so the $\theta$ and $\phi$ integrals are trivial. The final state is then $\sigma_{S B}^h=\m G_{S B}\left[\rho_S\otimes\kb{ah}{ah}_B\right]$, mimicking an encoding from a classical frame to quantum frame.

The $s_B\to\infty$ limit results in an unchanged decoherence map, but an infinite reference frame on $B$ in the final state. This final state can be interpreted as a recovery \eqref{eq-rec} from a finite reference frame $A$ to infinite (`classical') reference frame $B$, where the mixing on the system $\rho_S$ is the decoherence due to the initial encoding with the imperfect $A$ reference frame.

The simultaneous infinite limit of $s_A$ and $s_B\to\infty$ then describes a change of classical reference frame operation.

\subsection{$SU(2)$ Coherent states:  A direction indicator}

As an illustrative example of the effect of the choice of fiducial state, we consider using an $SU(2)$ coherent state to define a direction reference frame.  Such a state indicates a direction on the two-sphere and has rotational symmetry (it is invariant up to global phase) about this direction. These $SU(2)$ coherent states reside within a single irreducible representation $\m M^{(j)}$ of $SU(2)$ and transform under $SU(2)$, but with a $U(1)$ invariance corresponding to the rotation about the direction in which the state is pointing. The set of possible orientations of a direction indicator therefore has the structure of a coset space $SU(2)/U(1)$, rather than a group. Consequently, the results in this example take different forms to the previous examples. Even in classical cases or limits of frames using this coset space, we will see dephasing operations on quantum systems due to the $U(1)$ rotational symmetry \cite{BRST}.

For this example we will use the Euler angle parametrisation of $SU(2)$ \cite{BRST}, as it allows us to easily separate the $J_z$ rotations under which the coherent states are invariant up to global phase
\be
U(g)=U(\alpha,\beta,\gamma)=\Ee^{-\ii \alpha J_z}\Ee^{-\ii \beta J_y}\Ee^{-\ii \gamma J_z}\label{eq-SU2CSparam}
\ee
with $\alpha,2\beta,\gamma\in[0,2\pi]$ and $\di\mu(g)=\di \alpha\,\sin \beta\, \di \beta\, \di \gamma/8\pi^2$. The `identity' coherent state on irreducible representation with total spin $j$ is defined and denoted
$\ket{j;e}_{\rm CS}:=\ket{j,j}_\text{N}$
and the $SU(2)$-rotated state $\ket{j;g}_{\rm CS}\equiv\ket{j;(\alpha,\beta,\gamma)}_{\rm CS}$ is \cite{Perelomov}
\be
U(\alpha,\beta,\gamma)\ket{j,j}_\text{N}=\Ee^{-\ii\gamma j}\sum_{m=-j}^j{2j\choose j+m}^{\half}\cos^{j+m}\frac \beta2\sin^{j-m}\frac \beta2 \Ee^{-\ii\alpha m}\ket{j,m}_\text{N}=:\ket{j;g}_{\rm CS}.
\label{eq-su2csrotate}
\ee

The overlap of a rotated state with the identity coherent state is
\be
_{\rm CS}\bk{j;e}{l;g}_{\rm CS}=\  _\text{N}\!\bra{j,j}U(g)\ket{l,j}_\text{N}=\delta_{jl}\Ee^{-\ii(\alpha+\gamma)j} \cos^{2j}(\beta/2).\label{eq-su2csoverlap}
\ee
The $G$-twirl of \eqref{eq-su2csrotate} is $D_s^{-1}I_s$.

In this example the reference frame state size parameters $s_A$ and $s_B$ are given by the total spin $j$ of the coherent state. The measurement projectors \eqref{eq-proj} will consist of coherent states of the same sizes. The normalisation factors in the measurement \eqref{eq-M} are given by $D_{s}=2s+1$. The decoherence map is then
\begin{align}
\m F_S^{(A)}&=D_{s_A}\int\di\mu(g)\cos^{4s_A}(\beta/2)\m U_S(g^{-1})\nonumber\\
&=(2s_A+1)\int_0\frac{\di\alpha}{2\pi}\frac{\di\gamma}{2\pi}\sin\beta\frac{\di\beta}2 \cos^{4s_A}(\beta/2)\m R_S^z(-\gamma)\circ\m R_S^y(-\beta)\circ\m R_S^z(-\alpha)\notag\\
&=\m D_S\circ\left((2s_A+1)\int_0^\pi\sin\beta\frac{\di\beta}2 \cos^{4s_A}(\beta/2)\m R_S^y(-\beta)\right)\circ\m D_S.\label{eq-SU2CSF}
\end{align}
where $\m R_S^i(\theta)[\rho]:=\Ee^{-\ii \theta J_i}\rho\Ee^{\ii \theta J_i}$ is the superoperator for a unitary rotation of $\theta$ around the $i=y$ or $z$ axis and $\m D_S[\rho_S]=\int_0^{2\pi}\frac{\di\theta}{2\pi}R_z(\theta)\rho_S R_z(\theta)^\dag$ is dephasing noise on $\rho_S$.

\begin{figure}[h]
\centering
\ifpdf                                                %%%COMMENT FOR SUBMISSION
\includegraphics[height=6cm]{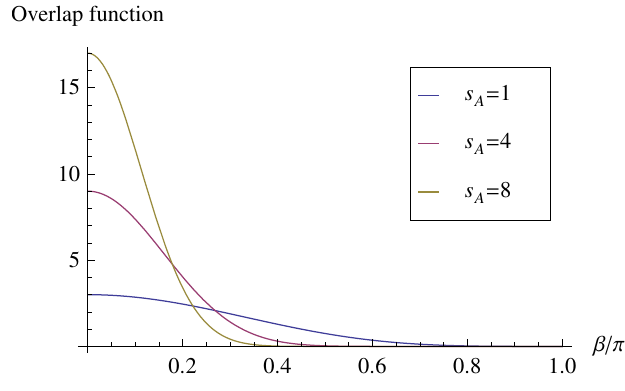}
\else                                                %%%COMMENT FOR SUBMISSION
\includegraphics[height=6cm]{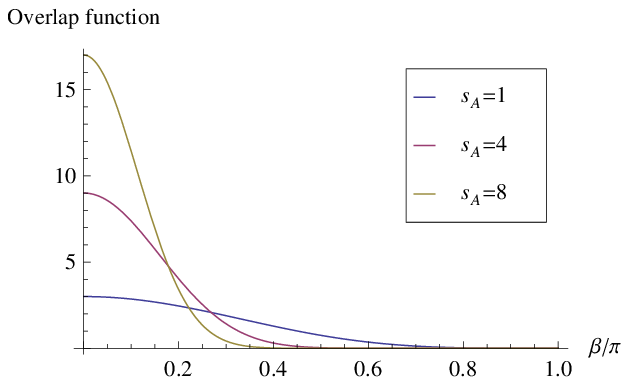}                                                %%%COMMENT FOR SUBMISSION
\fi                                                %%%COMMENT FOR SUBMISSION
\caption[Overlap function in the $SU(2)$ coherent state decoherence map (colour)]{The overlap function in $\beta$ for the $SU(2)$ coherent state decoherence map \eqref{eq-SU2CSF}. This depends on $\beta$ and is unconstrained in $\alpha$ and $\gamma$: they are rotationally symmetric on the two-sphere about the north pole (the identity), indicating we have mixing around circles of constant latitude. These overlap functions wrap around the longitude of the sphere from the north to south pole. We can see that the distribution becomes more tightly peaked near $\beta=0$ as $s_A$ increases.\label{fig-mixedSU2fCS}}
\end{figure}

This overlap function is plotted in \figref{fig-mixedSU2fCS} for choices of $s_A$. Note that the function is rotationally symmetric about the direction $\beta=0$ on the two-sphere.

\subsubsection{Classical limits of reference frame states}

We focus on the differences of this coset space example in the classical limits. From \cite{RowedeGuiseSanders,BRSandersT}, for large $j$, the overlap function \eqref{eq-su2csoverlap} can be approximated as
\be
\abs{_{\rm CS}\!\bk{j;e}{j;(\alpha,\beta,\gamma)}_{\rm CS}}^2=\cos^{4j}(\beta/2)\to \Ee^{-j\beta^2/2}.\label{eq-su2csoverlapdeltalim}
\ee
This distribution has a variance of $\sigma^2=1/2j$. Up to normalisation we thus have that in the $j\to\infty$ limit $\exp[-j\beta^2]$ approaches the delta function $\delta(\beta)$. Therefore, for the $s_A\to\infty$ limit, the overlap in the decoherence map \eqref{eq-SU2CSF} becomes essentially a delta function in $\beta$. Then, although other examples of this scenario in this limit indicate no mixing, for coherent state reference frames we instead have that the decoherence function $\m F_S^{(A)}=\m D_S[\rho_S]$ is dephasing noise on $\rho_S$.  The final state then has a uniform average over $z$-rotations of the system state $\rho_S$, i.e.
\be
\sigma_{S B}^h=\m E_{{\ket h_{\rm CS}}_B}\left[\m D_S(\m U_S(a^{-1})[\rho_S])\right].
\ee
As this limit takes the form of an encoding from classical frame $A$ to quantum frame $B$, it demonstrates that the direction indicator reference frame fundamentally cannot encode phases. Interestingly, the $a$ and $h$ rotations do not commute with the dephasing operator, so we cannot write this in the usual form as $\m E_{{\ket{ah}_{\rm CS}}_B}(\m D_S[\rho_S])$.

When $s_B\to\infty$ the decoherence map is unaffected and we have an infinite size $B$ reference frame, which indicates the decoherence that would occur due to decoding the state $\rho_S$  from the reference frame $A$. Even in the simultaneous limit $s_A,s_B\to\infty$ there is still dephasing noise. The $U(1)$ dephasing is merely an artifact of describing the $SU(2)/U(1)$ coset in a representation of $SU(2)$.

\section{Conclusions\label{sec-conclusion}}

In this chapter I investigated how the description of a state changes under a change of quantum reference frame in a static scenario. We did this by constructing a quantum operation which changes the quantum reference frame used to define a basis for another quantum system. We found that decoherence is in general induced on the quantum system due to the procedure. This decoherence is interpreted as a form of intrinsic decoherence due to a change of reference frame if one treats the frames as external objects with fundamental quantum uncertainties. Our results may provide insight into what form a relativity principle would take in such a scenario. A relativity principle would dictate how the descriptions of a physical system and its dynamics change upon a change to a new quantum reference frame. This is distinct to the `equivalence principle' as studied in \cite{AharonovKaufherr}, where the choice of reference frame used had an effect on internal relational measurements; i.e.\;no active change of quantum reference frame was made. Examples of the change of quantum reference frame procedure for $U(1)$ and $SU(2)$ reference frames were presented.

%\section*{Acknowledgements}
%MP thanks Maki Takahashi, Terry Rudolph, Peter Turner, David Jennings, Hans Westman, Iman Marvian, and Leon Loveridge for input and helpful discussions. \tblue{We thank an anonymous reviewer for helpful suggestions.} This research is supported by the Australian Research Council.

%% file: outlook.tex
\chapter{Summary and Outlook\label{ch-summary}}
In this thesis I explored three fundamental areas of quantum information theory. In the first area I defined the structure of the quantum information theory and the domain of applicability of localised fermion and photon qubits in curved spacetime. In the second I justified the form of the quantum observable for relativistic Stern--Gerlach measurement by modelling the interaction of a Dirac field with an electromagnetic field and recovering the quantum observable. These are aspects of relativistic quantum information theory. The third aspect regards the possible quantum nature of space. In this I studied the decoherence arising from the measurement of two quantum reference frames.

In this final chapter I will summarise and interpret the main results from each chapter of the thesis, and comment on the consequences and outlook of each chapter of work.

\subsection*{Localised qubits in curved spacetimes}

In this area of research the first accomplishment was to comprehensively define the domain of applicability in which localised one-particle states with two dimensional Hilbert spaces was a good approximation to the quantum field theory of particles. The one-particle qubits were obtained starting from the field equation for the particle, and by expanding orders of this equation using a high frequency limit, referred to as a WKB approximation. The particles we considered were the photon and massive fermion, so the corresponding field equations were the Maxwell and van der Waerden equations, respectively. The latter is the Dirac equation rewritten for a two-component spinor. We constructed physical arguments that fix the domain of applicability to a situation in which one begins with a single-particle localised qubit in a region of flat spacetime, and then one finishes with a single-particle localised qubit, even though in general the particle number in the curved region in between may not be well defined.

Once localisation is complete, the Hilbert space structure is carefully defined and derived from the field theory. The representation used is induced from the representation of the field and is manifestly Lorentz covariant. This requires a redefinition of unitarity of operators. With this the subsequent result was that these states evolved unitarily due to their motion through curved spacetime. Using covariant representations for the quantum states, the evolution differential equations were given by parallel transport, for geodesic paths, and in the case of massive fermions, by Fermi--Walker transport for general accelerated trajectories.

We also developed a unified formalism for the absolute phase accumulation $\Ee^{\ii\theta}$ of the quantum state as the qubit moves along a trajectory in curved spacetime. The phase term missing in the parallel transport and Fermi--Walker equations was obtained from the field equations in a WKB limit.  Combining this phase accumulation term with state rotation evolution we thus obtain the total evolution equations.

Finally we constructed quantum information theoretic protocols consistent with nonrelativistic quantum information theory, including a teleportation protocol. For localised qubits in curved spacetime, this protocol experiences the same issues as the protocol in special relativity, but without an automatic shared reference frame to use. We determined that the update map has no preferred plane of simultaneity, and that it is imperative for the two parties to share a reference frame.

The outlook and consequence for this chapter is that it provides theory for the basic experimental question of how photons and electrons can behave as localised qubits in realistic free space situations with gravity. The transport equation with phase accumulation, and the teleportation and entanglement protocol provides a fully relativistic quantum information theory of localised qubits. With this, we have a solid theoretical foundation for realistic experimental quantum information theory in gravitational fields, and can perform calculations of precession and phase accumulation around gravitational bodies, opening the way for long range quantum communication and sensitive measurements of gravitational quantities. By considering problems in the covariant formalism the chapter has provided a clearer picture and meaningful intuitive understanding of the evolution and measurement of localised qubits in curved spacetimes. The covariant formalism also reduces the possibility of applying results in invalid scenarios, or simply in the wrong frame, and results from general relativity can be used more directly.

\subsection*{Relativistic Stern--Gerlach measurement}

This research topic was focused on the question of relativistic spin measurement. We considered the specific operational model of a Stern--Gerlach measurement on a massive fermion approaching the apparatus with possibly relativistic speed. In a measurement of the spin of a massive fermion qubit, there are two observer frames: there is the rest frame of a measurement apparatus, which will see the qubit approaching at possibly relativistic velocity, and there is the rest frame of the qubit, in which the apparatus is approaching with the negative velocity. The scenario is physically distinct from a non-relativistic measurement, and the interaction between the apparatus and qubit is different. The question was then: what was the form of the operator that relates how the measurement apparatus is oriented in its rest frame to the spin eigenstates of the measurement, as described in the qubit rest frame?

In this chapter we provided a relativistic mathematical argument for the correct operator for relativistic Stern--Gerlach measurement. We did this by modelling a Stern--Gerlach measurement of a massive charged fermion in field theory. We obtained the same measurement interaction as obtained by considering relativistic transformations of the electromagnetic field between frames combined with the standard nonrelativistic argument regarding the quantisation and separation of spin packets in the Stern--Gerlach magnetic field.

The outcome of this research is that we now have a strong justification from field theory regarding the actual form of the operator for a Stern--Gerlach measurement of a relativistic spin qubit, and we can be confident about the behaviour of spin measurements in relativistic scenarios.

The consequences of the results are that there are results regarding relativistic spin in the literature which are based on other spin operators which are distinct to, and incompatible with, the relativistic Stern--Gerlach operator. These results regard non-violation of Bell inequalities of entangled spins with well-defined relativistic momentum. In these studies, the transformation of the chosen spin observable means that orientations chosen to determine Bell inequalities measure a different spin quantisation axis, returning different statistics. The Bell inequality violation can be recovered if the measured directions are modified to account for the transformation. However, an experiment testing this will likely be using a Stern--Gerlach technique to measure spin, in which case the calculations will have to be redone using the Stern--Gerlach observable in order to predict experimental results accurately.

An extension of the relativistic spin research is to study how spin-momentum wavefunctions look in different Lorentz frames (or equivalently how they transform under Lorentz boost). The spin operators measure different spin directions depending on the momentum of the qubit. Therefore a particle with a single spin, as defined in some way, but with a distribution in momentum can produce measurement statistics so that it looks like a mixed state. i.e.\;it appears the spin has decohered. Using this effect, one can look at Bell inequality non-violation due to momentum spread. This has been calculated using two spin operators, but not the Stern--Gerlach operator. The phenomenon still exists when using the Stern--Gerlach operator, but the transformation properties of the Stern--Gerlach operator are more complicated than other operators, so the quantitative result is more difficult to compute. From preliminary calculations it is clear that the form of the apparent decoherence is qualitatively different.

\subsection*{Changing quantum reference frames}

In this research I defined a quantum procedure to perform a change of quantum reference frame, and analysed the decoherence that resulted from the procedure. A quantum reference frame $A$ is a quantum state that can be used in place of an external reference frame. A quantum system $\rho_S$ which is using a quantum reference frame in this manner is said to be a `relational encoding' of $\rho_S$ using $A$. I defined a quantum procedure to take a quantum reference frame, $A$, being used in a relational encoding of $\rho_S$, and correlate its orientation with a second quantum reference frame $B$. The $A$ reference frame is then discarded by tracing over those degrees of freedom, resulting in a final state involving $\rho_S$ and a reference frame on $B$.  Since $A$ and $\rho_S$ were initially correlated, this final state is a relational encoding using $B$ of a modified system state $\rho_S'$. I found that due to the uncertainty in orientation of frame $A$, the change of quantum reference frame procedure results in a relational encoding of a partially decohered original system state $\rho_S$.

This change of frame procedure may also be described in terms of external frames rather than internal quantum frames. In this case the frames are external to the physical description and the decoherence would be described as an `intrinsic decoherence' on $\rho_S$ due to the uncertainties inherent in these external frames. I argued that the change of quantum reference frame procedure is a derivation of an example of an intrinsic decoherence due to changing frames in a quantum space, when the reference frames are internalised as quantum systems. This has been proposed as a semiclassical phenomenon of quantum gravity that arises due to the inherently quantum nature of space.

This change of quantum reference frame is one element of a relativity principle for quantum reference frames. The next step in characterising the relativity principle is to study the effects of the change procedure on the dynamics and conserved quantities defined relationally with respect to the quantum reference frames. The implications of the decoherence in this change of quantum reference frame procedure for quantum gravity should also be determined. It is possible that this model, or a modification of it, is useful for describing specific scenarios in which the quantum nature of space is proposed to lead to decoherence.

\subsection*{Conclusion}
The chapters in this thesis are all elements of research in quantum information theory in which I try to expand the domain of quantum theory into more fundamental regions. In some sense, all elements of research in this thesis are looking for semiclassical results in quantum gravity, as each combines quantum theory and relativity or quantum space in some way. The first two parts are about presenting the careful approximations required to produce a theory for experimentally accessible quantum mechanics amongst gravitational fields. The third uses quantum mechanics as a toy model for quantum space: internalising the parameters of space that are proposed to be inherently quantum mechanical into a quantum model in order to derive the proposed intrinsic decoherence as a consequence of measurement of these quantum parameters.

I hope that this research provides some useful tools for calculating predictions for experiments of new phenomena in semiclassical gravity or relativistic quantum information theory, and also provides some intuition for quantum mechanics and quantum phenomena in relativistic or quantum gravitational scenarios.

%% file: thesisarxiv.bbl
\newcommand{\etalchar}[1]{$^{#1}$}

%% file: LQiCST/QST_App.tex
%%%%%%%%%%%%%%%%%%%%%%%%%%%%%%%%%%%%%%%%%%%%%%%%%%%%%%%%%
\chapter{Appendix to Chapter 2\label{ch-app3}}
%%%%%%%%%%%%%%%%%%%%%%%%%%%%%%%%%%%%%%%%%%%%%%%%%%%%%%%%%
\section{Spinors and $SL(2,\mathbb{C})$}\label{secspinornotation}
%%%%%%%%%%%%%%%%%%%%%%%%%%%%%%%%%%%%%%%%%%%%%%%%%%%%%%%%%

In our analysis of qubits in curved spacetime it will be necessary to introduce some notation for describing spinors. A spinor is a two-component complex vector $\phi_{A}$, where $A=1,2$ labels the spinor components, living in a two-dimensional complex vector space $W$. We are going to be using spinors as objects that transform under  $SL(2,\mathbb{C})$, which forms a double cover of $SO^{+}(1,3)$. Hence, $W$ carries a spin-$\half$ representation of the Lorentz group. The treatment of spinors in this section begins abstractly, and ends with details specific to Dirac spinors. The material is based on \cite{Wald,Penrose,Bailin,DHM2010}.

%%%%%%%%%%%%%%%%%%%%%%%%%%%%%%%%%%%%%%%%%%%%
\subsection{Complex vector spaces}\label{complexvectorspace}
%%%%%%%%%%%%%%%%%%%%%%%%%%%%%%%%%%%%%%%%%%%%

Mathematically, spinors are vectors in a complex two-dimensional vector space $W$. We denote elements of $W$ by $\phi_A$. Just as in the case of tangent vectors in differential geometry, we can consider the space $W^*$ of linear functions $\psi: W\mapsto \mathbb{C}$, i.e.\; $\psi(\alpha\phi_1+\beta\phi_2)=\alpha\psi(\phi_1)+\beta\psi(\phi_2)$. Objects belonging to $W^*$, which is called the {\em dual space} of $W$, is written with the index as a superscript, i.e.\;$\psi^A\in W^*$.

Since our vector space is a complex vector space it is also possible to consider the space $\overline{W}^*$ of all {\em antilinear} maps $\chi:W\mapsto\mathbb{C}$, i.e.\;all maps $\chi$ such that $\chi(\alpha\phi_{1}+\beta\phi_{2})=\bar{\alpha}\chi(\phi_{1})+\bar{\beta}\chi(\phi_{2})$. A member of that space, called the {\em conjugate dual space} of $W$,  is written as $\chi^{A'}\in\overline{W}^{*}$. The prime on the index distinguishes these vectors from the dual vectors.

Finally we can consider the space $\overline{W}$ dual to $\overline{W}^*$, which is identified as the {\em conjugate space} of $W$. Members of this space are denoted as $\xi_{A'}$.

In summary, because we are dealing with a complex vector space in quantum mechanics rather than a real one as in ordinary differential geometry we have four rather than two spaces:
\begin{itemize}
\item the space $W$ itself: $\phi_A\in W$;
\item the space $W^*$ dual to $W$: $\psi^A\in W^*$;
\item the space $\overline{W}^*$ conjugate dual to $W$: $\chi^{A'}\in\overline{W}^*$;
\item the space $\overline{W}$ dual to $\overline{W}^*$: $\xi_{A'}\in\overline{W}$.
\end{itemize}
%
%%%%%%%%%%%%%%%%%%%%%%%%%%%%%%%%%%%%%%%%%%%%
\subsubsection{Spinor index manipulation\label{sec-geometricquantum}}
%%%%%%%%%%%%%%%%%%%%%%%%%%%%%%%%%%%%%%%%%%%%
There are several rules regarding the various spinor manipulations that are required when considering spinors in spacetime. Specifically, we would like to mathematically represent the operations of  complex conjugation, summing indices, and raising and lowering indices. The operation of raising and lowering indices will require additional structure which we will address later.

Firstly the operation of complex conjugation: In spinor notion the operation of complex conjugation will turn a vector in $W$ into a vector in $\overline{W}$. The complex conjugation of $\phi_{A}$ is represented as
$$
\overline{\phi_{A}} = \overline{\phi}_{A'}.
$$

We will also need to know how to contract two indices. We can only contract  when one index appears as a superscript and the other as a subscript, and only when the indices are either both primed or both unprimed, i.e.\; $\phi_A\psi^A$ and $\xi_{A'} \chi^{A'}$ are allowed contractions. Contraction of a primed index with an unprimed one, e.g.\;$\phi_A\chi^{A'}$, is not allowed.

The reader familiar with two-component spinors \cite{Wald,Penrose,Bailin} will recognise the index notation (with primed or unprimed indices) presented is commonly used in treatments of spinors. It should be noted however that this structure has little to do with the Lorentz group or its universal covering group $SL(2,\mathbb{C})$. Rather, this structure is there as soon as we are dealing with complex vector spaces and is unrelated to what kind of symmetry group we are considering. We will now consider the symmetry given by the Lorentz group.

%%%%%%%%%%%%%%%%%%%%%%%%%%%%%%%%%%%%%%%%%%%
\subsection{$SL(2,\mathbb{C})$ and the spin-$\half$ Lorentz group\label{spinhalfLG}}
%%%%%%%%%%%%%%%%%%%%%%%%%%%%%%%%%%%%%%%%%%%
The Lie group $SL(2,\mathbb{C})$ is defined to consist of $2\times2$ complex-valued matrices $\m L_A^{\ B}$ with unit determinant which mathematically translates into
\begin{eqnarray*}
\frac{1}{2}\epsilon_{CD}\epsilon^{AB}\m L_A^{\ C}\m L_B^{\ D}=1
\end{eqnarray*}
where $\epsilon^{AB}$ is the antisymmetric Levi--Civita symbol defined by $\epsilon^{12}=1$ and $\epsilon^{AB}=-\epsilon^{BA}$ and similarly for $\epsilon_{AB}$. It follows immediately from the definition of $SL(2,\mathbb{C})$ that the Levi--Civita symbol is invariant under actions of this group. If we use the Levi--Civita symbols to raise and lower indices it is important due to their antisymmetry to stick to a certain convention, more precisely: whether we raise with the first or second index. See e.g.\;\cite{Bailin} or \cite{Wald} for competing conventions.

The generators $G^{IJ}$ in the corresponding Lie algebra $\mathfrak{sl}(2,\mathbb{C})$ is defined by (matrix indices suppressed)
\begin{eqnarray*}
[G^{IJ},G^{KL}]=\ii\left(\eta^{JK}G^{IL}-\eta^{IK}G^{JL}-\eta^{JL}G^{IK}+\eta^{IL}G^{JK}\right)
\end{eqnarray*}
and coincides with the Lorentz $\mathfrak{so}(1,3)$ algebra. In fact, $SL(2,\mathbb{C})$ is the double cover of $SO^+(1,3)$ and is therefore a spin-$\half$ representation of the Lorentz group. Note also that the indices $I,J,K,L=0,1,2,3$ labelling the generators of the group are in fact tetrad indices. The Dirac $4\times4$ representation of this algebra is given by
\begin{eqnarray*}
S^{IJ}=\frac{\ii}{4}[\gamma^I,\gamma^J]
\end{eqnarray*}
where the $\gamma^I$ are the usual $4\times4$ Dirac $\gamma$-matrices. This representation is reducible, which can easily be seen if we make use of the Weyl representation of the Dirac matrices
\begin{eqnarray*}
\gamma^I=\begin{pmatrix}0&\sigma^I_{\ AA'}\\ \bar{\sigma}^{IA'A}&0\end{pmatrix}
\end{eqnarray*}
in which the generators become
\begin{eqnarray*}
S^{IJ} = \frac \ii4 \left[\gamma^I,\gamma^{J}\right]=\begin{pmatrix}(L^{IJ})_{A}^{\ \ B}&0\\ 0&(R^{IJ})^{A'}_{\ \ B'}\end{pmatrix}
\end{eqnarray*}
where
\begin{eqnarray*}
&(L^{IJ})_{A}^{\ \ B} =  \frac \ii4\left(\sigma^I_{\ AA'} \bar{\sigma}^{JA'B} - \sigma^J_{\ AA'}\bar{\sigma}^{IA'B} \right)\\
&(R^{IJ})^{A'}_{\ \ B'} = \frac \ii4\left(\bar{\sigma}^{IA'A}\sigma^J_{\ AB'} - \bar{\sigma}^{JA'A} \sigma^I_{\ AB'}\right).
\end{eqnarray*}
In this way the Dirac $4\times4$ representation decomposes into a left- and right-handed representation. Since primed and unprimed indices are different kinds of indices the ordering does not matter. However, if we want the spinors $\sigma^I_{\ AA'}$ and $\bar{\sigma}^{JA'A}$ to be the usual Pauli matrices it is necessary to have the primed/unprimed index as a row/column for $\bar{\sigma}^{JA'A}$ and {\it vice versa} for $\sigma^I_{\ AA'}$ \cite{Bailin}. Furthermore, $\sigma^I_{\ AA'}$ and $\bar{\sigma}^{JA'A}$ are in fact the same spinor object if we use $\epsilon_{AB}$ and $\bar\epsilon_{A'B'}$ to raise and lower the indices. Nevertheless, it is convenient for our purposes to keep the bar since that allows for a compact index-free notation $\sigma^I=(1,\sigma^i)$, $\bar{\sigma}^I=(1,-\sigma^i)$, where the $\sigma^i$ are (in matrix form) the usual Pauli matrices.

The Dirac spinor can now be understood as a composite object:
\begin{equation}
\Psi = \begin{pmatrix}\phi_{A} \\ \chi^{A'}\end{pmatrix}\label{eq-Diracspinor}
\end{equation}
where $\phi_A$ and $\chi^{A'}$ are left- and right-handed spinors respectively. In this chapter we take the left-handed component as encoding the quantum state. However, we could equally well have worked with the right-handed component as the result turns out to be the same.

Although $\epsilon_{AB}$ and $\bar\epsilon_{A'B'}$ are the only invariant objects under the actions of the group  $SL(2,\mathbb{C})$, the hybrid object $\bar{\sigma}^{IA'A}$ plays a distinguished role because it is invariant under the combined actions of the spin-1 and spin-$\half$ Lorentz transformations, that is
\begin{eqnarray*}
\bar{\sigma}^{IA'A}\rightarrow\Lambda^I_{\ J}\Lambda^{A}_{\ B}\bar\Lambda^{A'}_{\ B'} \bar{\sigma}^{JB'B}=\bar{\sigma}^{IA'A}
\end{eqnarray*}
where $\Lambda^I_{\ J}$ is an arbitrary Lorentz transformation and $\Lambda^A_{\ B}$ and $\bar\Lambda^{A'}_{\ B'}$ are the corresponding spin-$\half$ Lorentz boosts. $\Lambda^A_{\ B}$ is the left-handed  and $\bar\Lambda^{\ \;A'}_{B'}\ (=\bar\Lambda^{-1\, A'}_{\quad\ \ B'})$ the right-handed representation of $SL(2,\mathbb{C})$.

The connection between $SO(1,3)$ vectors in spacetime and $SL(2,\mathbb{C})$ spinors is established with the linear map $\bar{\sigma}^{IA'A}$, a hybrid object with both spinor and tetrad indices \cite{Wald}. The relation between a spacetime vector $\phi^I$ and a spinor $\phi_{A}$ is given by
\begin{eqnarray*}
\phi^{I} = \bar\sigma^{IA'A}\bar{\phi}_{A'}\phi_{A}.
\end{eqnarray*}
This relation can be thought of as the spacetime extension of the relation between $SO(3)$ vectors and $SU(2)$ spinors, i.e.\;this object is the Bloch 4-vector. This is in fact a null vector, and we can say that $\bar\sigma^{IA'A}$ provides a map from the spinor space to the future null light cone.

%%%%%%%%%%%%%%%%%%%%%%%%%%%%%%%%%%%%%%%%%%%%%%%%%%%%%%%%%%%%
\subsection{The geometric structure of the inner product}\label{appIP}
%%%%%%%%%%%%%%%%%%%%%%%%%%%%%%%%%%%%%%%%%%%%%%%%%%%%%%%%%

In order to turn the complex vector space $W$ into a proper Hilbert space we need to introduce a positive definite sesquilinear inner product. A sesquilinear form is linear in the second argument, antilinear in the first, and takes two complex vectors $\phi_A,\psi_A \in W$ as arguments. The nonlinearity in the first argument means that $\phi_A$ must come with a complex conjugation and the linearity in the second argument means that $\psi_A$ comes without complex conjugation. In order to produce a complex number we now have to sum over the indices. So we should have something looking like $\langle\phi|\psi\rangle=\sum\bar{\phi}_{A'}\psi_A$. However, we are not allowed to carry out this summation: both the indices appear as subscripts and in addition one comes primed and the other unprimed. The only way to get around this is to introduce some geometric object with index structure $I^{A'A}\in\bar{W}^*\otimes W^*$. The inner product then becomes
\begin{eqnarray*}
\langle\phi|\psi\rangle=I^{A'A}\bar{\phi}_{A'}\psi_A.
\end{eqnarray*}
In order to guarantee positive definiteness, the inner product structure $I^{A'A}$ should have only positive eigenvalues.

Now that we have defined an inner product structure we can state how the spinor index notation is related to Dirac bra-ket notation used in standard quantum theory. We can readily make the identifications
\begin{eqnarray*}
|\phi\rangle\sim\phi_A\qquad\langle\phi|\sim I^{A'A}\bar{\phi}_{A'}.
\end{eqnarray*}

In non-relativistic quantum theory, one would choose the inner product as $I^{A'A}=\delta^{A'A}$ where $\delta^{A'A}$ is the Kronecker delta. However,  a different structure arises from the inner product of the Dirac field in the WKB limit. To see this we begin with the conserved current $j^{\mu} =\bar\Psi(x)\hat\gamma^\mu\Psi(x)$. The net `flow' of this current through an arbitrary hypersurface forms the Dirac inner product, and is a conserved quantity. Now consider the Dirac inner product  between two 4-spinor fields $\Psi_1(x)$, $\Psi_2(x)$ in the Weyl representation \eqref{eq-Diracspinor}. We have
\begin{equation}
\int \bar\Psi_1(x)\hat\gamma^\mu\Psi_2(x)\   \di\Sigma_{\mu}=\int \bar{\sigma}^\mu_{\ A'A}\bar\chi^{A}_1(x)\chi^{A'}_2(x)+\bar{\sigma}^{\mu AA'}\bar\phi^1_{B'}(x)\phi^2_{A} (x) \ \di\Sigma_{\mu}
\label{covinproddemo}
\end{equation}
where the integration is over an arbitrary spacelike hypersurface $\Sigma$. If $n^{\mu}$ is the unit vector field normal to the hypersurface and $\di \Sigma$ is the induced volume element, we write $\di \Sigma^{\mu}= n^{\mu} \di \Sigma$. \eqref{covinproddemo} is further simplified by making use of the equations of motion $m\chi^{A'}=\ii\bar{\sigma}^{\mu A'A}D_\mu\phi_A$, where the covariant derivative reduces to $D_\mu\phi_A\approx k_\mu\phi_A$ in the WKB approximation. In this approximation we obtain
\begin{equation}
\int \bar\Psi_1(x)\hat\gamma^\mu\Psi_2(x)\   \di\Sigma_{\mu}\approx \int u^{1}_{\alpha} u^{2}_{\beta}\bar{\sigma}^{\alpha B'A} {\sigma}^\mu_{\ AA'}\bar{\sigma}^{\beta A'B}\bar\phi^{1}_{B'}(x)\phi^{2}_{B}(x)+\bar{\sigma}^{\mu AA'}\bar\phi^1_{A'}(x)\phi^2_{A} (x) \ \; \di\Sigma_{\mu}.
\end{equation}
If we further assume that $k^{1}_{\alpha}=k^{2}_{\alpha}$, i.e.\;the 4-momentum of the fields $\Psi_1$ and $\Psi_2$ in the WKB limit coincide, the inner product can be further simplified to
\begin{eqnarray}
\int \bar\Psi_1(x)\hat\gamma^\mu\Psi_2(x)\   \di\Sigma_{\mu} = \int 2 I_{u}^{A'A}\bar\varphi^{1}_{A'}(x)\varphi^{2}_{A} (x) u^{\mu}\ \di\Sigma_{\mu}
\label{covIPfrom DIP}
\end{eqnarray}
where we have made use of the identity \cite[Eqn (2.52) p16]{DHM2010}
\begin{equation}
\bar{\sigma}^{\alpha B'A}\sigma^\mu_{\ AA'}\bar{\sigma}^{\beta A'B} = g^{\alpha\mu}\bar{\sigma}^{\beta B'B}-g^{\alpha\beta}\bar{\sigma}^{\mu B'B}+g^{\beta\mu}\bar{\sigma}^{\alpha B'B}+\ii \epsilon^{\alpha\mu\beta\gamma}\bar{\sigma}^{\ B'B}_{\gamma}\label{posspinIP}.
\end{equation}
We therefore see that the inner product for the Weyl 2-spinor $I_{u}^{A'A}$, which we obtained in the WKB approximation, naturally emerges from the inner product of the Dirac field as the object contracting the spinor indices $A'$, $A$ at each point $x$ of the fields.\footnote{Note that the factor of 2 arises from differences in defining normalisation: the Dirac spinor is normalised by $\Psi(x)^\dag\Psi(x)\equiv1$ with $\Psi=(\phi,\chi)$, but the Weyl 2-spinor is normalised by $\phi(x)^\dag\phi(x)\equiv1$.}

%%%%%%%%%%%%%%%%%%%%%%%%%%%%%%%%%%%%%%%%%%%%%%%%%%%%%%%%%%%%%%%%%%%%%%%%
\section{Jerk and non-geodesic motion}\label{jerk}
%%%%%%%%%%%%%%%%%%%%%%%%%%%%%%%%%%%%%%%%%%%%%%%%%%%%%%%%%%%%%%%%%%%%%%%

We have seen that the transport of qubits as massive fermions is governed by the spin-$\half$ Fermi--Walker transport equation (\S\ref{secFermion}). One might then expect that the transport of qubits as polarisation of photons along non-geodesic null-trajectories should similarly be governed by a kind of Fermi--Walker transport. Transport of polarisation vectors for these {\it non-geodesic} null trajectories was developed by Castagnino \cite{Castagnino,SamuelNityananda,Jantzen}. However, these two proposals are mathematically distinct, and it is not clear to us which one is the correct one. Furthermore, both of these proposals involve the `jerk'  along the path, i.e.\;the time derivative of the acceleration, making the transport equation for non-geodesic paths look rather unpleasant.

It is easy to show that any transport of a polarisation vector along a null path must involve three or more derivatives of the trajectory $x^\mu(\lambda)$, i.e.\;involve one or more derivative of the acceleration $a^\mu(\lambda)$. From linearity and the requirement that the transport reduces to the parallel transport for geodesics we deduce that the transport must have the form $\frac{D^\text{NF}\phi^I}{D\lambda}=\frac{D\phi^I}{D\lambda}+T^I_{\ J}\phi^J=0$. We now show that no such choice of $T^I_{\ J}$ containing only the 4-velocity and the acceleration exists that preserves the orthogonality $\phi^Iu_I=0$ between the 4-velocity and the polarisation vector. We have that
\begin{eqnarray*}
0=\frac{\di}{\di\lambda}(\phi^Iu_I)=\frac{D}{D\lambda}(\phi^Iu_I)=\frac{D\phi^I}{D\lambda} u_I+\phi^Ia_I=-T^I_{\ J}u_I\phi^J+\phi^Ia_I.
\end{eqnarray*}
However, if we now assume that the transport contains at most the second derivative of $x^\mu(\lambda)$ (i.e.\; the velocity $u^I$ and the acceleration $a^I$) we deduce that $T^I_{\ J}=\alpha u^Iu_J+\beta u^Ia_J+\gamma a^Iu_J+\delta a^Ia_J$. But since $u^Ia_I\equiv0$ we see that $T^I_{\ J}u_I\phi^J\equiv0$ and we have thus deduced that $\phi^Ia_I=0$ for all trajectories and all polarisation vectors which is false. Therefore, we have a contradiction and we have to conclude that $T^I_{\ J}$ contains one or more derivatives of the acceleration $a^I$.

It is, however, not clear that it is appropriate to study transport of polarisation vectors along  non-geodesic null trajectories. Physically, non-geodesic paths of photons can only be achieved in the presence of a medium, in which case the photon trajectories will be timelike. In our approach a physically motivated way to obtain non-geodesic trajectories would be to introduce a medium in Maxwell's equations through which the photon propagates. Nevertheless, even without explicitly including a medium, it is easy to include optical elements such as mirrors, prisms, and other unitary transformations as long as their effect on polarisation can be considered separately to the effect of transport through curved spacetime.

%% file: CoQRF/CoQRF105_App.tex
\chapter{Appendix to Chapter 5\label{ch-app5}}
\section{Balanced Homodyne Detection of quantum phase references\label{app-BHD}}
Associated with \Ssref{sec-BHD}.

In this section we will make some connections of relational quantum measurements with experiment. Balanced Homodyne detection is a measurement technique in quantum optics in which two beams are incident on either side of a beamsplitter. The angle of incidence with the plane of reflection is $45^\circ$ so that reflected and transmitted beams are on two paths, but these cannot mix with the incident beams. The beams on the two transmission paths are then measured with photon counters, returning numbers of photons $n_A$ and $n_B$. Therefore the projected state is a simultaneous number eigenstate for each path. It has total photon number $2j=n_A+n_B$, and difference in photons $2m=n_A-n_B$ where $m=-j$ to $j$ in integer steps \cite{TycSanders}. The basic idea is that the outcome $m/j$ is related to the relative phase of the two beams. If $j$ is large, there are more outcome possibilities, admitting a greater resolution of relative phase.

We want to see whether balanced homodyne detection is a way to perform the POVM \eqref{eq-M} that measures relative orientation of the reference frames $A$ and $B$. If it is, it provides an immediately experimentally accessible way to study the change of quantum reference frame procedure for phase references.  Indeed, there exists a coherent state amplification scheme using balanced homodyne measurement \cite{Josse06}, which may be considered as a specific change of quantum phase reference operation, from one coherent state to an amplified coherent state.

The standard treatment of balanced homodyne detection is that one input is the quantum state with a phase to be measured, and the second input is a classical `local oscillator', providing the phase reference for the measurement \cite{TycSanders}. In the scenario suggested in this section, whereby two quantum phase references are directly measured in a single measurement, we would need to consider the general situation where each input state is of finite size. Also we want to analyse the possibility of measurement of two phase eigenstates, as well as two coherent states. Since this view of a balanced homodyne detection treats both input beams equally as quantum states, the interpretation of the measurement is then that it measures relative phase of the two optical states, and requires no phase reference to do so. Adapting results in \cite{TycSanders} we can analyse the large $s_A$ and $s_B$ limits of balanced homodyne measurements of coherent states and phase eigenstates. See also \cite{VogelGrabow93}.

\subsubsection*{Two coherent states}
From \cite{TycSanders} we have that the probability of $2j$ total photons and $2m$ difference in photons for two coherent states $\ket{s_A;a}_{\rm CS}$ and $\ket{s_B;b}_{\rm CS}$ is
\[
P_m^j=\Ee^{-s_A^2}\Ee^{-s_B^2}\frac{1}{(j+m)!(j-m)!}2^{-2j}\abs{s_A\Ee^{\ii a}-s_B\Ee^{\ii b}}^{2(j+m)} \abs{s_A\Ee^{\ii a}+s_A\Ee^{\ii b}}^{2(j-m)}.
\]
For two coherent states with equal amplitude $s$ and with phases $a$ and $b$, the measurement probabilities for $m$ are a function of $\cos^2[(b-a)/2]$:
\[
P^j_m=\Ee^{-2s}\frac{(2s^2)^{2j}}{(2j)!}\binom{2j}{j+m}\left[\cos^2\left(\frac{b-a}2\right)\right]^{j+m} \left[1-\cos^2\left(\frac{b-a}2\right)\right]^{j-m}.
\]
The magnitude of the relative phase is monotonically mapped to $m\in[-j,j]$. Larger $j$ gives better relative phase accuracy. As $s\to\infty$, the resolution becomes perfect.

From \cite{TycSanders} we have for large coherent state $\ket{s_A;a}_{\rm CS}$ the outcome probability
\be
P_m^j=\frac{\Ee^{-(2j-s_A^2)^2/2s_A^2}}{\sqrt\pi s_A^2} \abs{\Bigbk{x=\frac m{\sqrt j}}{\psi(b-a-\pi)}}^2\label{eq-appxP}
\ee
where $x$ are the eigenvalues of $\hat x=(\hat a+\hat a^\dag)/\sqrt 2$, so for two coherent states with one amplitude $s_A$ large we replace $\ket{\psi(b)}\to\ket{s_B;b}_{\rm CS}$ and use the position representation of a coherent state \cite[Ch.V]{cohen1977quantum} to obtain
\[
P^j_m=\frac{\Ee^{-(2j-s_A^2)^2/2s_A^2}}{\sqrt\pi s_A^2}(\pi\hbar)^{-\half}\exp[-\left(m/\sqrt j-s_B\cos(a-b+\pi)\right)^2].
\]
Again this maps $\cos(b-a)$ to $m$. Probability $P^j_m$ is sharply peaked about $j=s_A^2/2$ for $s_A$ large, so we obtain accurate phase measurement.

\subsubsection*{Coherent state and phase eigenstate}
The balanced homodyne detection worked well as a phase measurement for a coherent state $\ket{s_B,b}_{\rm CS}$ with a coherent state $\ket{s_A,a}_{\rm CS}$ (treated in this case as the `reference' oscillator) because the state is localised in the $x$-$p$ phase space (with a Gaussian probability distribution). If we instead were measuring a phase eigenstate $\ket{\psi(b)}=\ket{s_B;b}$ and a large coherent state $\ket{s_A;a}_{\rm CS}$, we can again use \eqref{eq-appxP}. Calculating $\ex x$ and $\Delta x$ for the phase eigenstate using $\hat x=(\hat a+\hat a^\dag)/2$ on Fock states \cite[Ch.V]{cohen1977quantum}, we have in the large $s_B$ limit
\be
\ex x=D_{s_B}^\half\frac23\cos b\qquad\text{and}\qquad(\Delta x)^2\approx\half\left(\frac32+\frac29D_B\cos^2b\right).
\label{eq-BHDPEvar}
\ee
As $B$ increases, the position variance is predominately determined by the phase $(b-a)$ of the state, and by the size of the Hilbert space the state has support on, $D_{s_B}=s_B+1$. By \eqref{eq-appxP} we have a mapping of $\cos(b-a)$ to $m$.

\subsubsection*{Two phase eigenstates}

Consider balanced homodyne detection of two phase eigenstates $\ket{s_A;a}$ and $\ket{s_B;b}$ with size $s_A$ and $s_B$. The beamsplitter does not change total photon number probability. Therefore the total probability of detection of $2j$ photons is
\begin{align*}
P^j=&(s_A+1)^{-1}(s_B+1)^{-1}(\min\set{j,s_A-j}+\min\set{j,s_B-j}+1)
\end{align*}
The probability grows from $D_{s_A}^{-1}D_{s_B}^{-1}$ at $2j=0$ linearly with $j$ to a plateau of $(\max\{s_A,s_B\}+1)^{-1}$ at $2j=\min\{s_A,s_B\}$ to $\max\{s_A,s_B\}$, then falling linearly with $j$ until probability is zero at $2j=s_A+s_B+1$. This plateau at moderate $j$ yields a lower average measurement accuracy than with two large coherent states. Modifying the derivation in \cite{TycSanders} that led to \eqref{eq-appxP}, the probability for $\ket{j,m}$ of BHD of two phase eigenstates is approximately in the form of an overlap of a position eigenstate with a phase eigenstate:
\[
P^j_m\approx D_{s_A}^{-1}D_{s_B}^{-1}\abs{j^{-\frac14}\bra{x=\frac m{\sqrt j}}\left(\sum_{k=\max\set{2j-s_A,0}}^{\min\set{s_B,2j}} \Ee^{\ii k(a-b-\pi)}\ket{k}\right)}^2.
\]
For $2j>s_A,s_B$, this superposition in the overlap looks like the $(s_B-s_A+2j+1)$ highest photon number components of a phase eigenstate $\ket{s_B;a-b-\pi}$. From \eqref{eq-BHDPEvar} we saw that the $x$ variance and expectation value of phase eigenstates depends on their cutoff and phase orientation. From our picture of phase eigenstates in phase space this overlap is the outer part of the pseudo-distribution of a phase eigenstate, without the inner part of the state. Since this becomes an isolated packet away from the origin, it allows some accuracy in correlation of a position measurement with the cosine of the phase. Notice however, as $j$ falls to $s_A/2$, the state becomes a complete phase eigenstate, so the overlap produces great inaccuracy mapping position to phase.

\subsubsection*{Balanced homodyne detection as a relational measurement}

In each case of study in this section, balanced homodyne detection provides a relational measurement of a parameter related to relative phase or quadrature. Since balanced homodyne detection projects to total photon number $j$ and a higher photon number offers more possible outcomes $m$, large coherent states are better suited to this type of measurement than phase eigenstates. Although the mapping between $m/j$ and relative phase is not linear, it becomes infinitely well resolved as the size of phase eigenstates or coherent states goes to infinity. However, only the relative phase modulo $\pi$ is measured. A final remark on the viability of balanced homodyne detection for use in the change of quantum reference frame procedure is that the homodyne detection destroys the state by absorbing the photons, so we would not be able to continue with the remainder of the change of reference frame procedure. Perhaps an extended optical setup could be utilised to produce an output state, similar to the balanced homodyne coherent state amplification scheme by Josse {\it et.\,al.\,} \cite{Josse06}.